\documentclass{aa}  

\usepackage[varg]{txfonts}
\usepackage[T1]{fontenc}
\usepackage{ae,aecompl}
\usepackage{natbib}	
 \bibpunct{(}{)}{;}{a}{}{,} 
\usepackage{graphicx}	
\usepackage{amsmath}	
\usepackage{amssymb}	
\usepackage{tikz,graphics,color,float,epsf,caption,subcaption}
\usepackage{rotating}
\usepackage{lscape} 
\usepackage{rotating}
\usepackage{xcolor}
\usepackage{calc}
\usepackage{hyperref}

\newcommand{\uniF}{{\rm erg} \; {\rm s}^{-1} {\rm cm}^{-2}}
\newcommand{\lsun}{\mbox{L}_\odot}
\newcommand{\rsun}{\mbox{R}_\odot}
\newcommand{\msun}{\mbox{M}_\odot}
\newcommand{\pab}{\mbox{Pa}\beta}
\newcommand{\brg}{\mbox{Br}\gamma}
\newcommand{\lacc}{L_{\rm acc}}
\newcommand{\macc}{\dot{M}_{\rm acc}}
\newcommand{\lstar}{L_\star}
\newcommand{\mstar}{M_\star}
\newcommand{\rstar}{R_\star}
\newcommand{\teff}{T_{\rm eff}}
\newcommand{\lbol}{L_{\rm bol}}

\usepackage{color}

\begin{document}

   \title{KMOS study of the mass accretion rate \\from Class~I to Class~II in NGC~1333 \thanks{Based on observations collected at the European Southern Observatory under ESO programme 0102.C-0679A.} \thanks{Reduced spectra of the sources described in Tables~\ref{jhktab1} and \ref{starparnoYSO} are only available in electronic form at the CDS via anonymous ftp to cdsarc.u-strasbg.fr (130.79.128.5) or via \url{http://cdsweb.u-strasbg.fr/cgi-bin/qcat?J/A+A/}.}}


\author{Eleonora Fiorellino\inst{1,2,3,4}
     \and Carlo F. Manara\inst{4}
     \and Brunella Nisini\inst{2}
     \and Suzanne Ramsay\inst{4}
     \and Simone Antoniucci\inst{2}
     \and Teresa Giannini\inst{2}
     \and Katia Biazzo\inst{2}
     \and Juan Alcalà\inst{5}
    \and Davide Fedele \inst{6,7}    }

\institute{Dipartimento di Fisica, Università di Roma `Tor Vergata' Via della Ricerca Scientifica 1, 00133, Roma, Italy\\
\email{eleonora.fiorellino@inaf.it}
  \and INAF-Osservatorio Astronomico di Roma, via di Frascati 33, 00078, Monte Porzio Catone, Italy
  \and INAF-IAPS, via del Fosso del Cavaliere 100, I-00133 Roma, Italy
  \and European Southern Observatory, Karl-Schwarzschild-Strasse 2, 85748 Garching bei München, Germany
  \and INAF-Osservatorio Astronomico di Capodimonte, via Moiariello 16, 80131 Napoli, Italy
  \and INAF-Osservatorio Astrofisico di Arcetri, L.go E.Fermi 5, 50126 Firenze, Italy
  \and INAF, Osservatorio Astrofisico di Torino, Via Osservatorio 20, 10025, Pino Torinese, Italy
}
   \date{Received ; accepted }

\abstract
   {The mass accretion rate ($\macc$) is the fundamental parameter to understand the process of mass assembly that results in the formation of a low-mass star. 
   This parameter has been largely studied in Classical T~Tauri (CTT) stars in star-forming regions with ages of $\sim 1 - 10$~Myr. 
   However, little is known about the accretion properties of young stellar objects (YSOs) in younger regions and early stages of star formation, such as in the Class~0/I phases.}
   {We present new near-infrared spectra of 17 Class~I/Flat and 35 Class~II sources located in the young ($<1$~Myr) NGC~1333 cluster, acquired with the KMOS instrument at the Very Large Telescope (VLT). 
   Our goal is to study whether the mass accretion rate evolves with age, as suggested by the widely adopted viscous evolution model, by comparing the properties of the NGC~1333 members with samples of older regions.
   }
   {For the Class II sources in our sample, we measured the stellar parameters (SpT, $A_V$, and $\lstar$) through a comparison of the IR spectra with a grid of non-accreting Class~III stellar templates. 
   We then 
   computed the accretion luminosity by using the known correlation between $\lacc$ and the luminosity of HI lines ($\pab$ and $\brg$). 
   For the Class I sample, where the presence of a large IR excess makes it impossible to use the same spectral typing method, we applied a procedure that allowed us to measure the stellar and accretion luminosity in a self-consistent way.     
   Mass accretion rates $\macc$ were then measured once masses and radii were estimated adopting suitable evolutionary tracks. 
   }
   {The NGC~1333 Class II sources of our sample have $\lacc \sim 10^{-4} - 1 $~$\lsun$ and $\macc \sim 10^{-11} - 10^{-7}$~$\msun$yr$^{-1}$. 
   We find a correlation between accretion and stellar luminosity in the form of $\log \lacc = (1.5 \pm 0.2) \log \lstar + (-1.0 \pm 0.1)$, and a correlation between mass accretion rate and stellar mass in the form of $\log \macc = (2.6 \pm 0.9) \log \mstar + (-7.3 \pm 0.7)$. 
   Both correlations are compatible within the errors with the older Lupus star-forming region, while only the latter is consistent with results from Chamaeleon~I. 
   The Class~I sample shows larger accretion luminosities ($\sim 10^{-2} - 10^{2}$~$\lsun$) and mass accretion rates ($\sim 10^{-9} - 10^{-6}$~$\msun$yr$^{-1}$) with respect to the Class~II stars of the same cloud.
   However, the derived mass accretion rates are not sufficiently high to build up the inferred stellar masses, assuming steady accretion during the Class~I lifetime. 
   This suggests that the sources are not in their main accretion phase and that most of their mass has already been accumulated during a previous stage and/or that the accretion is an episodic phenomenon. 
   We show that some of the targets originally classified as Class~I  through {\it Spitzer} photometry are in fact evolved or low accreting objects. 
   This evidence can have implications for the estimated protostellar phase lifetimes.  
   }
   {The accretion rates of our sample are larger in more embedded and 
   early stage YSOs. 
   Further observations of larger samples in young star-forming regions are needed to determine if this is a general result.
   In addition, we highlight the importance of spectroscopic surveys of YSOs to confirm their classification and perform a more correct estimate of their lifetime.}

\keywords{Accretion, accretion disks - Protoplanetary disks - stars: low-mass - Stars: pre-main sequence - Stars: protostars - Stars: variables: T Tauri, Herbig Ae/Be}
\maketitle
%

\section{Introduction}

The final mass of low-mass stars ($<2$~$\msun$) is mostly assembled during the protostellar phase. 
Therefore, it is during this phase that it is crucial to investigate how matter is accreted onto the forming star from the surrounding disk and envelope.
The widely adopted model for mass accretion in young stellar objects (YSOs) is the magnetospheric accretion model \citep[e.g.,][]{har16}. 
In this frame, the mass accretion rate not only regulates the final stellar mass build-up, but also the evolution of the protoplanetary disk, determining the initial conditions for planet formation \citep[e.g.,][]{man19}. 
In particular, the mechanism driving the evolution of disks is responsible for the transport of material through the disk, and thus, in the end, for the mass accretion onto the central star. 

According to the viscous evolution model, \citep[e.g.,][]{lyn74,har98} protostars follow a stage of vigorous accretion during the initial 10$^5$ yrs of life, that is during the Class~0 and I phases, when the protostar is still surrounded by its original envelope, while the mass accretion rate decreases significantly during the following pre-main sequence phase (Class~II, $\sim$ 10$^6$-10$^7$~yrs) during which the disk is dispersed \citep[e.g.,][]{erc17} and planetary systems form, until the source approaches the zero age main sequence (ZAMS), that is to say the Class~III phase.
This broad classification has been empirically based on the spectral index of the spectral energy distribution (SED) of YSOs in the range between 2 and 24~$\mu$m \citep{eva09}. However, this simple classification  might not reveal the complexity of such young stellar objects. For example, detailed 2D models \citep{whi03, rob06} show that, for a correct interpretation of the SED, the geometry of the system and its inclination angle need to be considered.

Observational verification of the above picture is still the subject of several studies. 
Most determinations of the mass accretion rate in YSOs have been obtained in classical T~Tauri (CTT) stars, in other words Class~II sources 
whose spectral energy distributions (SEDs) are mostly dominated by the stellar photosphere with little IR~excess due to the disk contribution. 
These sources are accessible to UV and optical observations; therefore, the accretion luminosity ($\lacc$) and mass accretion rates ($\macc$) can be directly measured from the UV-excess over the stellar photosphere caused by the accretion shock \citep[e.g.][]{cal04, her08, her09, rig12, ing13, alc14, fai15, rug18}. 
In addition, correlations between the $\lacc$ and the luminosity of optical and IR permitted lines, such as the Balmer, Paschen, and Brackett lines, have been found \citep{muz98,nat04,moh05,alc17}. 
Adopting these methods, several surveys of the accretion properties in YSOs of nearby star-forming regions with ages spanning between 1 and 10~Myr have been performed \citep[e.g., ][]{nat06, ant08, man12, bia12, ant14, ven14, ven19, alc14, alc17, bia14, man15, kim16, man17a, man20}, which have in particular addressed the connection between accretion rates and the stellar parameters.

Very little is known, however, on the accretion properties of younger and more embedded sources, such as the Class~I YSOs, whose large extinction, caused by their residual envelope, prevents observations in the UV and optical. 
These sources are best observed in the IR, although a lack of a good determination of extinction, which can reach values of up to $A_V = 40-50$~mag, can hamper a correct interpretation of even the IR emission lines.

One of the most intriguing open problems on the accretion properties of younger and more embedded YSOs is the luminosity problem: the luminosity of these sources is in general too low, by as much as an order of magnitude, with respect to that expected if accretion proceeds at the high mass accretion rates predicted by the standard steady-state collapse model of \citet{shu77}.
This problem was first noted by \citet{ken90} and later confirmed by the large  core-to-disk (c2d) {\it Spitzer} photometric surveys \citep[e.g.][]{eva09}. 
These IR observations suggest that most of the mass was accreted during the protostellar phase or in a non-steady frame \citep{eva09}, but direct accretion estimates are needed to confirm this. 
Assuming that the accretion decreases with time \citep[e.g.][]{har06}, studying the accretion rates in these young and poorly characterised objects can help us in understanding the luminosity problem itself, and the role of the Class~I stage in the overall picture of the accretion process.

Class~I objects are also interesting to study as they still have enough mass in their protoplanetary disk to allow planet formation \citep[e.g.,][]{tyc20}, contrary to Class~II objects \citep[e.g.,][]{man18}. 
Stellar properties of relatively large samples of Class~I sources were obtained by \citet{dop05} and \citet{con10}, who analysed the weak photospheric features of these objects which can be detected when they are not too `veiled' by a large  IR continuum excess. 

Mass accretion rate estimates were obtained on small samples of Class~I objects. 
In most studies, the luminosity of the Br$\gamma$ and Pa$\beta$ lines was used as a proxy for the accretion luminosity  \citep{muz98,nis05,ant08,ant11}.  
\citet{whi04} derived both the accretion and stellar parameters of about 15 Class~I stars through the analysis of their optical scattered light. 
These works highlight the large diversity of behaviour of the investigated Class~I objects. 
In fact, few studied sources show high levels of accretion, which is compatible with a scenario in which their bolometric luminosity is dominated by the accretion luminosity \citep[the so called accretion dominated young objects or ADYOs,][]{ant08}, while the majority of the samples have mass accretion rates comparable with those of Class~II stars of a similar mass. 
These results, however, refer to small and inhomogeneous samples of sources located in different environments. 
Hence  general conclusions on the accretion properties of young embedded sources with respect to more evolved objects in the same star-forming cloud can hardly be drawn from these studies. 
Nowadays, thanks to sensitive, high multiplex instrumentation in the IR, it is possible to perform more systematic surveys of embedded YSO populations in young star-forming clouds, thus contributing to the understanding of the  accretion process in the early stages of star formation.

In this work we present the results of a spectroscopic survey of YSOs in the young ($< 1$~Myr) NGC~1333 cluster, located at $293 \pm 22 \mbox{ pc}$ \citep[][which takes into account recent Gaia results]{ort18} in the Perseus cloud. This survey was performed with the K-band Multi Object Spectrograph \citep[KMOS, ][]{sha13}, at the ESO Very Large Telescope (VLT).

The Perseus complex is a suitable laboratory for the study of low mass star formation, because of YSOs at different evolutionary stages. 
The complex comprises several clusters, the largest being the IC~348 and NGC~1333 clusters \citep{sar79,kir06,eno07}. 
The prestellar cores and the YSO population of Perseus have been extensively investigated at different wavelengths: in the NIR and MIR \citep{red07,jor07,eva09, eno09, gut11, you15, mer17}, in the FIR and sub-millimetre \citep[][]{eno06,hat07a, hat07b,sad14, tyc20, pez20}, and at millimetre and radio wavelengths \citep{pec16, seg18, and19}. 



The NGC~1333 cluster shows the highest fraction of Class~I, in other words sources with $\alpha > 0.3$. 
Even if the age of the main clusters, IC~348 and NCG~1333, is still debated, there is a general agreement on the fact that star formation in NGC~1333 has taken place considerably later than in IC~348. 
In particular, the original work from \citet{sar79} estimated the ages of IC~438 and NGC~1333 to be about 1.3~Myr and less than 1~Myr, respectively. 
Those ages have been confirmed by the latest studies in which the age of NGC~1333 has been measured to be $<1$~Myr \citep{wil04,you15,yao18}. 

Optical and IR spectroscopy of YSOs in the NGC~1333 cluster has been performed by several authors and used to derive the stellar parameters, such as the spectral type, stellar luminosity and mass, of the members of the clusters \citep[the latest works are][]{win09,ito10,luh16,esp17}.
In particular, \citet{luh16} have reported the discovery of new members not previously observed with \emph{Spitzer}, thus providing  a brown dwarf catalogue of this cluster, later analysed by \citet{esp17}. 
Accretion properties of the NGC~1333 YSOs, however, have never been derived before. 

The aim of our work is to fill in this gap, providing the first homogeneous study of the accretion properties of a sample of Class~I and II sources in the NGC~1333 cluster through near-infrared (NIR) spectroscopy. 
The results of our survey are then compared with those derived in older regions to obtain information about the evolution of the accretion luminosity and the mass accretion rate with the age and the evolutionary stage of the YSOs. 

The paper is organised as follows. 
In Sect.~\ref{sect:obs} we present the seletion of our sample, the observations performed with KMOS and the data reduction steps; in Sect.~\ref{sect:emli} we report how we measured the Pa$\beta$ and Br$\gamma$ lines, which are the lines we used to derive the accretion luminosity; in Sect.~\ref{sect:ana2} and \ref{sect:ana1} we describe how we computed the stellar parameters and the accretion rates for Class~II and I YSOs, respectively; in Sect.~\ref{sect:disc} we discuss our results; finally, in Sect.~\ref{sect:con} we provide the summary and conclusions of this work. 

\section{Sample, observations, and data reduction} \label{sect:obs}

\subsection{The sample} \label{sect:sample}

\begin{table*}
 \center
 \caption{\label{jhktab1}The analysed sample of Class~II and Class~I/Flat YSOs.}
\resizebox{\textwidth}{!}{%
  \begin{tabular}{llllcccrccc}
  \hline
  \hline
ID$^{(a)}$&E09$^{(b)}$&2MASS name$^{(c)}$& Other name  &      RA$^{(d)}$     &    Dec$^{(d)}$                   &Class  & $\alpha_{\rm IR}^{(e)}$ &  J      &     H      &     K  \\
          &           &                  &             & ICRS (J2000)        & ICRS (J2000)           &       &                   & mag     &    mag     & mag   \\
    \hline
    \hline
 153 & 149  & J03284407+3120528 &  -                   & 03:28:44.08 & +31:20:52.82 &   II  & $-1.34$ &$14.24  \pm  0.03$ & $13.23 \pm 0.04$ & $12.63 \pm 0.03$\\
 156 & 153  & J03284782+3116552 &  -                   & 03:28:47.83 & +31:16:55.18 &   II  & $-1.10$ & $12.94  \pm  0.02$ & $11.76 \pm 0.02$ & $10.91 \pm 0.02$\\
 159 & 157  & J03285105+3116324 &  -                   & 03:28:51.07 & +31:16:32.50 &   II  & $-1.48$ & $13.29  \pm  0.02$ & $12.52 \pm 0.03$ & $12.12 \pm 0.03$\\
 162 & 160  & J03285213+3115471 &  -                   & 03:28:52.14 & +31:15:47.21 &   II  & $-1.34$ & $13.16  \pm  0.02$ & $12.47 \pm 0.03$ & $12.03 \pm 0.02$\\
 164 & 163  & J03285292+3111626 & ASR 46               & 03:28:52.91 & +31:16:26.53 &   II  & $-1.49$ & $13.62  \pm  0.03$ & $12.90 \pm 0.03$ & $12.48 \pm 0.03$\\
 165 & 164  & J03285392+3118092 &  -                   & 03:28:53.93 & +31:18:09.28 &   II  & $-1.29$ & $14.82  \pm  0.06$ & $12.27 \pm 0.04$ & $10.88 \pm 0.02$\\
 166 & 165  & J03285409+3136542 & ASR 42               & 03:28:54.08 & +31:16:54.33 &   II  & $-1.07$ & $13.03  \pm  0.02$ & $12.04 \pm 0.03$ & $11.60 \pm 0.02$\\
 167 & 166  & J03285461+3116512 &  -                   & 03:28:54.62 & +31:16:51.24 &   II  & $-1.08$ & $12.86  \pm  0.02$ & $11.19 \pm 0.02$ & $10.23 \pm 0.02$\\
 168 & 167  & J03285505+3116287 &  -                   & 03:28:55.07 & +31:16:28.83 &   II  & $-0.76$ & $13.58  \pm  0.03$ & $11.70 \pm 0.03$ & $10.68 \pm 0.05$\\
 173 & 172  & J03285663+3118356 &  -                   & 03:28:56.64 & +31:18:35.59 &   II  & $-1.14$ & $12.30  \pm  0.02$ & $10.70 \pm 0.02$ &  $9.70 \pm 0.02$\\
 174 & 173  & J03285694+3116222 &  -                   & 03:28:56.95 & +31:16:22.33 &   II  & $-0.74$ & $13.76  \pm  0.03$ & $11.92 \pm 0.03$ & $11.08 \pm 0.02$\\
 175 & 174  & J03285718+3115346 &  ASR 17              & 03:28:57.15 & +31:15:34.60 &   II  & $-0.71$ & $15.40  \pm  0.07$ & $13.98 \pm 0.05$ & $13.19 \pm 0.04$\\
 177 & 177  & J03285769+3119481 &  -                   & 03:28:57.70 & +31:19:48.10	&   II  & $-0.96$ & $13.04  \pm  0.03$ & $11.97 \pm 0.03$ & $11.38 \pm 0.02$\\
 178 & 178  & J03285811+3118037 & NAME SSS 107         & 03:28:58.10 & +31:18:03.75 &  III  & $-2.37$ & $12.81  \pm  0.02$ & $11.71 \pm 0.02$ & $11.34 \pm 0.02$\\
 180 & 180  & J03285824+3122021 &  -                   & 03:28:58.25 & +31:22:02.07	&   II  & $-1.07$ & $14.91  \pm  0.05$ & $13.49 \pm 0.04$ & $12.41 \pm 0.03$\\
 184 & 184  & J03285956+3121467 & LkHA 353             & 03:28:59.55 & +31:21:46.71 &   II  & $-1.11$ & $12.61  \pm  0.02$ & $11.26 \pm 0.02$ & $10.30 \pm 0.02$\\
 188 & 189  & J03290313+3122381 &  -                   & 03:29:03.14 & +31:22:38.08 &   II  & $-0.76$ & $13.72  \pm  0.03$ & $12.37 \pm 0.03$ & $11.32 \pm 0.02$\\
 192 & 194  & J03290386+3121487 &  -                   & 03:29:03.86 & +31:21:48.68 &   II  & $-1.06$ & $11.54  \pm  0.03$ & $10.14 \pm 0.03$ &  $9.22 \pm 0.02$\\
 194 & 198  & J03290466+3116591 &  -                   & 03:29:04.66 & +31:16:59.15 &   II  & $-0.46$ & $15.55  \pm  0.07$ & $13.91 \pm 0.04$ & $12.66 \pm 0.02$\\
 195 & 199  & J03290472+3111348 &  -                   & 03:29:04.73 & +31:11:34.90 &   II  & $-0.62$ & $<18.50$           & $<15.64$         & $14.45 \pm 0.08$\\
 197 & 201  & J03290518+3120369 & [HL2013]             & 03:29:05.18 & +31:20:36.90 &   II  & $-0.98$ & $16.42  \pm 0.19$  &($\dagger$)$16.84\pm0.05$&($\dagger$)$14.99\pm0.03$\\
 199 & 204  & J03290631+3113464 &  -                   & 03:29:06.32 & +31:13:46.42 &   II  & $-1.14$ & ($\dagger$)$19.02\pm0.03$ &  $14.70\pm0.06$ & $12.66\pm0.02$\\
 201 & 206  & J03290794+3122515 &  -                   & 03:29:07.94 & +31:22:51.51 &   II  & $-1.16$ & $13.00  \pm  0.03$ & $11.18 \pm 0.03$ & $10.19 \pm 0.02$\\
 203 & 208  & J03290908+3123056 &  -                   & 03:29:09.08 & +31:23:05.64 &   II  & $-0.33$ & $14.65  \pm  0.04$ & $12.94 \pm 0.04$ & $11.89 \pm 0.03$\\
 205 & 210  & J03290933+3121042 &  -                   & 03:29:09.33 & +31:21:04.20 &   II  & $-0.51$ & $16.42  \pm  0.11$ & $14.28 \pm 0.05$ & $13.15 \pm 0.03$\\
 209 & 214  & J03291046+3123348 &  -                   & 03:29:10.46 & +31:23:34.84 &   II  & $-1.41$ & $15.63  \pm  0.06$ & $13.82 \pm 0.04$ & $12.76 \pm 0.03$\\
 213 & 218  & J03291084+3116426 & ASR 23               & 03:29:10.82 & +31:16:42.68 &   II  & $-0.32$ & $15.65  \pm  0.07$ & $14.11 \pm 0.04$ & $13.04 \pm 0.03$\\
 219 & 225  & J03291312+3122529 &  -                   & 03:29:13.12 & +31:22:52.84 &   II  & $-1.33$ & $12.87  \pm  0.02$ & $11.12 \pm 0.03$ & $10.11 \pm 0.02$\\
 222 & 228  & J03291659+3123495 &  -                   & 03:29:16.59 & +31:23:49.53 &   II  & $-1.18$ & $13.25  \pm  0.02$ & $11.84 \pm 0.03$ & $11.18 \pm 0.02$\\
 224 & 230  & J03291683+3123251 & MBO 79               & 03:29:16.81 & +31:23:25.22 &   II  & $-0.57$ & $15.42  \pm  0.05$ & $14.32 \pm 0.04$ & $13.58 \pm 0.03$\\
 226 & 233  & J03291768+3122450 & Em* LkHA 270         & 03:29:17.68 & +31:22:45.01 &   II  & $-0.94$ & $9.97   \pm  0.02$ & $ 8.91 \pm 0.03$ &  $8.32 \pm 0.02$\\
 227 & 234  & J03291778+3119480 & ASR 80               & 03:29:17.78 & +31:19:48.00 &   II  & $-1.23$ & $14.80  \pm  0.04$ & $13.65 \pm 0.04$ & $12.99 \pm 0.03$\\
 233 & 240  & J03292155+3121104 &  -                   & 03:29:21.57 & +31:21:10.30 &   II  & $-1.24$ & $12.39  \pm  0.02$ & $11.72 \pm 0.03$ & $11.37 \pm 0.02$\\
 234 & 241  & J03292187+3115362 & Em* LkHA 271         & 03:29:21.87 & +31:15:36.22 &   II  & $-1.34$ & $11.18  \pm  0.02$ & $10.15 \pm 0.03$ & $ 9.50 \pm 0.02$\\
 235 & 242  & J03292317+3120302 & Em* LkHA 355         & 03:29:23.17 & +31:20:30.19 &   II  & $-1.03$ & $12.40  \pm  0.02$ & $11.65 \pm 0.03$ & $11.23 \pm 0.02$\\
 241 & 250  & J03292978+3121027 & -                    & 03:29:29.80 & +31:21:02.59 &   II  & $-1.15$ & $12.65  \pm  0.02$ & $11.62 \pm 0.03$ & $11.16 \pm 0.02$\\
     \hline
 147 & 142  & J03283875+3118068 & -                    & 03:28:38.78 & +31:18:06.59 & Flat  & $-0.25$& $<18.27$           &  $16.17 \pm 0.26$ & $14.01 \pm0.07$\\
 149 & 144  & J03283968+3117321 & -                    & 03:28:39.71 & +31:17:31.88 &  I    & $0.61$ &$<18.29$           &  $16.39 \pm 0.35$ & $13.66 \pm0.05$\\
 161 & 159  & J03285129+3117397 & -                    & 03:28:51.26 & +31:17:39.30 &  I    & $0.60$ &$16.35 \pm 0.17$   &  $15.18 \pm 0.17$ & $14.45 \pm0.04$\\
 181 & 181  & J03285842+3122175 & -                    & 03:28:58.43 & +31:22:17.51 &  I    & $0.82$ & 18.42              &  $14.80 \pm 0.09$ & $11.85 \pm0.04$\\
 183 & 183  & J03285930+3115485 & -                    & 03:28:59.32 & +31:15:48.71 & Flat  & $-0.08$ &$16.49 \pm 0.16$   &  $12.53 \pm 0.03$ & $10.44 \pm0.02$\\
 186 & 186  & J03290149+3120208 & -                    & 03:29:01.56 & +31:20:20.62 &  I    & $2.30$ & 15.75              &  $13.88 \pm 0.08$ & $10.88 \pm0.04$\\
 190 & 191  & J03290332+3123148 & -                    & 03:29:03.33 & +31:23:14.60 &  I    & $1.00$ & $17.25 \pm 0.23$   &  $15.83 \pm 0.14$ & $14.07 \pm0.05$\\
 191(*) &192& J03290378+3116038 & V512 Per, SVS 13  & 03:29:03.78 & +31:16:03.79 &  I       & $0.85$ & $12.39 \pm 0.01$   &  $10.42 \pm 0.01$ & $8.77  \pm0.01$\\
 196 & 200  & J03290493+3120385 & -                    & 03:29:04.95 & +31:20:38.40 & Flat    & $0.06$ & ($\dagger$)$18.43\pm0.05$&$14.60\pm0.14$& $12.97 \pm0.07$\\
 200 & 205  & J03290778+3121573 & [LAL96] 213          & 03:29:07.78 & +31:21:57.31 &  I    & $2.30$ & $<15.27$           &  $13.80 \pm 0.09$ & $10.43 \pm0.04$\\
 202 & 207  & J03290895+3122562 & -                    & 03:29:08.99 & +31:22:56.10 & Flat  & $-0.21$ & $16.05 \pm 0.14$   &  $13.42 \pm 0.06$ & $11.72 \pm0.03$\\
 204 & 209  & J03290907+3121291 & -                    & 03:29:09.10 & +31:21:28.69 &  I    & $1.05$ & $<15.57$           &  $<14.53$         & $12.97 \pm0.05$\\
 215 & 221  & J03291188+3121271 & -                    & 03:29:11.89 & +31:21:27.00 &  I    & $0.32$ & $<17.22$           &  $15.70 \pm 0.12$ & $12.82 \pm0.03$\\
 218 & 224  & J03291294+3118146 & -                    & 03:29:12.97 & +31:18:14.29 &  I    & $1.56$ & $<18.60$           &  $<17.77$         & $14.11 \pm0.05$\\
 231 & 238  & J03292003+3124076 & -                    & 03:29:20.06 & +31:24:07.49 &  I    & $0.34$ & $<17.16$           &  $14.70 \pm 0.07$ & $12.04 \pm0.03$\\
 232 & 239  & J03292044+3118342 &                      & 03:29:20.44 & +31:18:34.20 & Flat    & $-0.25$ & $14.40 \pm 0.04$   &  $12.01 \pm 0.03$ & $10.47 \pm0.02$\\
 237 & 245  & J03292409+3119576 & ASR 79               & 03:29:24.09 & +31:19:57.61 & Flat    & $-0.13$ & $15.52 \pm 0.07$   &  $14.24 \pm 0.06$ & $13.57 \pm0.04$\\   
\hline
\hline
  \end{tabular}
  }
  \begin{quotation}
  \textbf{Notes.} $^{(a)}$ID of the sources taken from \citet{you15} and adopted in this paper. 
  $^{(b)}$Corresponding ID in the \citet{eva09} paper. 
  $^{(c)}$2MASS name of the sources (the prefix 2MASS has been removed).
  $^{(d)}$Coordinates are the {\it Spitzer} coordinates from \citet{you15} catalog.  $^{(e)}${\it Spitzer} IR spectral index from \citet{you15}.
  \newline J, H and K photometry taken from the 2MASS catalogue except when indicated with ($\dagger$) (photometry from Luhman) and with ($^*$) (variable source, photometry taken on 12 march 2015, Giannini private communication).
  \end{quotation}  
  \end{table*}
  
\begin{figure}[t]
    \centering
    \includegraphics[width=\columnwidth]{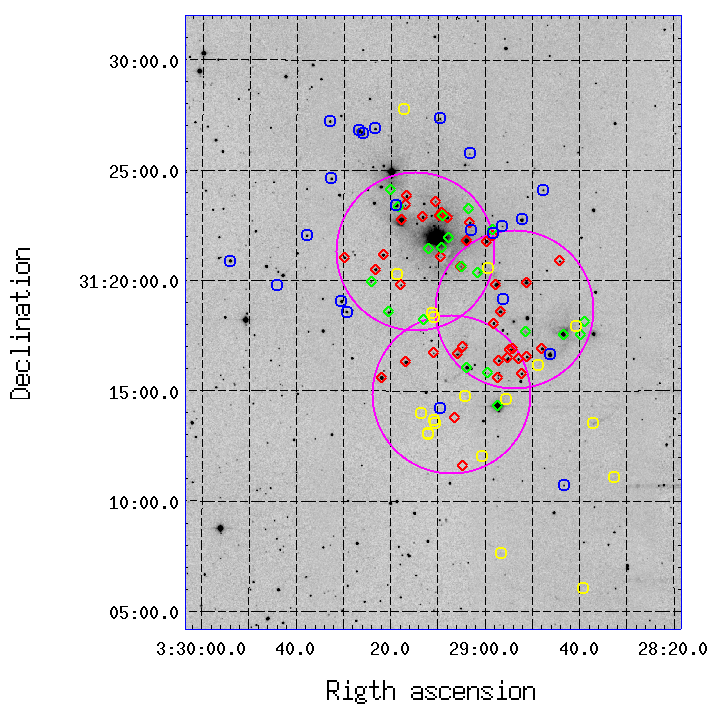}
    \caption{2MASS mosaic map of the NGC~1333 cluster achieved through the \emph{Image Mosaic Service} (\url{http://hachi.ipac.caltech.edu:8080/montage}). 
    Big magenta circles correspond to the KMOS pointings of 7.2$\arcmin$ diameter.
    Circles refer to the sample of YSOs with $m_K > 14.5$~mag. 
    The colour coding is the following: red (blue) and green (yellow) symbols indicate observed (not-observed) Class~II/III and I/Flat sources, respectively. 
    }
    \label{fig:obs}
\end{figure}

\begin{figure}
    \centering
    \includegraphics[width=\columnwidth]{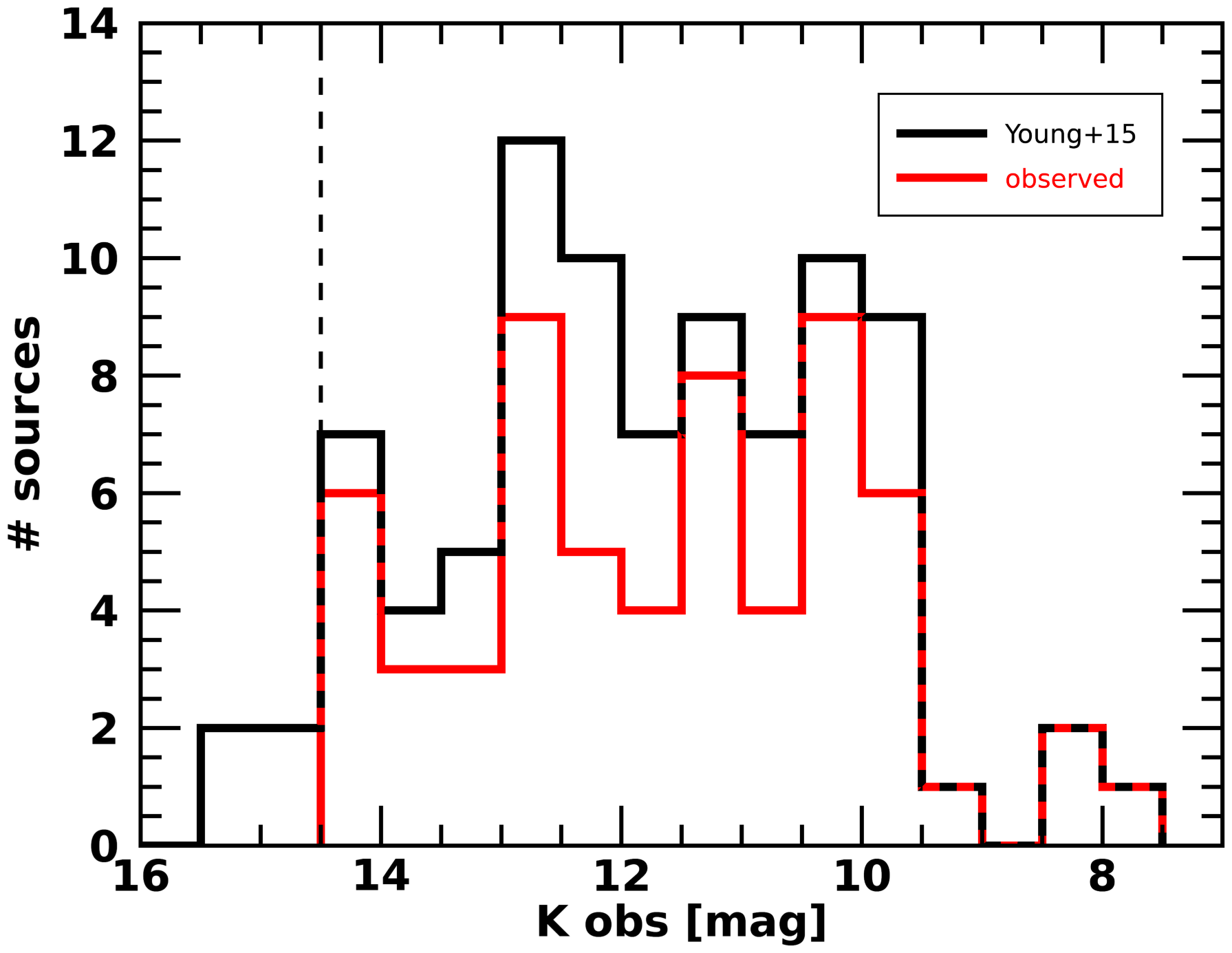}
    \caption{Histogram of the K-band magnitude of the {\it Spitzer} identified YSOs in the NGC~1333 cluster \citep[black,][]{you15} and the sample observed by KMOS (red). 
    The black-dashed vertical line represents the cut-off at 14.5~mag due to KMOS sensitivity.}
    \label{fig:histomk}
\end{figure}
We selected our sample from the catalogue provided by \citet{you15}, which contains 369 candidate YSOs in the Perseus cloud, classified using \emph{Spitzer} photometry up to 70~$\mu$m and 2MASS~photometry. 
In this catalogue 103 sources belong to the NGC~1333 cluster (40 Class~0+I/Flat, 58 Class~II, and 5 Class~III). 
To be able to observe those sources with KMOS, we selected only the targets which have a 2MASS counterpart (88 targets) and $m_K < 14.5$~mag, that is to say 84 targets (23 Class~I/Flat, 56 Class~II and 5 Class~III). 
We note that by considering sources that are bright in the NIR, we automatically removed any Class~0 from our sample. Therefore we will refer to the sources of our sample as Class~I objects. 

A prioritised list of targets was provided as the input catalogue to the KMOS ARM Allocator (KARMA)\footnote{\url{https://www.eso.org/sci/observing/phase2/SMGuidelines/KARMA.html}} software. 
The allocation of targets to arms follows the priority of the targets (Priority One is assigned to Class~I/Flat, Priority Two to Class~II, and Priority Three to Class~III) and the positional constraints due to the mechanical properties of the arms.
We identified three fields covering the largest number of sources (Fig. 1). Within these fields, KARMA selected 61 targets, i.e 60\%  of the initial sample. These targets are listed in Table~\ref{obs}. 

According to the classification given by \citet{you15}, 20 of them are Class~I/Flat, 39 are Class~II, and two are Class~III objects. 
Therefore, in total we successfully observed $50\%$ of the Class~I/Flat and $67\%$ of the Class~II sources. 
Our sample is 87\% and 70\% complete within $m_K < 14.5 \mbox{ mag}$ for Class~I/Flat and Class~II, respectively (for more details see Figs.~\ref{histomkclass1} and \ref{histomkclass2} in App.~\ref{app:obs}).

\subsection{KMOS observations}
The observations were carried out between October 2 and 31 2019 in service mode, with KMOS. 
KMOS is able to perform integral field spectroscopy in the NIR of 24 objects simultaneously within a $7.2 \arcmin$ patrol field. 
The targets were selected by 24 pick-off arms which place small mirrors in the focal plane to send a field of view of $2.8 \arcsec \times 2.8 \arcsec$ to an image-slicing  integral field unit (IFU). 

For each of the three fields, shown in Fig.~\ref{fig:obs},  we performed two observations, one optimised for bright targets ($ m_K < 11$~mag), and the other for the faintest ones ($11 \mbox{ mag} < m_K < 14.5$~mag). 
Each observation was performed with a dithering sequence (\emph{Freedither} mode) with one~cycle of a six~point dither pattern on target.
The dither step size is 0.72$\arcsec$ 
on right ascension (RA) and declination (Dec).
This pattern is interleaved with sky observations with an ABAA configuration for bright targets and ABA for dim targets, where A is the pointing on the target and B is the pointing on the sky. 
At each position of the pattern, we performed a long ($\sim 280-300$~s) or a short ($\sim 60-80$~s) exposure in order to have $\mbox{S/N} > 50$ for all the faint  and the bright  targets, respectively.  
Observations were taken, using this strategy for each J, H, and K band grating.
The resolving power at the band centre is about $\sim$~3600, $\sim$~4000, and $\sim$~4200, respectively. 

Three reference stars are acquired in each field.
A set of calibration images was taken in addition to each of the observations. 
This set consists of dark frames, flat fields, and an arc-lamp exposure. 
A G-type telluric standard star from the Hipparcos catalogue is observed in each field. 
Further observational details are reported in Appendix~\ref{app:obs}.

\subsection{Data reduction} \label{datared}
Data reduction was performed with the kmos-2.1.0 pipeline \citep{dav13} in combination with the ESOReflex pipeline, version 2.9.1 \citep{fre13}.  
The pipeline routines were used to calibrate and reconstruct the 3D data cubes from the raw data of the scientific targets and standard star.
The procedures used by the pipeline of KMOS are mostly standard for NIR spectroscopy: flat field correction, wavelength calibration, sky sybtraction, extraction of the spectra, telluric correction, and instrumental effects correction. 
We corrected the atmospheric transmission using a model of telluric absorption. 
This model was obtained using the molecfit code 
\citep{sme15} applied to the standard star spectrum, normalised for a black body function and the instrumental response. 
This telluric model was then applied to all the spectra of the targets to remove the telluric absorption effect. 
The reconstruction of the 3D data cubes was performed by interpolating from tables which provide the (x,y,z) location in the final cube for each pixel on the detector.

From the calibrated 3D data cubes, we then extracted the 1D spectra (see Appendix~\ref{app:spec}). 
In order to maximise the signal-to-noise (S/N) and be sure to avoid the contribution of spurious features in the spectrum, we extracted the spectra in a region of $0.6 \arcsec \times 0.6 \arcsec $ ($3 \times 3$ pixels), centred on the peak of the source.
This aperture is typically smaller than the airmass-corrected seeing, usually $\sim 0.8 - 1 \arcsec$ at the wavelength range and time of the observations.  
This assure us that the signal is dominated by stellar emission, minimising the background contribution. 
For those sources which are not well centred in the small IFU field of view, the spectrum has been extracted from a narrower aperture centred on the source peak. 
Since the KMOS spectro-photometry was not reliable, we decided to calibrate the flux scale using the 
2MASS photometry (see Table~\ref{jhktab1}\footnote{No errors on the spectral index values are provided by \citet{you15} and the SED fitting method used to derive these values does not account for photometric errors.}) for all targets. 

The median scale factors (2MASS/KMOS photometry) are 2.4, 2.7, and 2.6 in J, H, and K band, respectively, and they range between 0.94 and 5.10.
On the basis of the observed spectra, eight sources have not been considered as YSOs. As a result, we removed them from the analysis in this work (see Appendix~\ref{appnoyso}).

\section{Measurement of $\pab$ and $\brg$ lines} 
\label{sect:emli}

\begin{table}
\centering
\caption{\label{TabFlux23}Observed Fluxes}
\begin{tabular}{lcc}
\hline
\hline
ID  &  $F_{\small \pab}$     & $F_{\small \brg} $ \\
    & $10^{-16}\uniF$     & $10^{-16}\uniF$     \\
\hline		
\hline
153&$< 1.3          $&$< 1.1        $\\         
156&$  112 \pm  22  $&$  13 \pm 3 $\\         
159&$  7.8 \pm  2.3 $&$< 2.9        $\\             
162&$< 3.5          $&$< 2.5        $\\        
164&$< 2.3          $&$< 1.0        $\\   			
165&$< 1.3          $&$< 3.5        $\\         	  
166&$< 3.9          $&$< 3.3    	$\\            
167&$  87 \pm  7  $&$  40 \pm 7 $\\    	        
168&$  30 \pm  5  $&$  16 \pm 3 $\\       		
173&$  433 \pm  29  $&$  265 \pm 26 $\\       		
174&$< 2.4			$&$< 2.1        $\\            
175&$  1.0 \pm  0.2 $&$< 0.8    	$\\      		
177&$  9.3 \pm  3.0 $&$< 4.7      	$\\
178$^*$&$< 6.5			$&$< 1.8        $\\
180&$< 1.2			$&$  2.9 \pm 0.3$\\  		   	
184&$  479 \pm  59  $&$  163 \pm 15 $\\             
188&$  82 \pm  17   $&$  27 \pm 2 $\\         
192&$  430 \pm  36  $&$  190 \pm 63 $\\            
194&$< 1.0          $&$< 1.8        $\\
195&$< 4.3          $&$< 0.2        $\\       
197&$< 11           $&$< 3.6        $\\    
199&$< 3.7          $&$< 1.3        $\\     
201&$  26 \pm  3  $&$< 6.9        $\\              
203&$  20 \pm  4  $&$  5.1 \pm 1.3$\\           	
205&$< 1.0		    $&$< 0.9        $\\ 	 
209&$< 1.7		    $&$< 1.6        $\\
213&$  4.7 \pm  1.0 $&$< 0.4        $\\          	
219&$  32 \pm  7  $&$  14 \pm 3 $\\             
222&$  38 \pm  12   $&$  16 \pm 3 $\\         
224$^{**}$&$  1.2 \pm  0.3 $&$  2.5 \pm 0.8$\\          
226&$  379 \pm  51  $&$< 33	        $\\
227&$< 1.2          $&$< 0.9        $\\          
233&$< 8.0 			$&$< 3.9        $\\
234&$  1240 \pm  261$&$  439 \pm 27 $\\        
235&$  17 \pm  2  $&$< 3.6        $\\ 			 
241&$  < 6.3        $&$< 3.8        $\\
\hline		 				 				
147&$< 7.8 		    $&$< 0.5        $\\
149&$< 2.2		    $&$  1.3 \pm 0.3$\\
161&$< 2.1 		    $&$< 0.7        $\\
181&$< 3.3		    $&$  42 \pm 3 $\\
183&$< 0.7          $&$< 4.3        $\\
186&$< 49           $&$  31 \pm 5 $\\          
190&$< 1.7	        $&$< 0.4        $\\
191&$514 \pm 48  $&$959 \pm 28$\\
196&$< 63 	        $&$< 1.6        $\\
200&$< 51 	        $&$  79 \pm 16  $\\            
202&$  11 \pm 3   $&$  7.2 \pm 1.8$\\
204&$< 34 	        $&$  10 \pm 1 $\\
215&$< 14 	        $&$  6.6 \pm 1.4$\\
218&$< 3.0		    $&$< 0.7        $\\
231&$< 17  		    $&$  5.3 \pm 1.0$\\   
232&$  12 \pm 3   $&$  43 \pm 11  $\\ 
237&$< 1.3			$&$< 0.4        $\\
\hline
\hline
\end{tabular}
\begin{quotation}
    \textbf{Notes.} Fluxes are measured as explained in the text.  The horizontal line in the middle separates Class~II and III (above the line) from Class~I (below the line). $^*$Class~III target. $^{**}$Sub-luminous object.
\end{quotation}
 \end{table}

The brightest lines tracing the accretion process in the NIR are the $\pab$ at 1.28~$\mu$m and the $\brg$ at 2.16~$\mu$m lines. 
We measured the flux of these two emission lines in each target using the Image Reduction and Analysis Facility \citep[IRAF,][]{tod93}. 
The line flux was determined by simply summing the flux of the pixels contained between the continuum and the line.
We repeated the same measurement three times per line, considering the highest, lowest, and the middle position of the local continuum, depending on the local noise in the spectrum. 
We used the mean and the standard deviation of the measurements as the best estimate for the line flux and its error.
We considered a line detected when its $S/N>3$. 

For the non detected lines, we measured the root mean square (RMS) of the observed flux between 1.275~$\mu$m and 1.285~$\mu$m for $\pab$, and between 2.160~$\mu$m and 2.170~$\mu$m for $\brg$. 
Then we computed the upper limit of the line flux (F$^{upp}_{line}$) as in \citet{nis18}:
 \begin{equation}
  F^{upp}_{line} = 3\times \mbox{ RMS } \times (\lambda_{line}/R)
 \end{equation}
where 
$\lambda_{line}$ is the central wavelength of $\pab$ or $\brg$ and $R$ is the resolution of the instrument in the J or K band, respectively. 

As a result, 12 Class~II YSOs show both $\pab$ and $\brg$, seven show only $\pab$, and one shows only $\brg$; in three Class~I YSOs, we detect both $\pab$ and $\brg$, and in seven only $\brg$. 
Therefore $\sim 55\%$ of Class~II (20 out of 36) and $\sim 59\%$ of Class~I (ten out of 17) have at least one accretion line tracer above the threshold noise. 

Detected line fluxes and upper limits are reported in Table~\ref{TabFlux23}. 
Taking into account the uncertainty on the absolute flux due to the variability of CTT stars \citep[$\sim 0.5$~mag,][]{lor13}, the fluxes we measured could vary by about $50$\%.

\section{Analysis of the Class II sources} \label{sect:ana2}
In this section we present the analysis of the spectra of the Class~II sources. 
Source \#178 is a Class~III YSO, therefore is not included in the discussion. 
However, its $\pab$ and $\brg$ fluxes and stellar parameters are reported in Table~\ref{TabFlux23}, \ref{ext2} and \ref{starpar}, marked with an asterisk (*). 
The analysis performed consists in measuring the stellar parameters of the sources (i.e. spectral type, extinction, star luminosity, and mass) and their accretion parameters (accretion luminosity and mass accretion rate) from the luminosity of the $\pab$ and $\brg$ emission.

\subsection{Stellar parameter determination}
\label{sect:stelpar2}
\begin{table}[t]
\centering
\caption{\label{ext2} Extinction of the Class~II sample}
\begin{tabular}{lrrrr}
\hline
\hline
ID & $A_V^{CC}$ & $A_V^{L_{acc}}$  & $A_V^{SpT}$ & $A_V^L$\\
   & mag        & mag              & mag         & mag \\
\hline
\hline
  153 &   1.8       &  ...             &  4.5        & 4.1   \\
  156 &   2.2       &  0.0            &  4.0        & 3.8   \\
  159 &   0.3       &  ...             &  0.3        & 1.9   \\
  162 &   0.0       &  ...             &  0.0        & 0.4   \\
  164 &   0.0       &  ...             &  0.0        & 1.4   \\
  165 &  16.1       &  ...             & 17.0        &22.4   \\
  166 &   2.7       &  ...             &  2.7        & 3.1   \\
  167 &   7.7       &  4.5            &  8.5        & 8.6   \\
  168 &   9.9       &  ...             & 10.0        & 7.9   \\
  173 &   6.6       &  8.2            &  7.0        &7.2    \\
  174 &  10.9       &  ...             &  9.8        & 9.0   \\
  175 &   5.8       &  ...             &  5.7        & ...    \\
  177 &   2.7       &  ...             &  2.7        & 4.1   \\
  178$^*$ &   4.9       &  ...             &  3.0        & 3.8   \\
  180 &   3.5       &  ...             &  5.5        & 7.9   \\
  184 &   3.6       &  3.2            &  6.0        & ...  \\
  188 &   3.9       &  4.5            &  4.3        & ...  \\
  192 &   4.5       &  5.5            &  6.5        & 5.9 \\
  194 &   5.2       &  ...             &  8.8        & 10.7\\
  195 &  21.5       &  ...             & 15.0        & ...   \\
  197 &  17.7       &  ...             & 17.0        & 10.7   \\
  199 &  21.2       &  ...             & 28.0        & 30.7   \\
  201 &   9.6       &   ...            &  7.5        & 8.6   \\
  203 &   7.5       &   7.5           &  9.0        & 10.7\\
  205 &  12.5       &   ...            & 14.0        & 13.8   \\
  209 &   8.9       &   ...            & 11.7        & 11.4   \\
  213 &   5.3       &   ...            &  8.0        &  3.8  \\
  219 &   8.5       &   8.7           &  8.7        & 8.6\\
  222 &   6.7       &   3.0           &  5.8        & 5.9\\
  224$^{**}$ &   2.0       &   2.2           &  3.5        & 3.4 \\
  226 &   2.6       &   ...            &  3.0        & 2.9  \\
  227 &   3.3       &   ...            &  4.5        & 5.5  \\
  233 &   0.0       &   ...            &  0.5        & 1.0  \\
  234 &   1.7       &   4.5           &  3.0        & 1.8 \\
  235 &   0.0       &   ...            &  1.0        & 1.4   \\
  241 &   3.2       &   ...            &  3.1        & 2.9    \\
  \hline
  \hline
\end{tabular}
\begin{quotation}
 {\bf Notes.} Extinction values computed with the methods described in the text (see Sect.~\ref{sect:stelpar2}). 
 $A_V^{CC}$: visual extinction estimated from the colour-colour diagram, $A_V^{acc}$: visual extinction estimated imposing $L_{\rm acc}^{{\rm Pa}_{\beta}} = L_{\rm acc}^{{\rm Br}_{\gamma}}$. $A_V^{SpT}$: visual extinction estimated from SED fitting of the Spectral Type. $A^L$: extinction derived from the \citet{luh16} $A_J$ estimates, applying the \citet{car89} relationship. $^*$Class~III target. $^{**}$Sub-luminous object.
\end{quotation}
\end{table}
\begin{figure}
  \centering
  \includegraphics[width=\columnwidth]{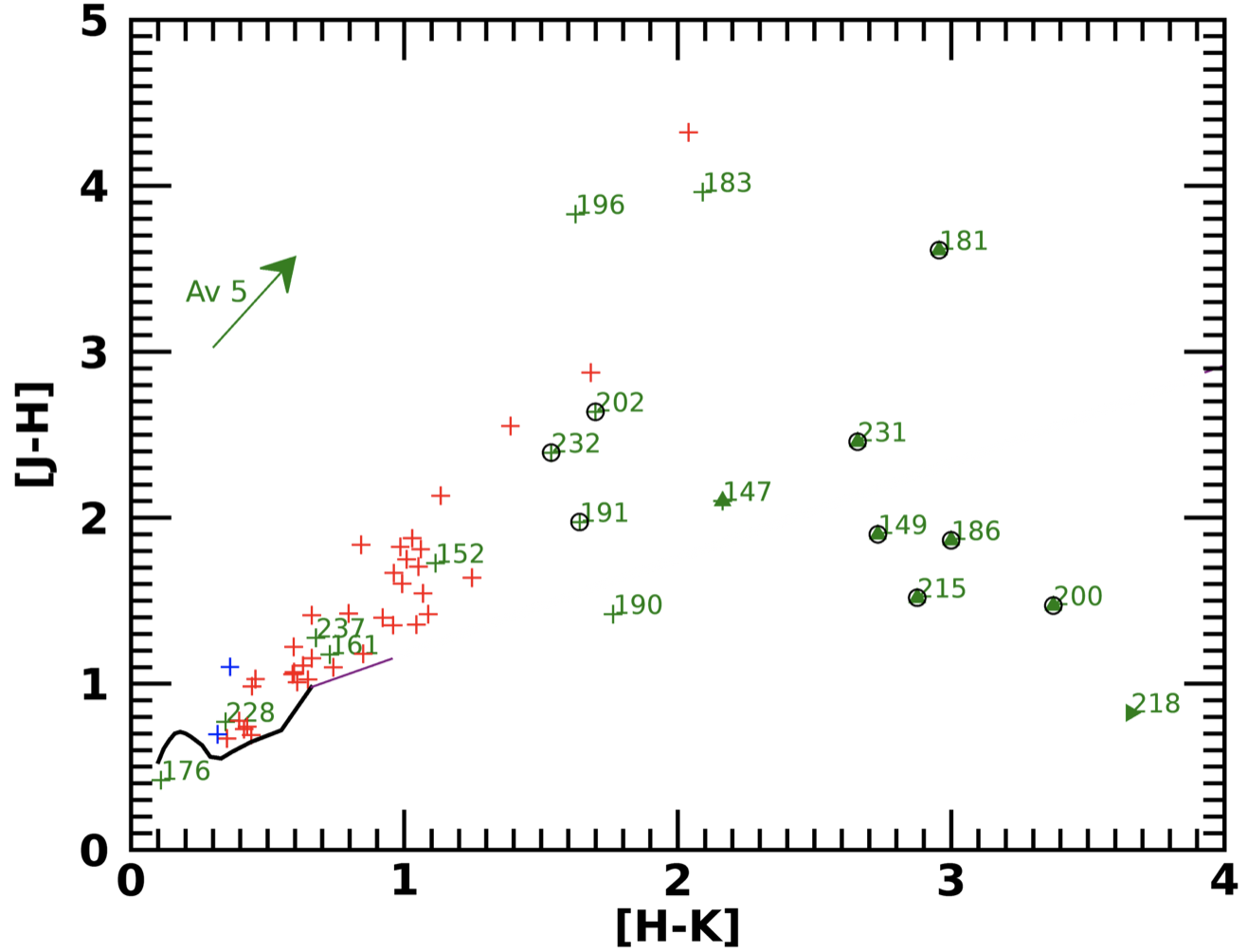}
  \caption{\label{colcol}Colour-colour diagram of our sample. 
  Class~I, II, and III YSOs are plotted as green, red, and blue crosses, respectively. 
  Triangles are lower limits. Class~I YSOs which show HI in emission are surrounded by a black circle.
  The main sequence { \citep[MS,][]{luh10} and CTT locus \citep{mey97}} are shown in black and purple, respectively.
  The green arrow in the left-top corner defines the reddening direction, following the \citet{car89} extinction law.
  Its length corresponds to $A_V =$~5~mag.}
\end{figure}
\begin{table*}[t]
 \centering
 \caption{\label{starpar}Stellar and accretion parameters of the Class~II sample}
  \begin{tabular}{lrllccllrr}
  \hline
  \hline
     ID&   $A_v$    & SpT$^{(a)}$   & SpT$_L^{(b)}$ &$T_{\rm eff}$ & log($\lstar/ \lsun$) & $\mstar$    &R$_\star$ & $\log{(\lacc/\lsun)}$ &  $\macc$\\
       &   mag   &       &         & K            &        &$\msun$      &$\rsun$   &  &$10^{-9} \msun \mbox{yr}^{-1}$\\   
 \hline
 \hline
153        &  4.5 &  M5.5  &     M6  &  2920    & $-1.17$      &  0.10	        &  1.02  &$< -2.90        $ & $< 0.06$ \\
156        &  4.0 &  M6.5  &   M6.5  &  2815    & $-0.72$      &  0.11	        &  1.84  &$ -1.95 \pm 0.36$ & $7.49 \pm 0.48$ \\
159        &  0.3 &  M5.5  &     M5  &  2920    & $-1.26$      &  0.10	        &  0.92  &$ -3.22 \pm 0.54$ & $0.22 \pm 0.02$ \\
162        &  0.0 &    M9  &   M7.5  &  2400    & $-1.29$      & ...	            &  1.31  &$< -2.84        $ &... \\
164        &  0.0 &  M5.5  &     M5  &  2920    & $-1.43$      &  0.10	        &  0.76  &$< -3.03        $ &$<  0.06$ \\
165        & 17.0 &    K7  &     K7  &  4020    &  0.16      &  0.70	        &  2.49  &$< -1.79        $ &$<  0.06$ \\
166        &  2.7 &    M4  &  M5.25  &  3190    & $-0.86$      &  0.21	        &  1.22  &$< -2.56        $ &$< 0.12$ \\
167        &  8.5 &    K7  &     M1  &  4020    & $-0.01$      &  0.72	        &  2.05  &$ -1.30 \pm 0.37 $ &$5.68 \pm 0.37$ \\
168        & 10.0 &    M0  &     M2  &  3900    & $-0.15$      &  0.61	        &  1.85  &$ -1.49 \pm 0.44$ &$3.91 \pm 0.30$ \\
173        &  7.0 &    M3  &   M1.5  &  3410    & $-0.04$      &  0.32	        &  2.75  &$ -0.59 \pm 0.34$ &$88.0 \pm 5.2$ \\
174        &  9.8 &    M4  &     M1  &  3190    & $-0.35$      &  0.25	        &  2.20  &$< -2.27        $ & $< 0.15$ \\
175        &  5.7 &  M6.5  &  -  &  2815    & $-1.52$      &  0.08($\dagger$)&  0.73   &$ -3.56 \pm 0.54$ &$0.10 \pm 0.01$ \\
177        &  2.7 &    M3  &   M3.5  &  3410    & $-0.83$      &  0.30	        &  1.11  &$ -2.87 \pm 0.51$& $0.20 \pm 0.02$ \\
178$^*$       &  3.0 &    M2  &   M0.5  &  3560    & $-0.68$      &  0.38	        &  1.21  & $< -2.87        $&$< 0.05$ \\
180        &  5.5 &  M9.5  &     M8  &  2330    & $-1.36$      & ...	            &  1.29  &$ -3.26 \pm 0.80$ &... \\
184        &  6.0 &    M2  &  -  &  3560    & $-0.26$      &  0.38	        &  1.96  &$ -0.75 \pm 0.36$& $36.5 \pm 2.3$ \\
188        &  4.3 &    M3  &  -  &  3410    & $-0.92$      &  0.30	        &  1.00  &$ -1.85 \pm 0.38$&  $1.87 \pm 0.14$ \\
192        &  6.5 &    M1  &     K6  &  3720    &  0.25      &  0.47	        &  3.22  &$ -0.70 \pm 0.35$& $54.5 \pm 3.3$ \\
194        &  8.8 &    M7  &     M6  &  2770    & $-1.23$      &  0.10	        &  1.06  &$< -2.58        $&$<  0.06$ \\
195        & 15.0 &  M4.5  &  -  &  3085    & $-1.67$      &  0.11	        &  0.51  &$< -2.60        $&$<  0.03$ \\
197        & 17.0 &  M6.5  &     M9  &  2815    & $-0.65$      &  0.11	        &  2.00  &$< -1.42        $&$<  0.81$ \\
199        & 28.0 &    M4  &     M4  &  3190    & $-0.40$      &  0.25	        &  2.07  &$< -1.69        $&$<  0.08$ \\
201        &  7.5 &  M3.5  &   M2.5  &  3300    & $-0.29$      &  0.28	        &  2.20  &$ -1.84 \pm 0.46$&  $4.53 \pm 0.36$ \\
203        &  9.0 &    M8  &     M6  &  2710    & $-0.86$      &  0.08($\dagger$) &  1.69  &$ -2.20 \pm 0.36$&  $5.32 \pm 0.34$ \\
205        & 14.0 &  M5.5  &   M5.5  &  2920    & $-0.97$      &  0.11	        &  1.28  &$< -2.50        $&$<  0.06$ \\
209        & 11.7 &  M6.5  &     M5  &  2815    & $-0.93$      &  0.10	        &  1.45  &$< -2.46        $&$<  0.11 $\\
213        &  8.0 &    M4  &     M2  &  3190    & $-1.31$      &  0.18	        &  0.73  &$ -2.57 \pm 0.49$&  $0.43 \pm 0.04$ \\
219        &  8.7 &    M4  &     M3  &  3190    & $-0.12$      &  0.25	        &  2.86  &$ -1.79 \pm 0.38$&  $7.4 \pm 0.5$ \\
222        &  5.8 &    M5  &     M3  &  2980    & $-0.62$      &  0.16	        &  1.84  &$ -1.94 \pm 0.40$&  $5.3 \pm 0.4$ \\
224$^{**}$        &  3.5 &    M3  &     M3  &  3410    & $-1.69$      &  0.24	        &  0.41  &$ -3.57 \pm 0.38$&  $0.02 \pm 0.001$ \\
226        &  3.0 &    M3  &     M4  &  3410    &  0.43      &  0.40	        &  5.74  &$ -0.95 \pm 0.42$& $52.8 \pm 4.5$ \\
227        &  4.5 &  M5.5  &  M5.75  &  2920    & $-1.40$      &  0.10	        &  0.78  &$< -3.16        $&$<  0.03$ \\
233        &  0.5 &    M5  &  M5.25  &  2980    & $-0.88$      &  0.14	        &  1.37  &$< -2.60        $&$<  0.26$ \\
234        &  3.0 &    K7  &     K4  &  4020    &  0.04      &  0.71	        &  2.17  &$ -0.55 \pm 0.34$& $34.3 \pm 2.1$ \\
235        &  1.0 &  M4.5  &  M4.75  &  3085    & $-0.81$      &  0.18	        &  1.38  &$ -2.78 \pm 0.50$&  $0.51 \pm 0.04$ \\
241        &  3.1 &    M4  &   M4.5  &  3190    & $-0.66$      &  0.22	        &  1.54  &$< -2.53        $&$<  0.18$ \\
    \hline     
    \hline
  \end{tabular}                                   
  \begin{quotation}
  \textbf{Notes.} $^{(a)}$ SpT derived by using Class~III templates (see Sect.~\ref{sect:stelpar2}), $^{(b)}$ SpT derived by \citet{luh16}.
{\small ($\dagger$) Mass derived using the \citet{bar15} isochrones, see Sect.~\ref{sect:ana1}. $^*$Class~III target. $^{**}$Sub-luminous object. }
  \end{quotation}  
  \end{table*}
  
\begin{figure}[t]
 \includegraphics[width=\columnwidth]{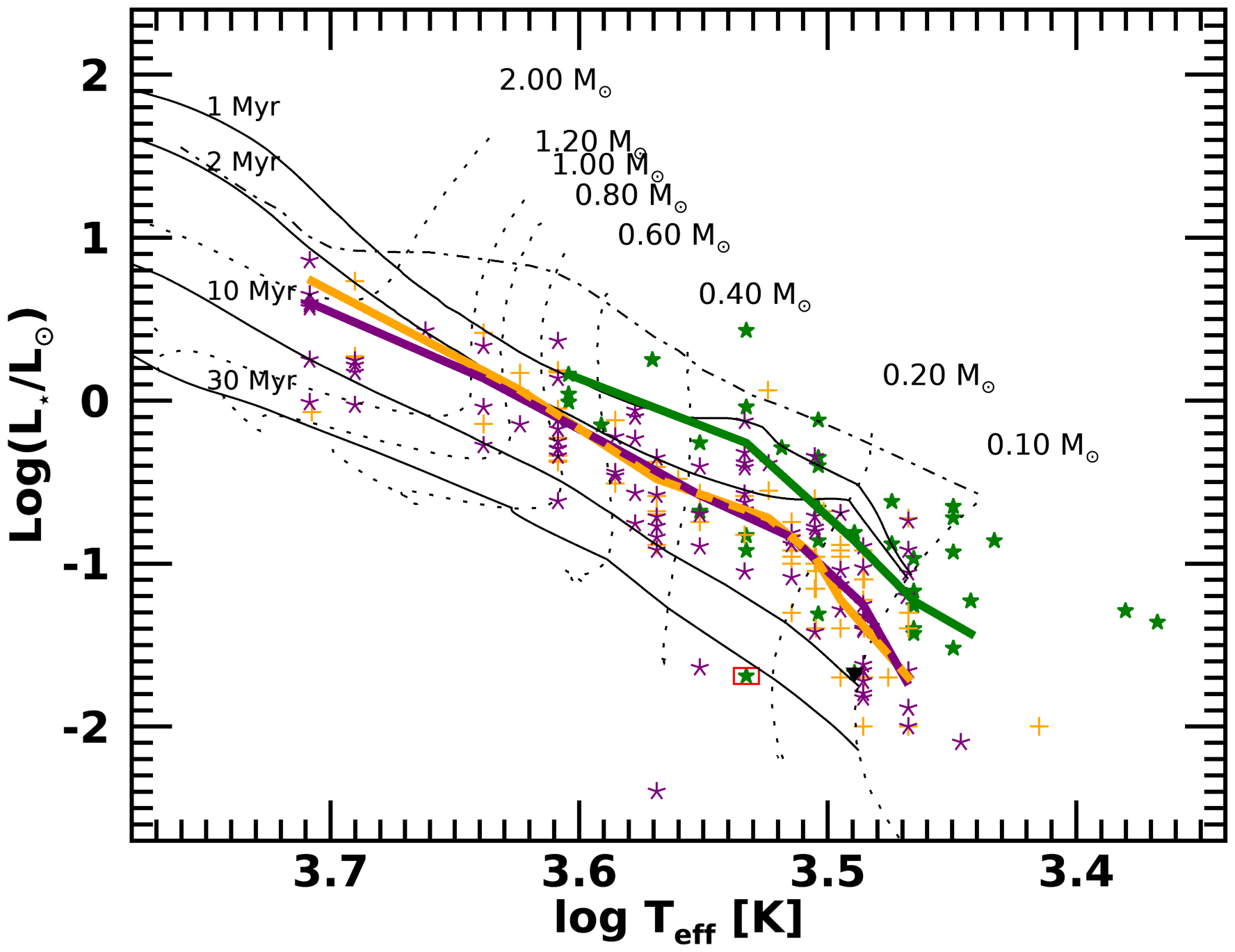}
  \caption{\label{hrdiag}HR diagram for the Class~II YSOs of NGC~1333 (green stars), Lupus (orange crosses) and Cha~I (purple asterisks).
  The green, orange, and purple lines correspond to the median of NGC~1333, Lupus, and Cha~I samples (sub-luminous objects are not included), respectively. 
  The birthline \citep{pal93} is plotted as a dot-dashed line.
  Black solid and dashed lines show the isochrones and evolutionary tracks by \citet{sie00}.}
\end{figure}
\begin{figure*}
    \centering
    \includegraphics[width=\textwidth]{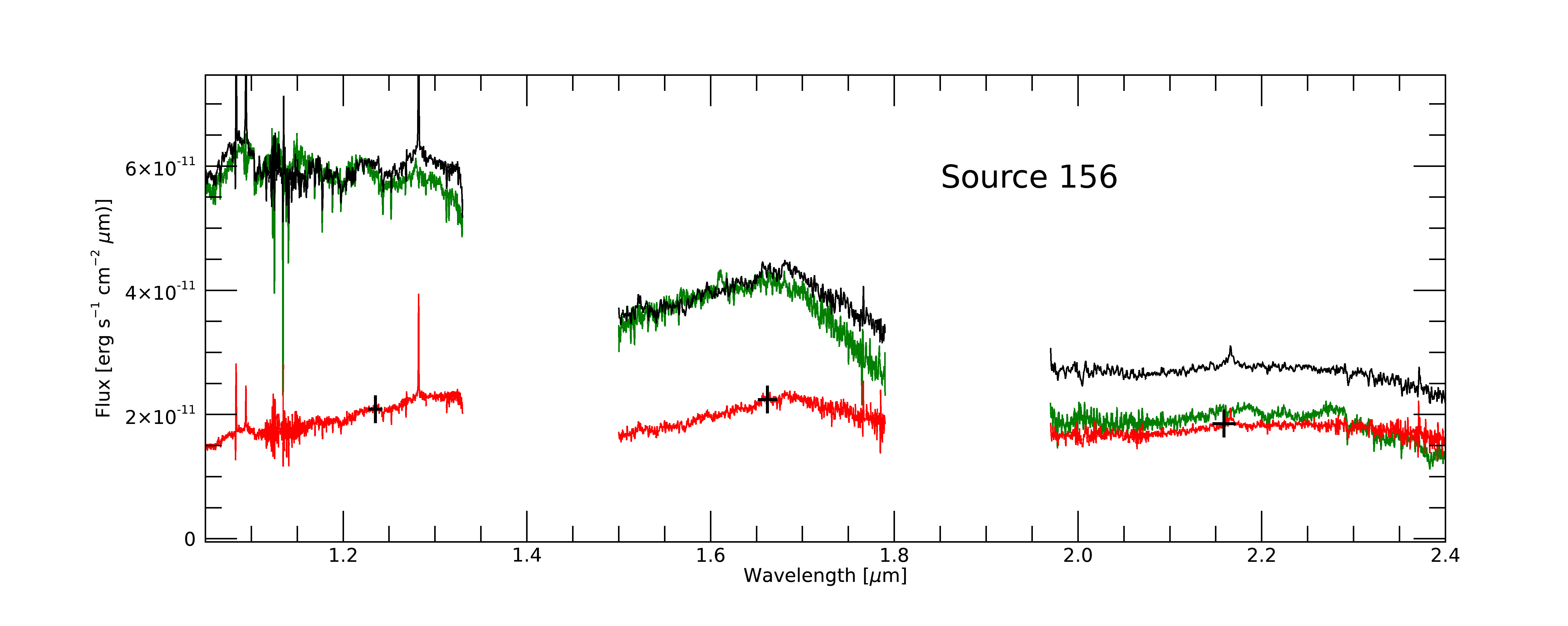}
    \caption{Example of the spectral typing result for Class~II source \#156. The observed spectrum is shown in red. The dereddened spectrum (black) matches the spectrum of an M6.5 Class~III (green) from \citet{man13}. The difference in the K band is due to the fact that source \#156 is accreting, while the Class~III template is not.}
    \label{fig:SpT}
\end{figure*}

To determine the spectral type~(SpT) and the extinction~($A_V$) of each target, we compared our spectra with a grid of photospheric templates consisting of observed spectra of non-accreting Class~III YSOs \citep{man13, man17b}, which cover spectral types from G4 to M9.5, with a typical step of one spectral sub-class for G and K types, and 0.5 spectral sub-classes for the M type. 
For each source, we estimated the SpT and $A_V$ by varying the template and the extinction, and by matching the shape and the molecular features (mostly H$_2$O in H band) of our data with the ones of the Class~III template 
\citep[see App.~B of ][for a list of NIR molecular bands]{man13}. 
This is a procedure typically used in the literature \citep[e.g.,][]{alc14,man15}. 
To compare the templates to KMOS spectra, we reddened and normalised them within a window of 0.02~$\mu$m around $\lambda = 1.60$~$\mu$m, where no molecular features are present.
The comparison was made by eye, taking into account the match with the molecular bands and the overall shape of the spectrum in the J and H bands. 
We did not consider the K band for matching due to the IR excess expected in this band from to the disk emission. 
An example of a spectral typing result is shown in Fig.~\ref{fig:SpT}.

The J band of the KMOS spectra is in general noisier than the other bands, and for several sources the S/N is too low to compute the scale factor in this band. 
We verified that the spectral typing changes at most within one sub-Class and one mag in $A_V$, scaling the template flux to the flux of our data in the J or H band for the sources whose J band is not too noisy. 
Therefore, we decided to use H band for the overall sample.
This method can provide some small degeneracies in SpT and $A_V$, because the presence of certain molecular bands is related to the spectral class. 
We adopted $\pm 1$ sub-Class and $\pm 1$ magnitude for the uncertainties of spectral type and extinction, respectively.

We compareed our SpT and $A_V$ with the ones estimated by \citet{luh16} for the sources in common, finding a general agreement and verifying that different methods give consistent results.
We discuss this comparison more in detail in Appendix~\ref{app:luh}.

We further estimatedd $A_V$ with other methods and compared the results. 
The first method is based on the use of the [J~-~H]~versusd~[H~-~K] colour-colour diagram. 
This is shown in Fig.~\ref{colcol} together with the observed sample and the extinction-vector, 
which indicates the direction of the colour difference of a star extincted by $A_V$= 5~mag according to the \citet{car89} extinction law, with the standard value $R_V=3.1$.  
We estimated the $A_V$ from the colour difference between the position of the observed targets and the CTTS locus \citep{mey97}, following the extinction-vector. 

Another method is to estimate $A_V$ from the accretion luminosity obtained with tracers at different wavelengths. 
Indeed, as we show in the next section, for those sources in which both $\pab$ and $\brg$ are detected, we can estimate the value of extinction that makes $L_{\rm acc}^{{\rm Pa}_{\beta}} = L_{\rm acc}^{{\rm Br}_{\gamma}}$.

Table~\ref{ext2} summarises $A_V$ estimates derived from all methods. 
The $A_V$ derived from the comparison with the reddened spectral templates are generally consistent with those derived from the colour-colour diagram within 2~mag for the majority of sources. 
Larger differences occur in a few cases, including sources \# 165, \#180, \#197, and \#213. 
The extinction derived from the colour-colour diagram is on average smaller than that estimated through the other methods, which can be explained by the presence of scattered light in the source environment.

In the following, we refer to $A_V$ as the extinction derived by comparing the spectra with the reddened templates ($A_V^{ \rm SpT}$).
The SpT and $A_V$ for each Class~II YSO is reported in Table~\ref{starpar}. 

Knowing the SpT, we can then estimate the star luminosity ($\lstar$) from the observed magnitudes corrected for the extinction and assuming a bolometric correction ($BC$).
We used bolometric correction values in the $J$ band ($BC_J$) of $5-30$~Myr stars from Table~6 of \citet{pec13} for spectral types down to M5, and from \citet{her15} for SpT from M5.5 to M7.5. 
From the same works, we also took the effective temperature ($\teff$) corresponding to the estimated SpT. 
In several cases, the SpT class or sub-class of our targets is not present in these works. 
In such cases we obtained $BC$ and $\teff$ by interpolating with the available SpTs.

Then, $\lstar$ is given by: 
 \begin{equation}
  \log \left(\frac{\lstar}{\lsun}\right) = 0.4[M_{\rm bol,\odot}-M_{\rm bol}]
  \label{eq:lstar}
 \end{equation}
where $M_{{\rm bol},\odot}$ is the bolometric magnitude of the Sun \citep{mam15}, and the bolometric magnitude of a source is $M_{\rm bol}= m_J - 5 \log (d/10 \mbox{[pc]}) + BC_J$ where $m_J$ is the extinction corrected magnitude. 
The typical uncertainty in computing the luminosity in this way is of 0.2~dex in $\log{\lstar/\lsun}$, taking into account the uncertainty of the observed magnitude and of the SpT. 
 
After deriving $\lstar$ and $\teff$ for each source, we used the evolutionary tracks by \citet{sie00} to determine the corresponding $M_\star$. 
Since \citet{sie00} models are available down to 0.1~$\msun$, for objects with lower masses, we used the \citet{bar15} evolutionary tracks instead. 
As investigated in \citet{alc17} and \citet{man17b}, the results of both stellar parameters and accretion rates are compatible within each other by using different evolutionary models. 
The range of stellar masses derived for our sample is  $0.08 - 0.72$~$\msun$, with 0.1~dex in $\log({\mstar/\msun})$ of uncertainty (see Table~\ref{starpar}).

We plotted the locus of the NGC~1333 stars in the HR diagram in Fig.~\ref{hrdiag}. 
For comparison, in this diagram, we also plotted the values for Lupus and Chamaeleon~I (Cha~I) samples taken from \citet{alc19} and \citet{man19}, respectively. 
We note that there is one target (\#224) below the 30~Myr isochrone. This is highlighted in Fig.~\ref{hrdiag} with a red box.
The spectrum of this target shows HI emission lines and features typical of a star as young as the others in the sample. Therefore, we suggest that this target is a candidate sub-luminous object, where the low apparent luminosity can be due to the presence of an edge-on disk which occults the central star \citep[e.g.,][]{alc14}. 

By looking at the median stellar luminosity of the sources, we observe that the Class~II targets in  NGC~1333 have a systematically higher stellar luminosity than the targets in the Lupus and Cha~I regions. 
This is an indication that, as expected and in agreement with previous estimates of this cluster age, NGC~1333 is younger than Lupus and Cha~I.

Assuming that the central object emits as a black-body, we computed the stellar radius:
\begin{equation}
    \rstar = \frac{1}{2\teff^2} \sqrt{\frac{\lstar}{\pi \sigma}}
    \label{rad*}
\end{equation}
where $\sigma$ is the Stefan–Boltzmann constant. 
The mean uncertainty on the radius is about $\pm 0.6$~$R_\odot$.
All the stellar parameters are listed in Table~\ref{starpar}. 

\subsection{Accretion rate estimate} \label{sect:acc}
As discussed in the introduction, the accretion luminosity ($\lacc$) of the sources can be measured by adopting the relationships that connect $\lacc$ with the luminosity of the $\pab$ and $\brg$ lines ($L_{\rm line}$). 

We computed the line luminosity as $L_{\rm line}=4\pi d^2 F_{\rm line}$, where $F_{\rm line}$ is the extinction corrected line flux and $d$ is the distance to the NGC~1333 region.
Then, we estimated the accretion luminosity by using the relations from \citet{alc17}
 \begin{equation}
 \log \left( \frac{\lacc}{\lsun}\right) = a \log\left( \frac{L_{\rm line}}{\lsun} \right) + b
 \label{eqLacc}
 \end{equation}
where $a=1.06 \pm 0.07$ and $b=2.76 \pm 0.34$ for $\pab$, and $a= 1.19 \pm 0.10$ and $b= 4.02 \pm 0.51$ for $\brg$.
The value of the accretion luminosity for each Class~II source is reported in Table~\ref{starpar}.
For the targets observed twice, we considered the mean value of the two $\lacc$ estimates. 
We computed the uncertainty on the accretion luminosity by propagating the error of the coefficients and $L_{line}$, obtaining a typical error of 0.4~dex in $\log \left ( \lacc / \lsun \right)$.
When there is no detection of $\pab$ nor $\brg$, we report in Table~\ref{starpar} the larger $3 \sigma$ upper limit.
The accretion luminosity in our sample is $-3.57 \le \log \left( \lacc/\lsun \right) \le -0.55$. 

From the derived $\lacc$, $\mstar$, and $\rstar$, we computed the mass accretion rate $\macc$
using the relation
 \begin{equation}
 \label{eqmacc}
  \macc \sim \left(1 - \frac{\rstar}{R_{\rm in}}\right)^{-1} \frac{\lacc \rstar}{G \mstar}
 \end{equation}
 where $R_{\rm in}$ is the inner-disk radius which we assume to be $R_{\rm in} \sim 5 R_\star$ \citep{har98}.
By propagating the uncertainties, we estimated the error on $\macc$, finding an average value of $\Delta\macc/\macc = 0.08$. 
The mass accretion rate of our sample ranges between $\sim 10^{-11}$ and $10^{-7}$$\msun$yr$^{-1}$. 
The derived $\macc$ are listed in Table~\ref{starpar}. 

\section{Analysis of Class~I YSOs} \label{sect:ana1}
Class~I objects have a large residual envelope that significantly contributes to the NIR emission and cannot be separated from the photospheric and disk contribution. 
Consequently, for those sources, it is not possible to adopt the same classification method we applied to the Class~II sample. 
Indeed, estimating the extinction from the colour-colour diagram (Fig.~\ref{colcol}) for this sample is not possible because half of the sources have upper limits in J magnitudes and, in principle, we cannot assume that the CTTS locus is also valid for Class~I YSOs.
Therefore, for the observed Class~I objects, we adopted a different method, already applied by \citet{ant08}, based on a procedure which we describe in the next section.

\subsection{Method} \label{sect:met}
The self-consistent method we used is based on the computation of the stellar luminosity in two equations which both involve extinction. 
Firstly, the stellar luminosity can be computed by assuming that the bolometric luminosity of YSOs is due to the sum of the accretion and the stellar contribution 
\begin{equation}
\label{eq:lum}
    \lstar = \lbol - \lacc
\end{equation}
As shown in Sect.~\ref{sect:stelpar2}, the stellar luminosity depends on the extinction and the bolometric correction. 
In Class~I sources, we also need to take the contribution of the emission into account due to the envelope and the disk, which we have considered negligible in the H-band of Class~II sources. 
This contribution is generally described with the veiling parameter $r = F_{\rm ex}/F_\star$, representing the excess emission above the photosphere.
Therefore for Class~I sources for which we used the $\brg$ line flux, in Eq.~\ref{eq:lstar}, the bolometric absolute magnitude can be given as follows
\begin{equation}
\label{eq:mbol}
M_{\rm bol} = BC_K + m_K + 2.5\log({1 + r_K}) - A_K - 5 \log (d/10 \mbox{pc})
\end{equation}
Secondly, as discussed in Sect.~\ref{sect:acc}, it is possible to compute the accretion luminosity from the flux of HI emission lines (Eq.~\ref{eqLacc}) which depends on the extinction as well.

The observed parameters that enter into Eqs.~\ref{eq:lum} and \ref{eq:mbol} are the $\brg$ flux, the observed K-band magnitude ($m_K$), the bolometric luminosity, and the distance, while assumptions on the K-band veiling ($r_K$) and on the age of the object, which set the spectral type and thus the bolometric correction ($BC_K$) in Eq.~\ref{eq:mbol}, need to be made. 
If all these parameters are determined, then Eqs.~\ref{eq:lum} and \ref{eq:mbol} only depend on $A_K$, so that we can find the value of the extinction which consistently provides the same $\lstar$ from the two computations.

For our analysis, we assumed that our Class~I sources are located between the birthline, as defined by \citet{pal93}, and the 1~Myr isochrone of the \citet{sie00} models. 
Appendix~\ref{app:hrdiagrI} displays the corresponding evolutionary tracks and how the sources are displayed on them.
Once the previous quantities were determined, the other stellar and accretion parameters, namely $M_\star$, $R_\star$, and $\dot{M}_{\rm acc}$, were derived as described in Sect.~\ref{sect:stelpar2}.

\subsection{Results}
Our sample comprises 17 confirmed Class~I/Flat sources (see Table~\ref{TabFlux23}). 
In only ten of them at least the $\brg$ line was detected at minimum. 
We notice that there is only one Flat source (namely source \#202) in our sample that shows accretion features.
We analysed only sources with at least one HI line detection.

Our adopted method relies on a good determination of the source bolometric luminosity. 
For this reason, we computed $\lbol$ by integrating the observed spectral energy distributions (SEDs) of the sources including, in addition to NIR and {\it Spitzer} photometry, also more recent \emph{Herschel} photometric measurements and millimetre measurements, when available. 
Details on the construction of SEDs construction are given in Appendix~\ref{app:sed}, while the derived $\lbol$ for the Class~I accretors are listed in Table~\ref{starparI}. 

\begin{figure}
    \centering
    \includegraphics[width=\columnwidth]{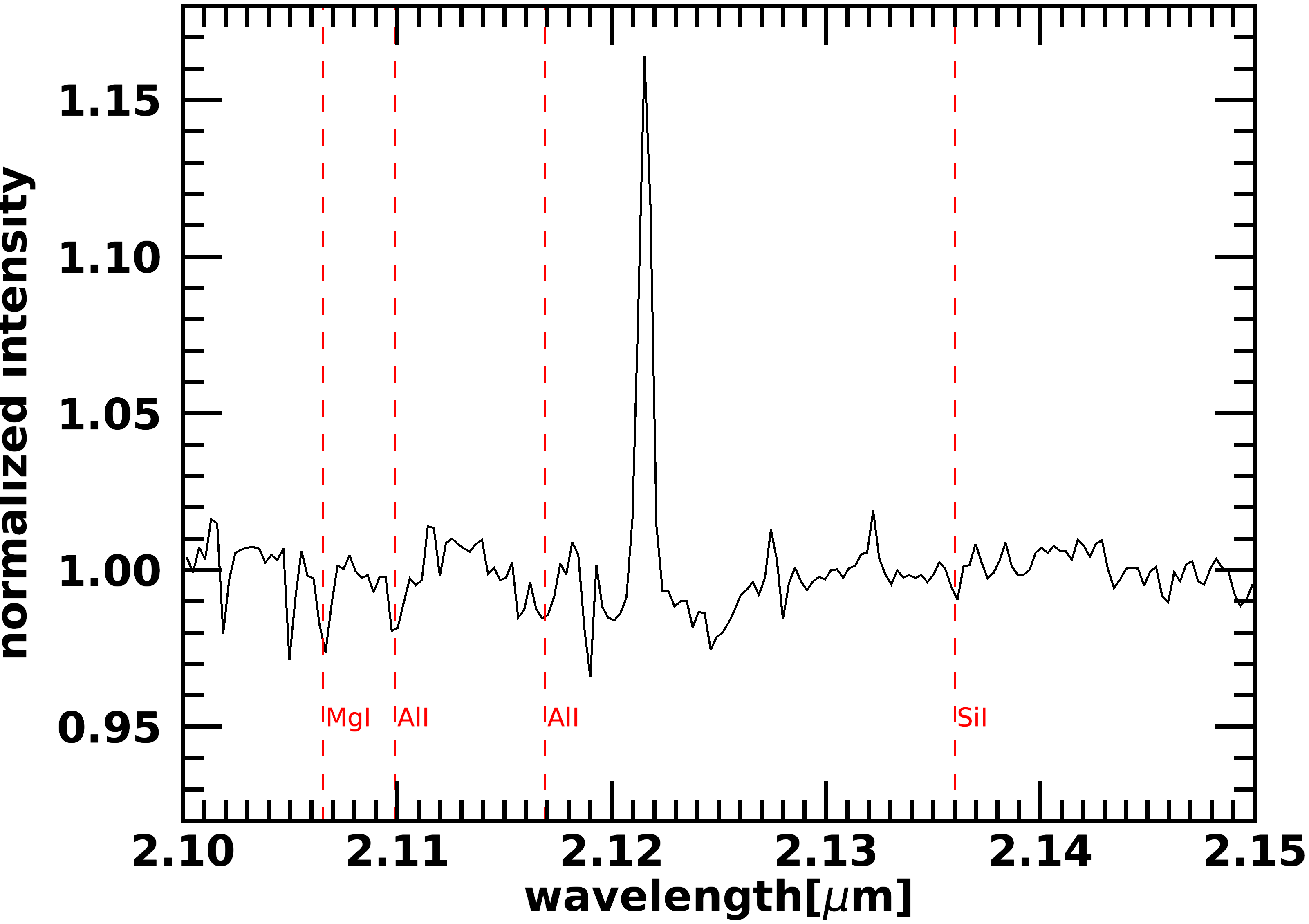}
    \caption{K-band spectrum of SVS13~A, normalised to the continuum.
    Red-dashed lines correspond to photospheric lines as labelled.
    }
    \label{fig:191}
\end{figure}

As discussed in the previous section, the method also assumes a value for the K-band veiling of our sources. 
The effect of the veiling is to fill the photospheric absorption lines, decreasing their equivalent widths and making them undetectable for extreme veiling values. 
From the inspection of the K-band spectra of our ten accreting sources, we see that photosperic lines typical of late-type stars, such as the Na~doublet at 2.209~$\mu$m or the Al~I at 2.117~$\mu$m \citep[see e.g.][]{nis05,dop05}, are detected in four sources (namely \#202, \#215, \#231, and \#232), while in the other ones such lines are either very weak or not detected. 
For Class~I sources where photospheric features have been detected, we assume $r_K$ values ranging between 1 and 3, as estimated in the samples of Class~I objects in the literature \citep[e.g. ][]{dop05, con10}. 
The same works estimate $r_K \gtrsim 5$ in sources where photospheric lines are not or are barely detected. 
In the bright source \#191, where the continuum was measured at the highest S/N, weak photospheric lines of Al and Mg are detected at a $5 \sigma$ level (Fig.~\ref{fig:191}).
Comparing the equivalent width of these lines with templates of stars with SpT between K7 and G1, we estimate that the veiling should be between 3 and 6. 
To be conservative, we explored a range of veiling between 3 and 8 for the source \#191 and the other six sources where photospheric lines could not be detected.
The results of our analysis are shown in Table~\ref{starparI},  where we report the range of values for the stellar and accretion parameters that we derived from our assumptions.

\begin{table*}
 \centering
 \caption{\label{starparI}Stellar and accretion parameters of the Class~I sample.}
\resizebox{\textwidth}{!}{%
 \begin{tabular}{lccccccccccc}
 \hline
 \hline
ID     & age	 & $r_K$  & $\lbol$& A$_V$	  & $\lacc$.	 & $\lstar$	&$T_{\rm eff}$& SpT	   & $\mstar$	  & R$_\star$	& $\macc$	\\
       &	 &	  & $\lsun$&mag 	 & $\lsun$	& $\lsun$      & K	     &  	  & $\msun$	 & $\rsun$     & $10^{-8} \msun$~yr$^{-1}$	    \\
\hline
\hline
  149  &    BL   &3-8	  & 0.31   & 44.5 - 51.0  & 0.05-0.10  & 0.25-0.30  &2818-2754  & M7.5       & 0.13-0.11  &2.2-2.1  & 3.2-8.3 \\
       &    1Myr &	  &	   & 43.0 - 49.5  & 0.04-0.08  & 0.26-0.31  &3075-3038  & M5.5       & 0.20-0.17  &1.8-1.6  & 1.4-3.2 \\
 &&&&&&&&&&&\\
  181  &    BL   &3-8	  & 1.11   & 31.5-34.5    & 0.60-0.86  & 0.25-0.51  & 3020-2754   & M6-M7.5	 & 0.19-0.11  &2.5-2.1  & 32-70 \\
       &    1Myr &	  &	   & 31.0-34.0    & 0.57-0.81  & 0.30-0.54  & 3218-3075   & M4.5-M5.5& 0.25-0.20  &2.1-1.8  & 19-31 \\
 &&&&&&&&&&&\\
  186  &    BL   &3-8	  &18.2    & 45.0-51.0   & 2.46-5.08  & 13.1-15.7  & 5248-5248    & G9.5     & 2.84	  &4.5      & 16-33 \\
       &  1Myr   &	  &	   & 45.5-52.0   & 2.62-5.7  & 12.5-15.6  & 5019-4942	 & K1-K1.5  & 3.03-2.80 &4.9-4.5  & 18-38 \\
    &&&&&&&&&&&\\   
  191  &    BL   &3-8	  &58.8    &33.0-38.0	  & 16.4-29.9  &28.9-42.5   &5754-5623 & G5.5-G7   & 3.07-2.95  &5.7-5.3  &120-220  \\
       &  1Myr   &	  &	   &34.0-39.0	  & 18.5-33.7  &25.1-40.4   &5395-5173 & G9.5-K0   & 3.53-3.28  &2.1-1.8  &19-31 \\
  &&&&&&&&&&&\\
  200  &    BL   &3-8	  & 7.63   &36.0-41.0  & 2.19-4.01  & 3.62-5.44  &3981-3802   & K8-M0.5   & 0.70-0.53  &4.7-3.9  & 60-120 \\
       &   1Myr  &	  &	   &34.0-39.5  & 1.72-3.34  & 4.29-5.91  &4683-4547   & K3.5-K4.5 & 1.90-1.52  &3.5-3.1  & 13-28 \\
  &&&&&&&&&&&\\
  202  &    BL   &1-3	  & 0.23   &17.0-22.5  & 0.02-0.04  & 0.19-0.21  &2754-2754   & M7.5	   & 0.11	  &2.1         &1.7-3.3 \\
       &  1Myr   &	  &	   &15.5-21.5  & 0.02-0.04  & 0.20-0.22  &3002-3002   & M6	   & 0.15	  &1.4         &0.7-1.4 \\
&&&&&&&&&&&\\
  204  &    BL   &3-8    & 0.51   &38.0-41.0  & 0.24-0.34  & 0.17-0.27  &2754-2754   & M7.5	   & 0.11	  &2.1         &19-28 \\
       &  1Myr   &	  &	   &36.5-40.5  & 0.20-0.32  & 0.19-0.31  &3075-3002   & M5.5-M6    & 0.20-0.15    &1.8-1.4     &7-13 \\
  &&&&&&&&&&&\\
  215  &    BL   &1-3	  & 0.12   &20.0-25.5  & 0.02-0.03  & 0.09-0.11  &2754-2754   & M7.5	   & 0.11	  &2.1         &1.4-2.6 \\
       &  1Myr   &	  &	   &19.0-24.0  & 0.01-0.03  & 0.09-0.11  &2928-2928   & M6.5	  & 0.10	 &1.0	      &0.6-1.2 \\
 &&&&&&&&&&&\\
  231  &    BL   &1-3	  & 0.59   &27.5-34.0  & 0.03-0.07  & 0.52-0.56  &3020-3020   & M6	   & 0.19	  &2.5         &1.6-3.6 \\ 
       &  1Myr   &	  &	   &26.0-32.5  & 0.03-0.06  & 0.53-0.56  &3289-3218   & M4-M4.5    & 0.28-0.25    &2.2-2.1     &0.8-1.9 \\
  &&&&&&&&&&&\\
  232  &    BL   &1-3	  & 0.70   &14.0 - 20.0  & 0.07-0.15  & 0.55-0.63  &3090-3020	& M5.5-M6   & 0.22-0.19  &2.5  &3.6-8.1 \\
       & 1Myr	 &	  &	   &12.5-18.5  & 0.06-0.13  & 0.57-0.64  &3289-3289   & M4	  & 0.28	 &2.2 &2.0-4.2 \\

\hline
\hline
 \end{tabular} 
 }
 \begin{quotation}
   \textbf{Notes:} the first two parameters of the table (the age and the K-band veiling, $r_K$) are assumed. 
   The bolometric luminosity is derived in Appendix~\ref{app:sed}, while for all the other 
   parameters the range of values derived from our method, as described in Sect.~\ref{sect:met} is reported. 
   BL denotes the birthline of \citet{pal93}.  
   
 \end{quotation}
\end{table*}

All Class~I YSOs but three (namely sources \#186, \#191, and \#200) are late-type stars (K-M), with stellar parameters (mass and luminosities) in the range of those estimated for the Class~II sources of the region. 
Three sources are, however, more massive and luminous. 
In particular, source \#191 (SVS13 A) and \#186 have masses in the range of $2.5-3.5$~$\msun$, and stellar luminosities between 12~$\lsun$ and 40~$\lsun$.
These results are due to the high bolometric luminosities of the two sources and also to the derived accretion luminosity, which can account for a maximum of $60\%$ of $\lbol$. 

The derived extinction is high in all the sources, and it ranges between 12 and 51 mag.
Finally, the accretion luminosity and mass accretion rate range are $10^{-2} - 10^2$~$\lsun$ and $10^{-9} - 10^{-6}$~$\msun/$yr, respectively.
We discuss this last issue in Sect.~\ref{sect:disc1} and the peculiarity of source \#191 in Sect.~\ref{vvser}.

\subsection{Upper limits on $\lacc$ in Class~I with no line detection} \label{sect:noHIlines}
For the seven sources which show no HI emission line, we derived the upper limits from the $3\sigma$ upper limits of the $\brg$ line (see Table~\ref{TabFlux23}), applying a reasonable estimate of the extinction, based on the spectral features such as IR-excess and photospheric lines.
Sources \#147, \#190, \#196 and \#237 show small or no IR excess. 
Therefore, it is possible to estimate $A_V$ and SpT by comparing their spectra with Class~III templates, as we did for the Class~II sample (see Sect.~\ref{sect:ana2}). 
Sources \#161 and \#183 show a some IR excess, but also photospheric lines in the K-band, as MgI, AlI, and SiI.
For these sources we assumed the mean of the values we found as extinction for the low-veiling sources ($A_V = 20$~mag).
Source \#218 shows a large IR excess and no photospheric absorption lines. 
For this source we assume the typical extinction value we find for the highly veiled sources ($A_V = 40$~mag).
The derived upper limits for the above sources under these assumptions are listed in Table 6.

\begin{table}
 \centering
 \caption{\label{noHIclI}Upper limits of the Class~I sources with no HI emission lines}
 \begin{tabular}{lcccccc}
 \hline
 \hline
ID  & $\lbol$ & SpT & $A_V$ & $\lacc$ & SpT$_L$ & $A_V^L$	\\
    & $\lsun$ &     & mag   & $\lsun$ &         &\\
\hline
\hline
147 & 0.08 &  M5.5 & 30.0 & < 0.0025     & M9      & 33.0\\
161 & 0.06 &  -    & 20.0  & < 0.0009 & - & -\\
183 & 1.20 &  -    & 20.0  & < 0.0094 & - & -\\    
190$^*$ & 0.15 &  M0.5 & 11.5  & < 0.0002 & - & - \\
196 & 0.20 &  M7.5 & 16.0  & < 0.0019 & M7.5 & 16.5\\
218 & 1.28 &  -    & 40.0  & < 0.0121 & - & -\\
237$^*$ & 0.06 &  M4.5 & 5.0   & < 0.0001 & M2 & 5.86  \\
\hline
\hline
 \end{tabular} 
 \begin{quotation}
   \textbf{Notes:} ($^*$) Sources whose bolometric luminosity was estimated as a lower limit, because we have no photometry over the peak, see Fig.~\ref{fig:sedI}.
   
 \end{quotation}
\end{table}

\subsection{The case of SVS~13A} \label{vvser}
The most luminous source of our sample, source \#191, is the well known variable Class~I object SVS~13A (or V512~Per), driving the chain of the Herbig Haro objects HH7-11. 
The source is part of a multi-object system composed of SVS~13A, B, and C \citep{loo00}. 
SVS~13A itself is a close radio binary separated by 0.3$\arcsec$ \citep[VLA~4A and 4B][]{ang00}, while another radio object has been identified  at about 5$\arcsec$ \citep[SVS13~A2][]{tob16}. 
The NIR source, responsible for the HH7-11 outflow has been associated with VLA~4B \citep{hod14,lef17}.

SVS~13A had a burst of about 1.5~mag in the K-band during 1990 (up to a K-mag of about 8~mag), followed by a slow decline in which it never reached the pre-burst value of about 9.5~mag \citep{hod14} .
In this paper, we use photometric values that were taken in 2015 (T. Giannini, private communication) where the K-band magnitude was 8.7~mag, which is about 0.5~mag weaker than the 2MASS value taken in 2000. 
The quick rise and slow decline of brightness resemble the characteristics of FUor-like objects; however, the small amplitude of the burst and the spectrum rich in emission lines exhibited by this source is more typical of EXor-type outbursts, as noted by  \citet{hod14}. 
High resolution spectral images of this source \citep{hod14} have identified an atomic micro-jet surrounded by a series of expanding bubbles, whose dynamical time suggests that the most recent of them can be associated with the photometric outburst from 1990.  

Our analysis suggests that SVS~13A is an intermediate mass star of $\sim 3$~$\msun$, accreting at a rate of a few $10^{-6}$~$\msun {\rm yr}^{-1}$. 
This mass accretion rate is in line with the upper values of $\macc$ derived on Herbig stars of a similar mass \citep[e.g.][]{wic20}. 
We exclude that this result could be affected by the binarity of the source. 
First of all, the companion to VLA~4B is not detected at NIR wavelengths, as testified by the high spatial resolution observations of \citet{hod14}. 
And also, observations in the I band performed with AstraLux at Calar Alto did not detect the companion, indicating that the difference in brightness with respect to the primary should be at least 3 mag (F. Comeron, private communication). 
Secondly, although at 3.6 cm, VLA~4A and B have comparable brightness, VLA~4B has a much steeper SED at shorter wavelengths and it is about five times brighter than VLA~A at 1.4 mm \citep{ang04,lef17}. 
This suggests that the contribution from VLA 4A to the source bolometric luminosity should be negligible. 

From spatially resolved observations of the [FeII] micro-jet of this source, \citet{dav11} estimated  a mass ejection rate of $\dot{M}_{\rm loss} = 1.4 \times 10^{-6}$~$\msun {\rm yr}^{-1}$, which would imply a $\dot{M}_{\rm acc} / \dot{M}_{\rm loss} \sim $ 1, that is to say about a factor of $5 - 10$ smaller than expected. 
A factor of a few larger mass accretion rate would be derived in our analysis only by assuming extreme values of the veiling in the K-band, in other words larger than 20. 
However, the detection of weak photospheric features in the K-band spectrum of SVS13~A clearly indicates that the IR~excess of this source is not extreme, and very high veiling values are ruled out. 
More likely, the $\dot{M}_{\rm loss}$ derived from the micro-jet gives a value of the mass ejected averaged in time; therefore, it takes into account periods of higher accretion activity of the source.

\section{Discussion} 
\label{sect:disc}

\subsection{Accretion properties of Class~II sources } \label{sect:disc2}

Previous studies on CTTs in different star-forming clouds have shown that a correlation exists between accretion properties ($\lacc$ and $\macc$) and properties of the central star ($\lstar$ and $\mstar$), see for example \citet{hil92}, \citet{moh05}, \citet{nat06}, \citet{ven14}, \citet{har16}, \citet{alc17}, \citet{man17b}, \citet{rug18}, \citet{aru19}, and \citet{bia19}. 
These studies have been mainly concentrated on relatively old clouds (i.e. $2 -10$~Myr). 
Here, we want here to investigate these correlations for our sample of Class~II stars in NGC~1333, and how the results in this cluster are related to those of older regions.

\subsubsection{Accretion luminosity}
\begin{figure}[t]
\includegraphics[width=\columnwidth]{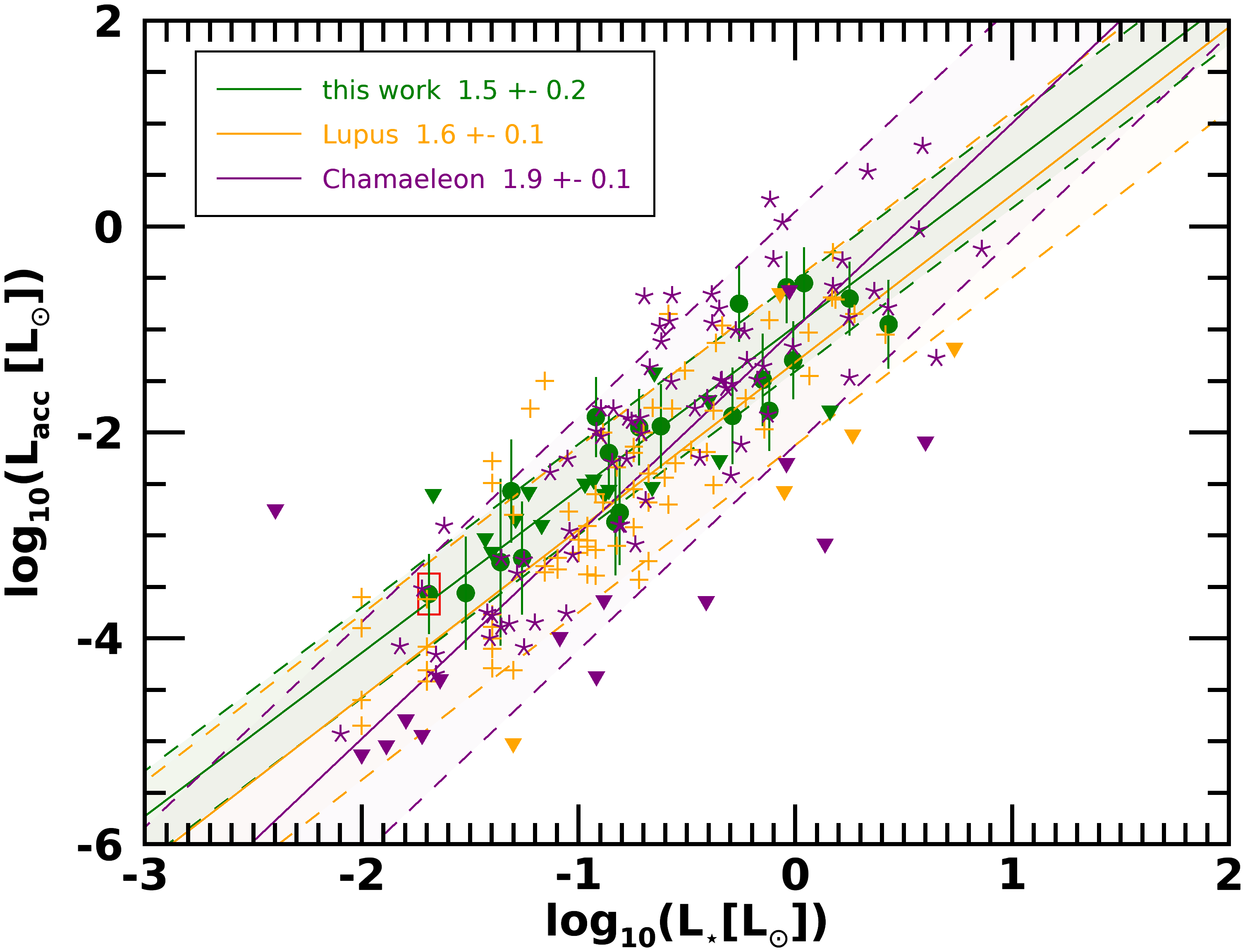}
\caption{\label{lacclstar}Accretion luminosity as a function of stellar luminosity for the NGC~1333 Class~II sample (green filled circles). 
Orange and purple stars represent Lupus and Cha~I sources, respectively. 
Upper limits are displayed as triangles. 
Solid lines show the best fit for each star-forming region as specified in the legend. 
The standard deviation from the best fit is shown with dashed lines. The subluminous target is highlighted with a red box.}
\end{figure}
\begin{figure}[t]
    \centering
    \includegraphics[width=\columnwidth]{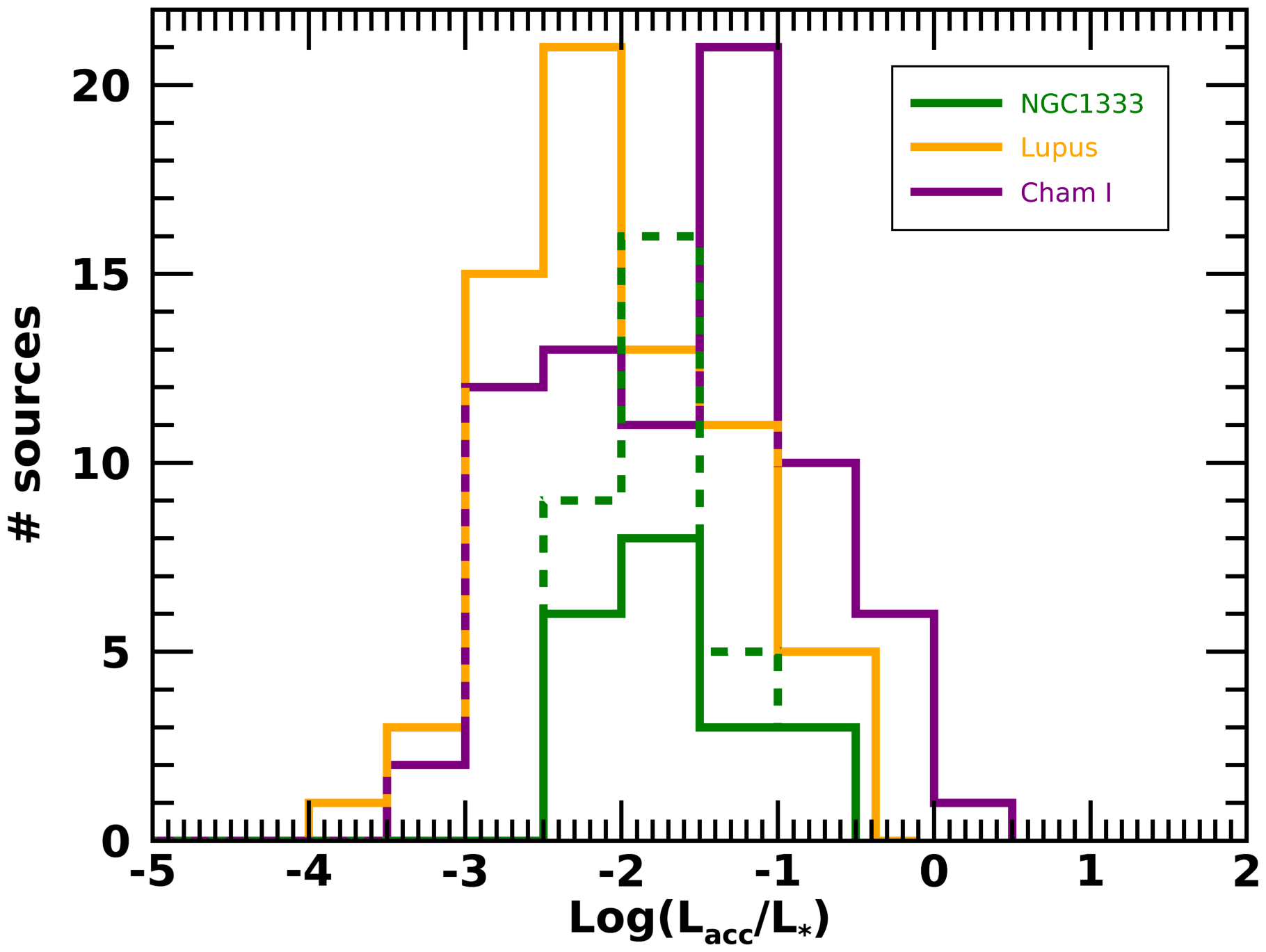}
    \caption{Histograms of the luminosity ratio $\lacc/\lstar$ of NGC~1333, Lupus, and Cha~I samples are shown in green, orange, and purple, respectively. The green-dashed line corresponds to the histogram of NGC~1333, including the upper limits.}
    \label{fig:histoLacc}
\end{figure}

Fig.~\ref{lacclstar} shows the distribution of $\lacc - \lstar$ for our sample. 
For comparison, we report on the same plot the values relative to the samples of the Lupus and Chamaeleon~I clouds \citep[][rescaled for Gaia distances by \citealt{alc19,man19}]{alc17,man17a}.
The figure highlights that, for the stars of NGC~1333, $\lacc$ correlates with $\lstar$ in a similar way as the sources of the other clouds, roughly following a linear slope in a logarithmic scale.
Using the IDL routine $\rm linmix\_err$, which takes into account errors on both axes and the presence of upper limits \citep{kel07}, a linear fit through the data points 
\begin{equation}
\log \lacc = \gamma \log \lstar + \beta
\end{equation}
yields $\gamma = 1.5 \pm 0.2$ and $\beta = -1.0 \pm 0.1$, with a standard deviation of $\sigma = 0.3 \pm 0.1$. 
The same values for the fit were obtained excluding subluminous object \#224.
In order to compare this relationship with those of the Lupus and Chamaeleon~I data, we applied the same fitting procedure to these data, considering upper limits, finding the following fit parameters: $\gamma_{\rm Lup} = 1.6 \pm 0.1$, $\beta_{\rm Lup} = -1.3 \pm 0.1$ with a standard deviation of $\sigma_{\rm Lup} = 0.6 \pm 0.1$, and $\gamma_{\rm ChaI} = 1.9 \pm 0.1$, $\beta_{\rm ChaI} = -1.0 \pm 0.1$, with a standard deviation of $\sigma_{\rm ChaI} = 0.8 \pm 0.1$. 
These values are consistent with the parameter fits given by \citet{alc17} and \citet{man17b}, before correcting the luminosities for the {\it Gaia} distance. 
The best fit relationships, with their uncertainties, are displayed in Fig.~\ref{lacclstar} for comparison.

We note that the fit through the data of NGC~1333 has a larger uncertainty, by up to a factor of two in the slope, with respect to the other regions, which is mainly due to the lower number of data points and higher fraction of upper limits. 
This figure also shows that the NGC~1333 and the Lupus samples have a similar slope of the $\lacc$ versus $\lstar$ correlation, while the fit through the Cha~I data is steeper, although compatible within the uncertainties. 

To compare the distribution of $\lacc$ among the three regions, we plotted the histograms of the luminosity ratio $\lacc / \lstar$ in Fig.~\ref{fig:histoLacc}. 
We note that, similarly to the Lupus and Cha~I samples, in NGC~1333 there are no sources where the accretion luminosity is larger than the stellar luminosity. 
This result is in agreement with previous studies, and is theoretically justified by the reasonable assumption that at this evolutionary stage the disk and star lifetimes coincide, as explained in \citet{cla06} and \citet{til08}. 
In addition, the NGC~1333 histogram is similar in shape to the Lupus one and peaks at a roughly higher $\lacc / \lstar$ value, that is to say  $\log{\lacc / \lstar} $ between -2.3 and -1.8. 

We note that the NGC~1333 sample has a lower spread in $\lacc / \lstar$ with respect to Lupus and Cha~I (Fig.~\ref{fig:histoLacc}).  
We think this is mostly due to the smaller number of sources in the NGC~1333 sample and to the lower sensitivity of our observations in detecting low $\lacc$ values. 
To better discuss this latter point, we also consider the histogram of the $\lacc/\lstar$ upper limits in Fig.~\ref{fig:histoLacc} (green dashed line). 
This shows that the derived upper limits are not stringent enough to confirm or disregard the possibility that the stars of the NGC~1333 cluster do indeed have a smaller spread.
Therefore, we cannot exclude the possibility that this small spread is related to the younger age of the cluster with respect to Lupus and Cha~I. 
A similar effect of an increasing $\lacc$ spread with age is indeed predicted by the protostellar evolution model of \citet{pad14}, in which the infall rates from turbulent clouds drive the accretion rates from the collapse to the pre-main sequence (PMS) phase. 

In conclusion, our analysis suggests that the relationship between $\lacc$ and $\lstar$ in NGC~1333 is similar to that of older regions, although with a smaller spread in the data. 
Our smaller statistical sample, however, does not allow us to infer whether this small spread is true and due to evolutionary effects or to a limited sensitivity.

\subsubsection{Mass accretion rate}
\begin{figure}[t]
 \includegraphics[width=\columnwidth]{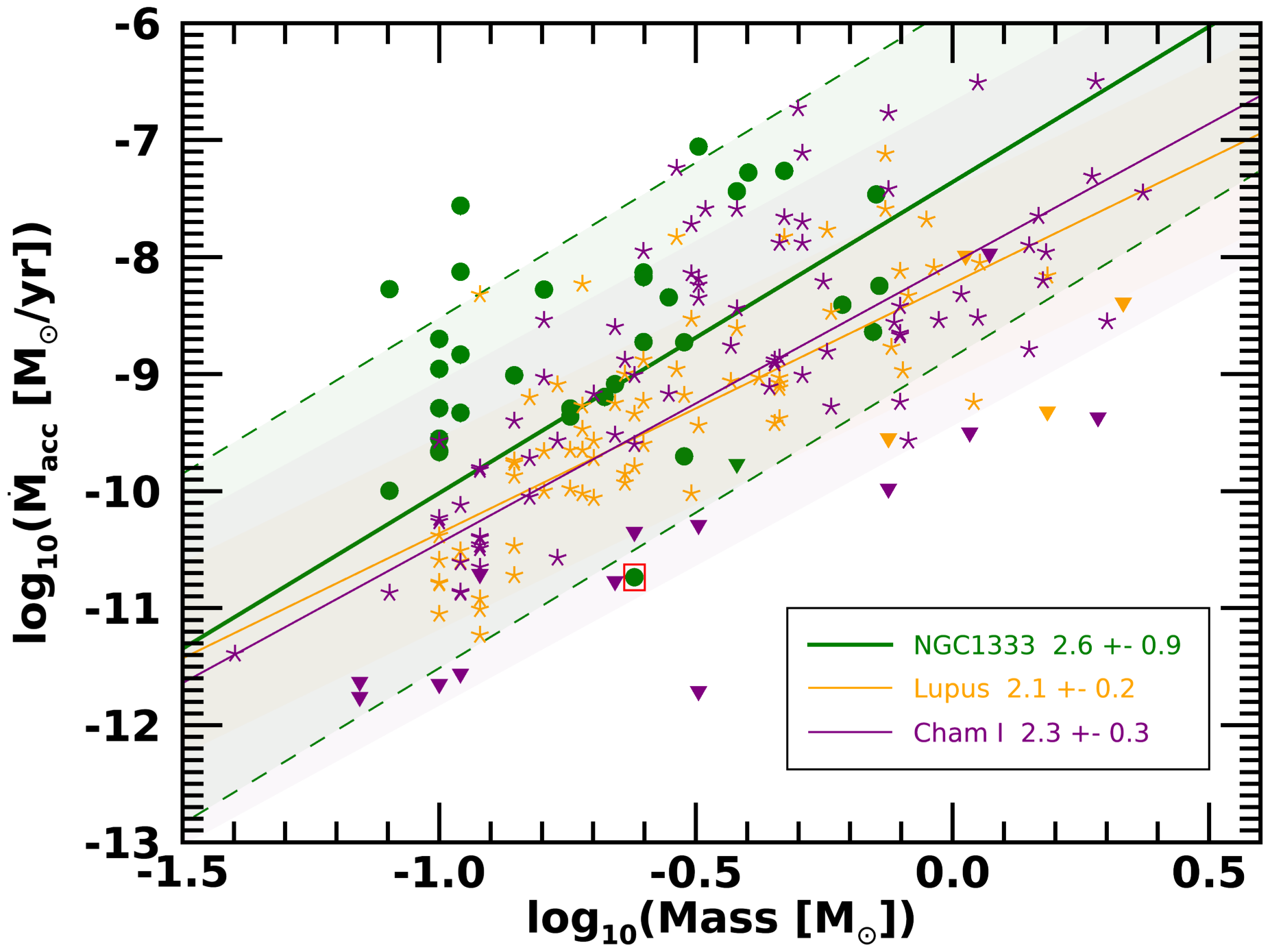}
  \caption{\label{maccs00} 
  Mass accretion rate as a function of stellar mass for the NGC~1333 Class~II sample (green). 
  Detections are displayed as filled circles, while upper limits are shown as triangles. 
  Orange and purple stars represent Lupus and Cha~I sources, respectively. 
  Solid lines show the best fit for each star-forming region as specified in the legend, and the standard deviation from the best fit is shown with dashed lines. The subluminous target is highlighted with a red box. 
  }
\end{figure}

The relation between the mass accretion rates versus the stellar mass for the NGC~1333 sources is shown in Fig.~\ref{maccs00} together with the objects in the Lupus and Cha~I samples, for comparison. 
Similarly to what we have done for the $\lacc - \lstar$ relationship, we performed a linear fit considering the upper limits on $\dot{M}_{\rm acc}$ in the following form:
\begin{equation}
  \log \macc = \eta \log \mstar + \zeta
\end{equation}

The linear fit of NGC~1333 results in $\eta = 2.6 \pm 0.9$ and $\zeta = -7.3 \pm 0.6$ with a standard deviation of $\sigma = 1.1 \pm 0.1$. 
By repeating the fit excluding the subluminous source \#224, we obtain the same values for the fit. The only difference is in the fact that the standard deviation decreases to a value of $\sigma = 0.9 \pm 0.1$. 
We used this value of the standard deviation in the following. The slope obtained for the sample in the NGC~1333 region is compatible within the error with the slope obtained here for both Lupus and Cha~I ($\eta_{\rm Lup} = 2.1 \pm 0.2$ and $\eta_{\rm ChaI} = 2.3 \pm 0.3$).
The fit obtained for the sample in NGC~1333 appears slightly shifted towards higher $\macc$ values with respect to the other clouds (whose intercepts are $\zeta_{\rm Lup} = -8.2 \pm 0.1$ and $\zeta_{\rm ChaI} = -8.0 \pm 0.1$). 
The standard deviation of the best fit for Lupus and Cha~I are $\sigma_{\rm Lup} = 0.6 \pm 0.1$ and $\sigma_{\rm ChaI} = 1.0 \pm 0.1$, respectively.
We can thus state here that in this plot the Cha~I linear fit is compatible, within the error, with the fit of the other two regions. 

\begin{figure}[t]
    \centering
    \includegraphics[width=\columnwidth]{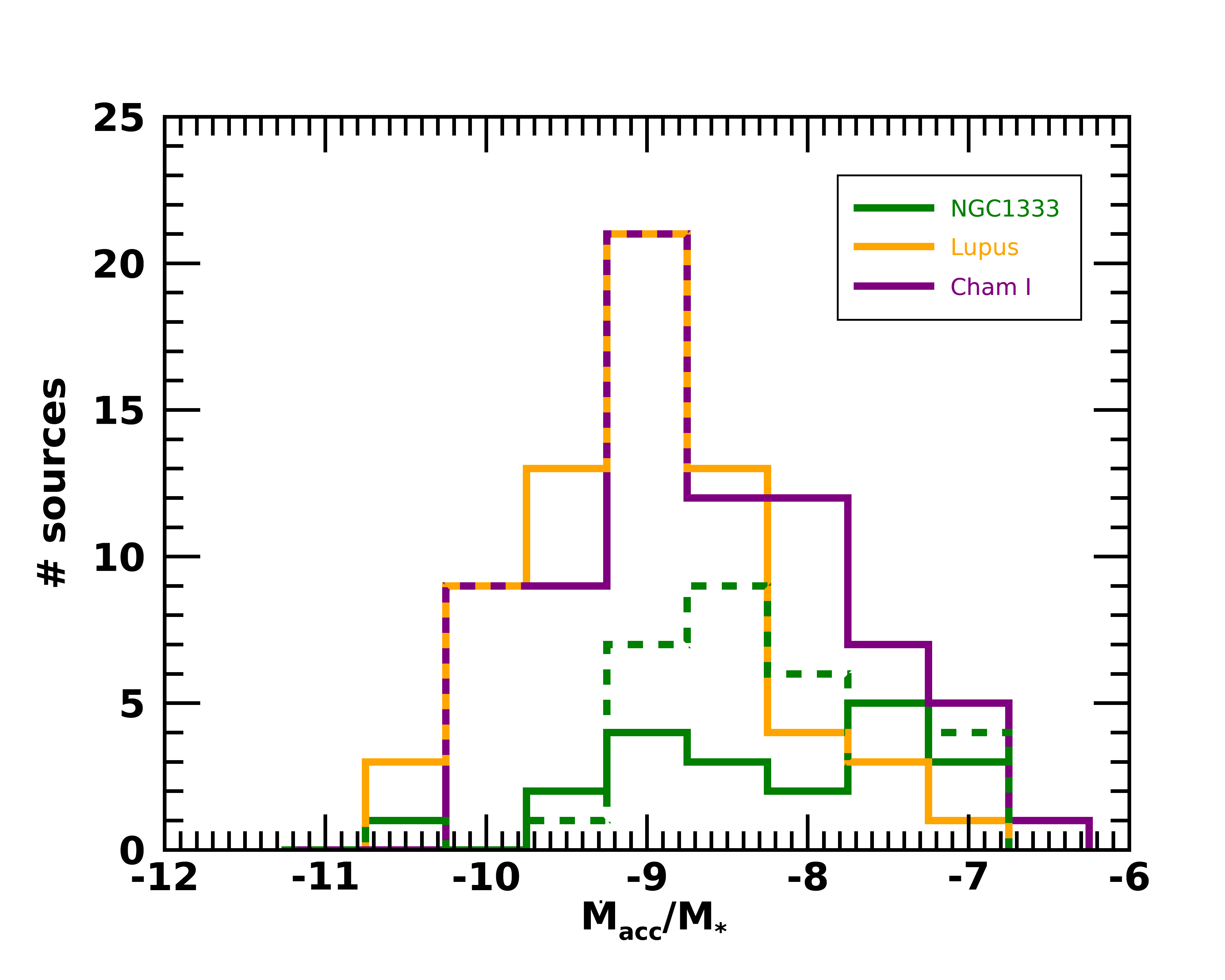}
    \caption{Histograms of the ratio $\macc/\mstar$ for NGC~1333, Lupus, and Cha~I samples are shown in green, orange, and purple, respectively. The green-dashed line corresponds to the histogram, which also includes upper limits.}
    \label{fig:histoMacc}
\end{figure}

We also note that the slope of the linear fit for NGC~1333 is driven by the large number of upper limits for  $\log \mstar <-0.6$. 
In fact most of the detections are located above the linear relationship. 
This is better shown in Fig.~\ref{fig:histoMacc}, which displays the distribution of the $\macc / \mstar$ values in the three clouds. 
In NGC~1333, these values show a spread in the range~$\sim 10^{-11} - 10^{-7}$, without a defined peak. 
This range is in agreement with the accretion values of the CTT sample investigated by \citet{nat06} in $\rho-$Ophiucus ($\sim 1$~Myr). 
However, more recent observations set a better and nearer estimate of the distance of this region, and the corresponding decrease in the luminosity implies an older age of this cloud, as recently confirmed by \citet{esp20}, which set the $\rho-$Ophiucus age at about $\sim 6$~Myr. Indeed, more recent estimates of $\rho-$Ophiucus accretion rates provided by \citet{rig11}, \citet{man15} are smaller by an order of magnitude than the NGC~1333 CTT sample. 

When comparing the NGC~1333 Class~II sample with Lupus and Cha~I regions,  
the $\macc / \mstar$ ratio ranges in the same interval, peaking at $\sim 10^{-9}$.
Although the lack of low $\macc / \mstar$ values in NGC~1333 can be caused by the limited sensitivity of our IR observations with respect to the optical surveys of Lupus and Cha~I, this histogram suggests that in NGC~1333 there is a larger fraction of sources with high $\macc / \mstar$ with respect to the older clouds. 

\citet{alc17} and \citet{man17b} suggest that the log $\macc \; {\rm  vs. }\log \mstar$ relationship in Lupus and Cha~I can be better reproduced by a double-slope power law, as the higher mass ($> 0.2$~$\msun$) stars show a flatter distribution of $\macc$ as a function of $\mstar$ with respect to the sources at a lower mass.
Moreover, theoretical models by \citet{vor09} predicted a bimodal distribution to be due to the action of large gravitational instabilities in the more massive disks, limiting the disk masses and the corresponding mass accretion rates.  

Here, we can investigate if a similar bimodal distribution is present in our NGC~1333 sample by addressing the role of possible evolutionary effects at the origin of such a distribution.  
Unfortunately, due to the large number of upper limits for $\mstar < 0.2$~$\msun$, the statistics that we have is not sufficient to draw firm conclusions. 
We note, however, that all the $\macc$ values measured in sources with masses of $\mstar < 0.2$~$\msun$ (seven values) are above the fitted single linear relationship, and are higher than the corresponding $\macc$ of the Cha~I and Lupus in the same mass range.

This result therefore suggests that a change in the slope at the lower masses is not present in the sample of Class~II of NGC~1333. 
A similar result is found by \citet{man15} for the very low mass stars in $\rho-$Ophiucus, which is a star-forming region younger than Lupus and Cha~I. 
A possible interpretation of this absence of the bimodal distribution can be that disks evolve with different timescales around stars with different masses, as already noticed by \citet{man15}. 
In particular, the mass accretion rate of very low mass stars ($\mstar < 0.2$~$\msun$) evolve faster than solar-type stars. 
This is in agreement with the models by \citep{har06} and is opposite to what was predicted by \citet{ale06}. 
In conclusion, our analysis of the mass accretion rate as a function of the stellar mass gives indications of larger $\macc/\mstar$ values, on average, for the sample of YSOs in NGC~1333 with respect to the older regions Lupus and Cha~I, despite the poor statistics and the upper limits on the low mass regime that prevent us from drawing firm conclusions.


\subsection{Accretion properties of the Class~I sample } \label{sect:disc1}

According to the viscous disk accretion model, the mass accretion rate is high during the first $3 \times 10^5$~yr, after which it declines with time \citep{har98,ale09,gor09}. 
Therefore we would expect that objects in the Class~I stage, which are assumed to be younger than Class~II sources,  would show higher accretion rates with respect to the latter. 
In the following we discuss whether our results are in agreement with this prediction or not through the comparison of the accretion properties of Class~I and II objects of the same cloud. 

\begin{figure}
\centering
\includegraphics[width=\columnwidth]{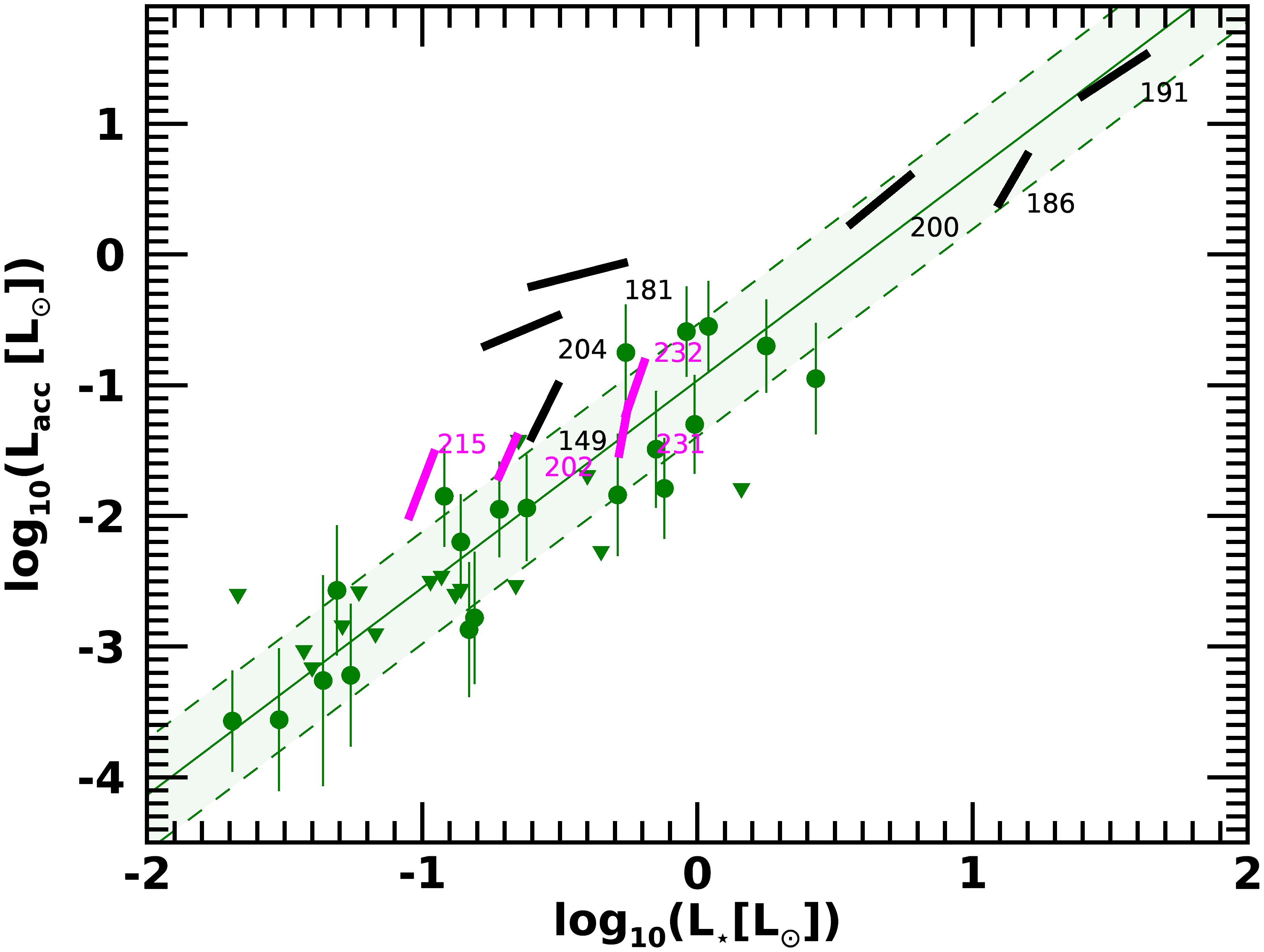}
\caption{\label{lacclstar12}Accretion luminosity as a function of stellar luminosity for NGC~1333. 
The Class~II sample is shown as in Fig~\ref{lacclstar}; Class~I sources with veiling between 1 and 3 are plotted in magenta, while those with veiling between 3 and 8 are in black.}
\end{figure}

\begin{figure}
 \includegraphics[width=\columnwidth]{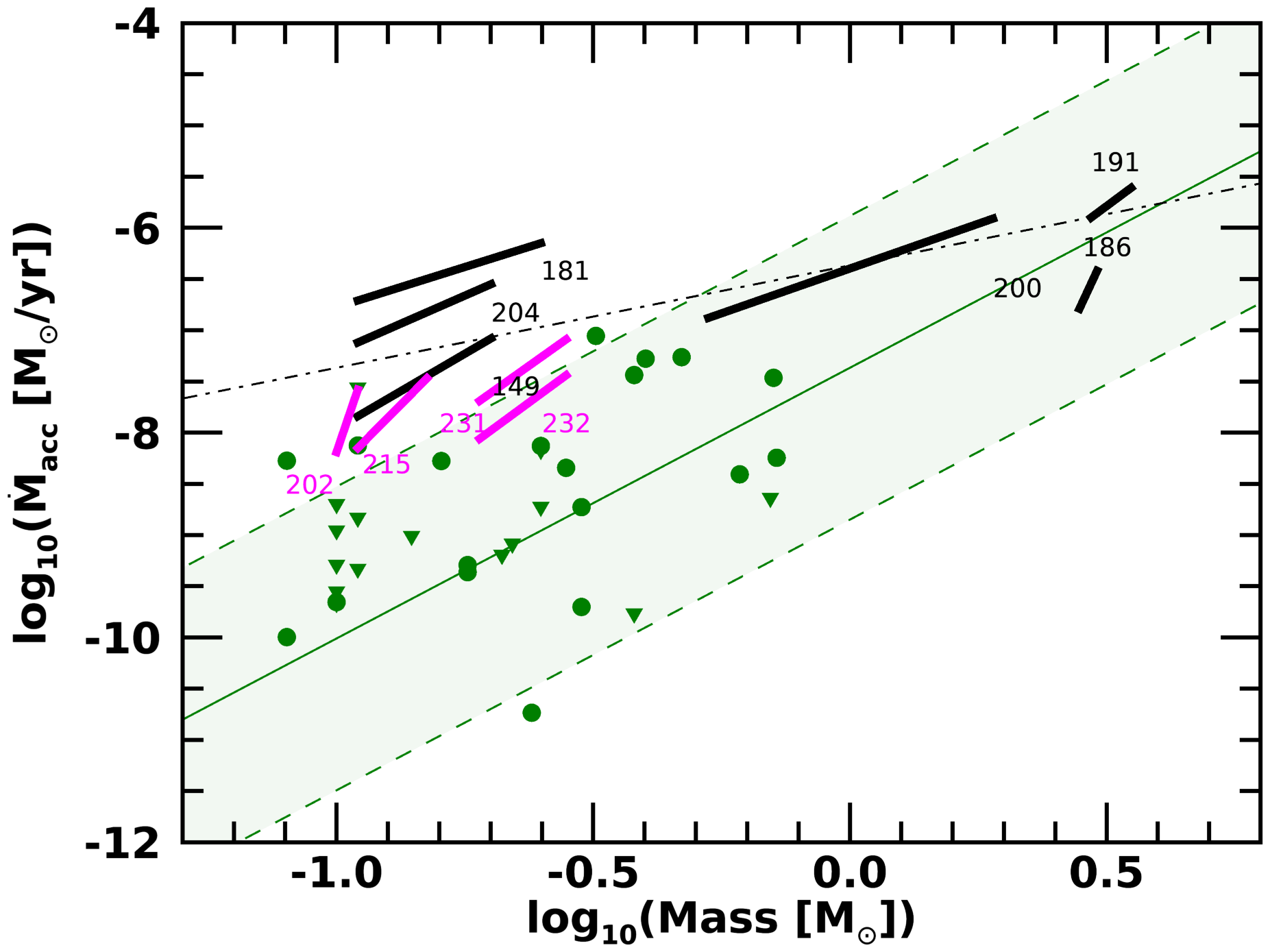}
  \caption{\label{maccs0012}Mass accretion rate versus stellar mass of the NGC~1333 cluster. 
  Class~II sample is shown as in Fig~\ref{maccs00}; Class~I sources with veiling between 1 and 3 are plotted in magenta, and the ones with veiling between 3 and 8 are in black. 
  The black dashed-dotted line represents a linear relation with a unitary slope.
  }
\end{figure}

Fig.~\ref{lacclstar12}~and~\ref{maccs0012} show the distributions of $\lacc - \lstar$ and $\macc - \mstar$, respectively, for both the Class~I and the Class~II samples of NGC~1333.
As discussed in Sect.~\ref{sect:ana1}, the inferred spread of stellar and accretion parameters for Class~I objects is large, and it is caused by the assumptions made on the possible range of veiling and age for these sources. 
Consequently, the Class~I results in these plots are not displayed as individual data points with errors, but rather as areas covering the possible range of values. 
The object number is also indicated to facilitate the identification of the different sources. 
The colour of each source is related to the veiling.
Sources with $1<r<3$ are displayed in magenta, while highly veiled ($r>3$) sources are in black.
Less veiled sources show a lower accretion luminosity than the Class~I YSOs with $r > 3$. 
In also considering the Class~II sources, which have a smaller veiling in general, our results suggest that the more embedded and veiled the objects, the higher the accretion luminosity is, as already noted in previous works \citep[e.g.,][]{cal04, fis11}.

Fig.~\ref{lacclstar12} shows a diversity of behaviour among the various sources. 
The stars with a mass $> 0.6$~$\msun$ and the  sources \#231 and \#232 are consistent with the relationship found for the Class~II objects, within the associated uncertainty.
On the contrary, the other sources have $\lacc$ appreciably higher than Class~II sources of the same stellar luminosity, up to a factor of 2. 
\\ \\
The spread in age that we assume for the Class~I analysis is at the origin of the even higher uncertainty in the range of possible $\macc$ and $\mstar$ values which is displayed in Fig.~\ref{maccs0012}. 
A wider spread in the stellar mass is present in the Class~I sample, with respect the CTTs. 

Given the small number of sources and and the large uncertainty on the parameters, no linear fit was performed on this sample. 
Nevertheless, we can infer that low-mass sources have, on average, mass accretion rates higher than Class~II YSOs of the same mass, up to more than two orders of magnitudes. 
In fact, the distribution of values  is compatible with a relationship $\macc \propto \mstar$, which is indicated in the plot with a dashed line. \citet{har06} suggested that this form of relationship is valid in the limit of pure irradiation with pure viscous heating, for stars with a mass up to 2~$\msun$. 

The degeneracy of the parameters for source \#200 does not allow us to infer if this is a low mass source with a mass accretion rate of the order of 10$^{-6}$~$\msun$yr$^{-1}$ or a 2~$\msun$ source with a factor almost 100 lower value of $\macc$. 
On the contrary, the mass accretion rates of the two more massive  stars \#186 and \#191 are quite constrained between 10$^{-7}$ and a few times 10$^{-6}$~$\msun$yr$^{-1}$, respectively, that is to say in agreement with the relationship found for the Class~II sources (assuming that this relation is also valid at higher masses). 
The mass accretion rates of these two sources are consistent with the upper values found by \citet{wic20} in Herbig stars of similar masses, as already noted for source \#191 in Sect. 5.4.  
\\ \\

\begin{figure}
    \centering
    \includegraphics[width=\columnwidth]{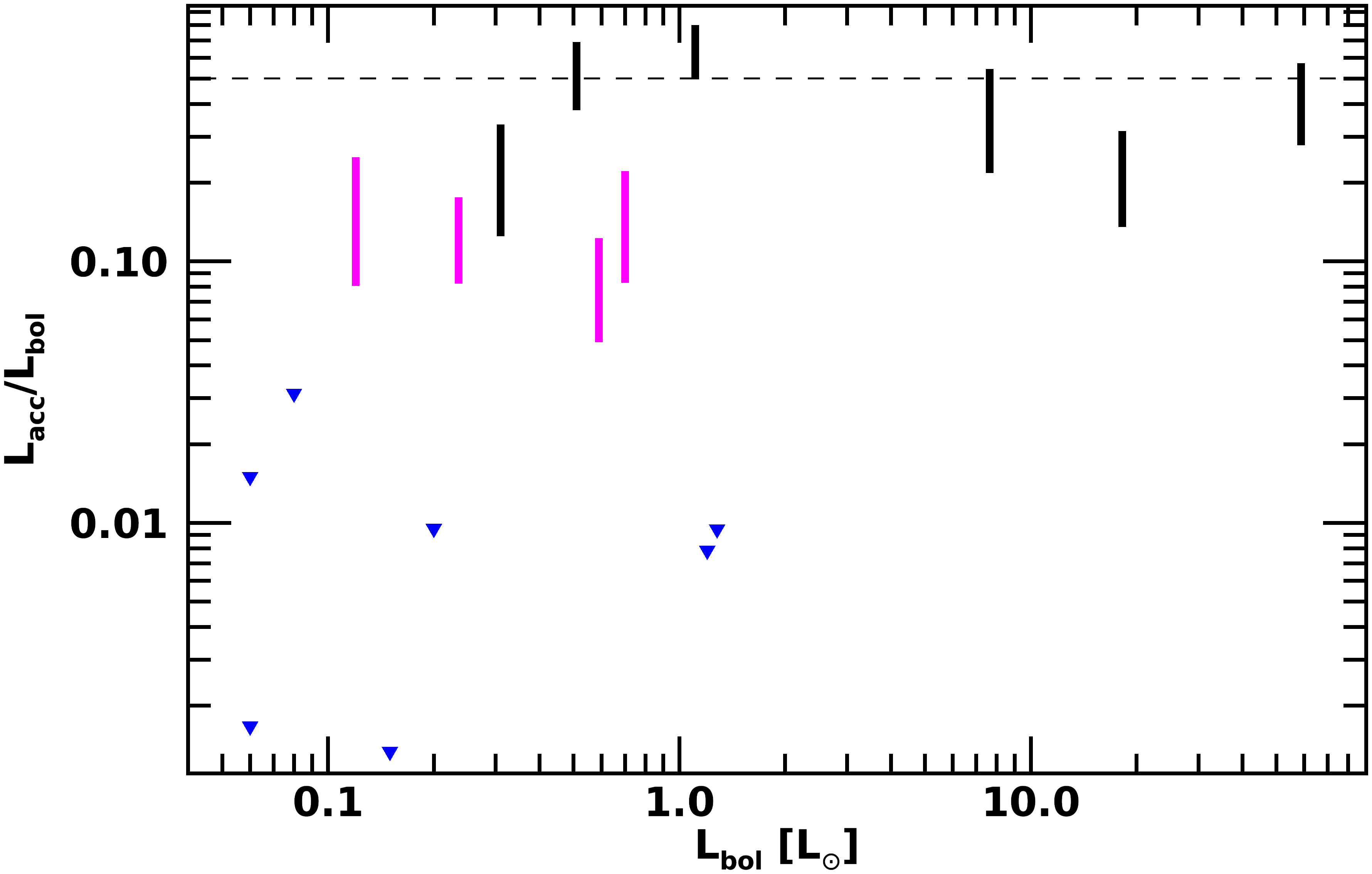}
    \caption{$\lacc / \lbol$ ratio is plotted as a function of the bolometric luminosity of our Class~I sample, low-veiled sources are plotted in magenta and high-veiled ones are in black.
    Blue triangles are the upper limits for the Class~I/Flat YSOs which show no HI detection.
    The horizontal dashed line represents the accretion luminosity fraction above which a source is in the main accretion phase.}
    \label{fig:LaccLbolMain}
\end{figure}

We therefore conclude that the accretion luminosities and mass accretion rates derived for the Class~I YSOs are, on average, higher than those of Class~II stars of the same mass. 
However, this finding does not necessarily imply that the sources we have investigated are in their main accretion phase, that is the phase in which a protostar is accreting most of its mass and its luminosity is dominated by the accretion luminosity. 
Previous studies on the accretion properties of Class~I sources have shown that indeed only a small fraction of them have $\lacc/\lbol>0.5$ \citep[e.g.,][]{whi04,nis05,ant08}. 
Fig.~\ref{fig:LaccLbolMain} shows $\lacc/\lbol$ as a function of the bolometric luminosity for our sample. 
We can see that, although the large uncertainty on the $\lacc$ value, only up to four out of the ten Class~I sources of our sample (i.e. $\sim 22\%$) might have their bolometric luminosity dominated by accretion. 

This evidence suggests that, assuming a steady-state condition, the objects considered should have already accreted most of their mass. 
This can be tested from our estimated values for the mass accretion rates in these sources. 
Assuming a lifetime of $\tau_I = 0.54$~Myr \citep{eno09, dun14} for the Class~I phase and that the mass accretion rate remains constant during this time, we can easily find that the sources in our sample should have accreted only 4$\% - 35\%$ of their mass during this phase. 
Consequently, either they have acquired most of their mass already during the Class~0 phase, or the accretion process 
proceeded in a non-steady framework.

This result is in agreement with the so-called luminosity problem.
In an attempt to solve the luminosity problem, different models have been developed during the last years to try to overcome the limitations of the standard Shu collapse model. 
The main feature of such models is the assumption that the accretion is not a steady process but it is characterised by variability of the mass accretion rate. 
In particular, \citet{dun12} propose a collapse model in which the accretion declines with time but has also short-term variability and episodic bursts caused by disk gravitational instability and fragmentation. 
In their simulations, \citet{dun12} foresee that the duration of the embedded phase depends on the final stellar mass produced, while the mass accretion rate, averaged over the full lifetime, is independent of the stellar mass. 
The range of mass accretion rates that we estimate for our sources, that is from  $10^{-8}$ to $10^{-6}$~$\msun \; {\rm yr}^{-1}$, would be in agreement with those predicted by these models in quiescent periods; while during bursts, $\macc$ increases up to three orders of magnitude.
Our derived values are also not in contradiction with the possibility that $\macc$ does not correlate with $\mstar$, given the associated uncertainty on the estimated values.
The fact that none of our sources are observed in bursts is expected, as their occurrence is estimated to be once every 10$^4$~yrs during the embedded phase \citep{sch13er}. 

A different model proposed to solve the luminosity problem is the one by \citet{pad14}. 
This model assumes that the accretion rate, during both the embedded and the Class~II phases, is controlled mainly by the mass infall from the large-scale envelope and the turbulent cloud, and not by the disk, whose role is only to mediate accretion. 
Following the evolution of the infall rates during the formation of a protostar, \citet{pad14} find that they are comparable with accretion rates inferred from protostellar luminosities or measured in pre-main sequence stars. 
Interestingly, in these simulations, large variations in the infall rates during the protostellar evolution are found, due to the stochastic nature of turbulent flow and not to accretion bursts. 
Consequently, large spreads of accretion rates are predicted at any age and for any mass, with infall rate values again being consistent with the mass accretion rate measured in our sample. 
\\ \\

\subsection{Class I with no line detection} \label{weak_acc}
We have so far discussed the results found on the ten Class~I sources of our sample for which we were able to give an estimate of the accretion properties. 
Seven other sources do not show any detected line emission in their spectra, which suggests that they should be low accretors, as shown by the upper limits on the accretion luminosity derived in Sect.~\ref{sect:noHIlines}. 
In this section, we discuss the characteristics of these `candidate weak accretors'.

Tab.~\ref{jhktab1} shows that four of them are classified as Flat, with $-0.25 \leq \alpha_{\rm IR} \leq 0.06$, and three are classified as Class~I, with $0.60 \leq \alpha_{\rm IR} \leq 1.56 $. Thus, the IR spectral index of this sub-sample does not have a peculiar trend with respect the Class~I strong accretors. 

Most of these sources have a SED (see Fig.~\ref{fig:sedI}) similar to the other accreting Class~I YSOs of the sample. However, the SEDs of sources \#161 and \#237 decrease at wavelengths longer than 2~$\mu$m and increase again at $\lambda > 5$~$\mu$m. This might suggest that the NIR source is not the counterpart of the {\it Spitzer} source. 

Also, by inspection at the colour-colour diagram in Fig.~\ref{colcol}, we note that all these sources lie in the same region of the diagram in which the Class~II sample is located. 
In addiction, because of the lack of IR excess in the K band (see Fig.~\ref{fig:spec_classI_non_acc1}), in four of these sources, we are able to determine the spectral type and the stellar parameters in the same way we have done it for the Class~II sample (see Sect.~\ref{sect:ana2}). Indeed, these sources lie in the same region of the Class~II sample.

Moreover, because Class~I protostars are defined by the condition that $\lacc$ dominates their luminosity, in Fig.~\ref{fig:LaccLbolMain}, we plotted the corresponding $\lacc/\lbol$ upper limits for a comparison with the other Class~I objects. 
The derived upper limits suggest that these sources have $\lacc/\lbol$ more than two orders of magnitude lower than those of the Class~I targets which show $\brg$ lines. 
This suggests that the observational classification based on SEDs' spectral index value (Class~0, I, II, III) does not match with the corresponding evolutionary `Stage' (e.g., Stage 0, I, II, III), which instead refers to a source's true physical nature \citep{jay01,eno09, dun14}. 
We conclude that this sample is composed of Class~I YSOs in the evolutionary stage, Stage~II. 
Spectra with typical features of more evolved objects than Class~I, such as photospheric absorption lines, were also found by \citet{car16} for a sample of sources classified by {\it Spitzer} as YSOs, meaning that the spectroscopic follow-up is probably crucial to determine the physical nature of the evolutionary stage of YSOs.

\subsubsection{Impact on the lifetime estimates of Class~I/Flat sources}
As we have seen in previous sections, the Class~I sample contains strong accretors which show strong HI emission lines, and more evolved YSOs whose accretion, if any, is weak, hence, not detected over the continuum.
It is however instructive to highlight the results found on the global sample of embedded objects. 

From the original sample of YSOs in NGC~1333, we observed 20 Class~I/Flat sources, that is $\sim$ 55\% of all the sources of these classes identified in NGC~1333 in \citet{you15}. 
Of these 20 sources, ten are strong accretors, seven are YSOs with an evolutionary stage compatible with Class~II, and three have been disregarded as YSOs on the basis of their spectra 
(see Appendix~\ref{appnoyso}). 
In conclusion, we derive that only 85\%,  17 out of 20 of our investigated Class~I sample, are accreting YSOs, and only 50\% are strong accretors. 
Assuming that the same fraction applies to the sources of the \citet{you15} sample that we have not observed, we might conclude that  the number of embedded accreting stars is smaller than estimated from the photometric information alone. 
This evidence can have implications for the estimated protostellar embedded phase lifetime. 
In fact, \citet{eva09} derive the lifetime of the Class~I and Flat phases from the number of Class~I/Flat relative to the Class~II sources, assuming that the Class~II lifetime is of the order of $2 \pm 1$~Myr. 
Our results show, however, that the contamination of non-accreting sources, likely more evolved objects, in the sample of embedded YSOs identified through {\it Spitzer} can be high, and consequently the lifetime of the Class~I/Flat phase might be overestimated by up to a factor of two. 

\section{Summary and conclusions} \label{sect:con}
In this paper we present the analysis of NIR VLT/KMOS spectra of 52 young stellar objects in the $<1$~Myr old NGC~1333 cluster in the Perseus star-forming region. 
Our goal is to investigate the evolution of accretion properties of YSOs from Class~I to Class~II.

The stellar ($A_V$, $\lstar$, $\mstar$, $\rstar$) and accretion ($\lacc$, $\macc$) parameters of the sample have been derived from the NIR spectra. 
For the Class~II sample, we derived $A_V$ and SpT through a comparison with a grid of non-accreting Class~III stellar templates. 
We then computed the accretion luminosity by using the known correlation between $\lacc$ and the luminosity of HI lines ($\pab$ and $\brg$). 
For the Class~I sample, where the presence of a large IR excess in the K band makes use of the same spectral typing method impossible, we used a procedure that allowed us to measure the stellar and accretion luminosity in a self-consistent way, based on a range of veiling values.     
Mass accretion rates $\macc$ were then measured once stellar masses and radii were estimated adopting suitable evolutionary tracks. This work represents the first unbiased study of the accretion properties of 17 Class~I YSOs belonging to the same cloud. 
In particular, we were able to measure the mass accretion rates of ten sources of our sample, where HI lines have been detected, while we set upper limits for the remaining seven sources.
We compared our results of the Class~II NGC~1333 sub-sample with the samples of CCTs in the older Lupus and Chamaeleon~I regions \citep{alc17,man17a}.

The main results of this analysis are the following:
\begin{itemize}
    \item The accretion luminosity of the Class~II sample correlates with the stellar luminosity, as found for the other regions.
    The linear fit in the log plane is $\log \lacc = (1.5 \pm 0.2) \log \lstar + (-1.0 \pm 0.1)$, which is in agreement within the errors with the relationship found on the Lupus sources and shallower with respect to the Cha~I relation. 
    The mass accretion rates of the Class~II sample correlate with the stellar masses, following the relationship: $\log \macc = (2.6 \pm 0.9) \log \mstar + (-7.3 \pm 0.7)$, in agreement within the error with both Lupus and Cha~I relationships. 
    However, at variance with what is found for the $\log \macc - \log \mstar$ distribution in Lupus and Cha I, we do not find evidence for a double power-law trend in NGC~1333. 
    If confirmed, this might suggest that the double-slope found in older regions is due to a faster evolution of the $\macc$ in the low mass regime.
    
    \item The accretion luminosity of the Class~I sample ($\sim 10^{-2} - 10^{2}$~$\lsun$) is higher by up to two orders of magnitude with respect to values found on the Class~II sample ($\sim 10^{-4} - 10^{0}$~$\lsun$). 
    Similarly, the mass accretion rate of the Class~I sample ($\sim 10^{-9} - 10^{-6}$~$\msun$yr$^{-1}$) is higher than the one found on the Class~II stars up to two orders of magnitudes ($\sim 10^{-11} - 10^{-7}$~$\msun$yr$^{-1}$). 
    We find that the higher accretion rates are associated with the sources with the larger IR veiling. 
    Given the large uncertainty on the derived parameters, due to the assumptions needed by our method, and the low number of Class~I targets, we are not able to define whether a correlation exists between the stellar parameters and $\lacc$ or $\macc$. 
    
    \item We computed the $\lacc/\lbol$ ratio for the Class~I sample, finding that for the majority of the sources (13 out of 17) this ratio is smaller than 0.5. 
    This indicates that for most of the sources the bolometric luminosity is not dominated by the accretion luminosity and thus the sources are not in their main accretion phase. 
    This result is in agreement with the well known luminosity problem, and suggests that those objects have already accreted most of their mass or the accretion occurs in a non-steady fashion.
    \item We observed 55\% of the Class~I/Flat YSOs listed in the \citet{you15} catalogue of the NGC~1333 cluster.
    The spectroscopic follow-up of them shows that three are not YSOs and seven shows no HI emission lines. 
    These results highlight the importance of more extended spectroscopic surveys of YSOs in different star-forming clouds in order to confirm their classification and infer a more solid estimate of their lifetimes and of the star formation efficiency.
\end{itemize}

In conclusion, our analysis suggests that early stage YSOs have accretion rates on average higher than Class II stars of the same cloud. However, such values are still not sufficiently high to be responsible for the assembly of most of the stellar mass during the source lifetime, giving support to theories predicting that mass accretion does not occur in a steady fashion. 
To provide more solid conclusions, observations of larger samples of YSOs in young star-forming regions with sensitive IR instrumentation, as will be made possible with James Webb Space Telescope (JWST) in the near future, are highly demanded.

\begin{acknowledgements}
This work  has  been supported  by the project PRIN-INAF-MAIN-STREAM 2017 “Protoplanetary disks seen through the eyes of new-generation instruments” and by the European Union's Horizon 2020 research and innovation programme under the Marie Sklodowska-Curie grant agreement No 823823 (DUSTBUSTERS). We also acknowledge the support by INAF/Frontiera through the ‘Progetti Premiali’ funding scheme of the Italian Ministry of Education, University, and Research and by the Deutsche Forschungs-Gemeinschaft (DFG, German Research Foundation) - Ref no. FOR 26341/1 TE 1024/1-1. 
E. Fiorellino was supported by an ESO studentship during the preparation of this work.

\end{acknowledgements}

\bibliographystyle{aa}
\bibliography{bibyso.bib} 

\appendix
\section{Technical details about observations}
\label{app:obs}

In this section, we report some technical details of each OB. 
In Table~\ref{obs}, the name of each OB (Col.~1) and the date and time when the OB was observed (Col.~2) are reported. 
The exposure time on target is the product of the DIT time in seconds (Col.~3), times the number of DIT (NDIT, Col.~4), times the number of exposures in the A position (i.e. 'on sky', see Sect.~\ref{sect:obs}, Col.~5). 
The airmass (Col.~6) and the seeing corrected for the airmass (Col.~7) are reported as an average value. 
Col.~8 shows the number of KMOS arms allocated to science targets for each OB.

\begin{figure}[t]
 \includegraphics[width=\columnwidth]{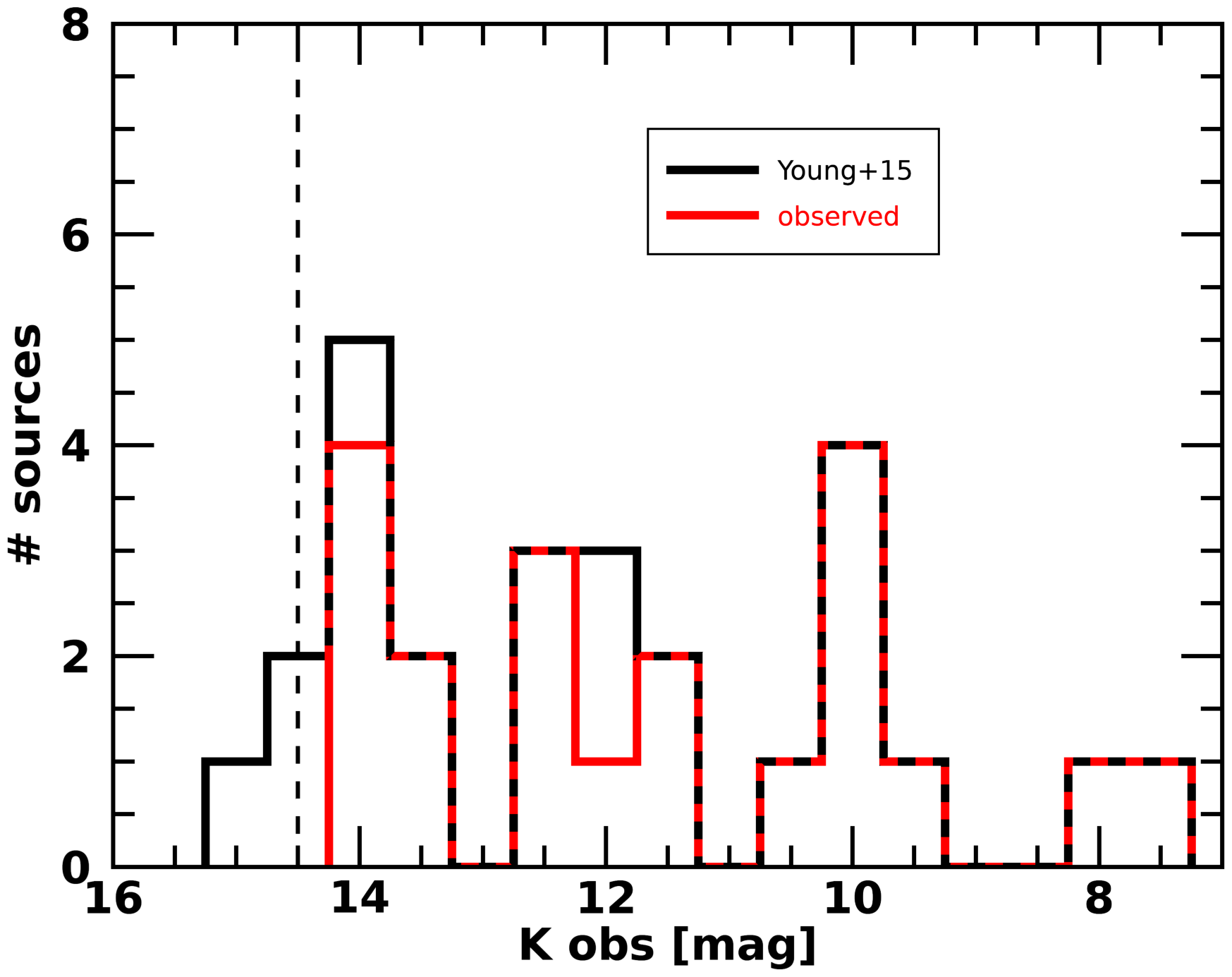}
  \caption{\label{histomkclass1}Histogram of the K-band magnitude of the {\it Spitzer} identified Class~I in the NGC~1333 cluster \citep[black,][]{you15} and the sample observed by KMOS (red). 
    The black-dashed vertical line represents the cut-off at 14.5~mag due to KMOS sensitivity.}
\end{figure}

\begin{figure}[t]
 \includegraphics[width=\columnwidth]{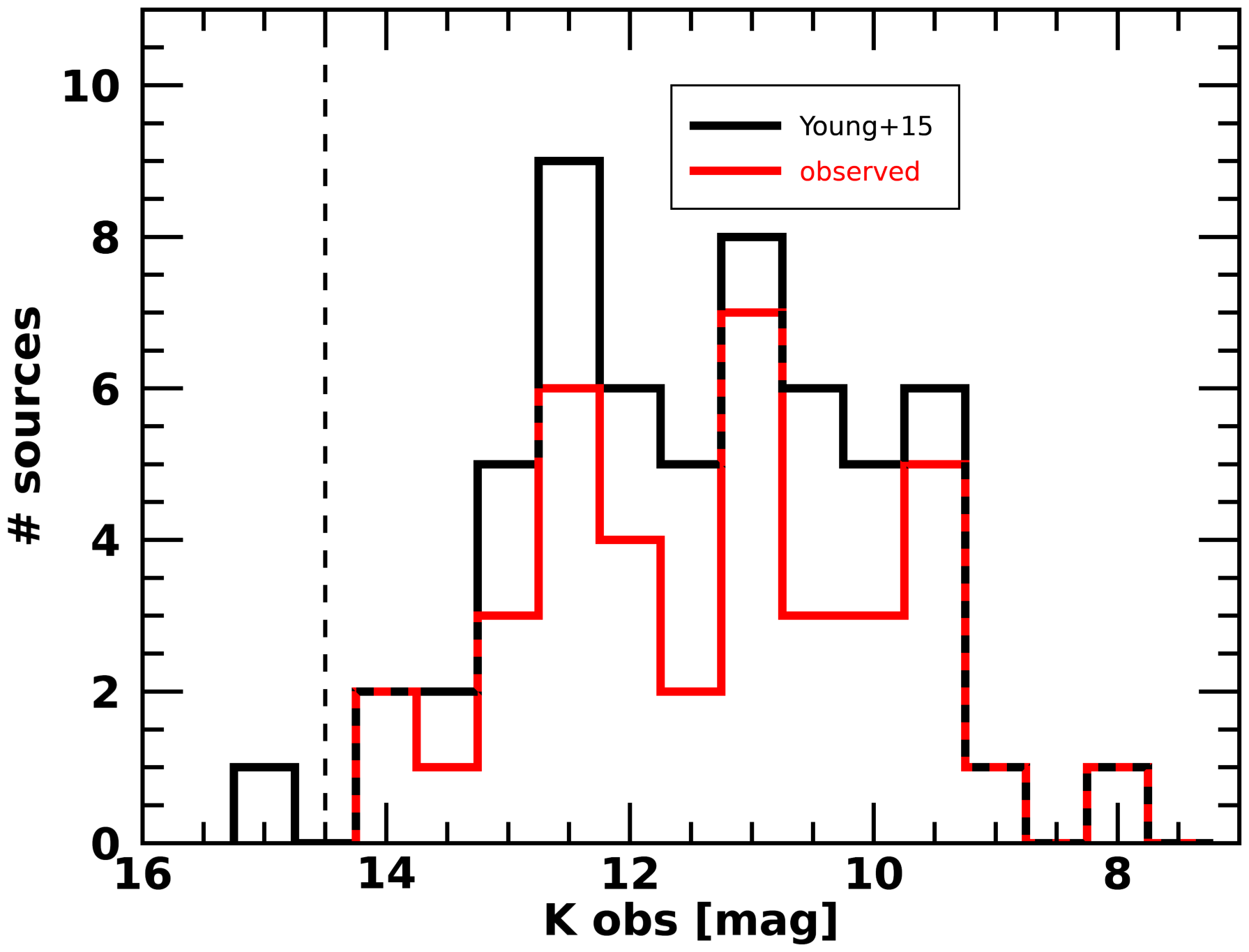}
  \caption{\label{histomkclass2}Histogram of the K-band magnitude of the {\it Spitzer} identified Class~II in the NGC~1333 cluster \citep[black,][]{you15} and the sample observed by KMOS (red). 
    The black-dashed vertical line represents the cut-off at 14.5~mag due to KMOS sensitivity.}
\end{figure}

In the OB named 1bright, the following sources were observed: \#184, \#188, \#192, \#197, \#200, \#201, \#203, \#204, \#205, \#208, \#209, \#218, \#219, \#222, \#226, \#228, \#231, \#232, \#233, \#235, \#237, and \#241.
In the OB named 1dim, the following sources were observed: \#180, \#181, \#186, \#188, \#190, \#196, \#200, \#202, \#203, \#204, \#205, \#215, \#218, \#224, \#227, \#231, \#232, and \#237.
In the OB named 2bright, the following sources were observed: \#152, \#153, \#156, \#159, \#160, \#161, \#162, \#164, \#166, \#173, \#174, \#177, \#178, \#180, \#186, \#191, \#194, and \#196.
In the OB named 2dim the following sources were observed: \#147, \#149, \#153, \#161, \#165, \#168, \#174, \#175, \#183, \#197, and \#198.
In the OB named 3bright, the following sources were observed: \#167, \#168, \#175, \#176, \#183, \#195, \#198, \#199, \#213, \#221, \#223, and \#234.
In the OB named 3dim, the following sources were observed: \#164, \#194, \#195, \#199, \#213, and \#221.
\\ \\
We also added information about completeness for each class, showing the histograms of Class~I (Fig.~\ref{histomkclass1}) and Class~II (Fig.~\ref{histomkclass2}) sub-samples.

\begin{table*}
\center
\caption{\label{obs} Observing log}
 \begin{tabular}{l|c|ccc|ccc}
 \hline
  OB Name  & DATE-OBS            & DIT TIME & NDIT & Nexp & airmass & seeing & arms\\
           &                     &    [s]   &      & &         &[$\arcsec$]& \\
  \hline
  \hline
  1brightJ & 2018-10-24T05:52:38 & 20       & 4    &   18     & 1.8  & 0.90  & 22 \\
  1brightH & 2018-10-24T04:46:46 & 20       & 4    &   18     & 1.9  & 0.78  & 22 \\
  1brightK & 2018-10-03T08:40:36 & 15       & 4    &   18     & 1.9  & 1.11  & 22 \\
  1dimJ    & 2018-10-12T07:45:10 & 140      & 2    &   12     & 1.9  & 1.07  & 18 \\
  1dimH    & 2018-10-23T06:14:43 & 145      & 2    &   12     & 1.8  & 0.89  & 18 \\
  1dimK    & 2018-10-09T06:36:18 & 145      & 2    &   12     & 1.8  & 0.99  & 18 \\
  2brightJ & 2018-10-23T05:16:11 & 20       & 4    &   18     & 1.8  & 0.89  & 18 \\
  2brightH & 2018-10-23T05:36:30 & 20       & 4    &   18     & 1.8  & 0.83  & 18 \\
  2brightK & 2018-10-23T05:55:55 & 15       & 4    &   18     & 1.8  & 0.91  & 18 \\
  2dimJ    & 2018-10-31T04:16:23 & 150      & 2    &   12     & 1.9  & 0.79  & 11 \\
  2dimH    & 2018-10-24T06:44:00 & 150      & 2    &   12     & 1.8  & 0.93  & 11 \\
  2dimK    & 2018-10-09T07:50:17 & 150      & 2    &   12     & 1.8  & 0.98  & 11 \\
  3brightJ & 2018-10-24T06:24:09 & 20       & 4    &   18     & 1.8  & 0.91  & 12 \\
  3brightH & 2018-10-24T05:12:31 & 20       & 4    &   18     & 1.8  & 0.93  & 12 \\
  3brightK & 2018-10-10T08:27:54 & 15       & 4    &   18     & 2.0  & 0.98  & 12 \\
  3dimJ    & 2018-10-31T05:47:49 & 150      & 2    &   12     & 1.8  & 0.88  & 6 \\
  3dimH    & 2018-10-25T04:57:18 & 150      & 2    &   12     & 1.9  & 0.93  & 6 \\
  3dimK    & 2018-10-25T06:02:41 & 150      & 2    &   12     & 1.8  & 0.91  & 6 \\
  \hline 
 \end{tabular}
 \begin{quotation}

\end{quotation}
\end{table*}

\section{Observed Spectra}
\label{app:spec}
We report here the complete JHK spectra and the individual spectra  of $\pab$ and $\brg$ lines for our sample.

\begin{figure*}
    \centering
    \includegraphics[width=\textwidth]{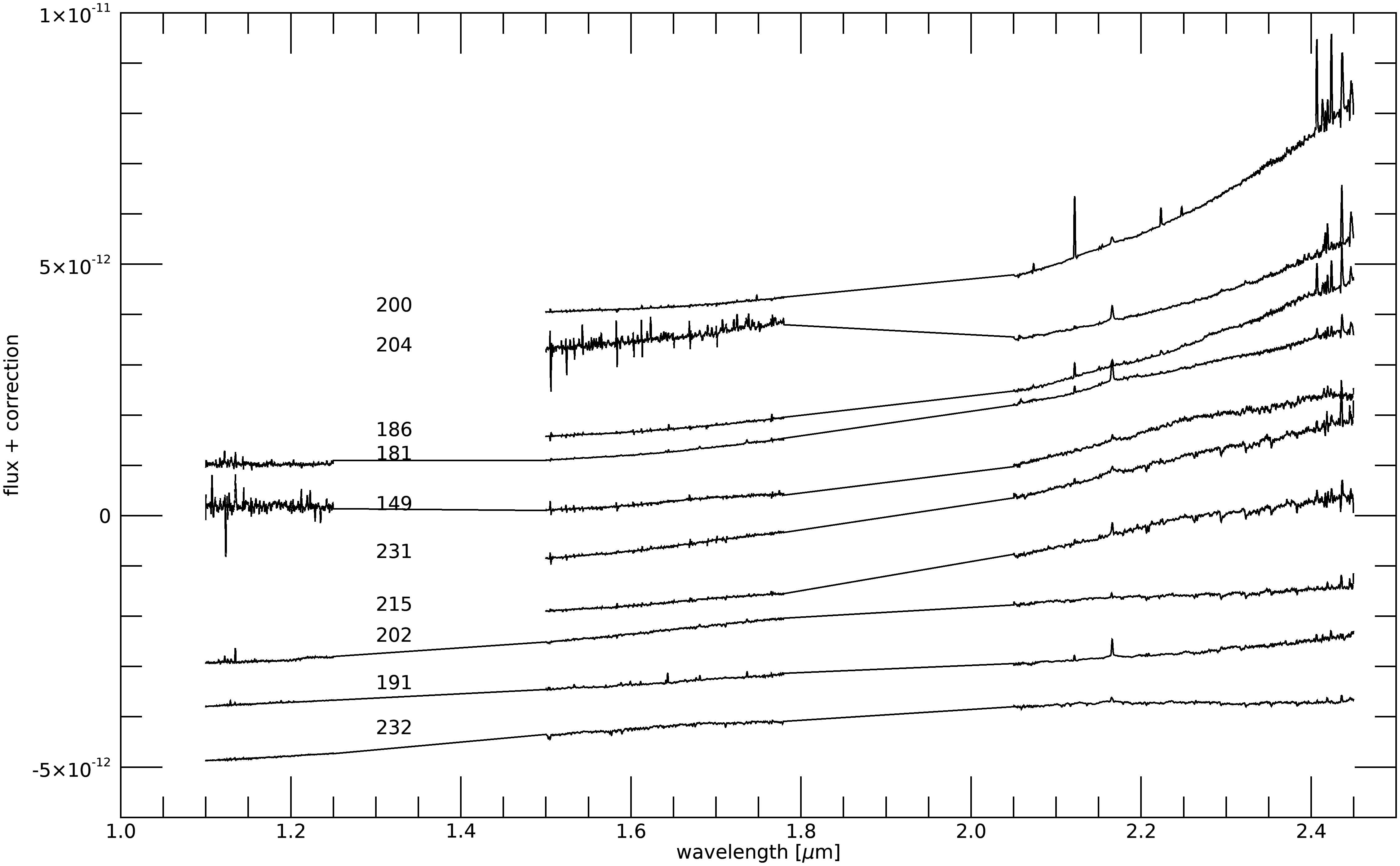}
    \caption{J, H, and K band spectra of Class~I YSOs which show at least one HI line. All the spectra are suitably shifted in flux for clarity.}
    \label{fig:spec_classI_acc1}
\end{figure*}

\begin{figure*}
    \centering
    \includegraphics[width=\textwidth]{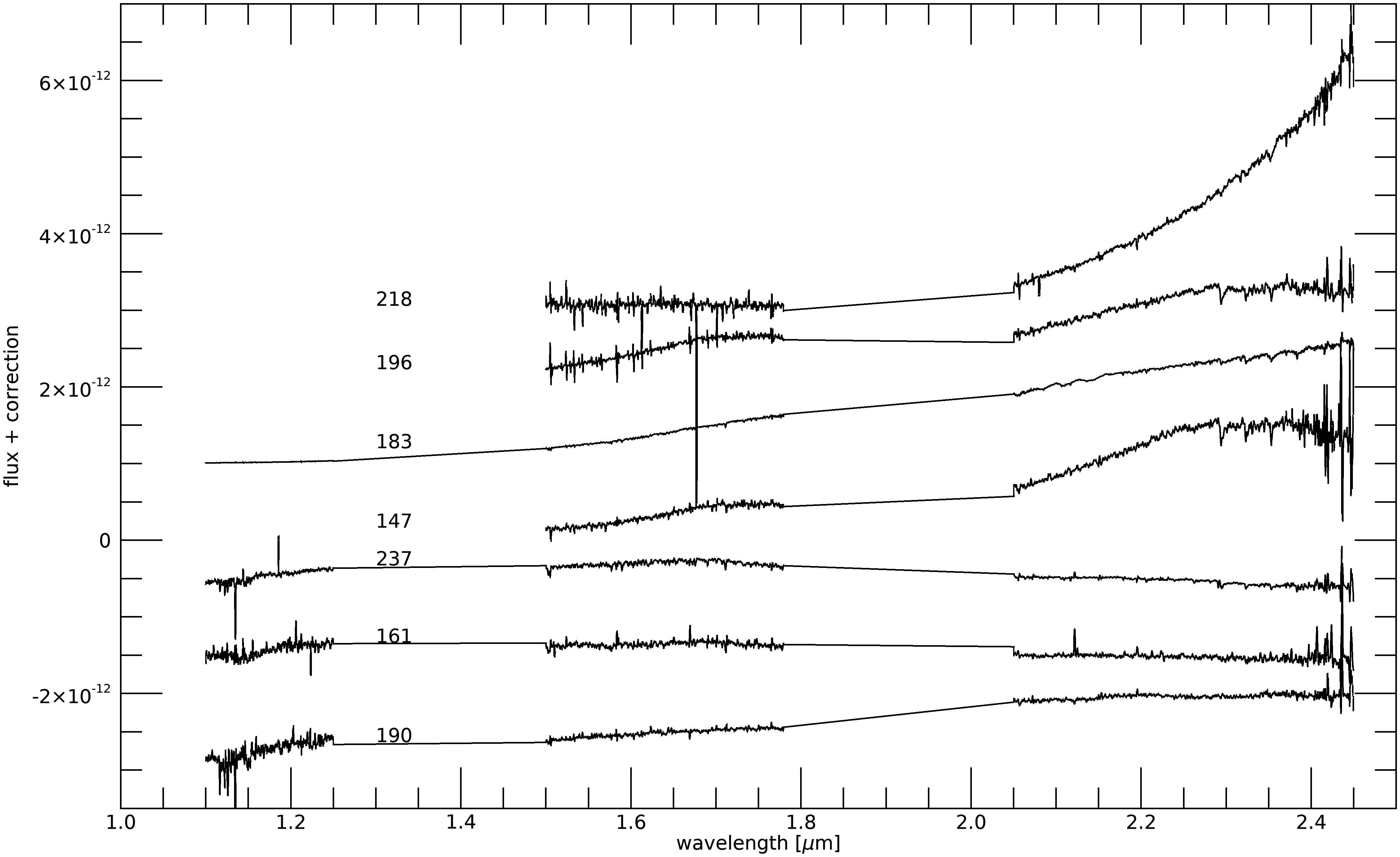}
    \caption{J, H, and K band spectra of Class~I YSOs which show no HI line. All the spectra are suitably shifted in flux for clarity.}
    \label{fig:spec_classI_non_acc1}
\end{figure*}

\begin{figure*}
    \centering
    \includegraphics[width=\textwidth]{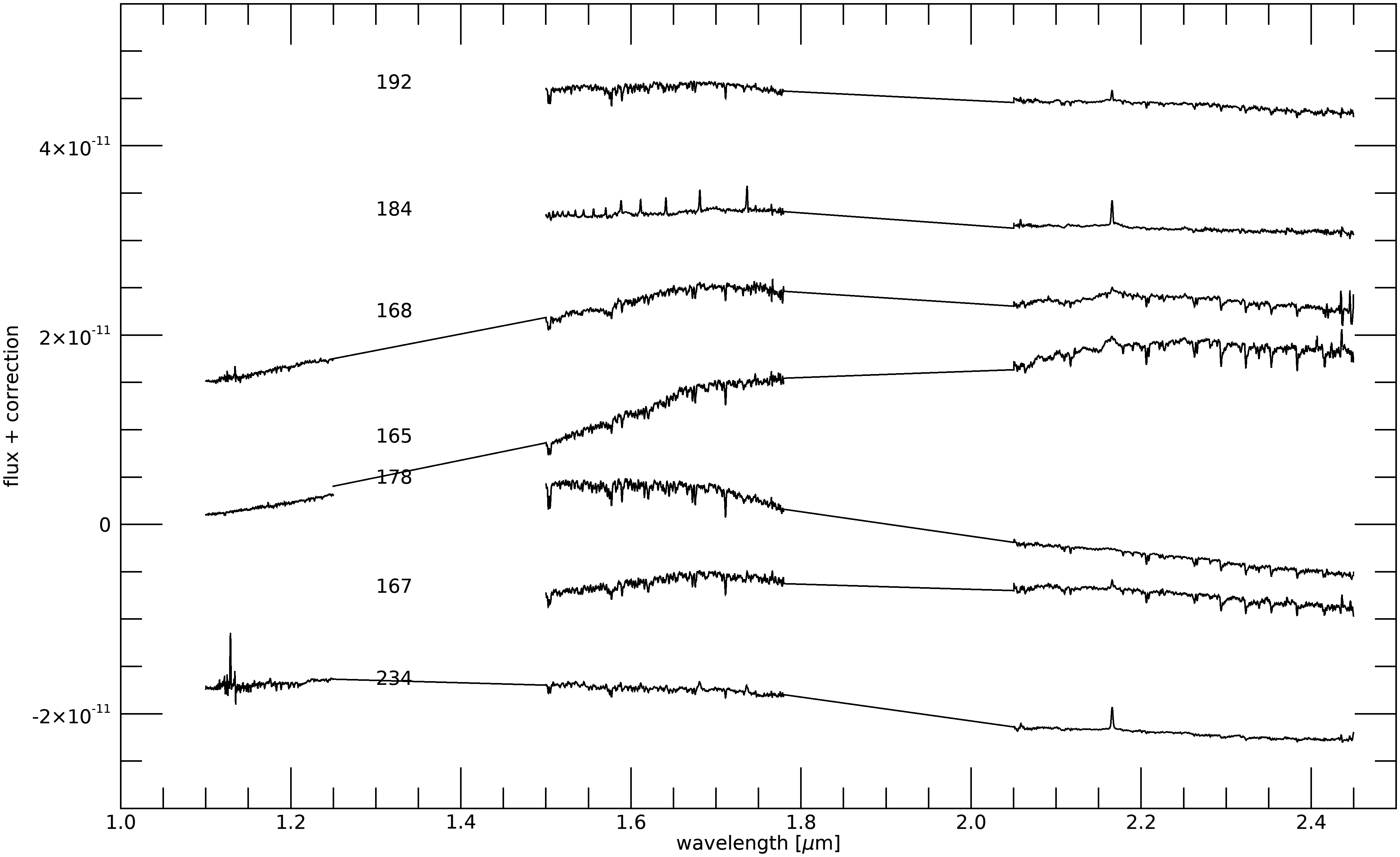}
    \caption{J, H, and K band spectra of Class~II YSOs with SpT from K7 to M2. All the spectra are suitably shifted in flux for clarity.}
    \label{fig:spec_classI_acc2}
\end{figure*}

\begin{figure*}
    \centering
    \includegraphics[width=\textwidth]{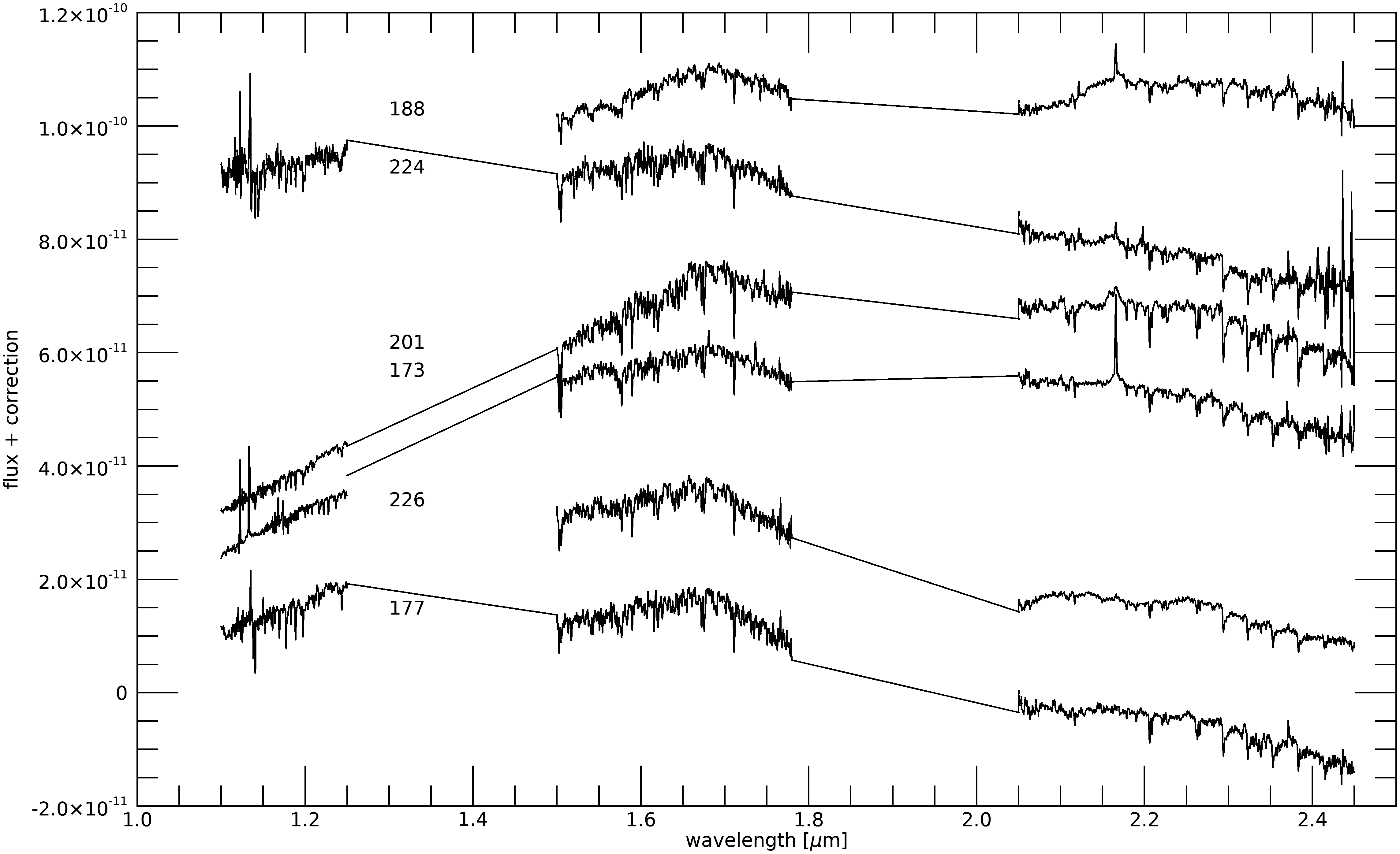}
    \caption{J, H, and K band spectra of Class~II YSOs with SpT M3 and M3.5. All the spectra are suitably shifted in flux for clarity.}
    \label{fig:spec_classI_acc3}
\end{figure*}

\begin{figure*}
    \centering
    \includegraphics[width=\textwidth]{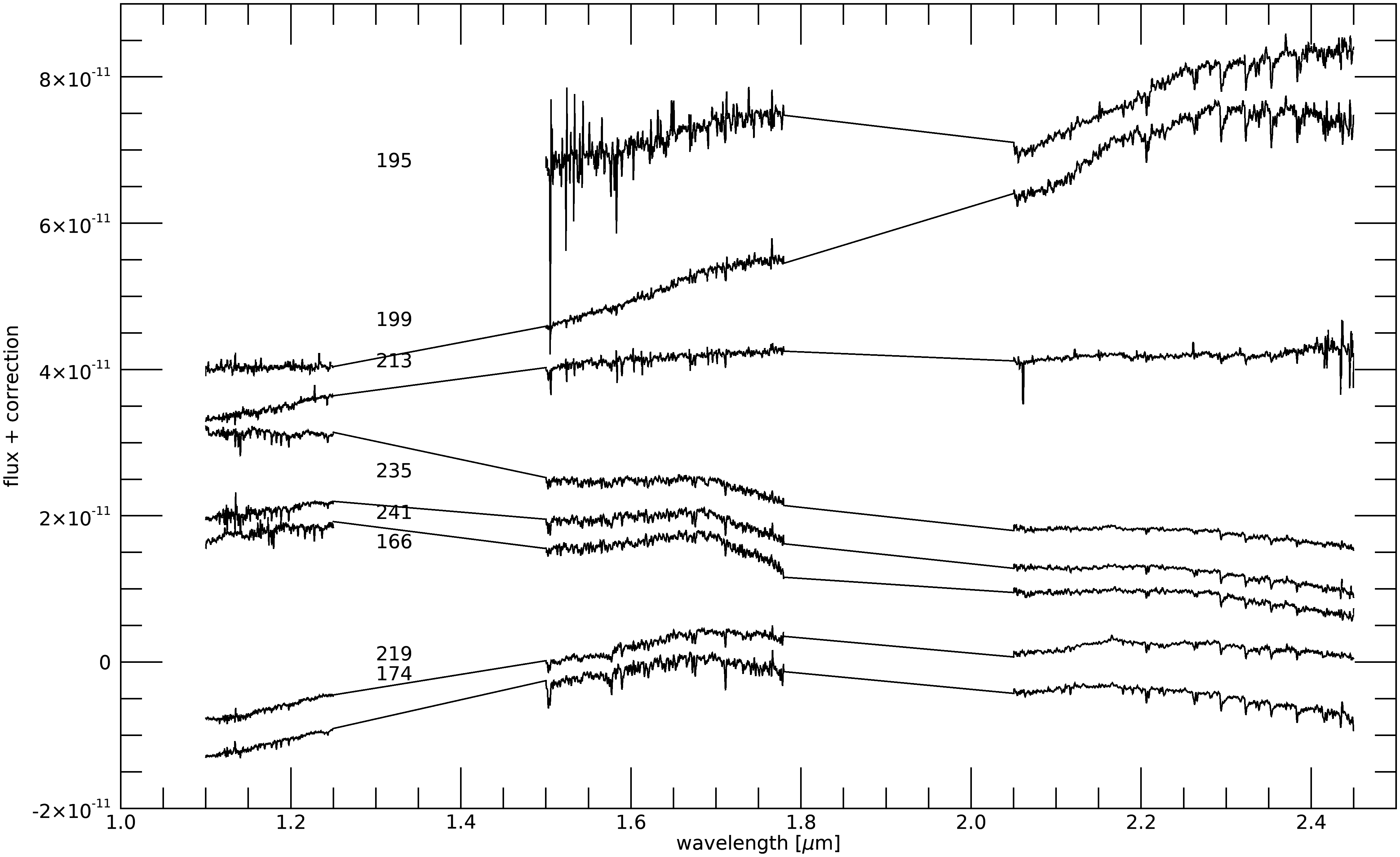}
    \caption{J, H, and K band spectra of Class~II YSOs with SpT M4 and M4.5. All the spectra are suitably shifted in flux for clarity.}
    \label{fig:spec_classI_acc4}
\end{figure*}

\begin{figure*}
    \centering
    \includegraphics[width=\textwidth]{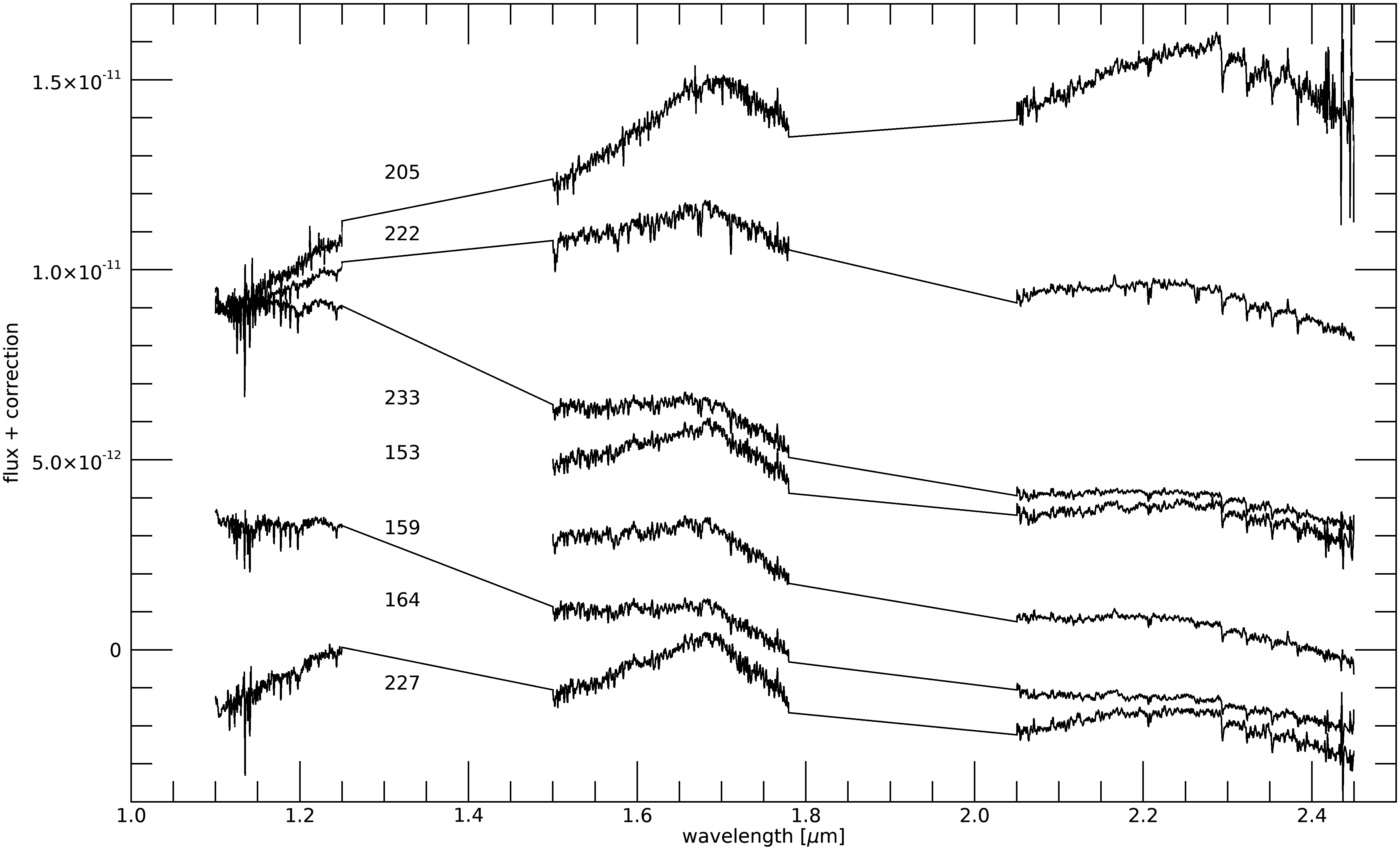}
    \caption{J, H, and K band spectra of Class~II YSOs with SpT M5 and M5.5. All the spectra are suitably shifted in flux for clarity.}
    \label{fig:spec_classI_acc5}
\end{figure*}

\begin{figure*}
    \centering
    \includegraphics[width=\textwidth]{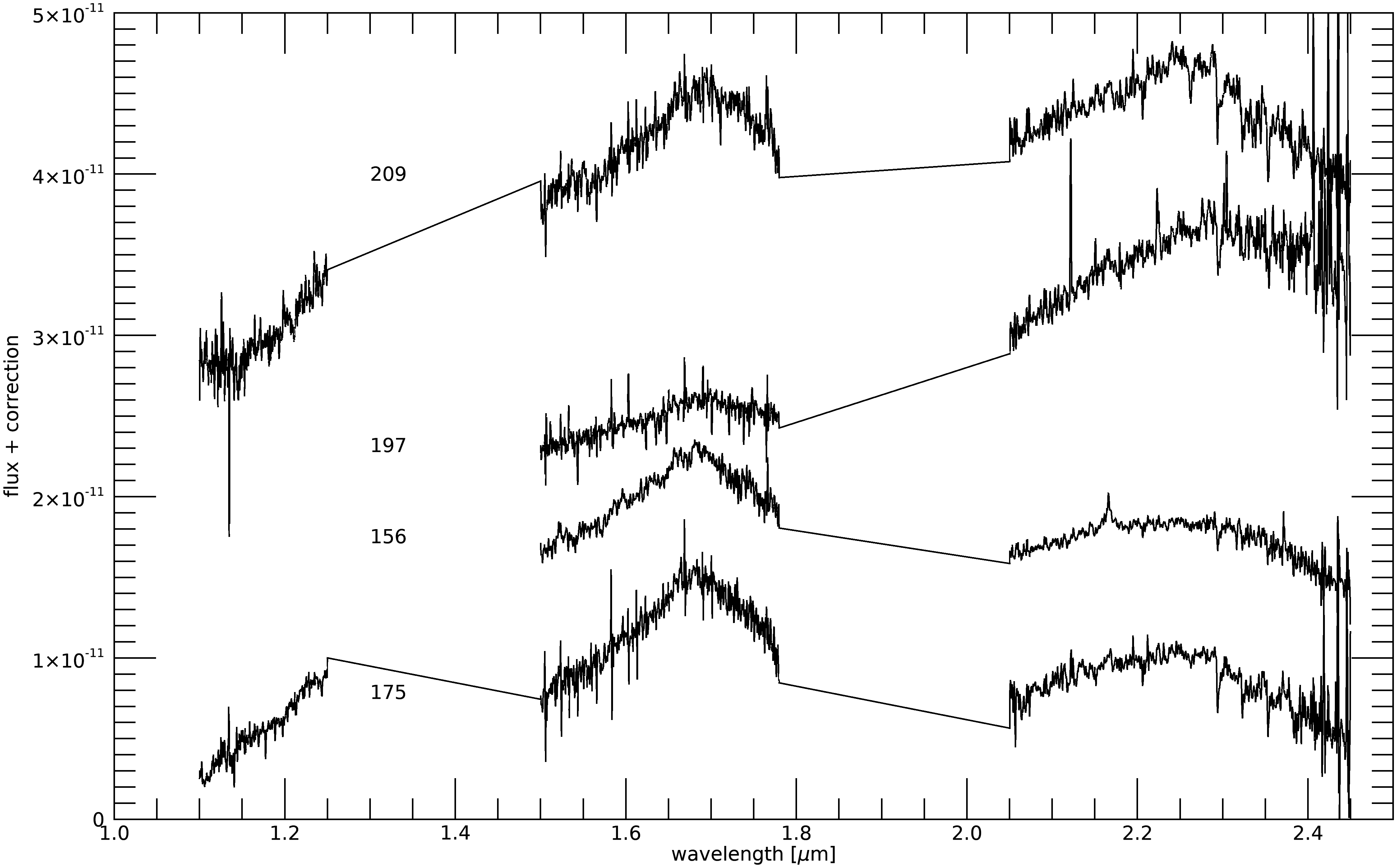}
    \caption{J, H, and K band spectra of Class~II YSOs with SpT M6 and M6.5. All the spectra are suitably shifted in flux for clarity.}
    \label{fig:spec_classI_acc6}
\end{figure*}

\begin{figure*}
    \centering
    \includegraphics[width=\textwidth]{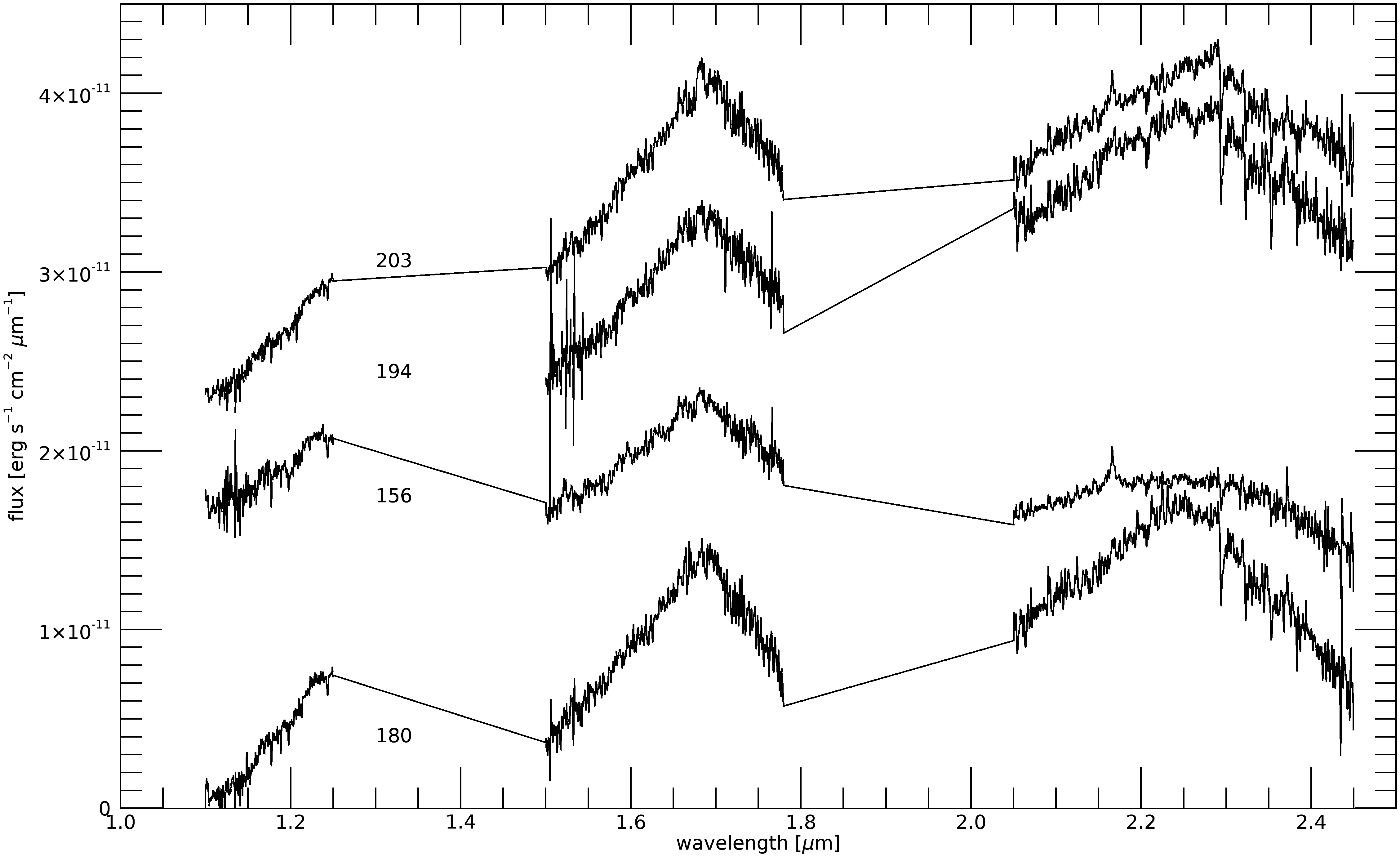}
    \caption{J, H, and K band spectra of Class~II YSOs with SpT from M7 to M9.5. All the spectra are suitably shifted in flux for clarity.}
    \label{fig:spec_classI_acc7}
\end{figure*}

Fig.~\ref{fig:spec_classI_acc1} shows the spectra of the Class~I YSOs for which we detect at least one HI line in emission. 
Fig.~\ref{fig:spec_classI_non_acc1} shows the spectra of the Class~I YSOs for which no HI line is detected. 
Figs.~\ref{fig:spec_classI_acc2}, \ref{fig:spec_classI_acc3}, \ref{fig:spec_classI_acc4}, \ref{fig:spec_classI_acc5}, \ref{fig:spec_classI_acc6}, and \ref{fig:spec_classI_acc7} show the Class~II YSOs spectra from early K to late M spectral type. 
We shifted the spectra in flux by a suitable amount, to optimise the visualisation of the YSOs spectra in the figure.

\begin{figure*} 
\centering
 \begin{subfigure}{\textwidth}
 \centering
 \includegraphics[width=0.2\textwidth]{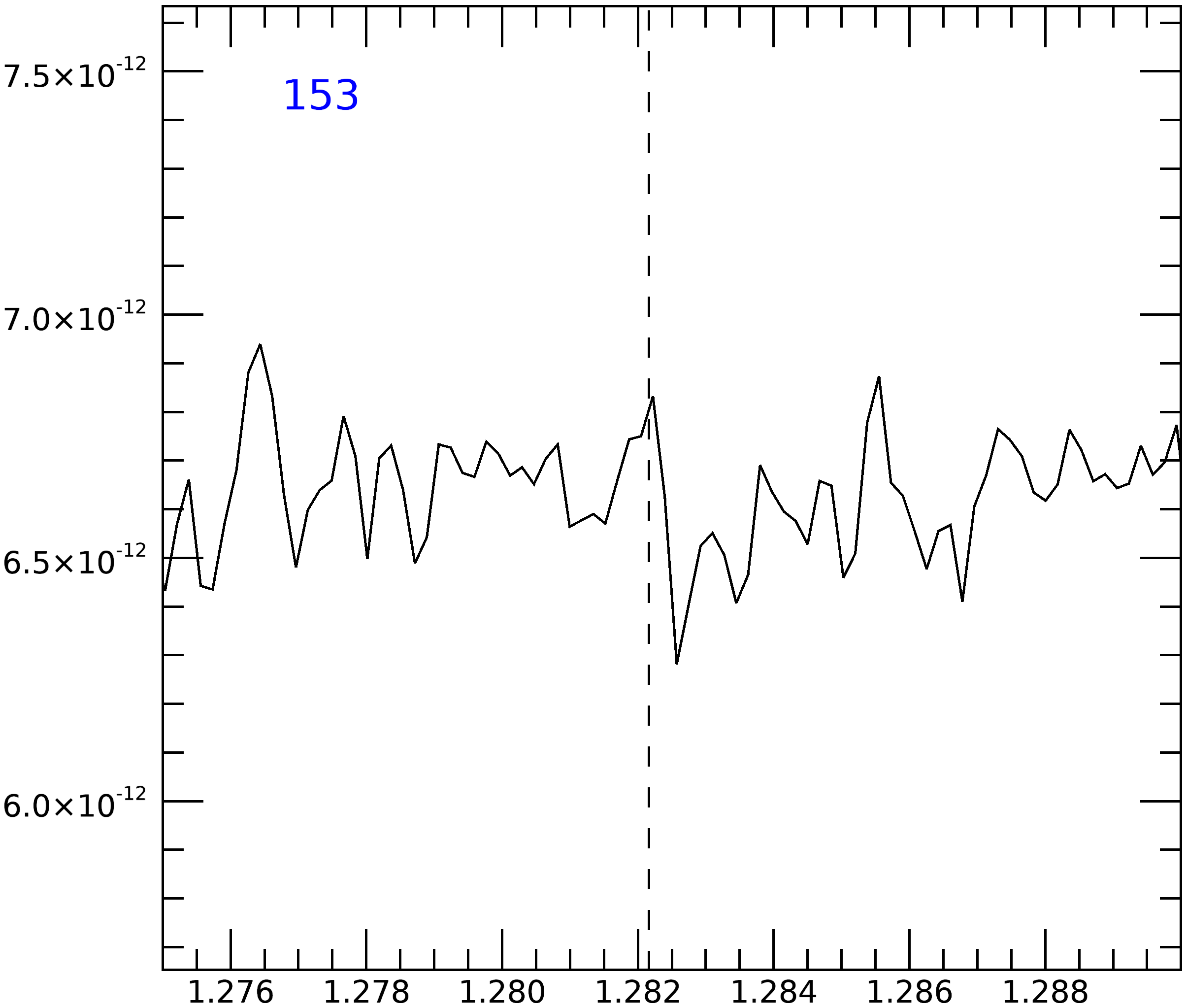}%
 \includegraphics[width=0.2\textwidth]{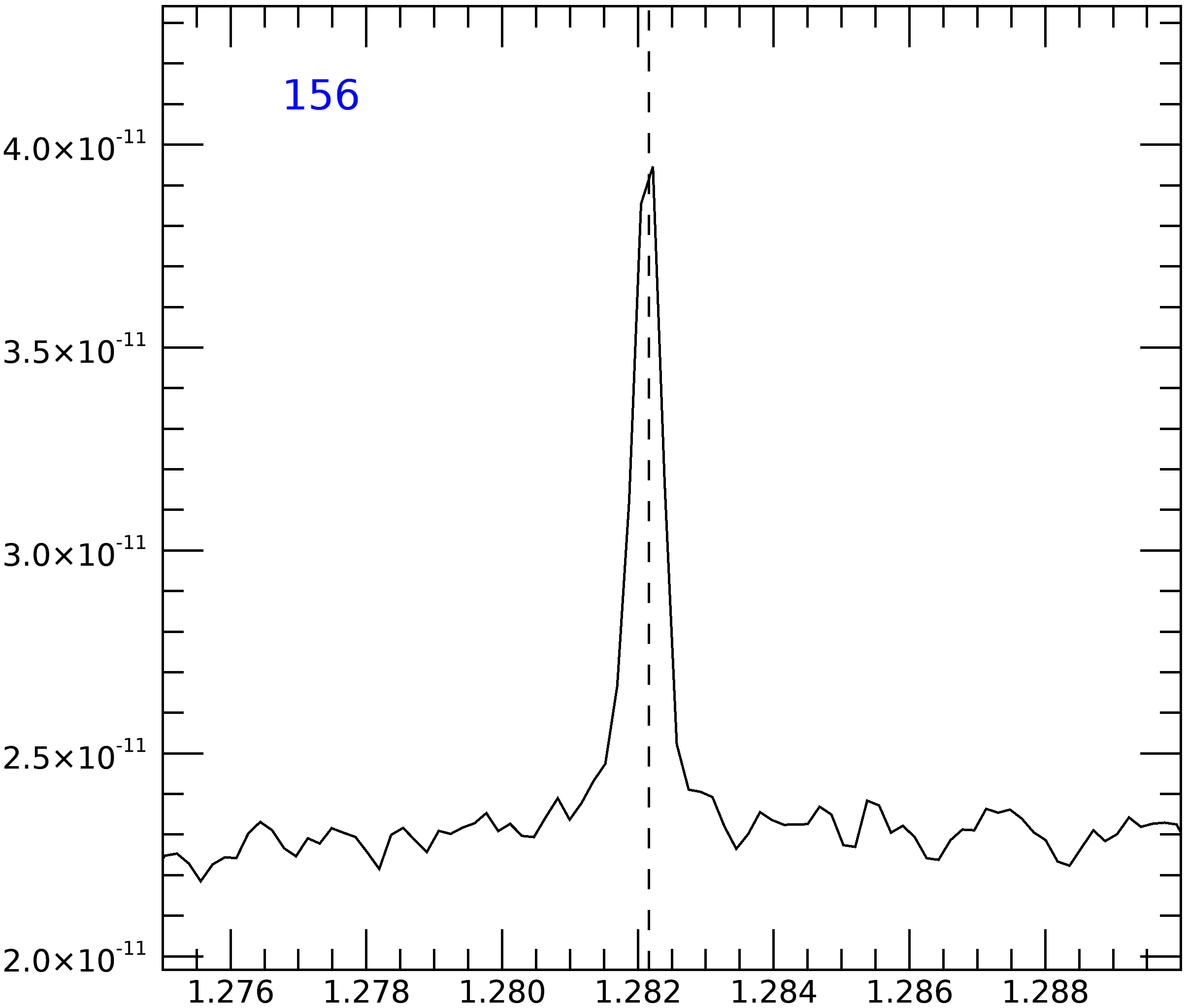}%
 \includegraphics[width=0.2\textwidth]{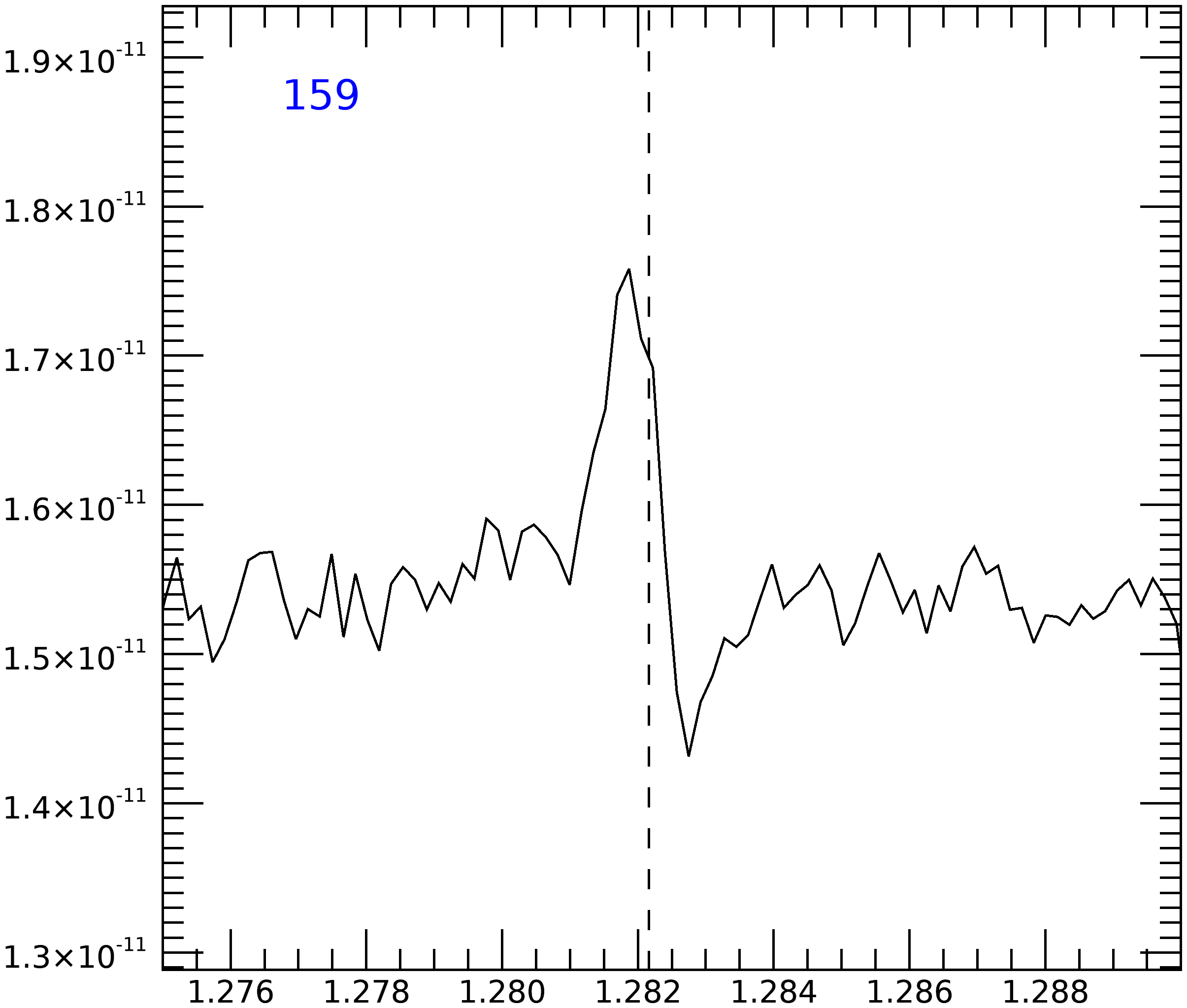}%
 \includegraphics[width=0.2\textwidth]{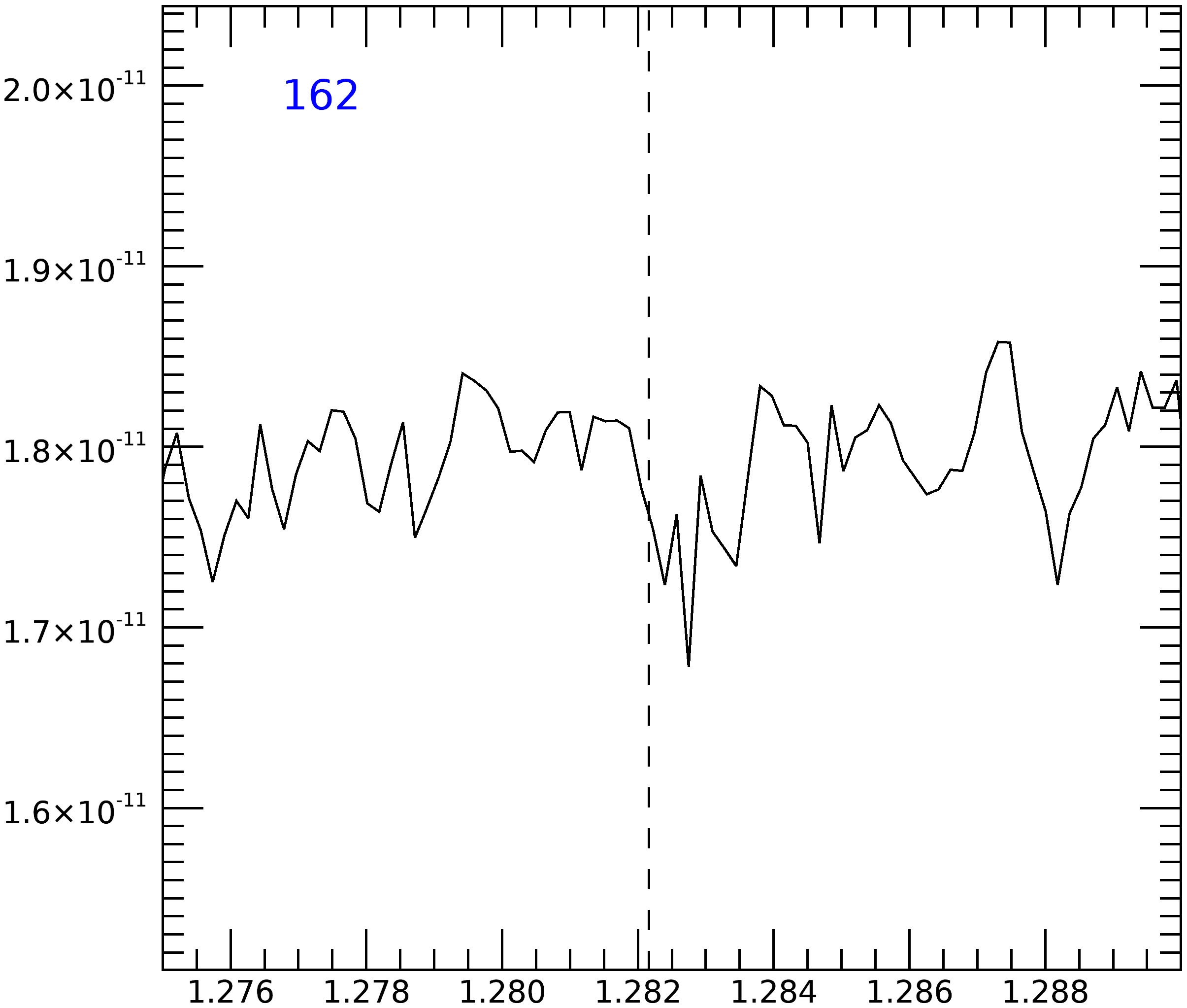}%
 \includegraphics[width=0.2\textwidth]{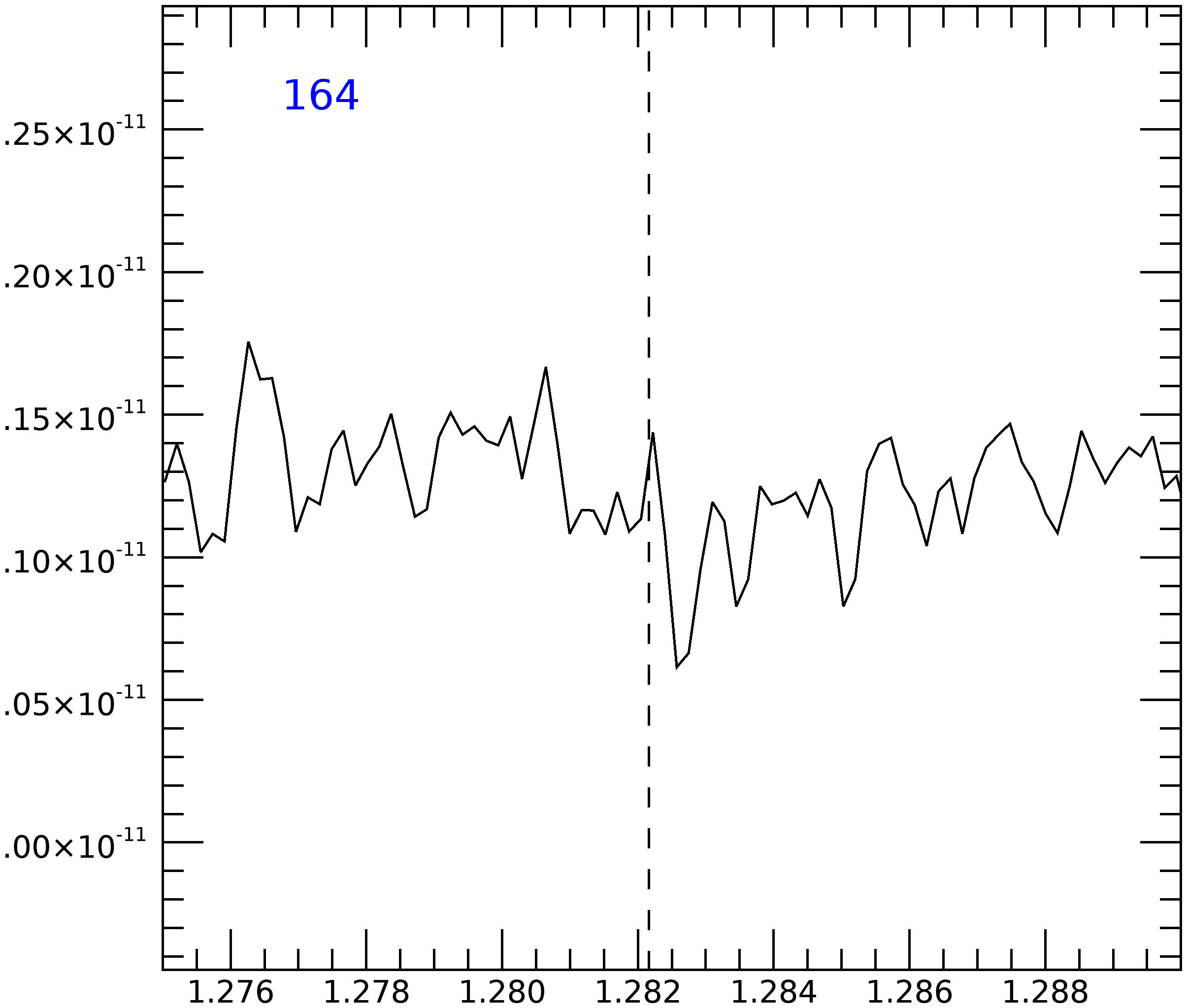}%

 \includegraphics[width=0.2\textwidth]{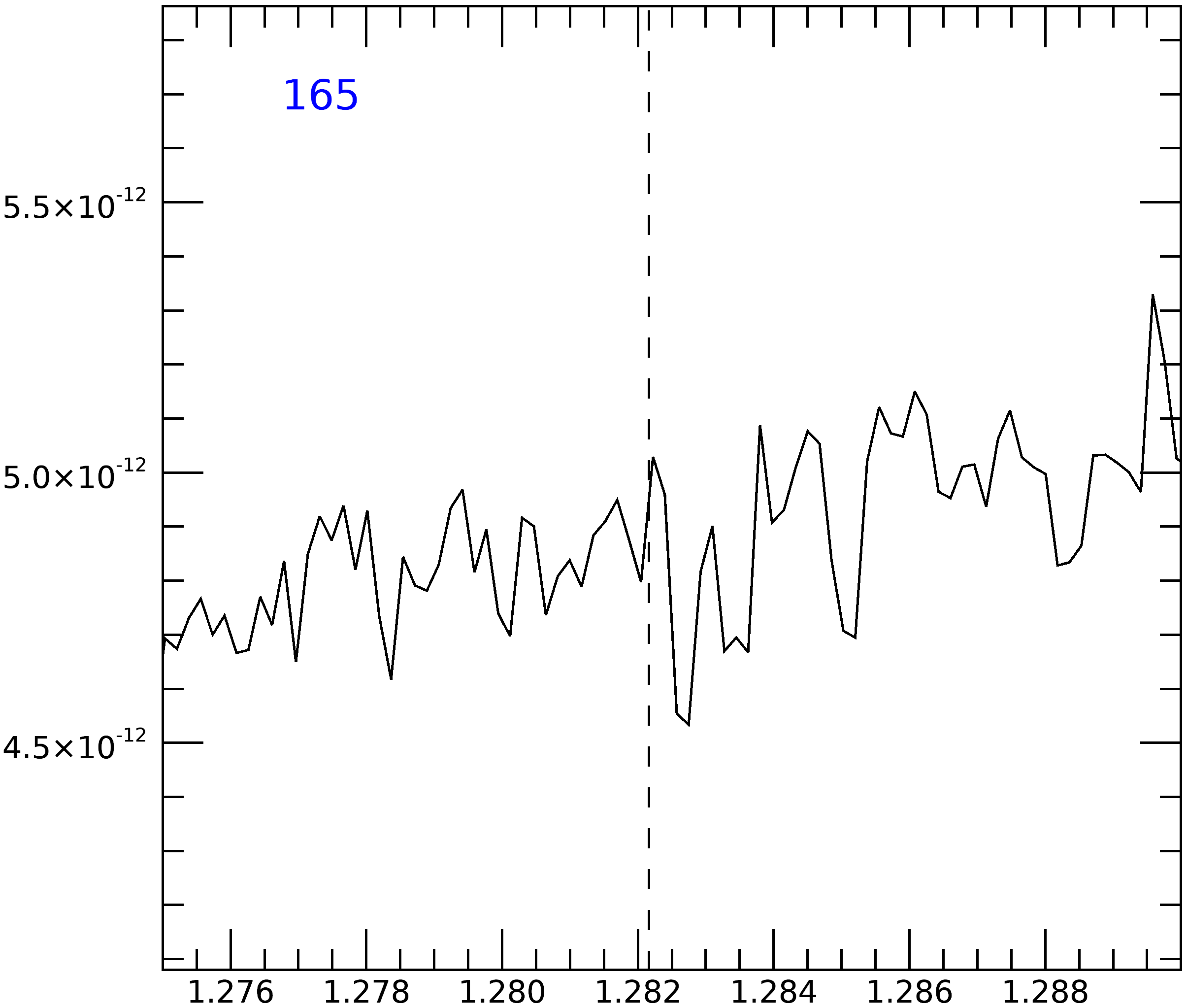}%
 \includegraphics[width=0.2\textwidth]{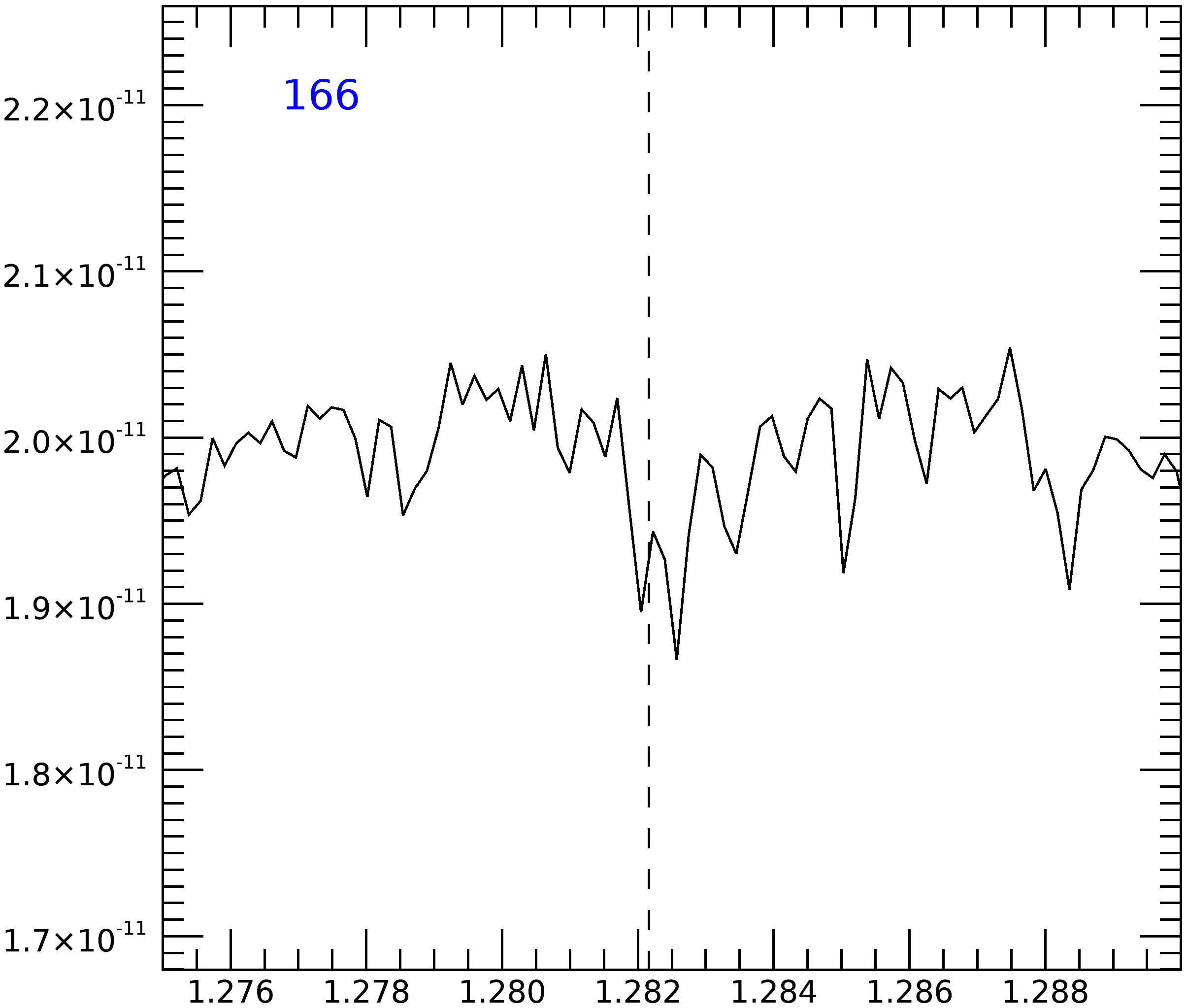}%
 \includegraphics[width=0.2\textwidth]{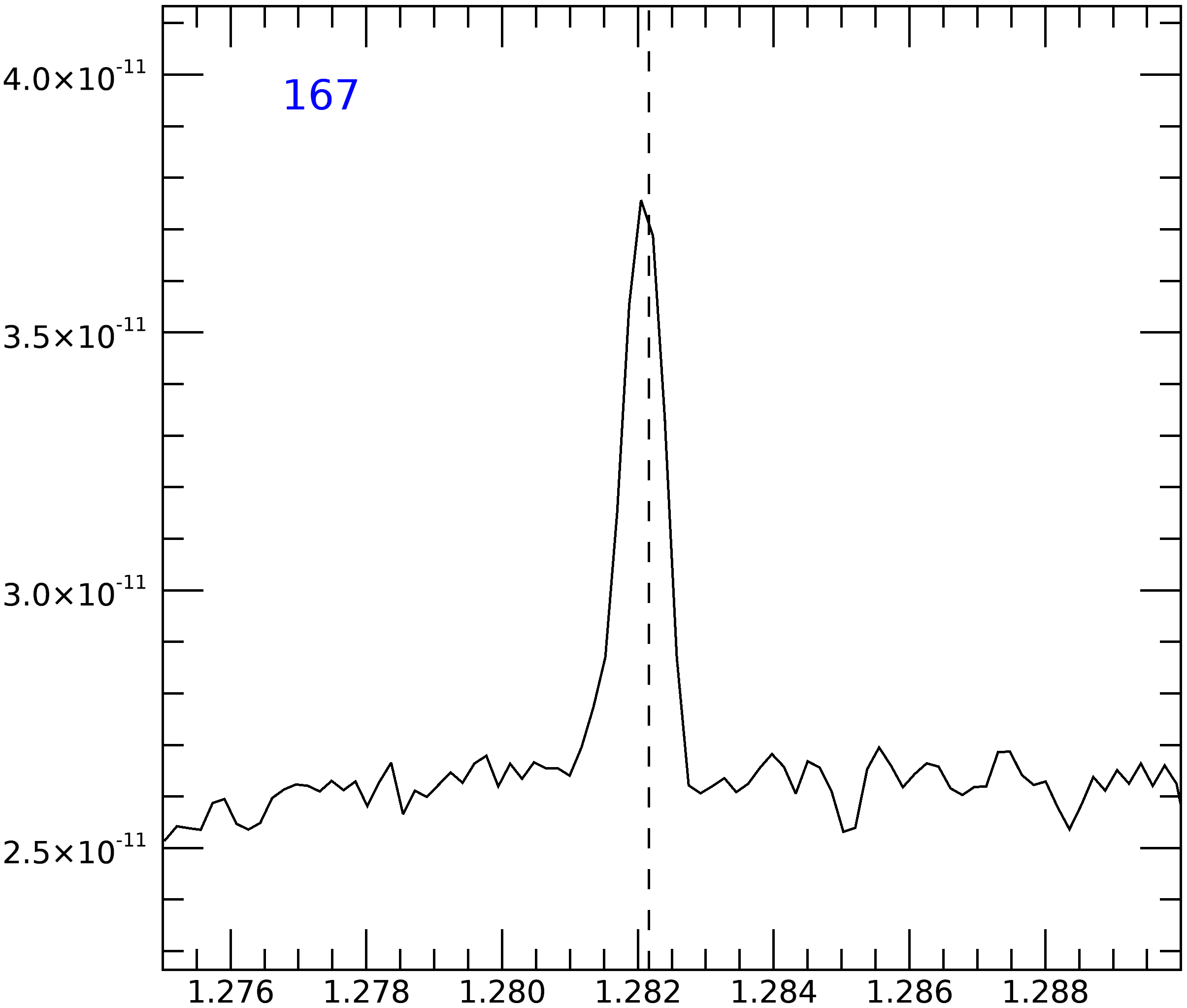}%
 \includegraphics[width=0.2\textwidth]{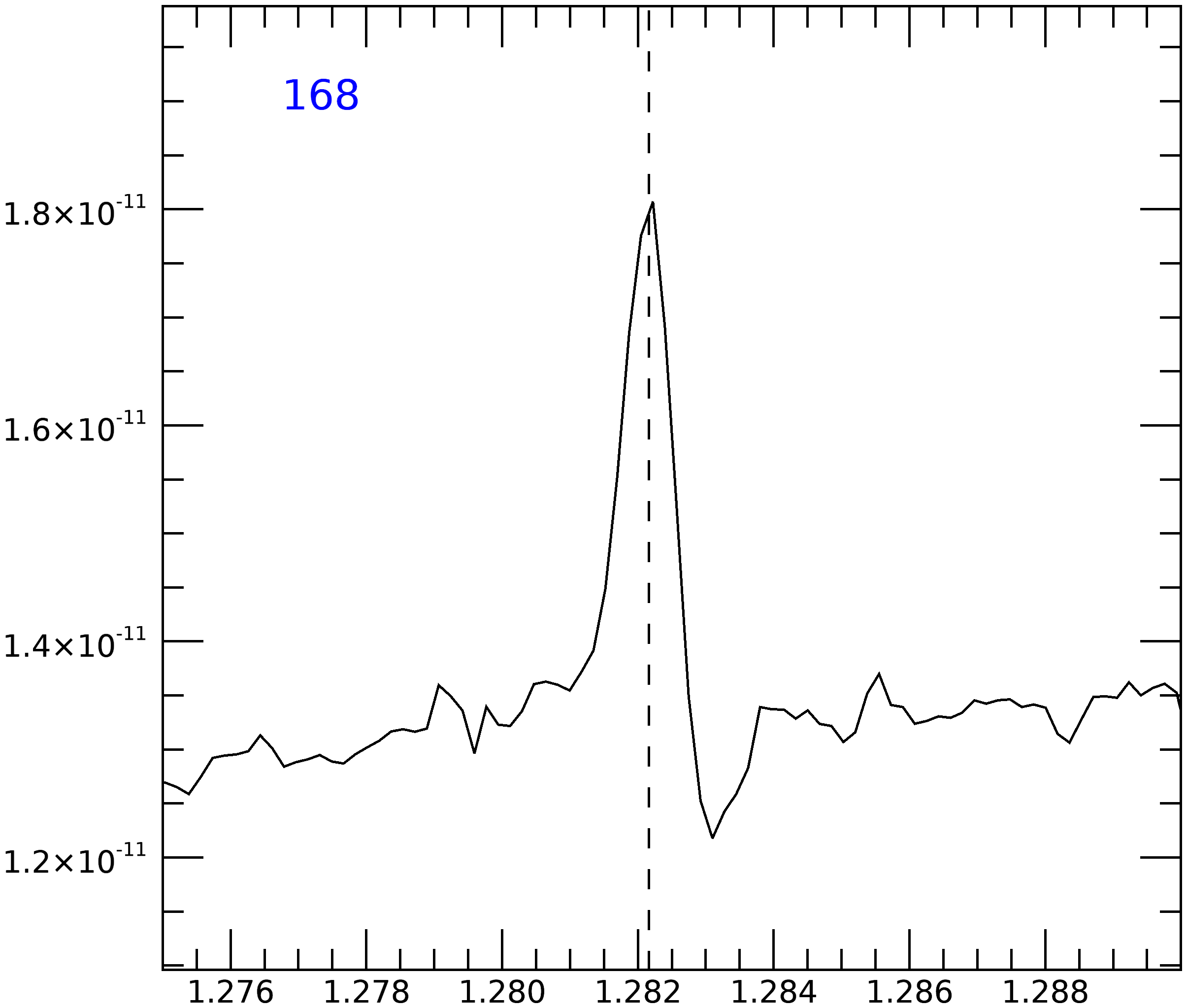}%
 \includegraphics[width=0.2\textwidth]{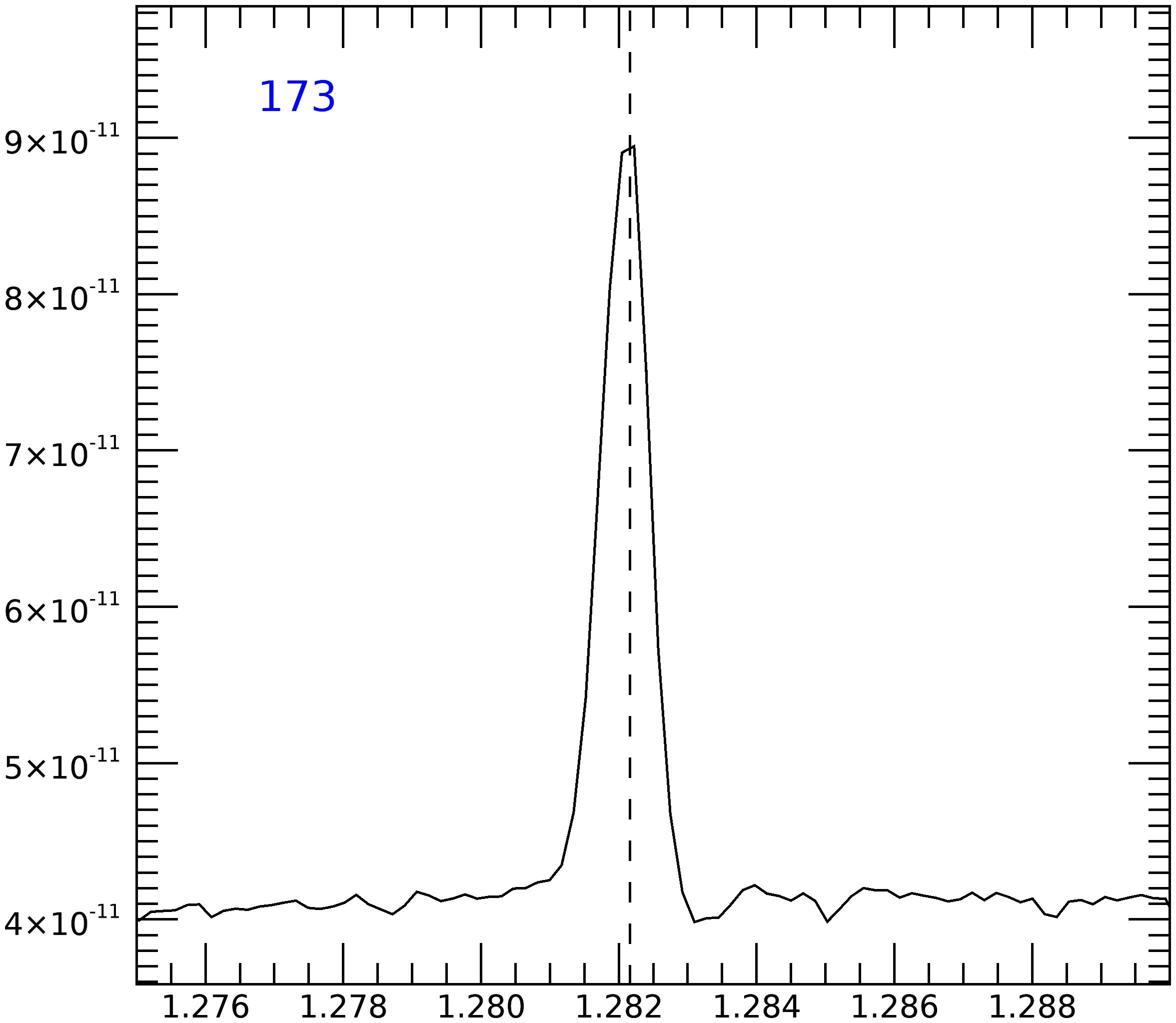}%

 \includegraphics[width=0.2\textwidth]{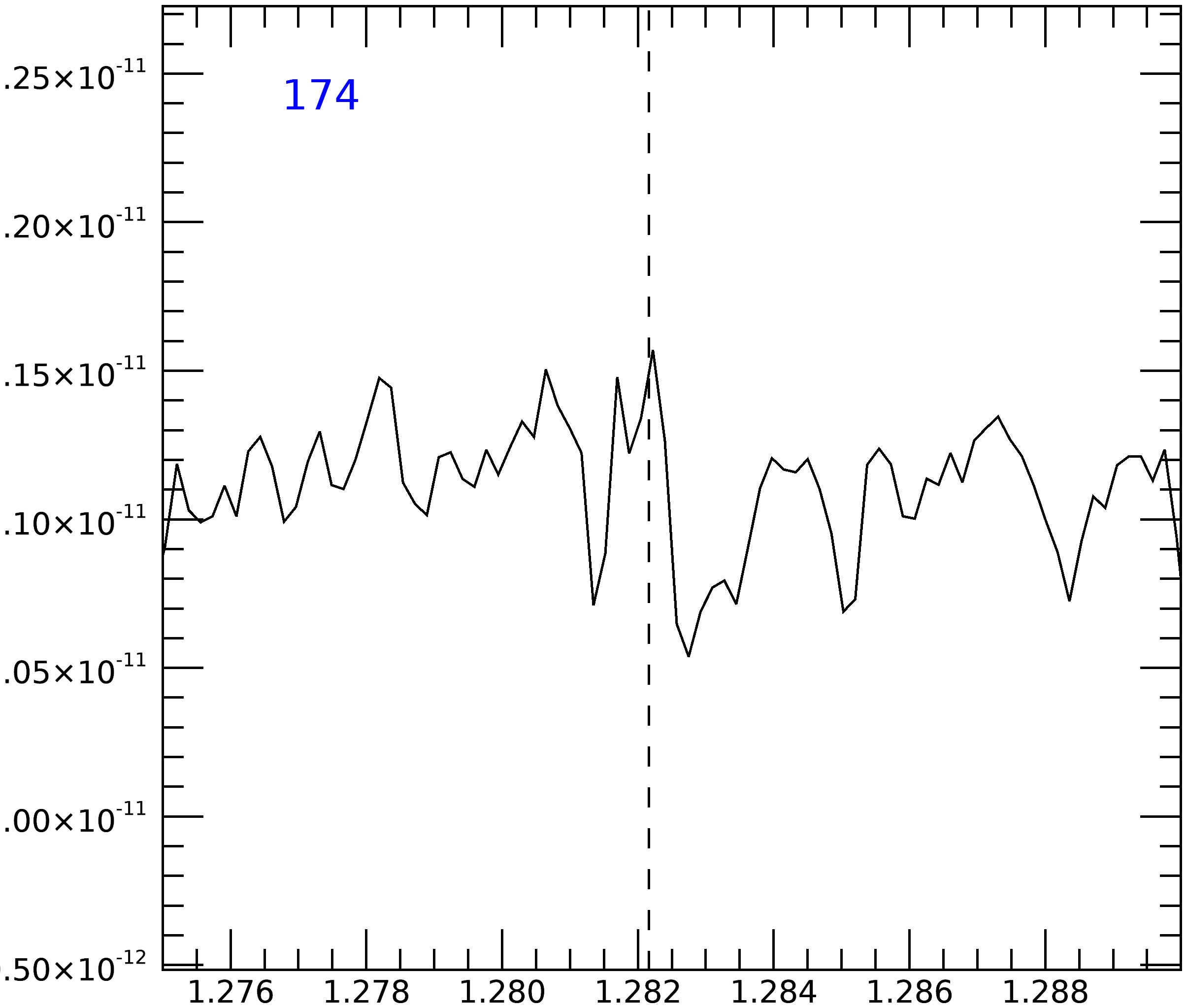}%
 \includegraphics[width=0.2\textwidth]{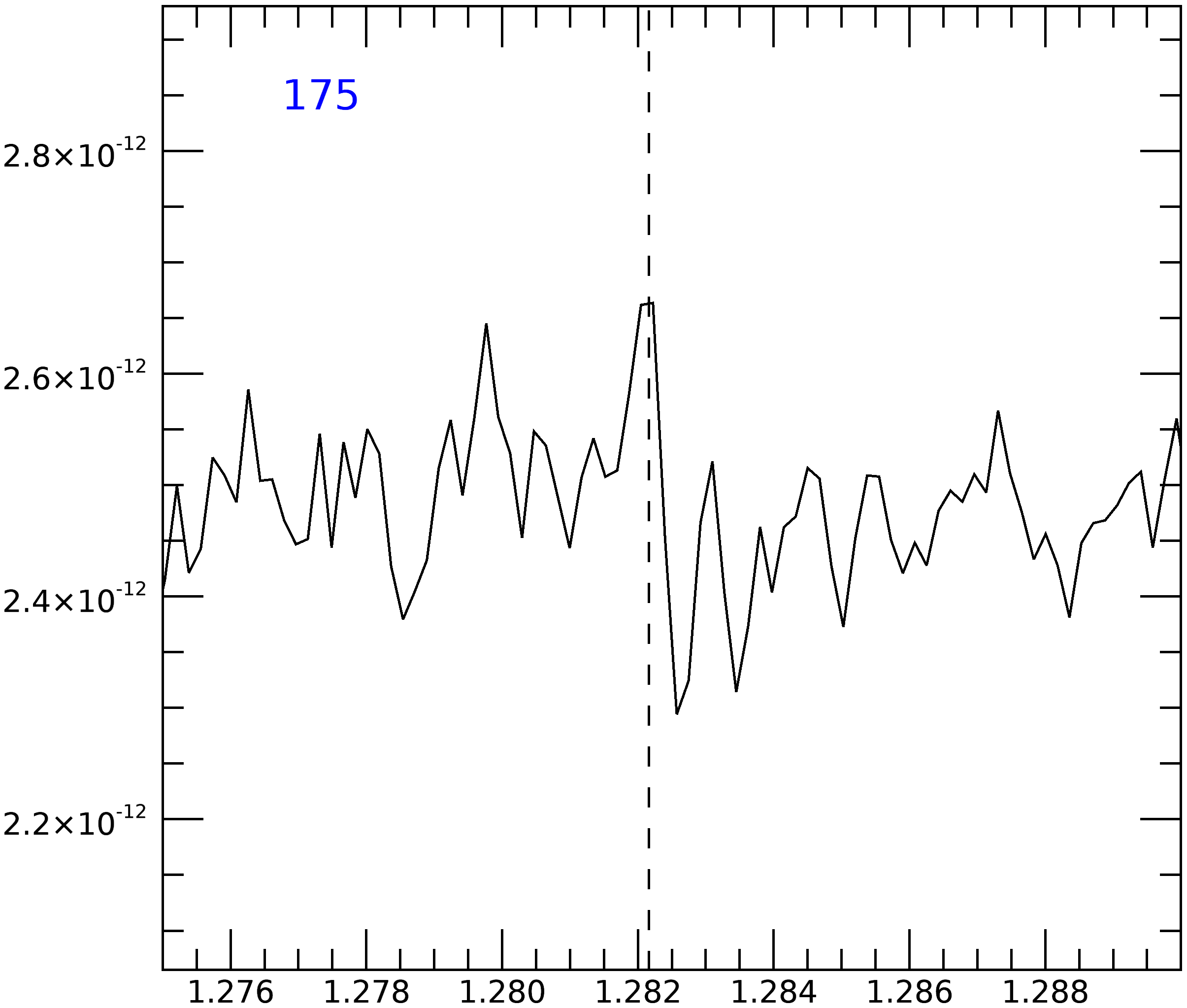}%
 \includegraphics[width=0.2\textwidth]{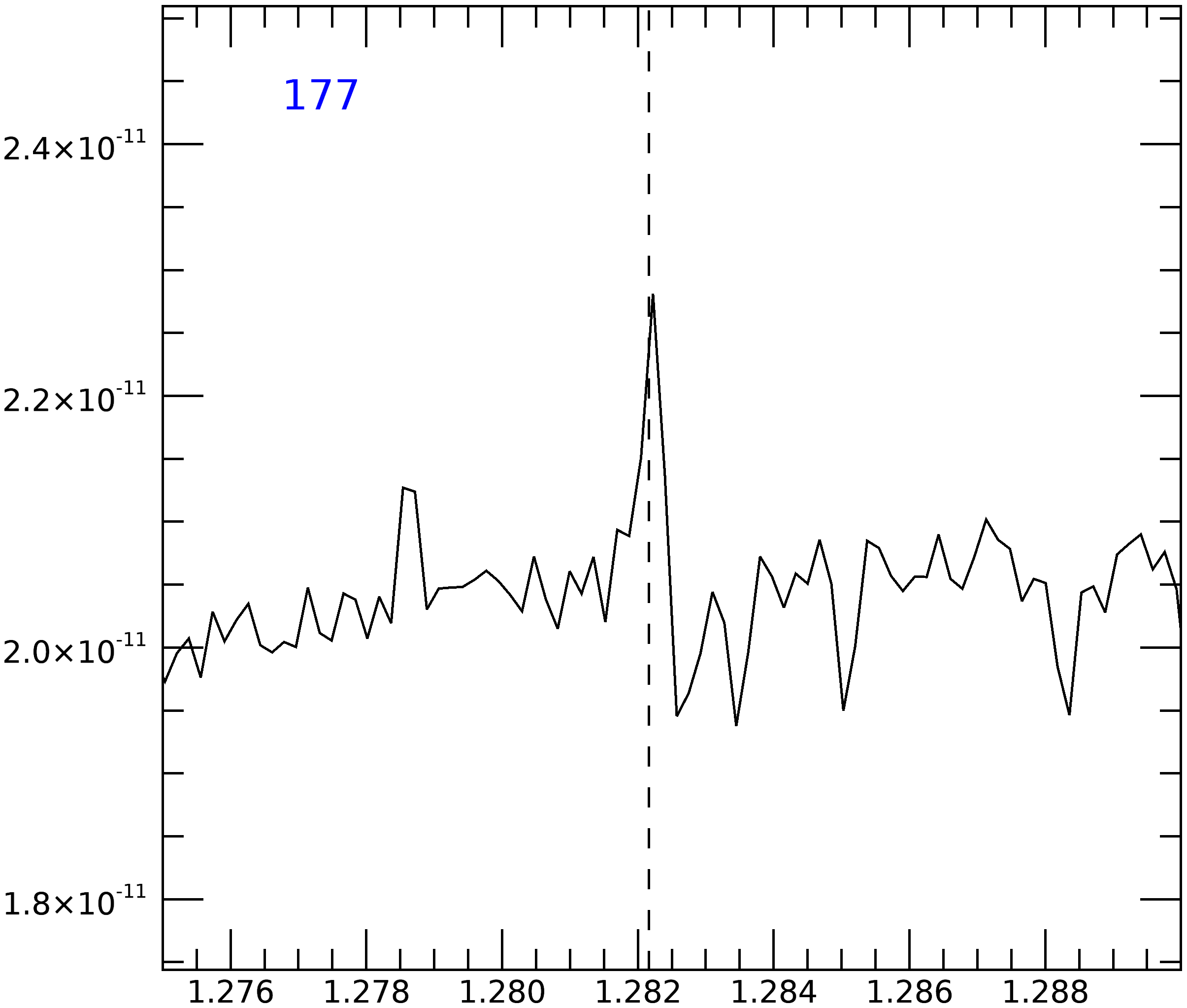}%
 \includegraphics[width=0.2\textwidth]{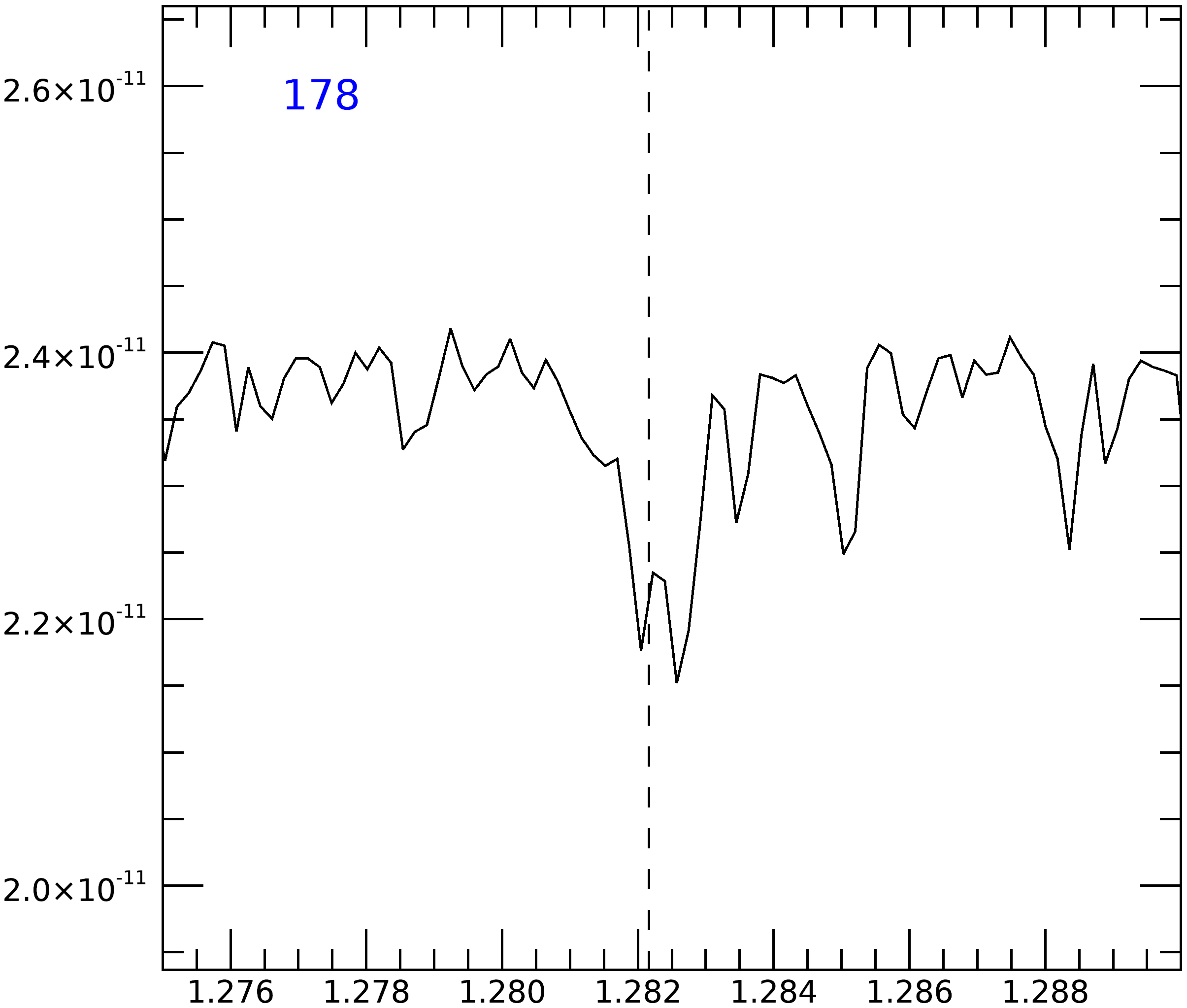}%
 \includegraphics[width=0.2\textwidth]{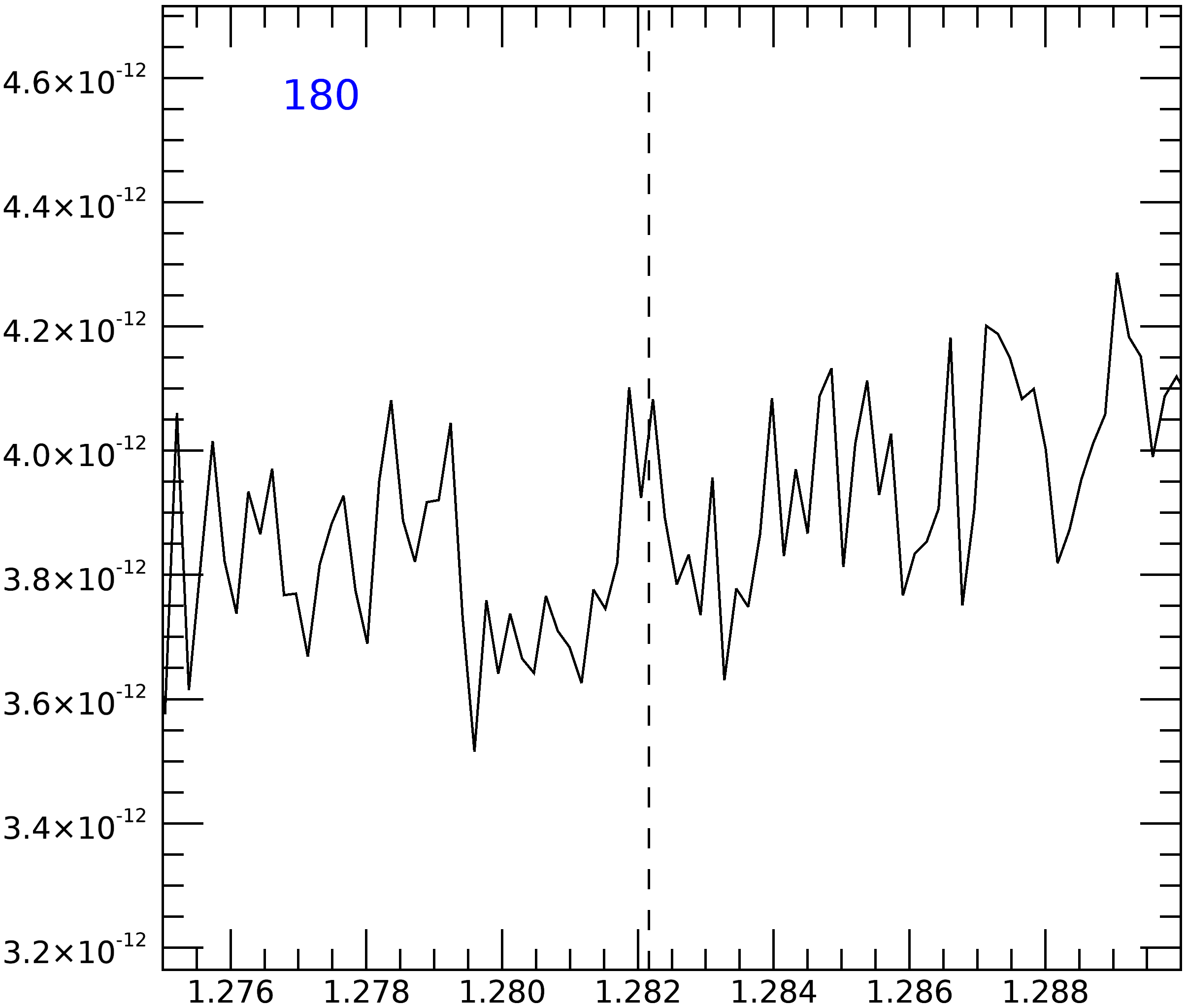}%

 \includegraphics[width=0.2\textwidth]{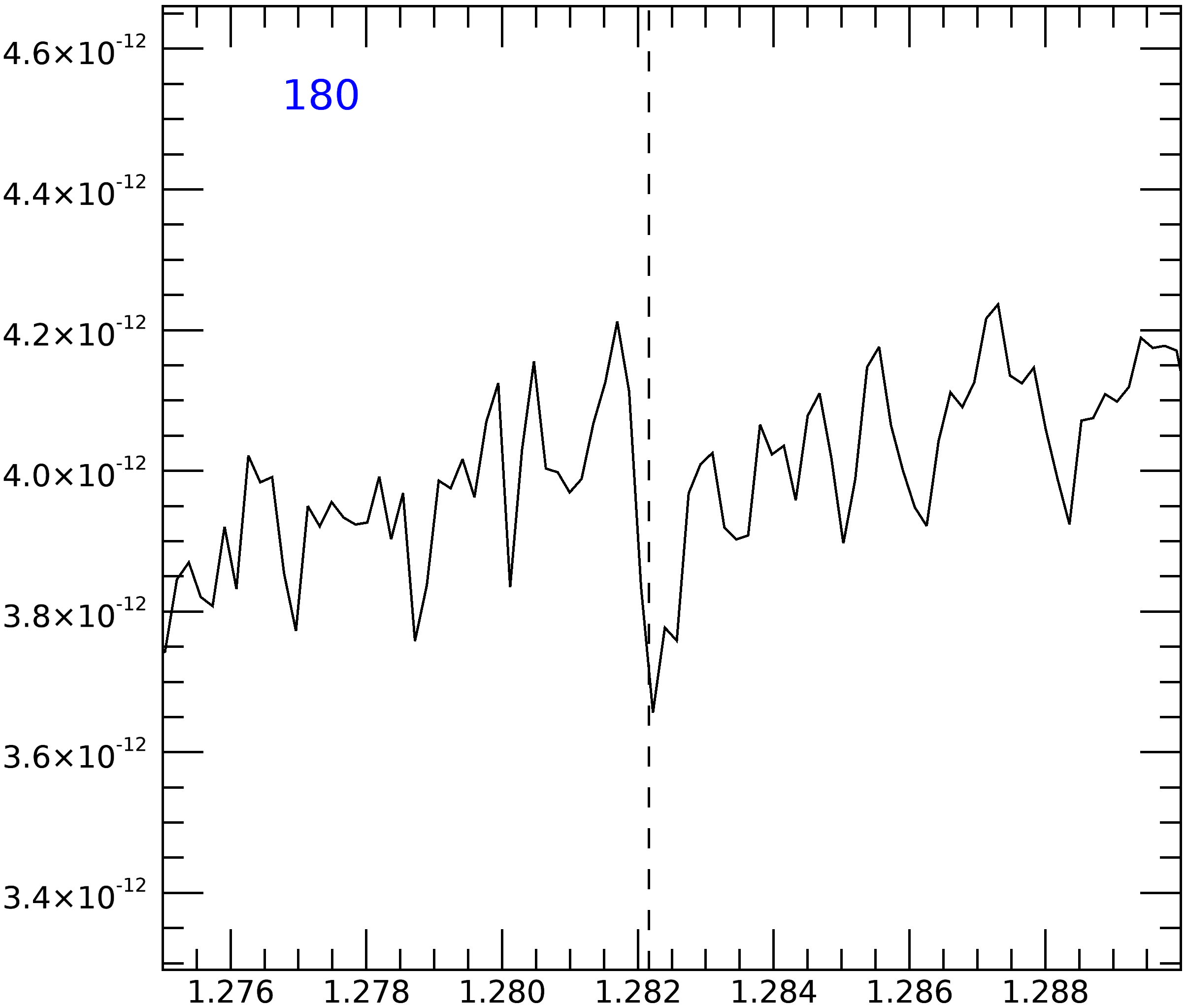}%
 \includegraphics[width=0.2\textwidth]{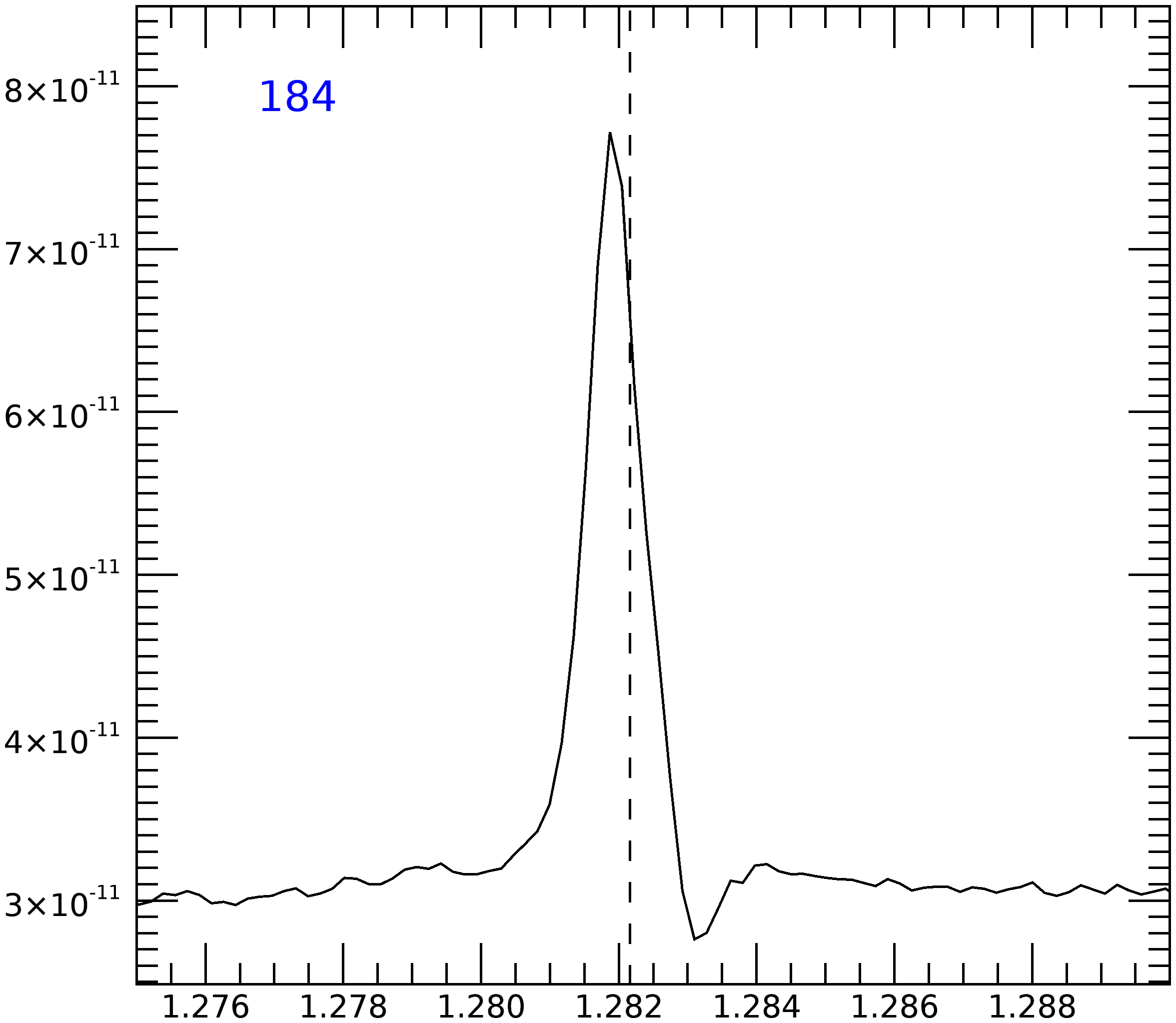}%
 \includegraphics[width=0.2\textwidth]{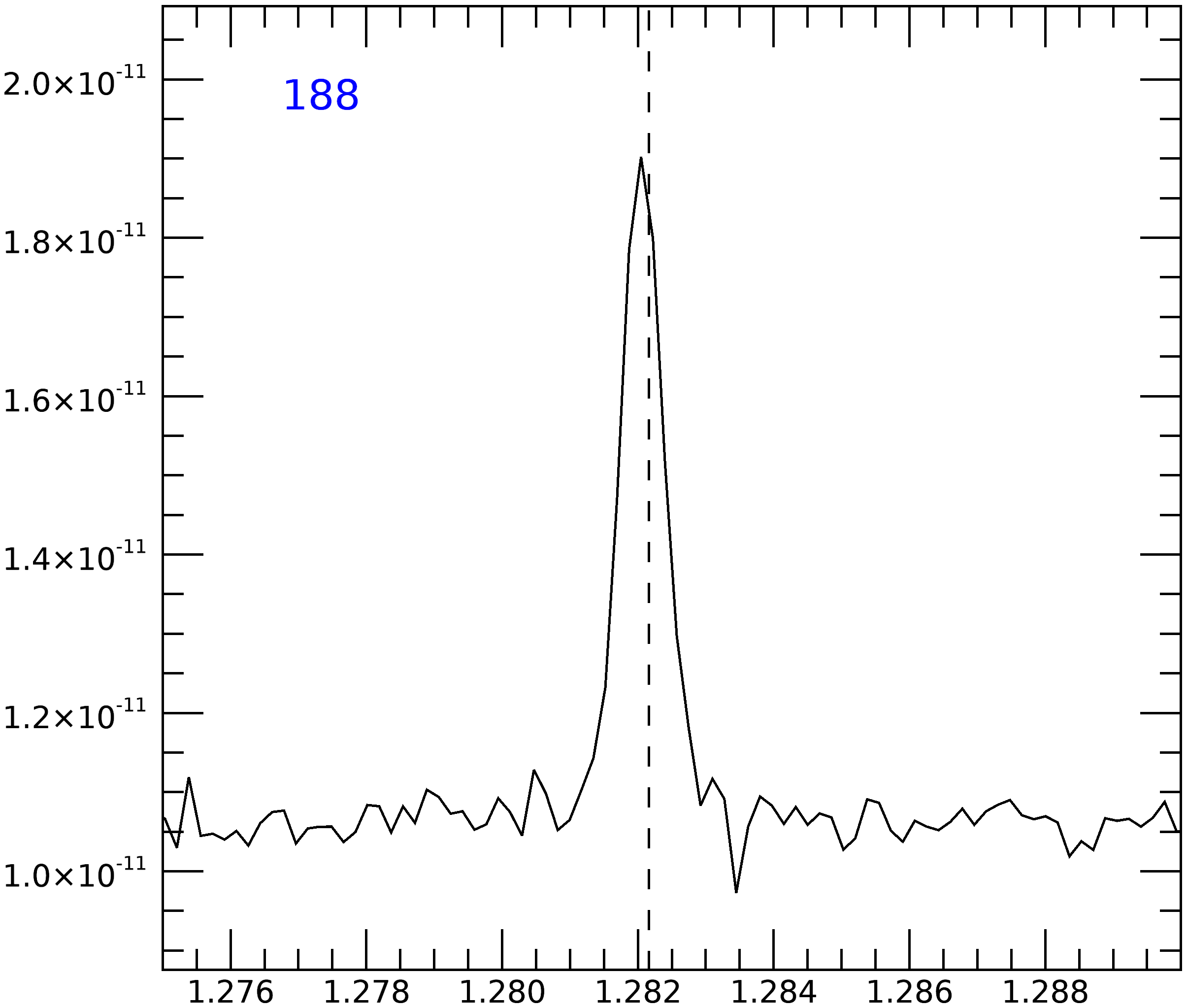}%
 \includegraphics[width=0.2\textwidth]{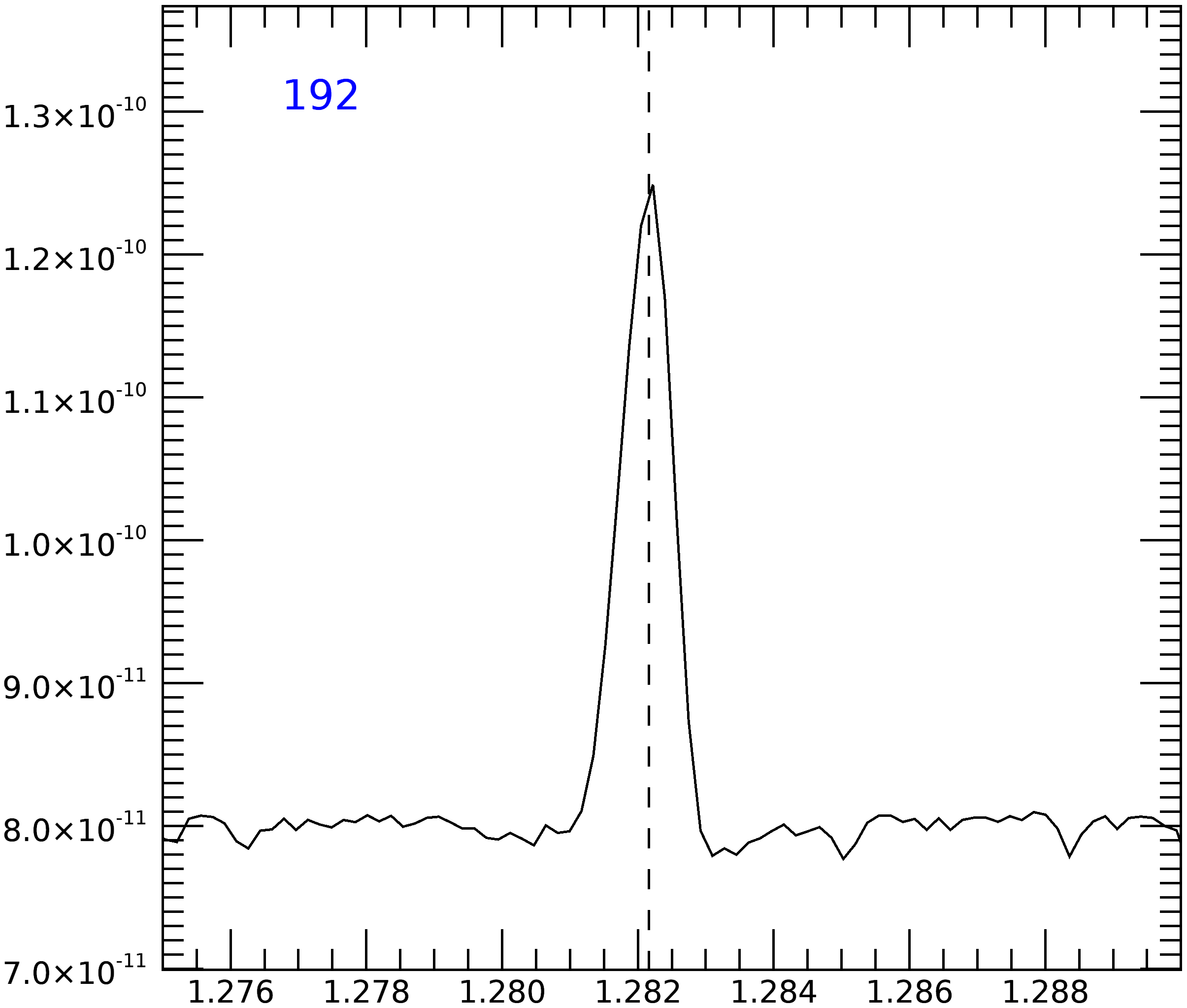}%
 \includegraphics[width=0.2\textwidth]{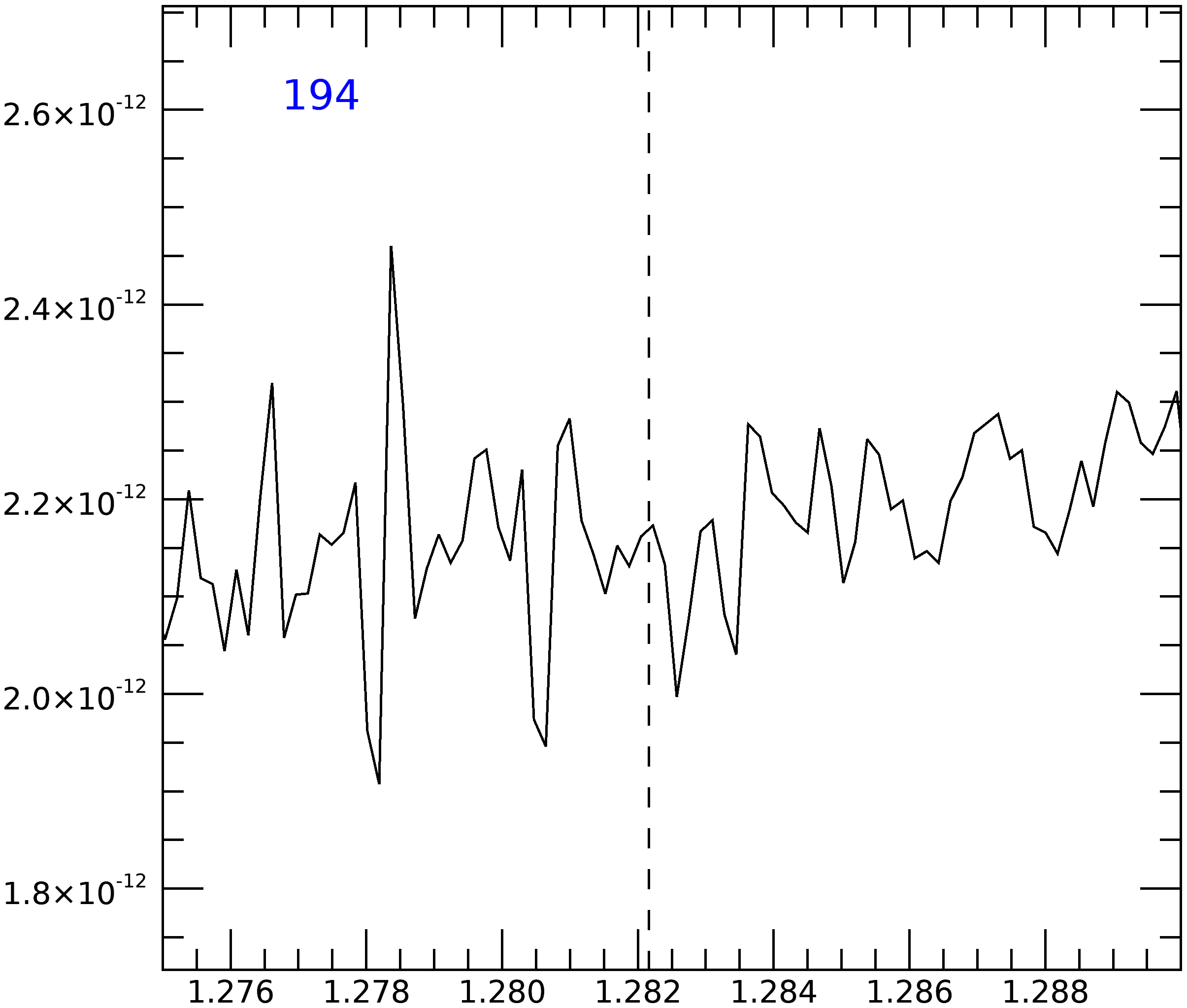}%
 
 \includegraphics[width=0.2\textwidth]{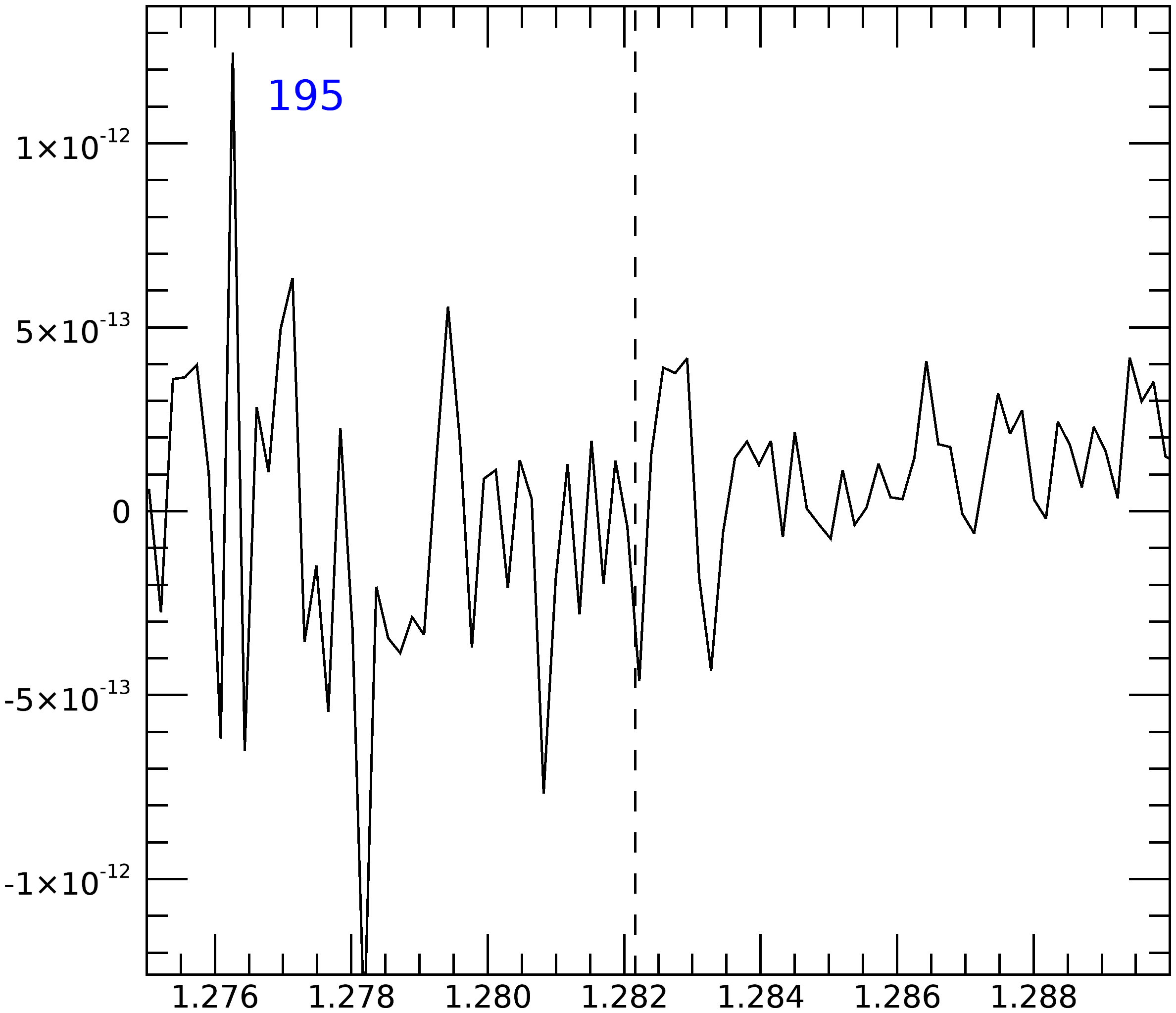}%
 \includegraphics[width=0.2\textwidth]{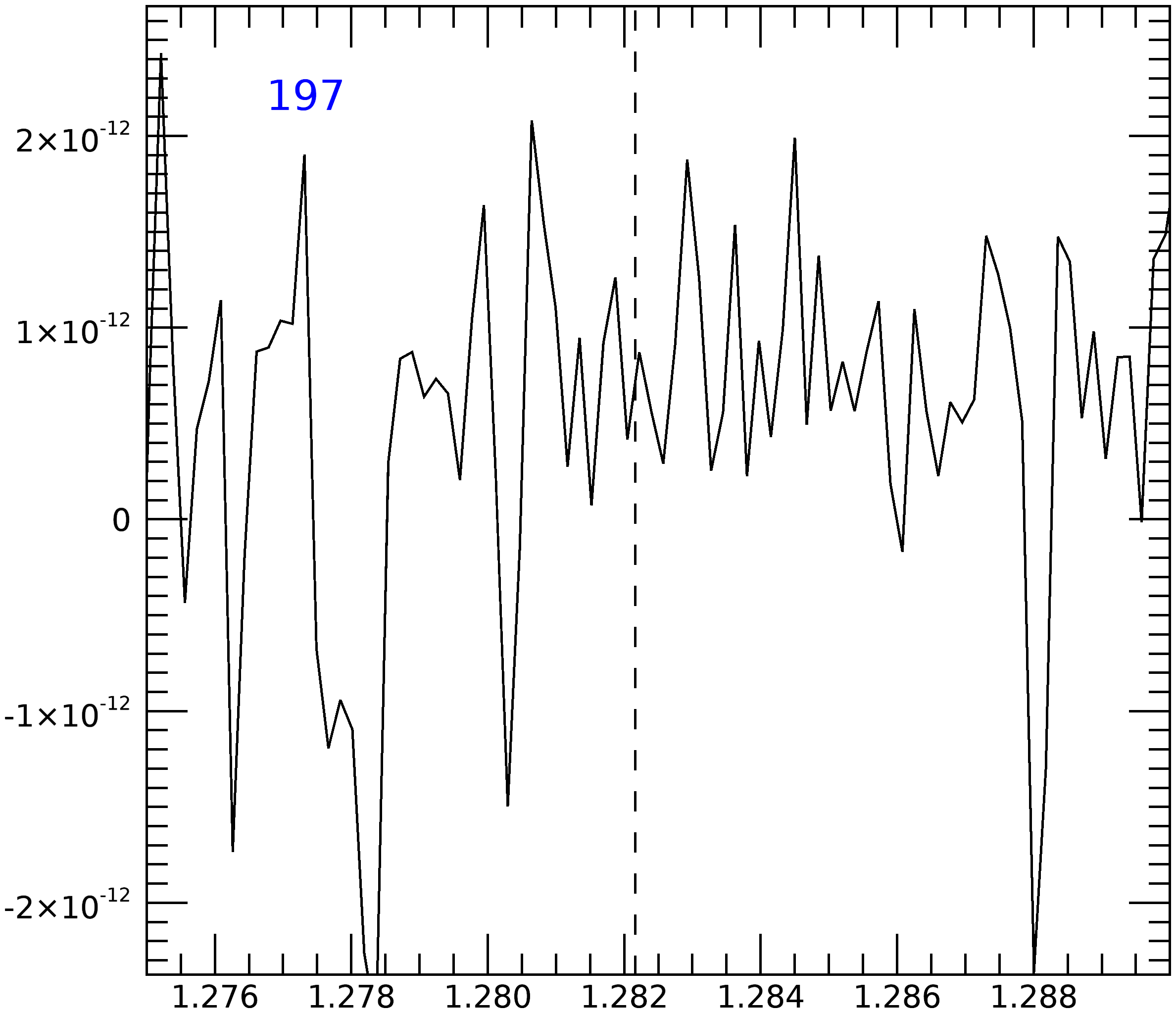}%
 \includegraphics[width=0.2\textwidth]{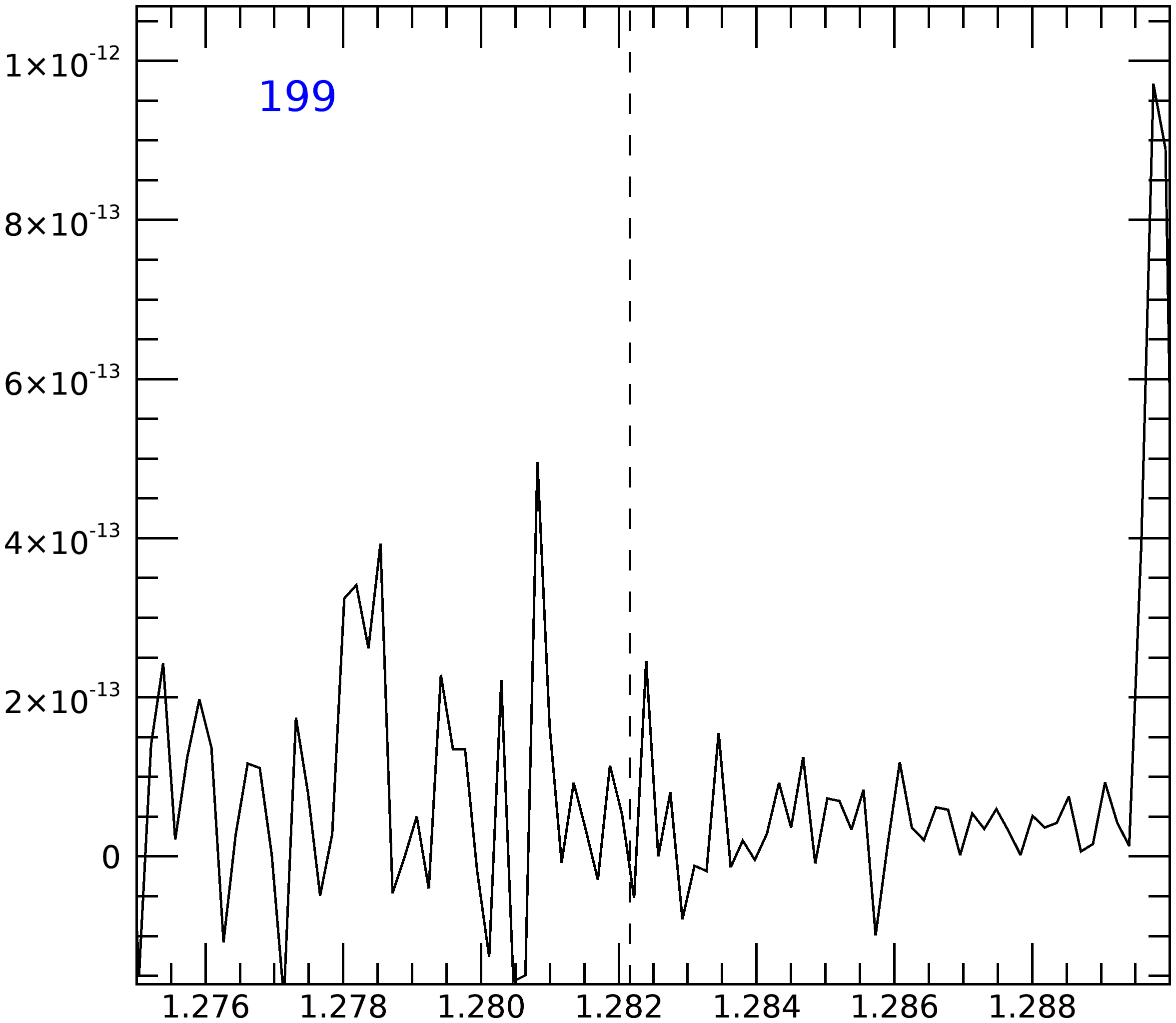}%
 \includegraphics[width=0.2\textwidth]{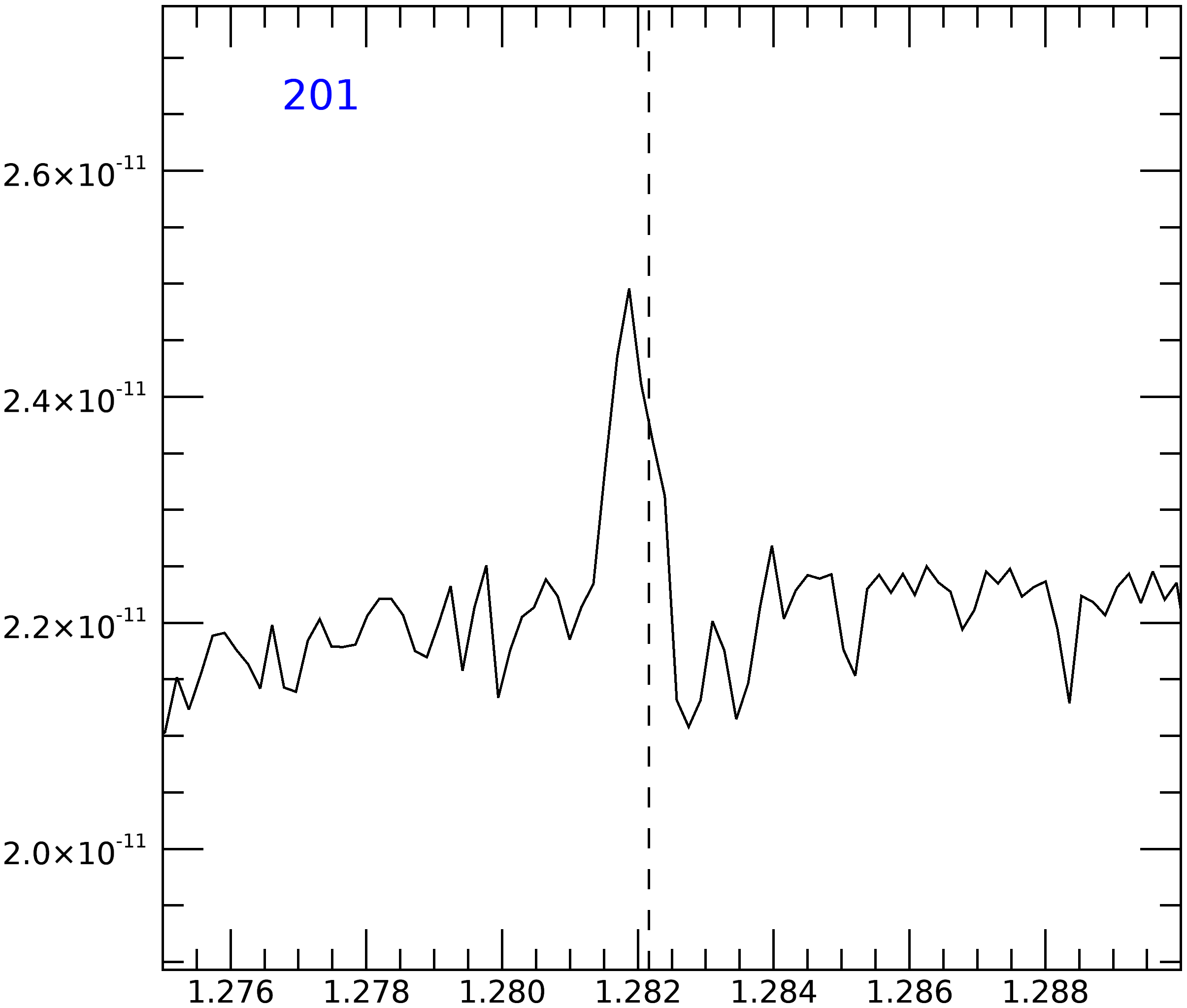}%
 \includegraphics[width=0.2\textwidth]{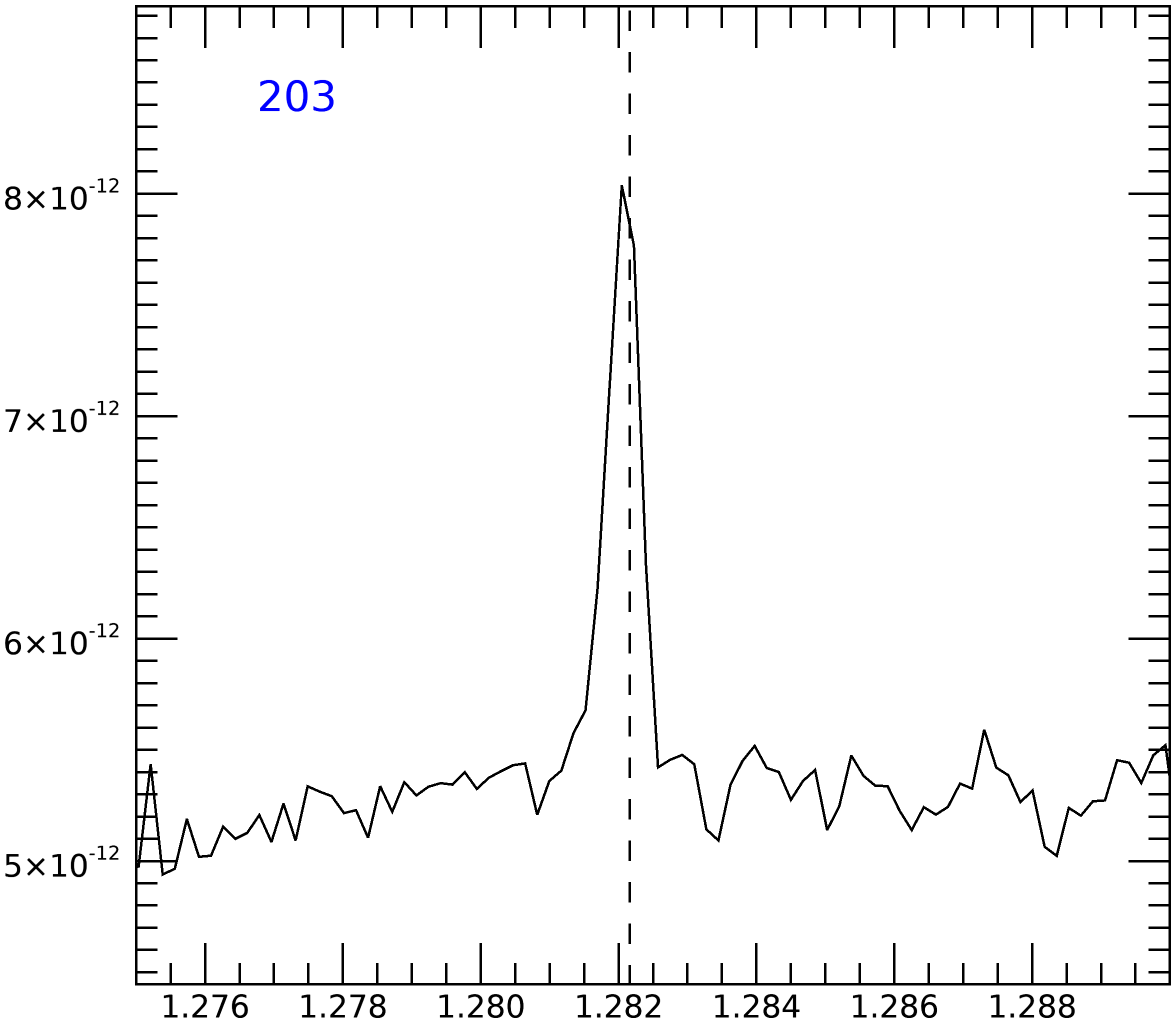}%
 
 \includegraphics[width=0.2\textwidth]{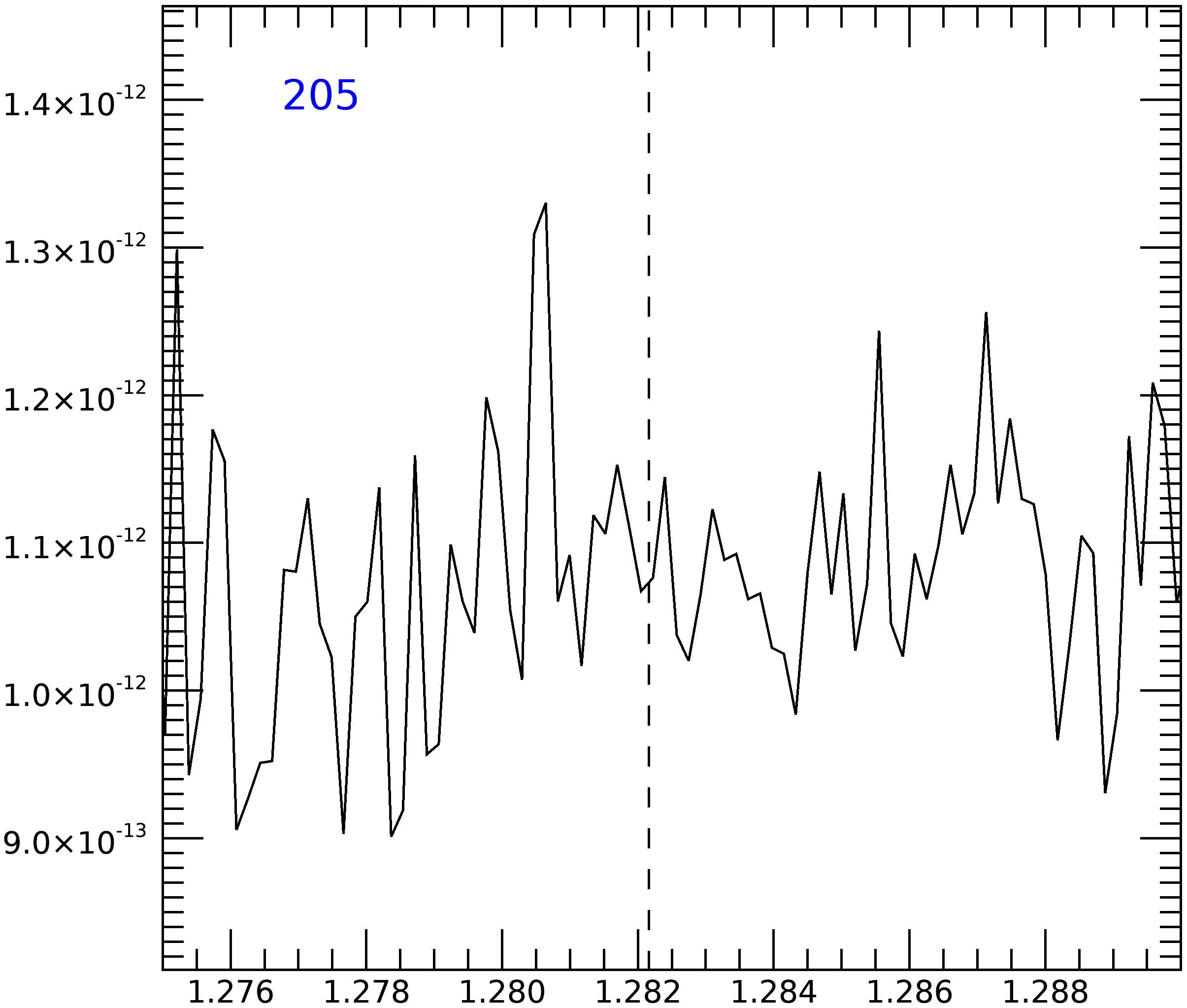}%
 \includegraphics[width=0.2\textwidth]{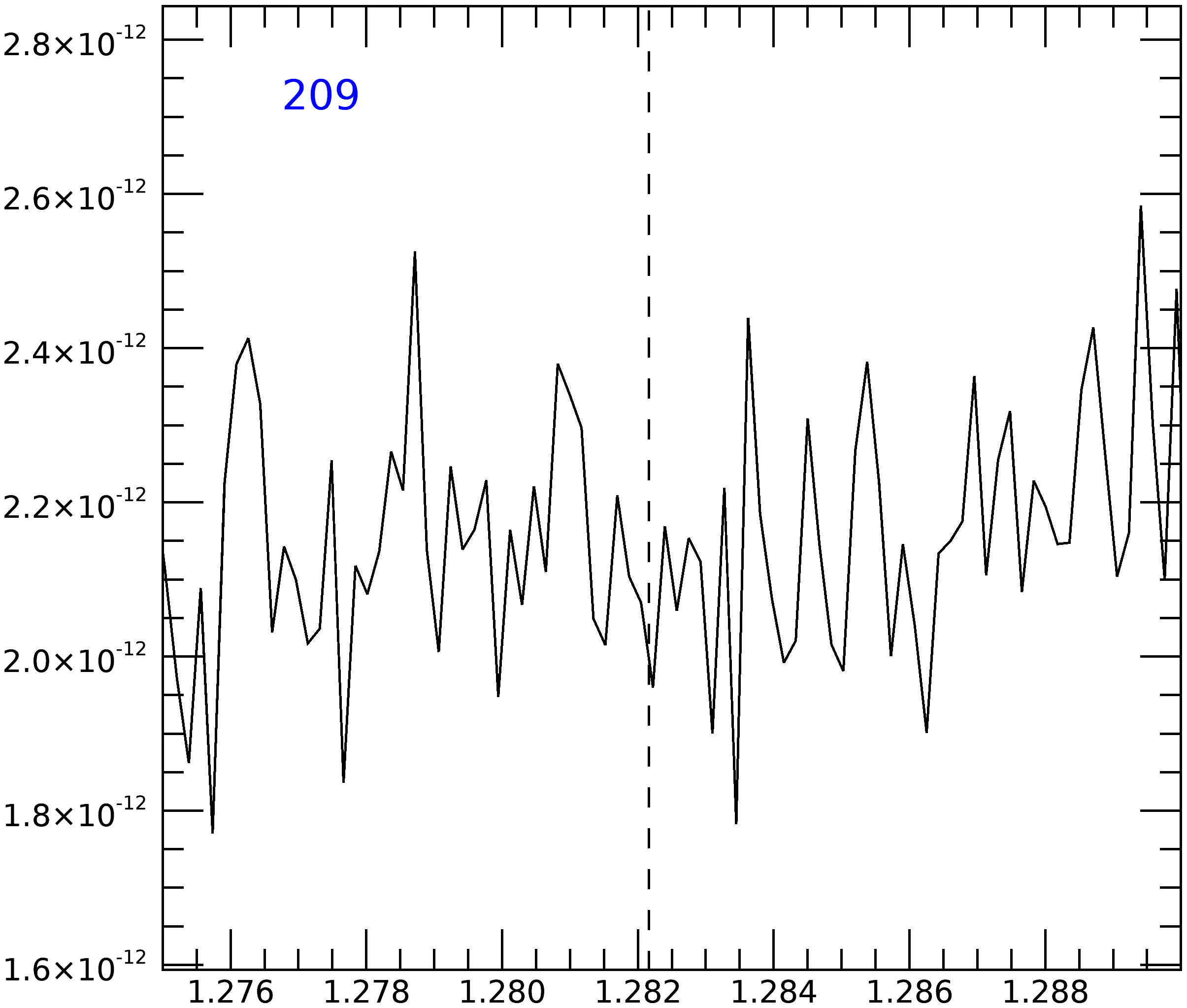}%
 \includegraphics[width=0.2\textwidth]{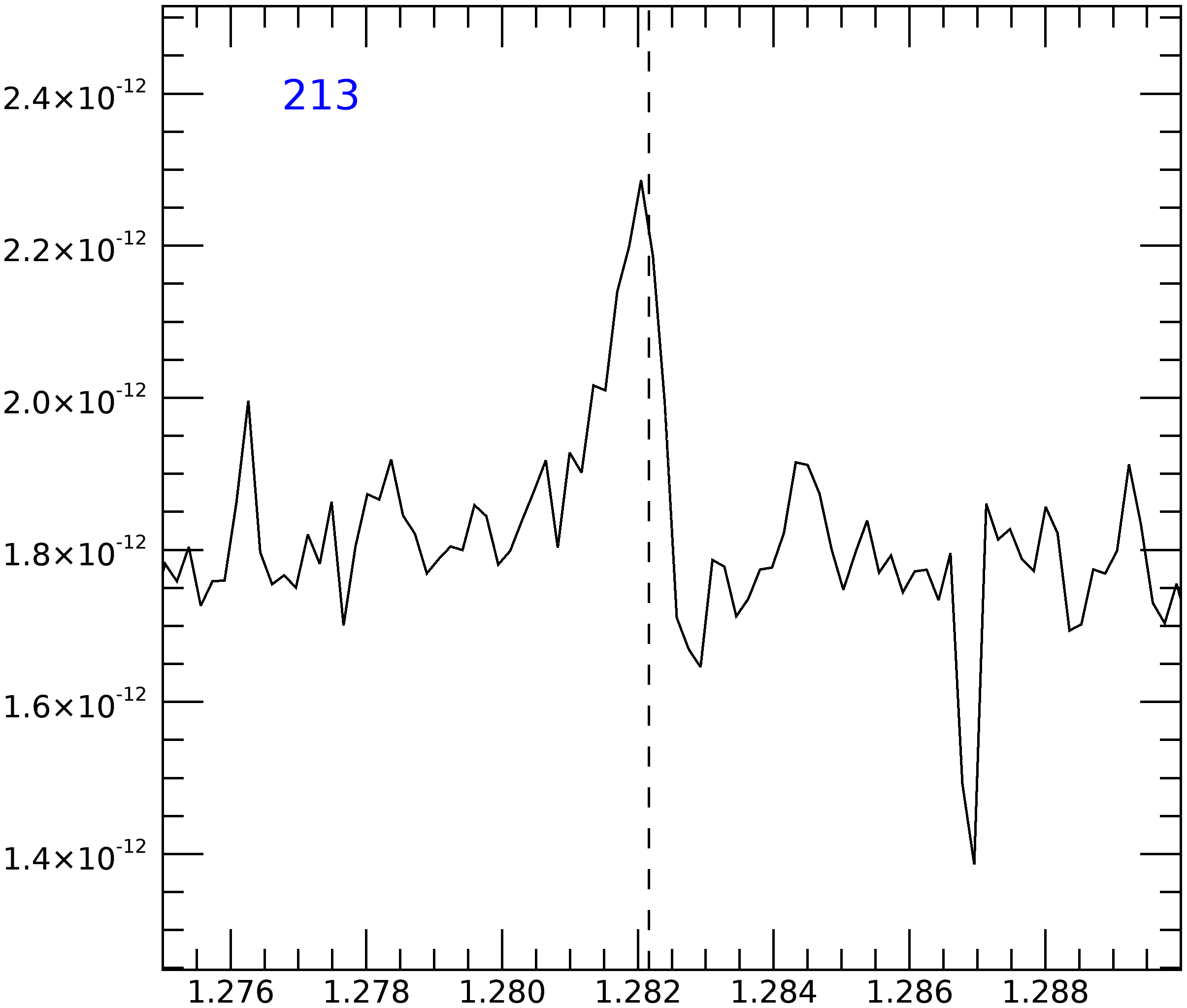}%
 \includegraphics[width=0.2\textwidth]{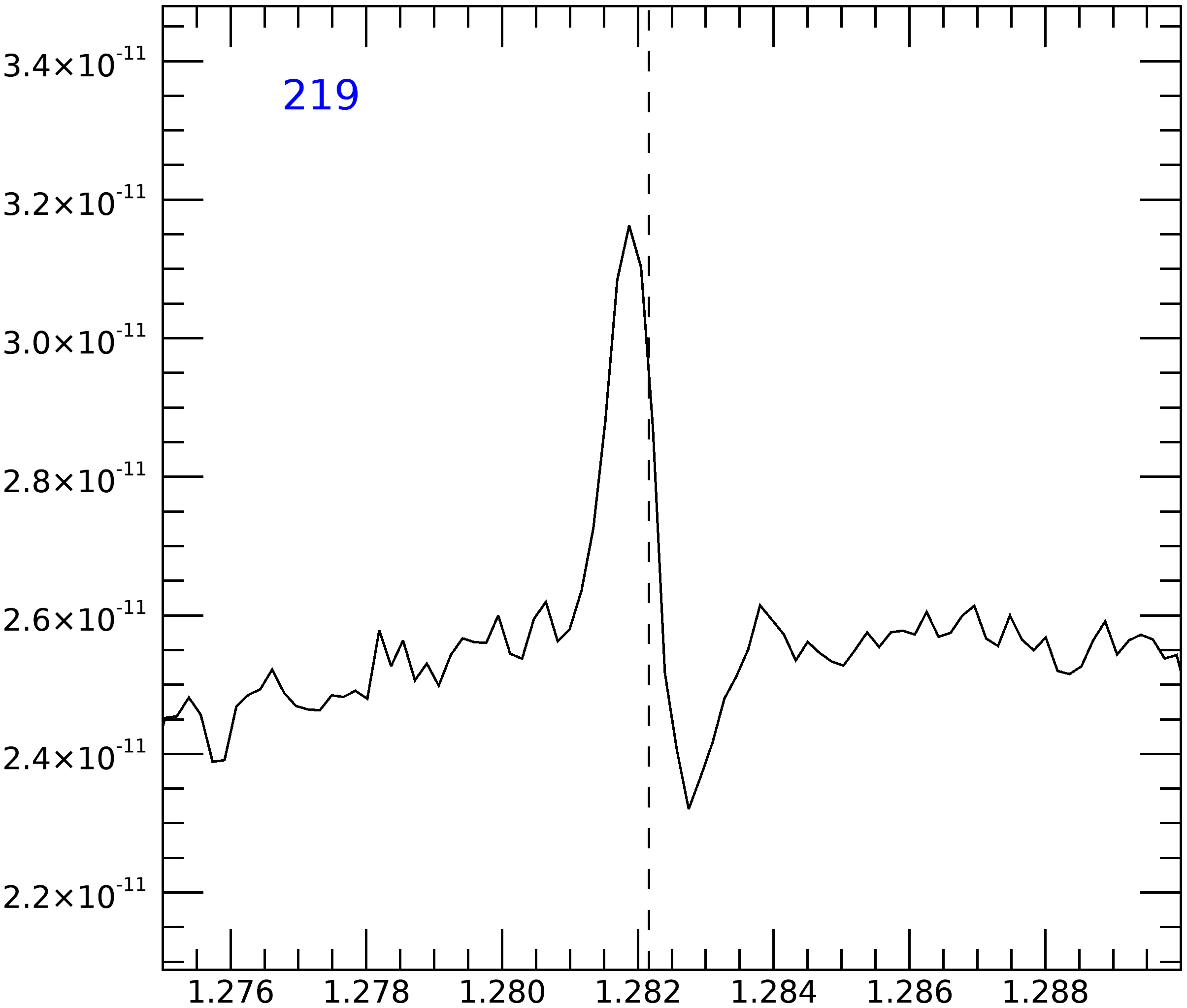}%
 \includegraphics[width=0.2\textwidth]{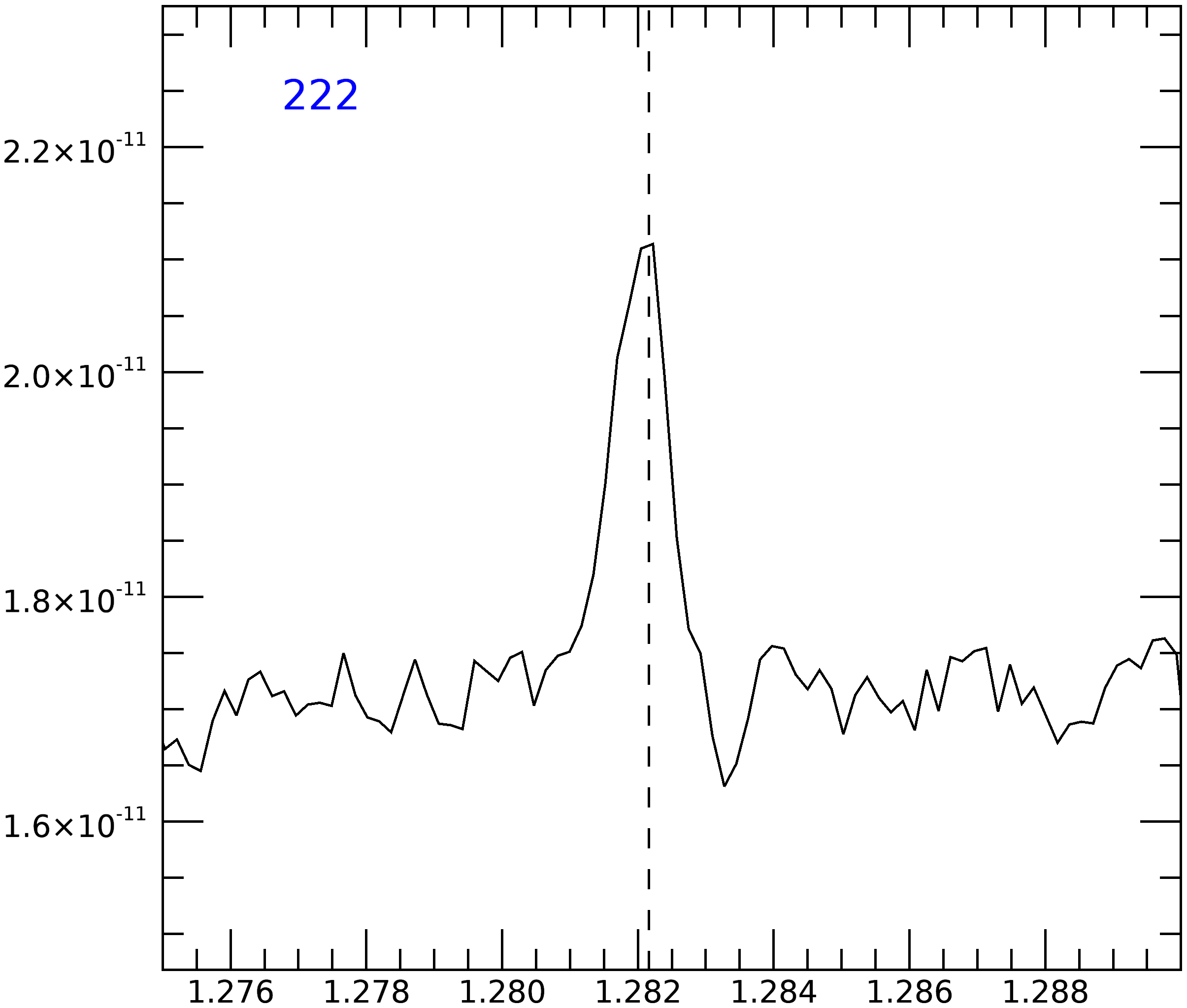}%

 \includegraphics[width=0.2\textwidth]{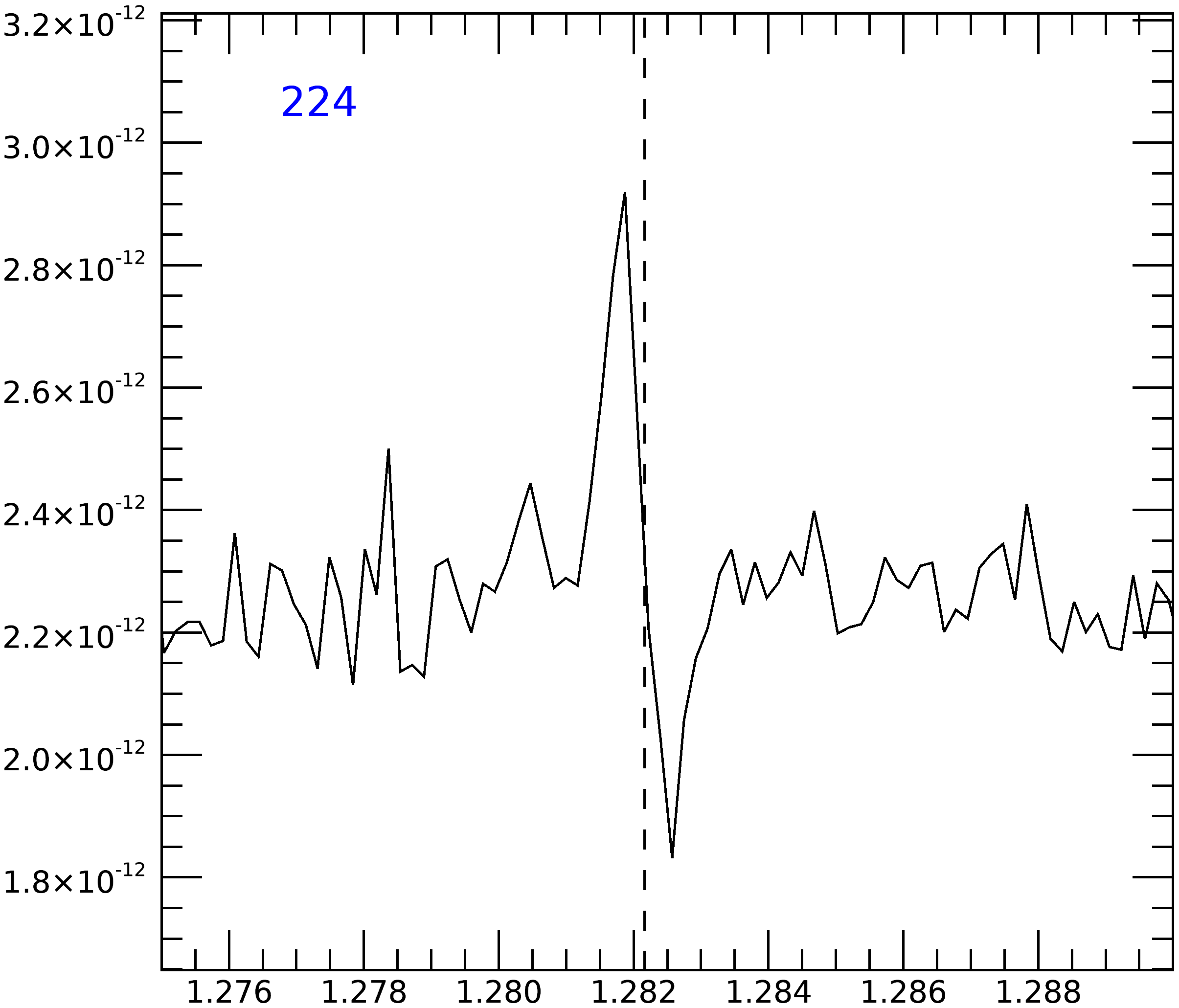}%
 \includegraphics[width=0.2\textwidth]{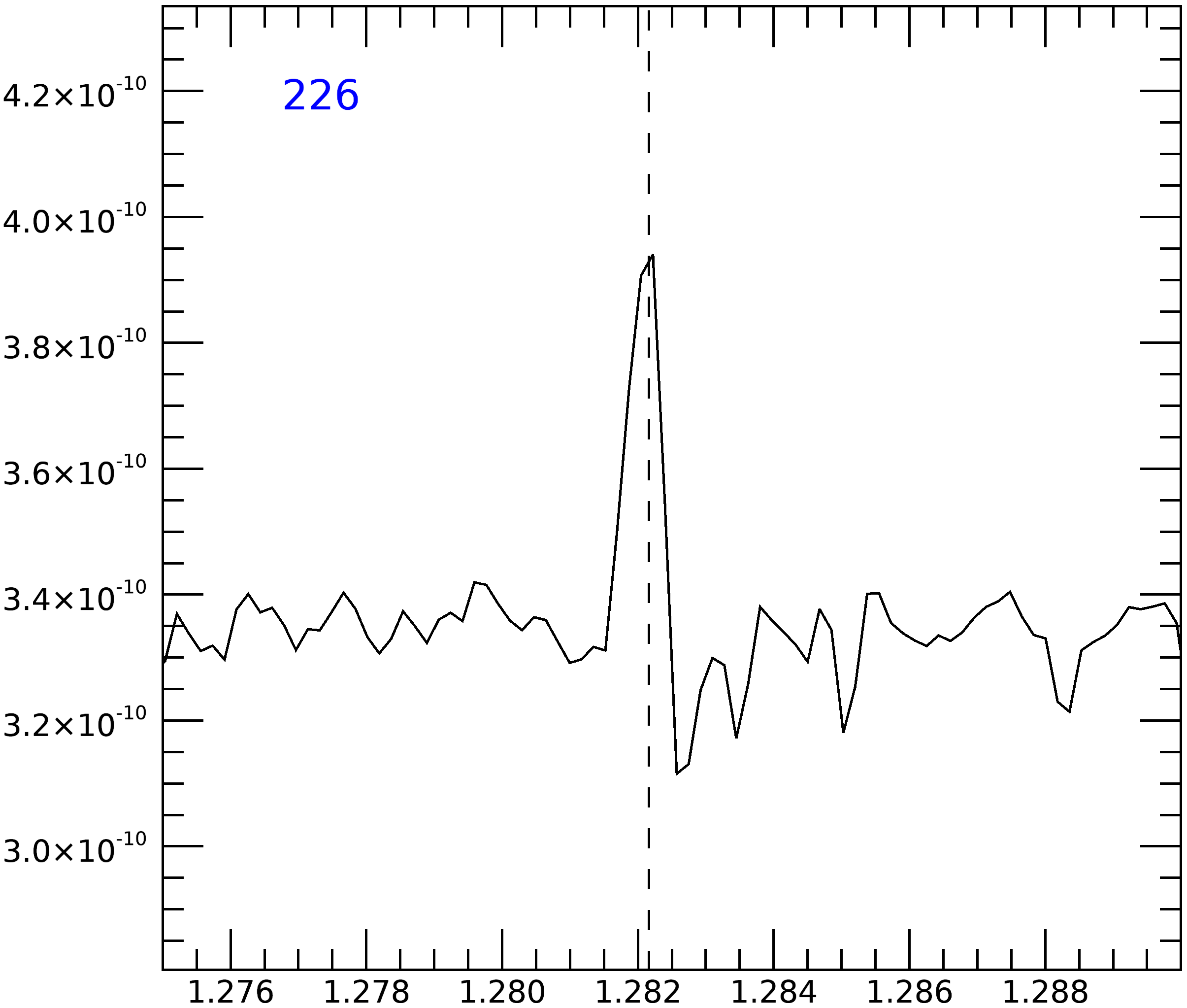}%
 \includegraphics[width=0.2\textwidth]{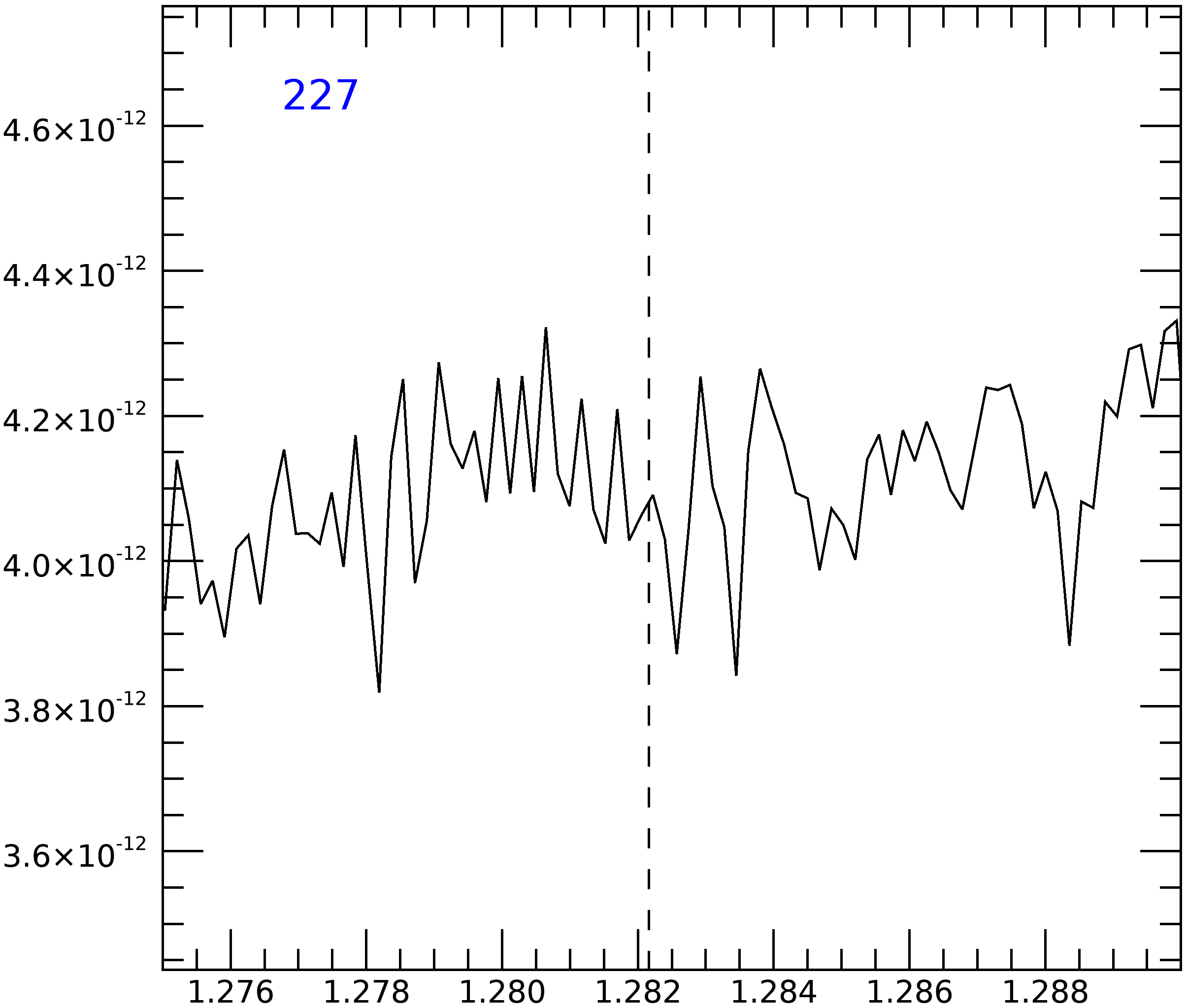}%
 \includegraphics[width=0.2\textwidth]{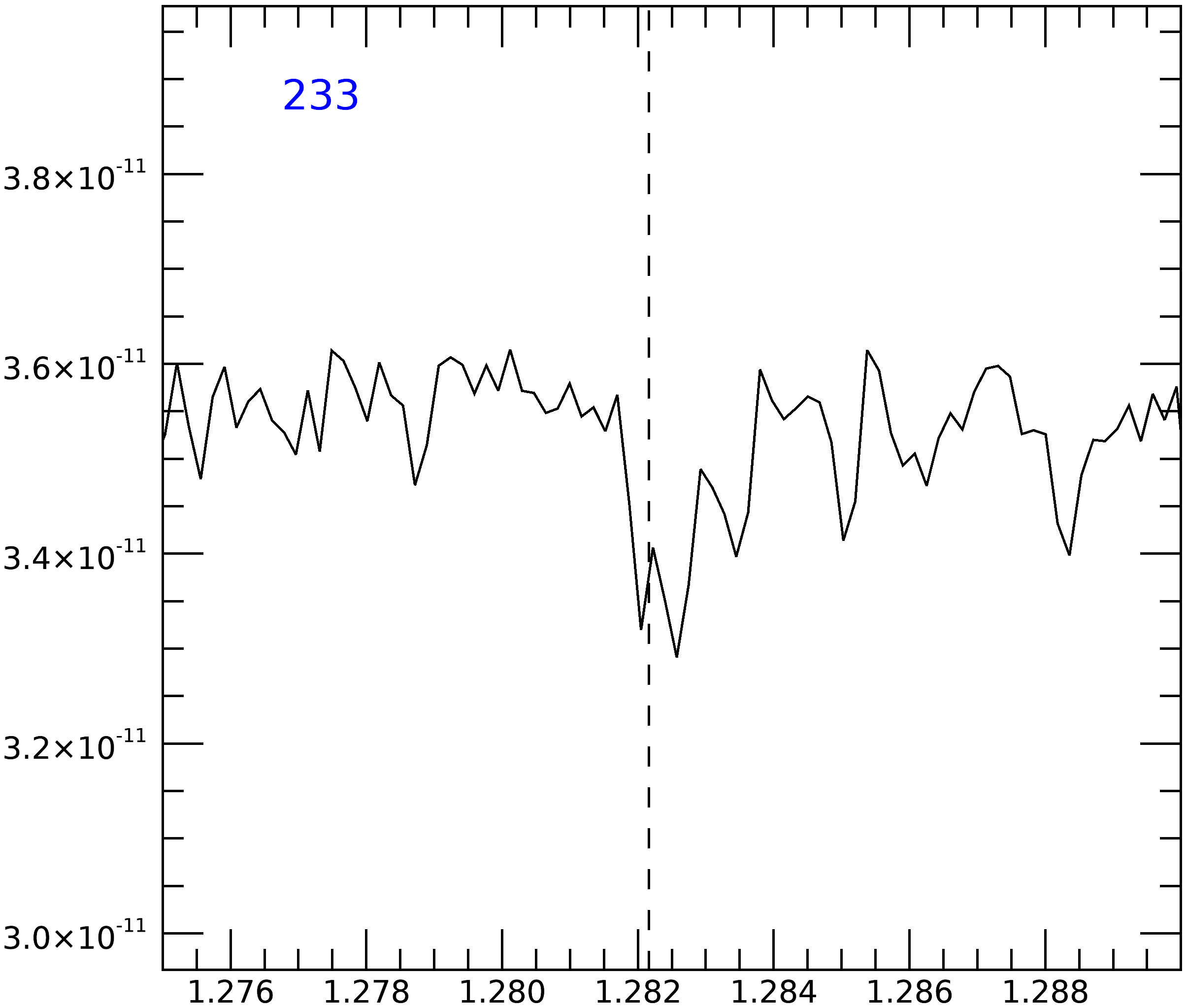}%
 \includegraphics[width=0.2\textwidth]{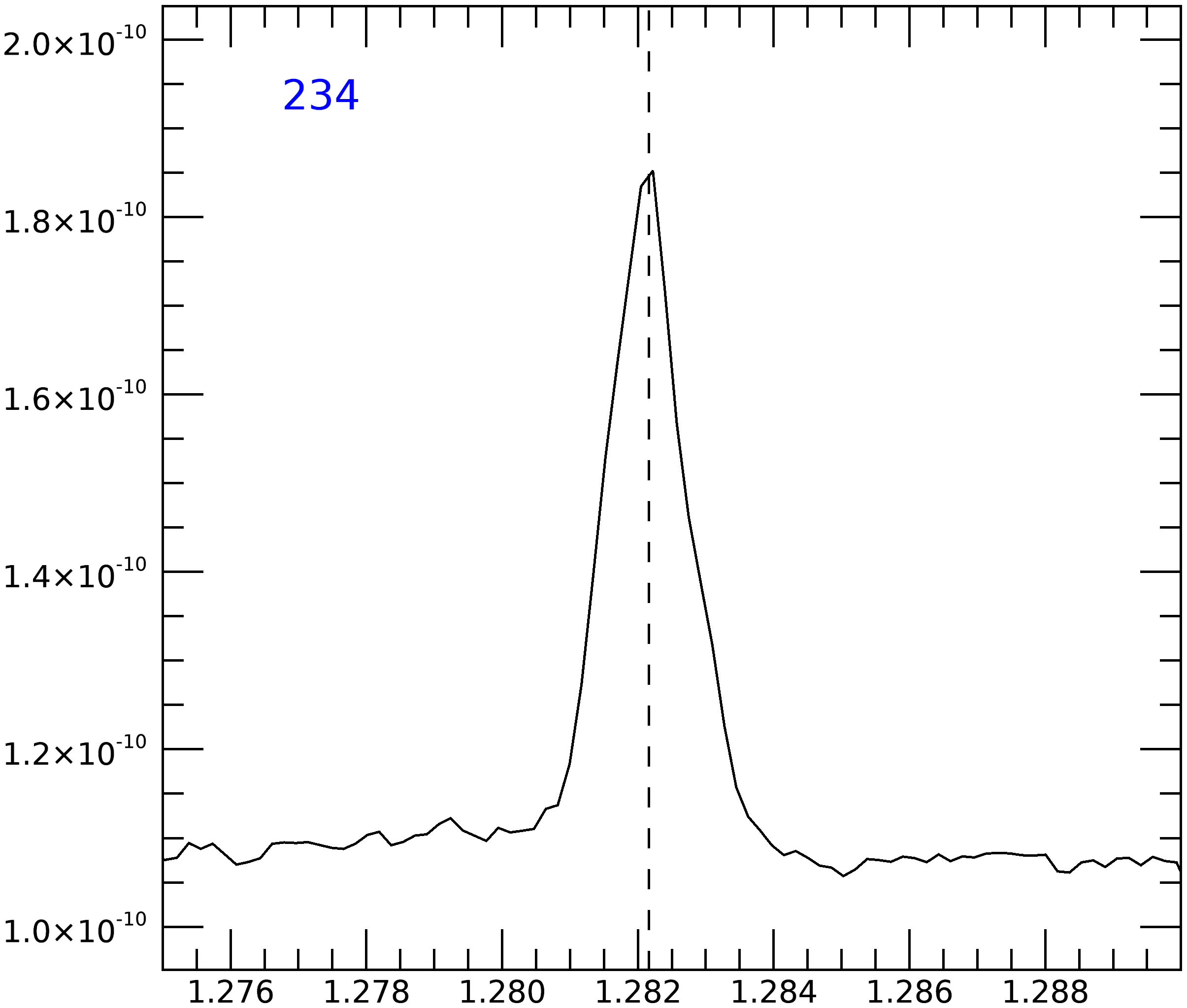}%

 \end{subfigure}
 \caption{\label{fig:lines2a} $\pab$ lines of Class~II objects. The flux is in erg~s$^{-1}$cm$^{-2}\mu$m$^{-1}$. 
 }
\end{figure*}

\begin{figure*} 
 \centering
 \begin{subfigure}{\textwidth}
 \centering

 \includegraphics[width=0.2\textwidth]{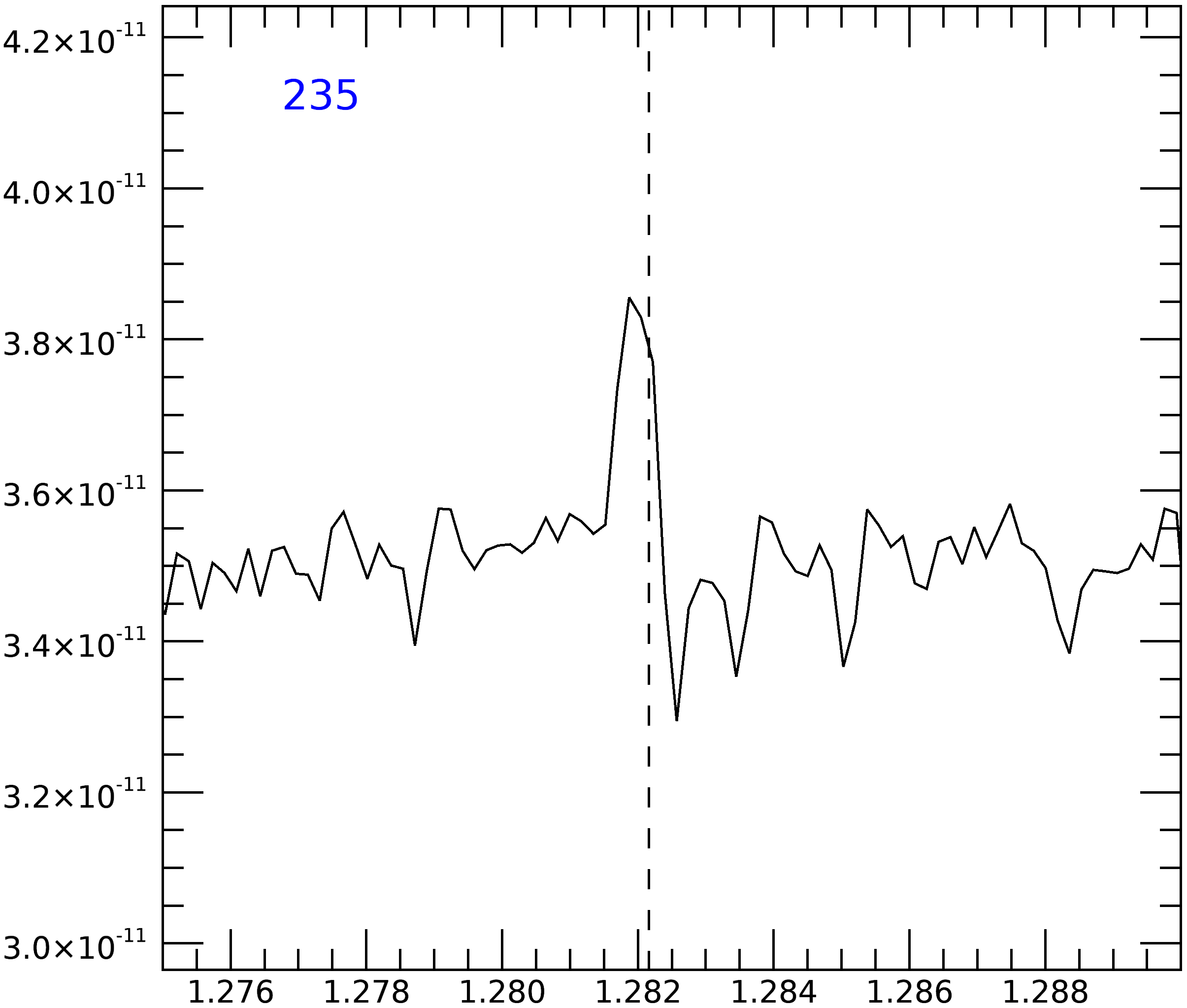}%
 \includegraphics[width=0.2\textwidth]{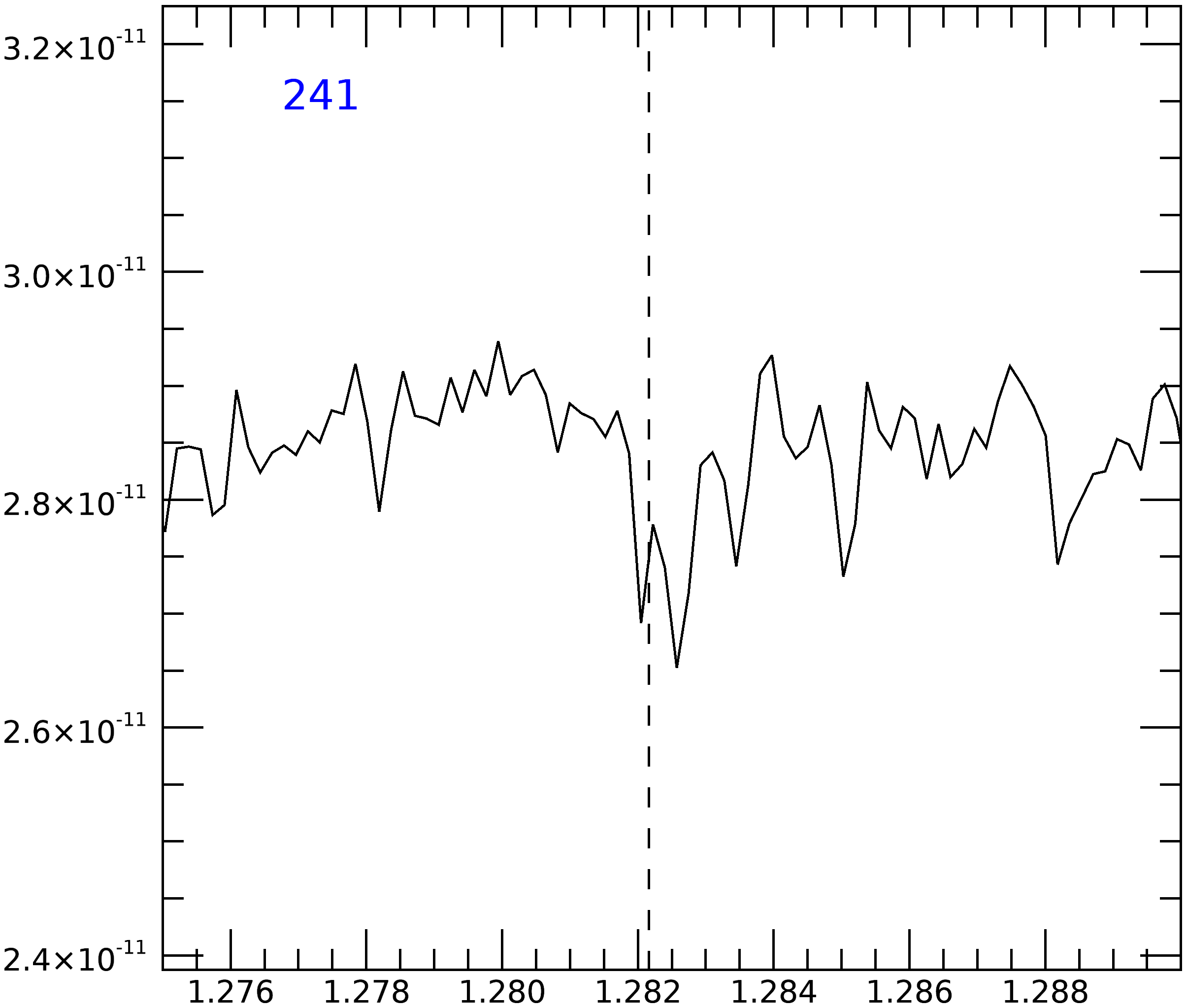}%

 \end{subfigure}
 \caption{\label{fig:lines2b}$\pab$ lines of Class~II objects. The flux is in erg~s$^{-1}$cm$^{-2}\mu$m$^{-1}$.} 
\end{figure*}

\begin{figure*} 
\centering
 \begin{subfigure}{\textwidth}
 \centering
 \includegraphics[width=0.2\textwidth]{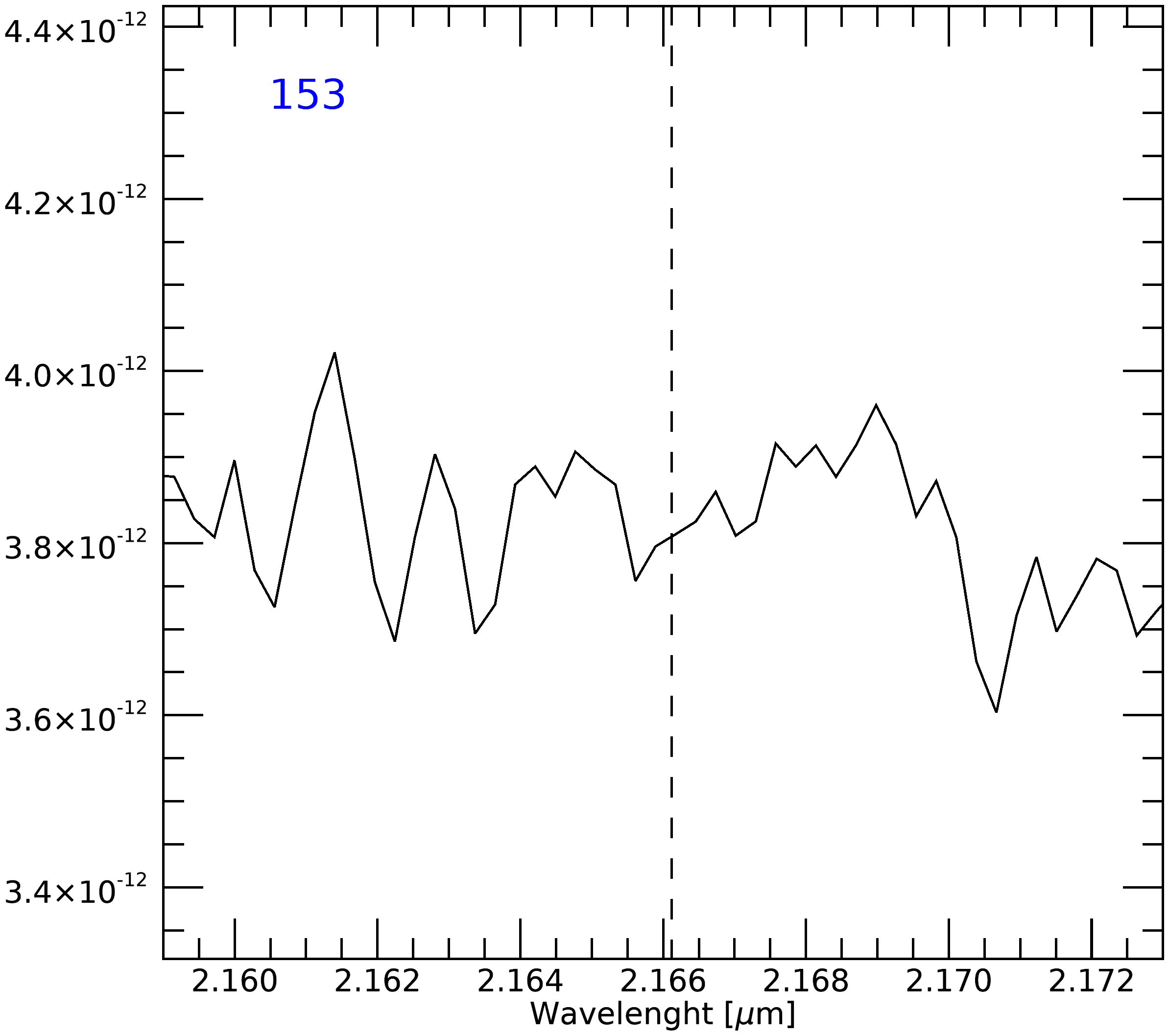}%
 \includegraphics[width=0.2\textwidth]{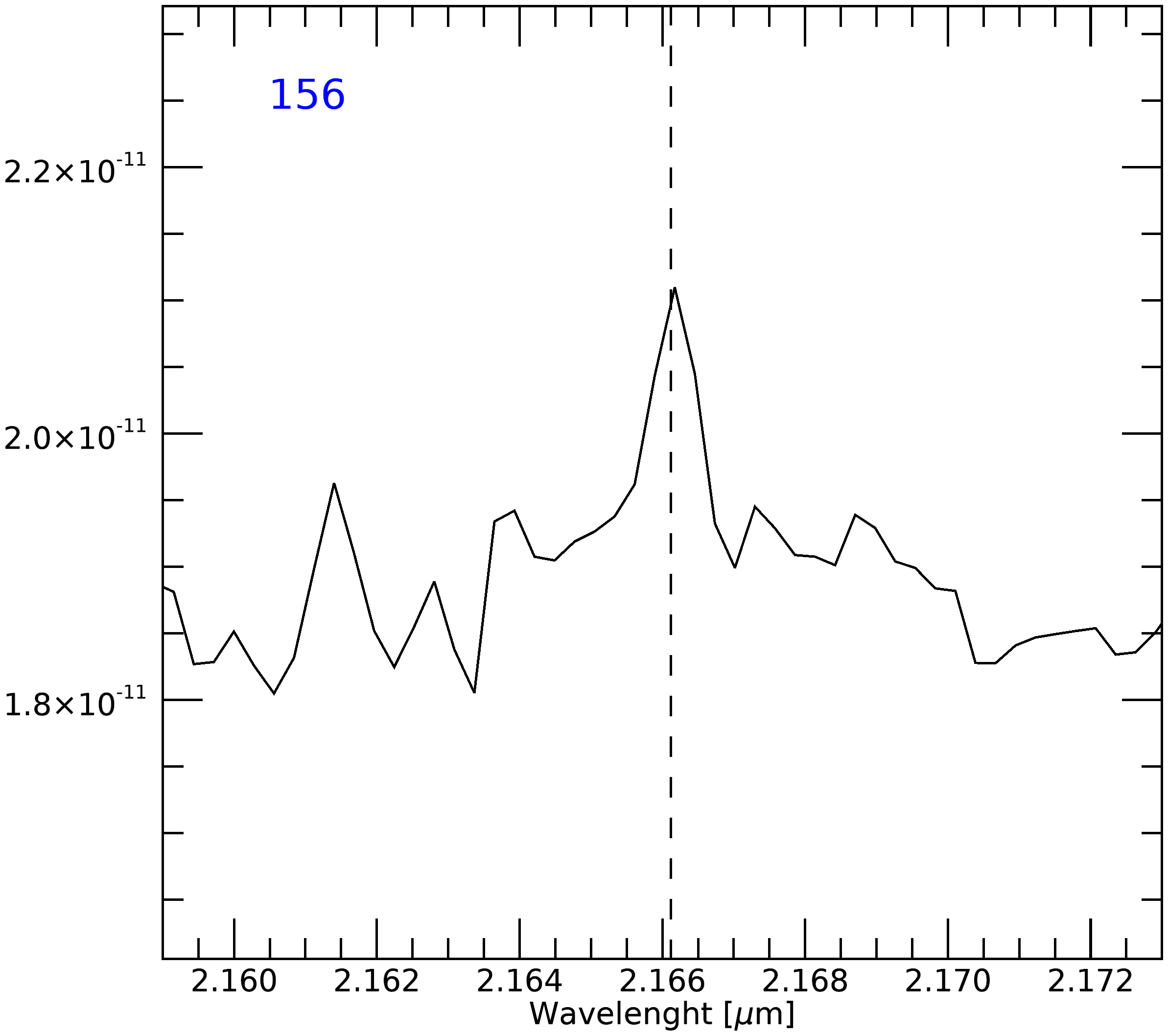}%
 \includegraphics[width=0.2\textwidth]{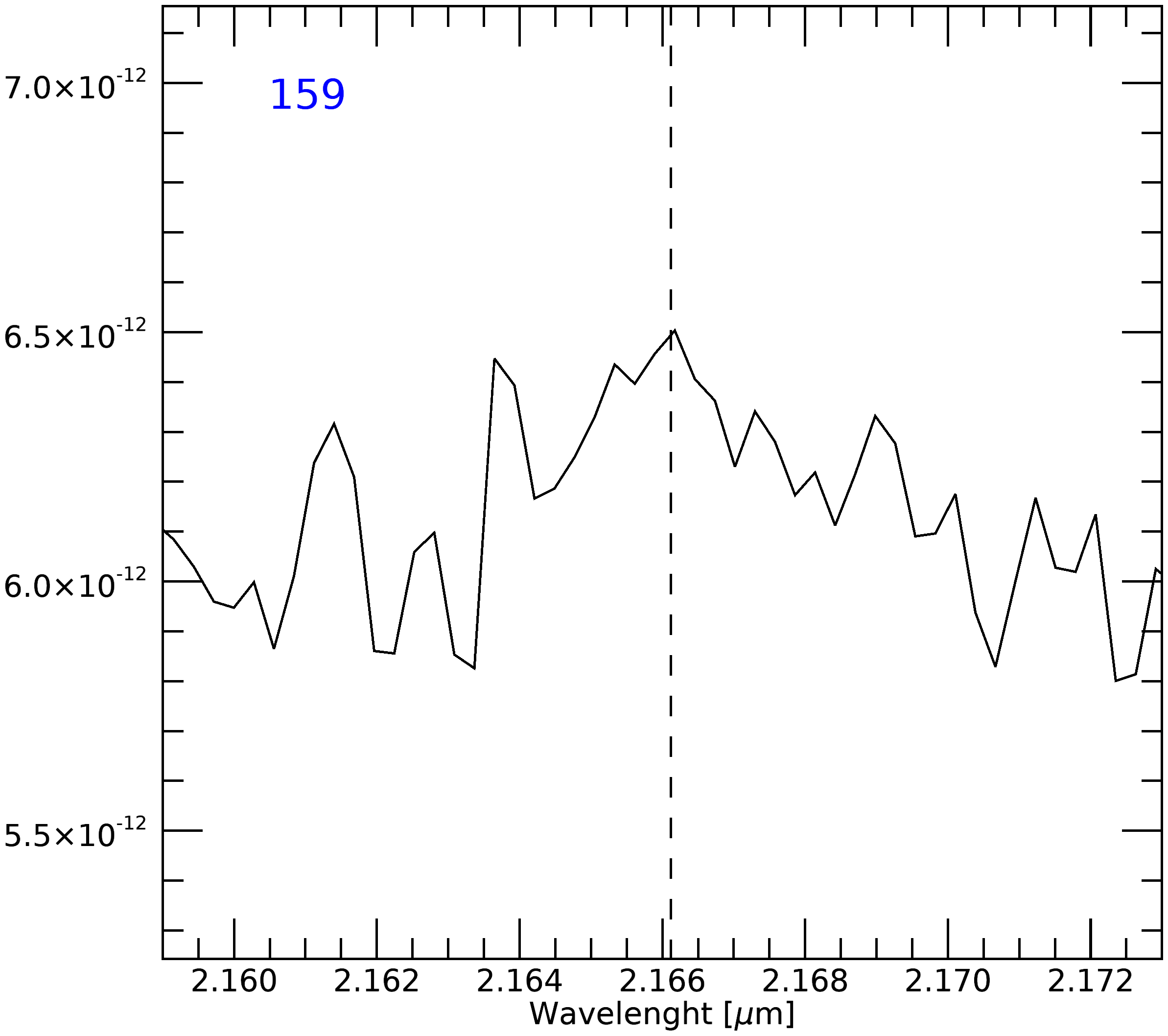}%
 \includegraphics[width=0.2\textwidth]{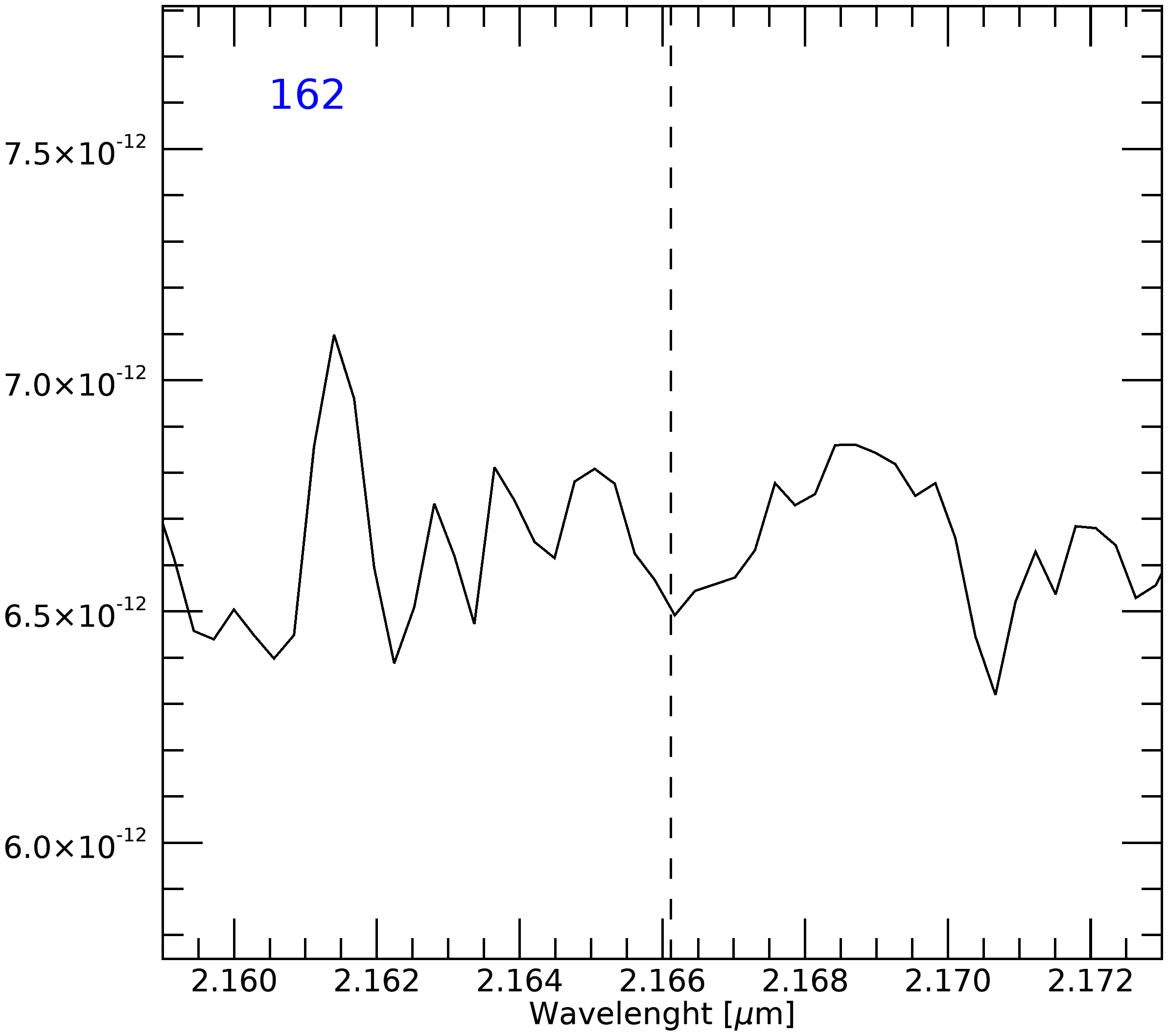}%
 \includegraphics[width=0.2\textwidth]{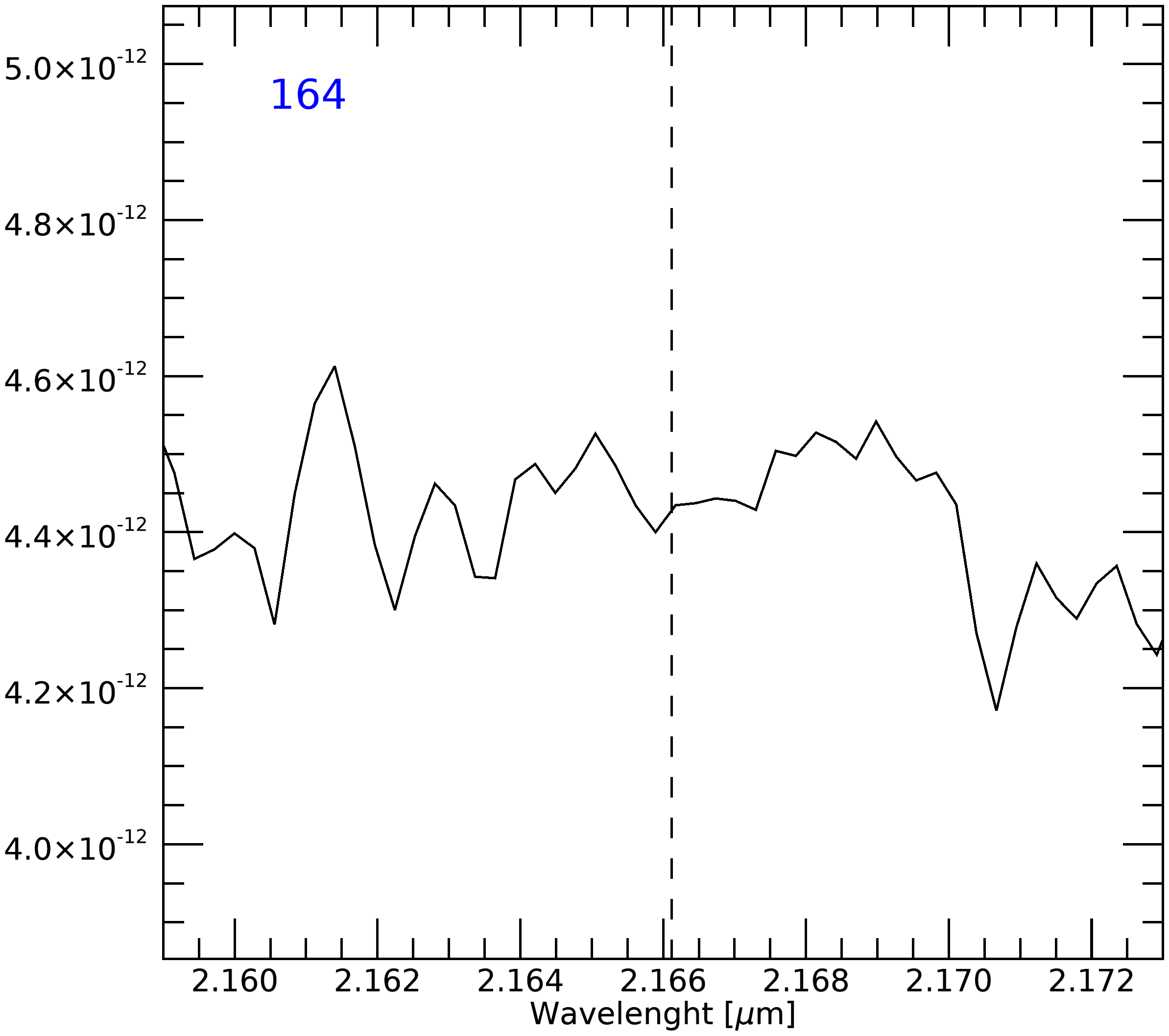}%
 
 \includegraphics[width=0.2\textwidth]{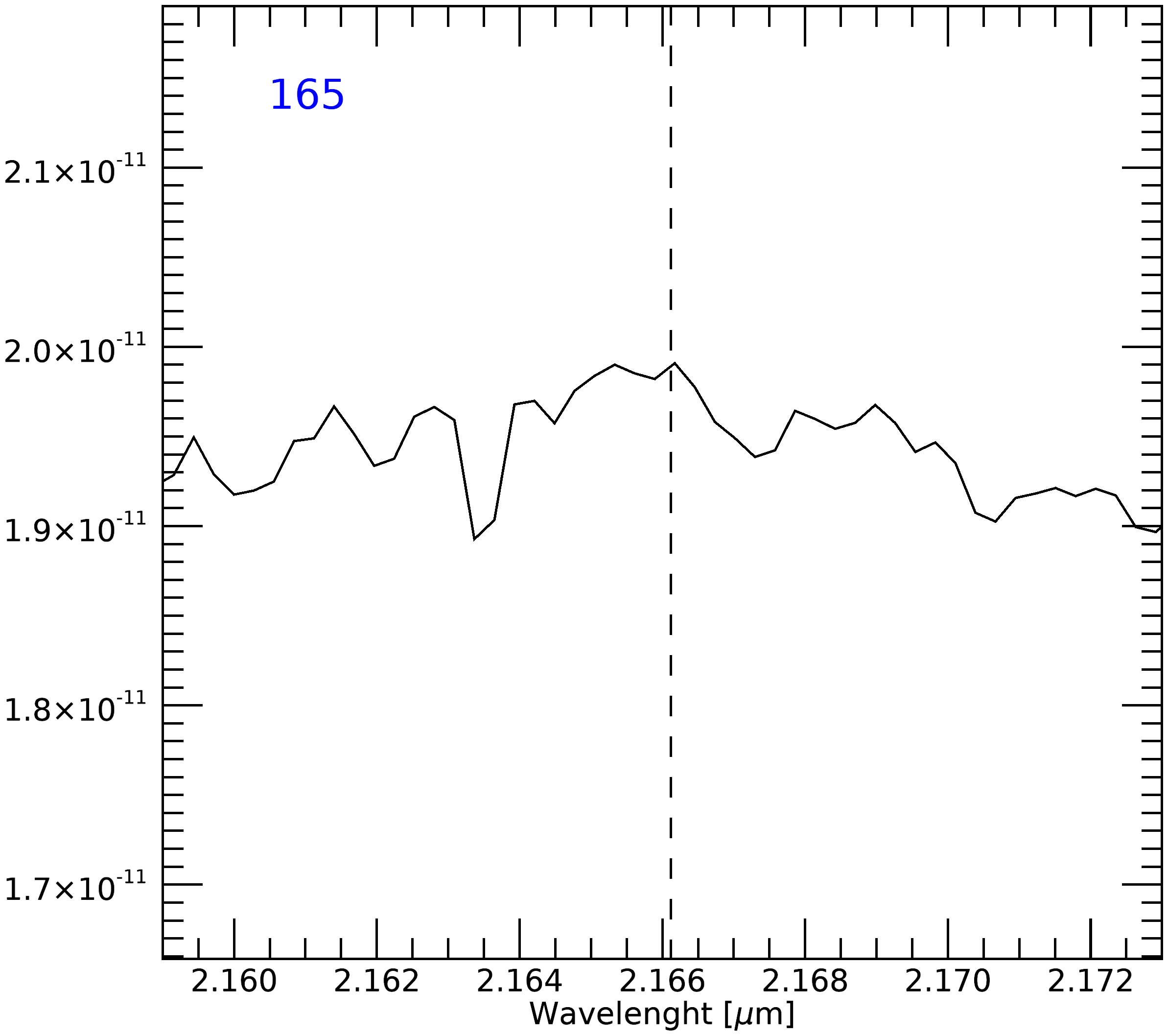}%
 \includegraphics[width=0.2\textwidth]{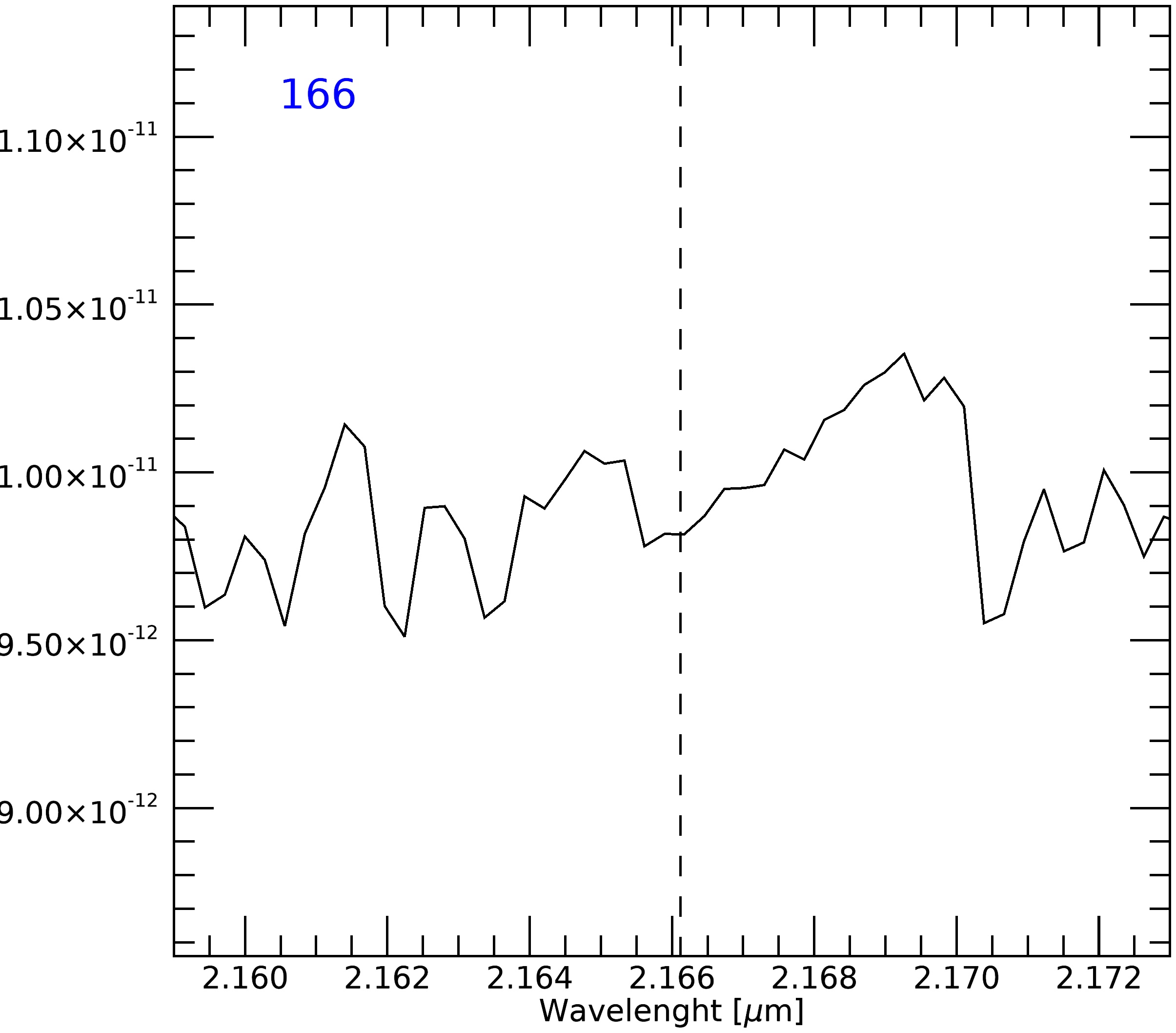}%
 \includegraphics[width=0.2\textwidth]{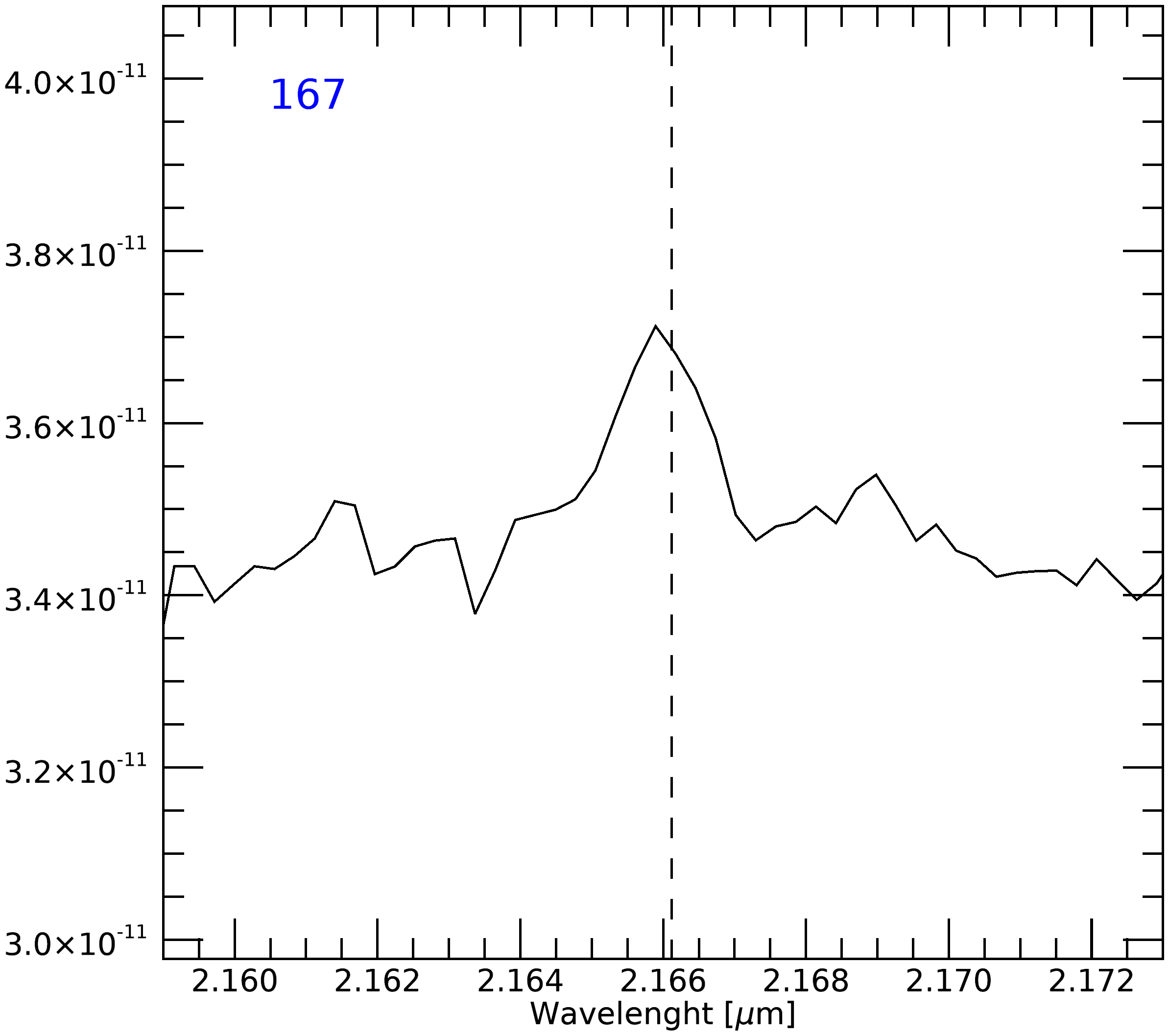}%
 \includegraphics[width=0.2\textwidth]{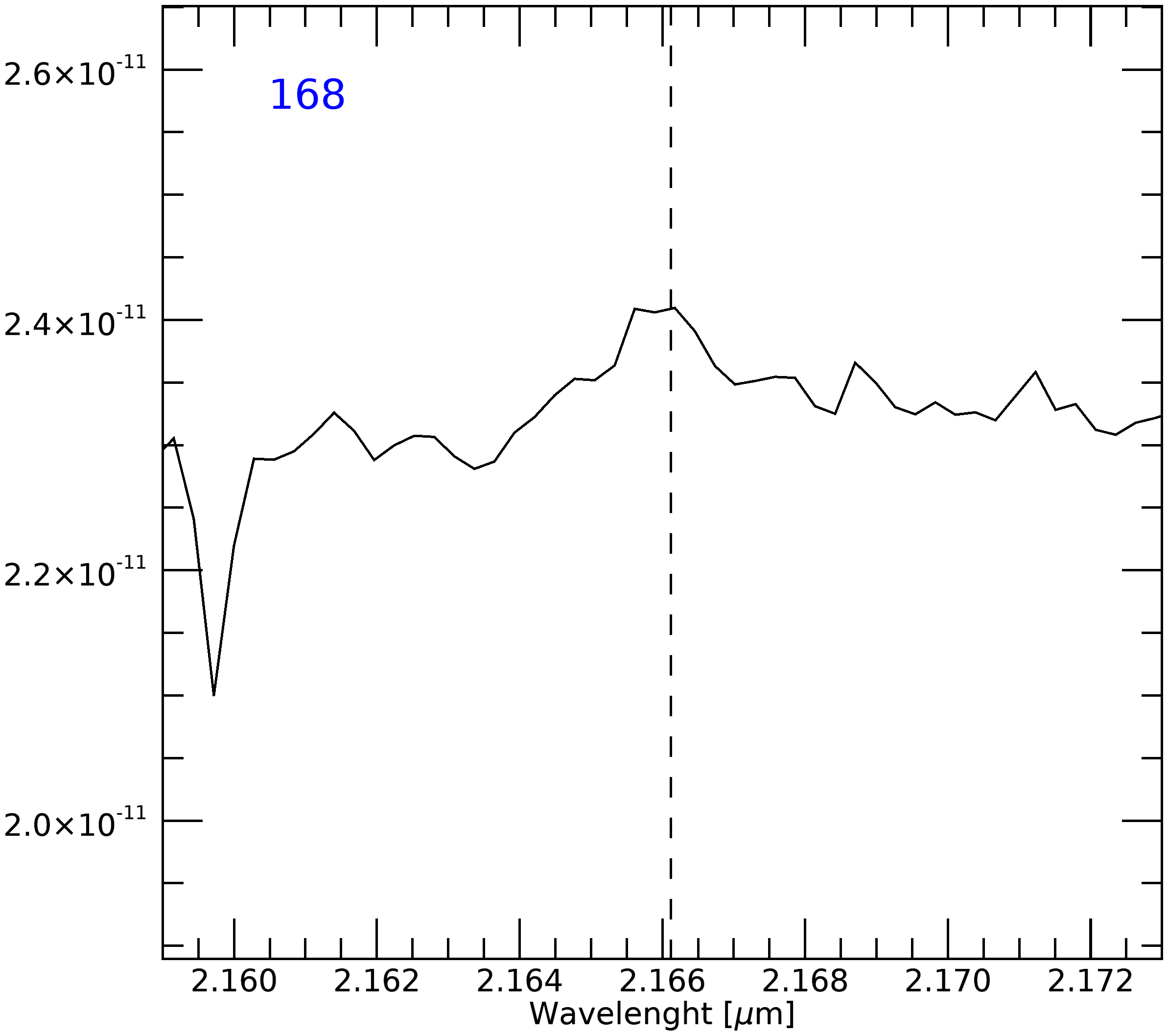}%
 \includegraphics[width=0.2\textwidth]{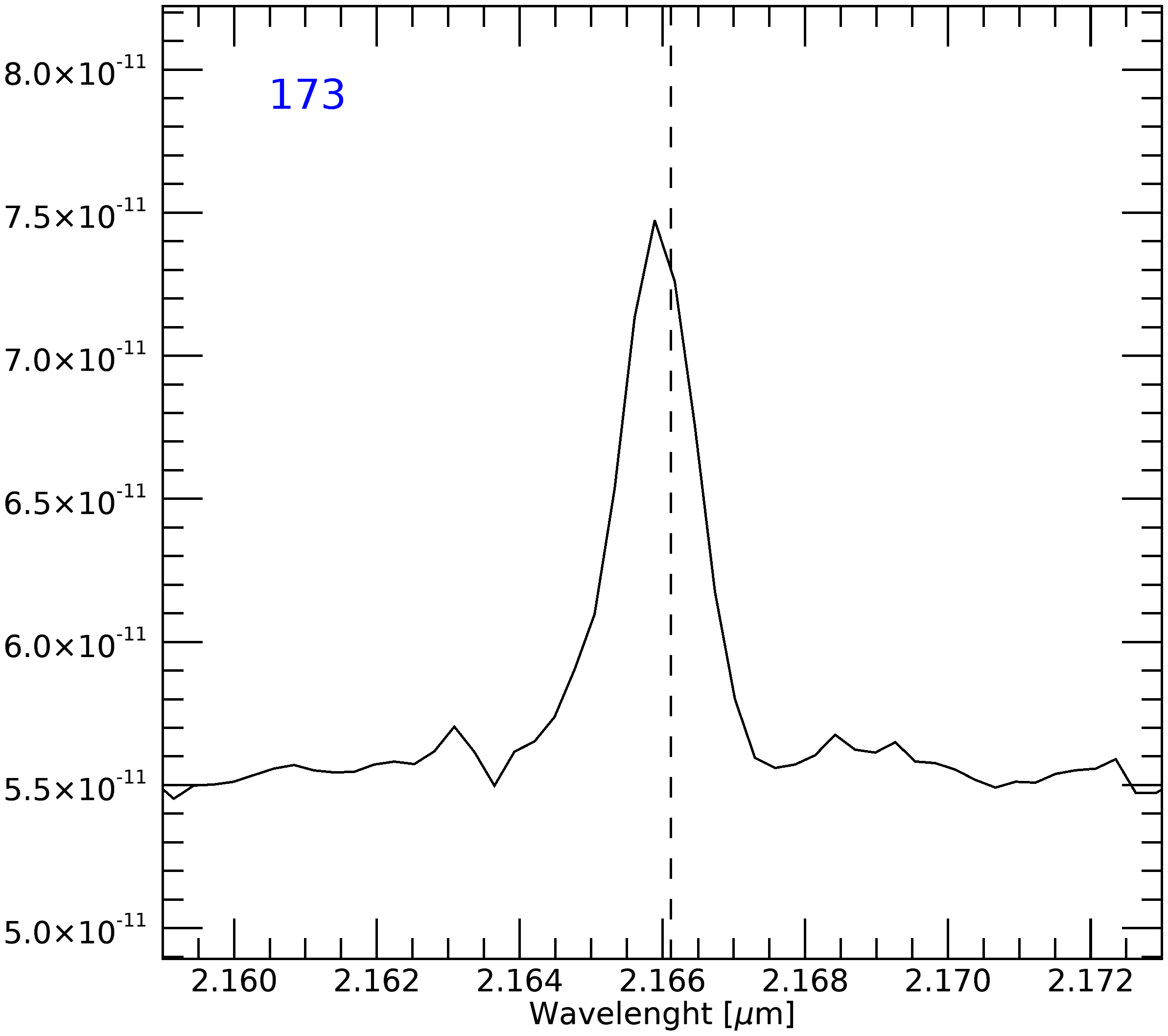}%
 
 \includegraphics[width=0.2\textwidth]{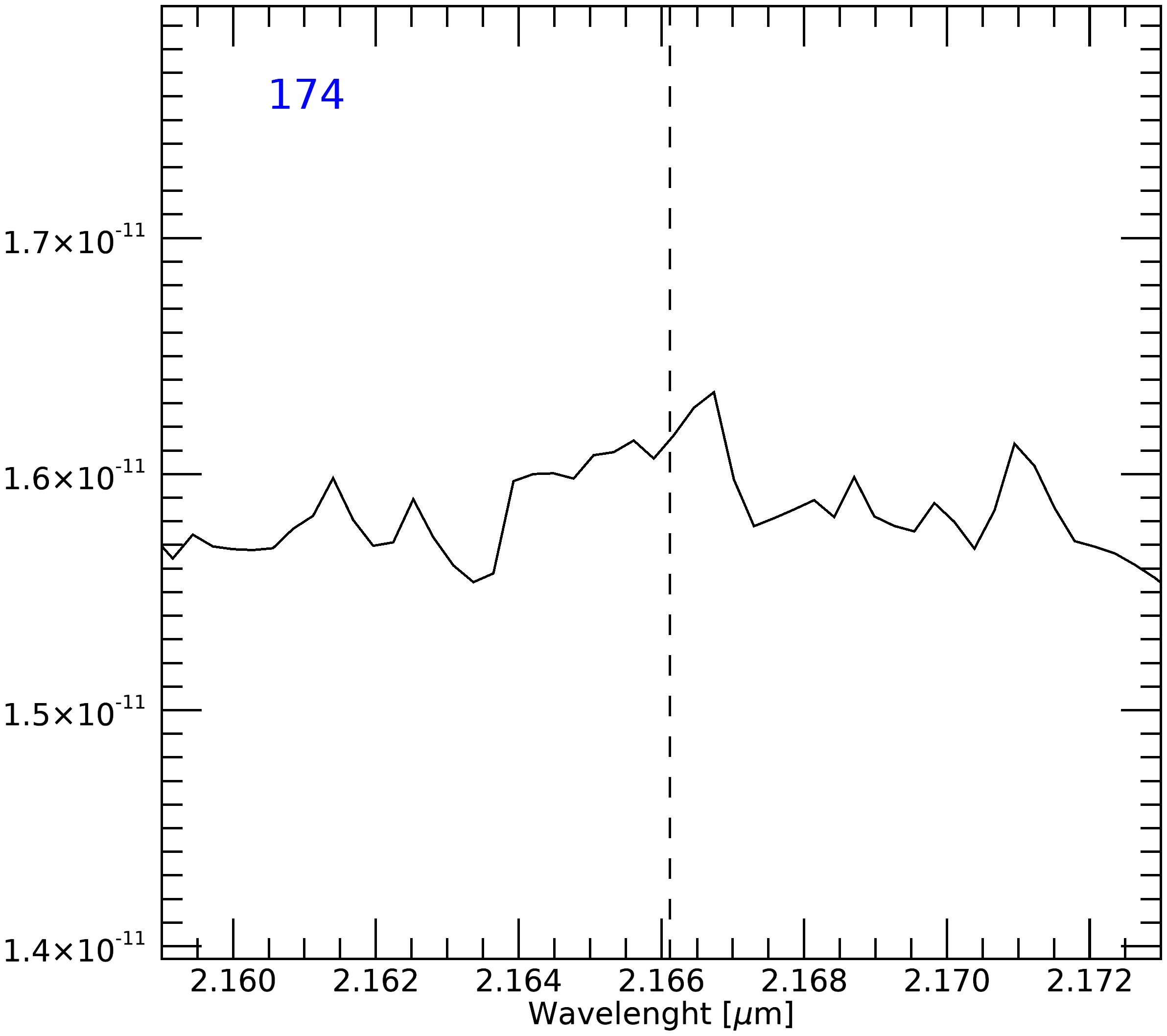}%
 \includegraphics[width=0.2\textwidth]{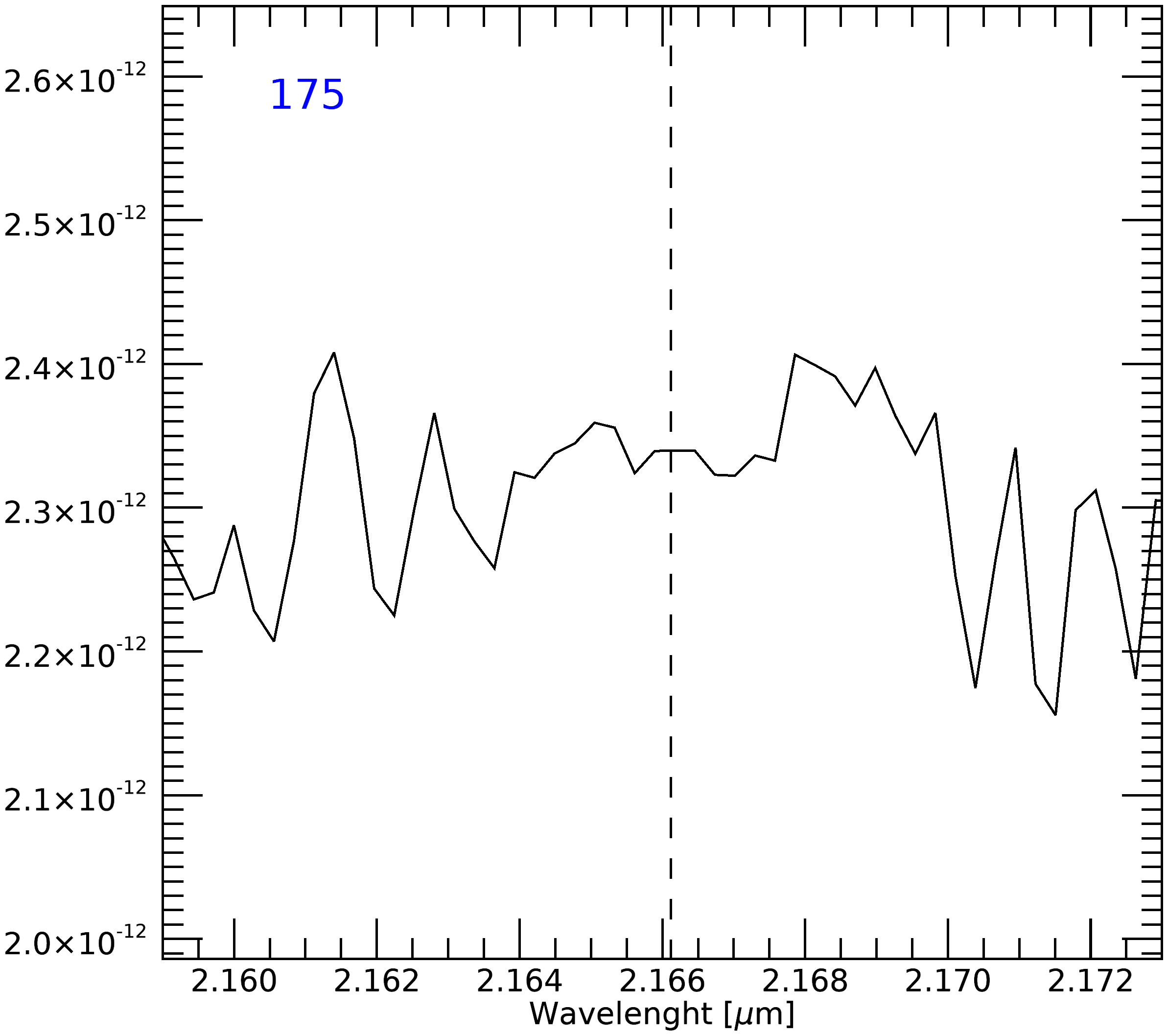}%
 \includegraphics[width=0.2\textwidth]{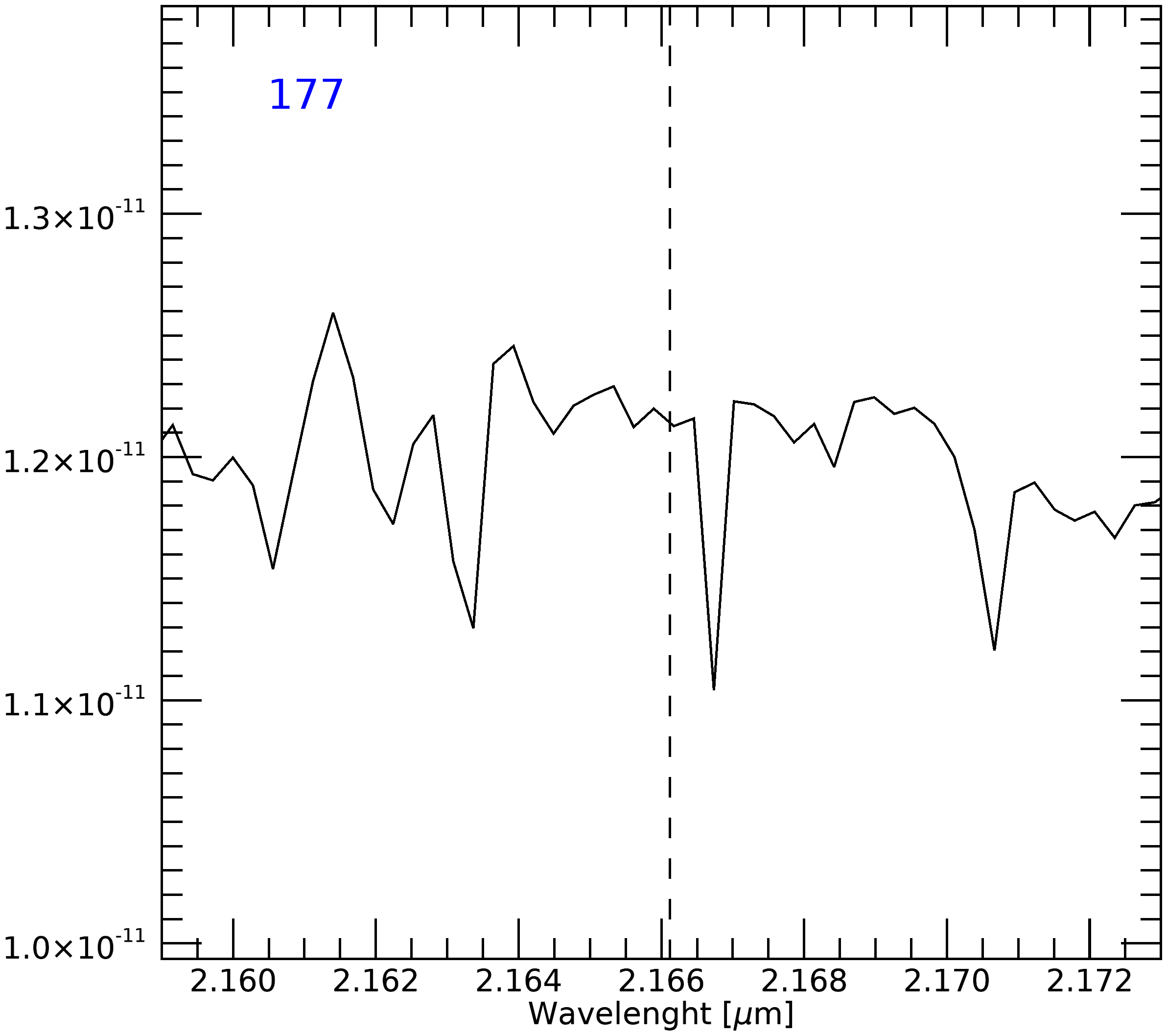}%
 \includegraphics[width=0.2\textwidth]{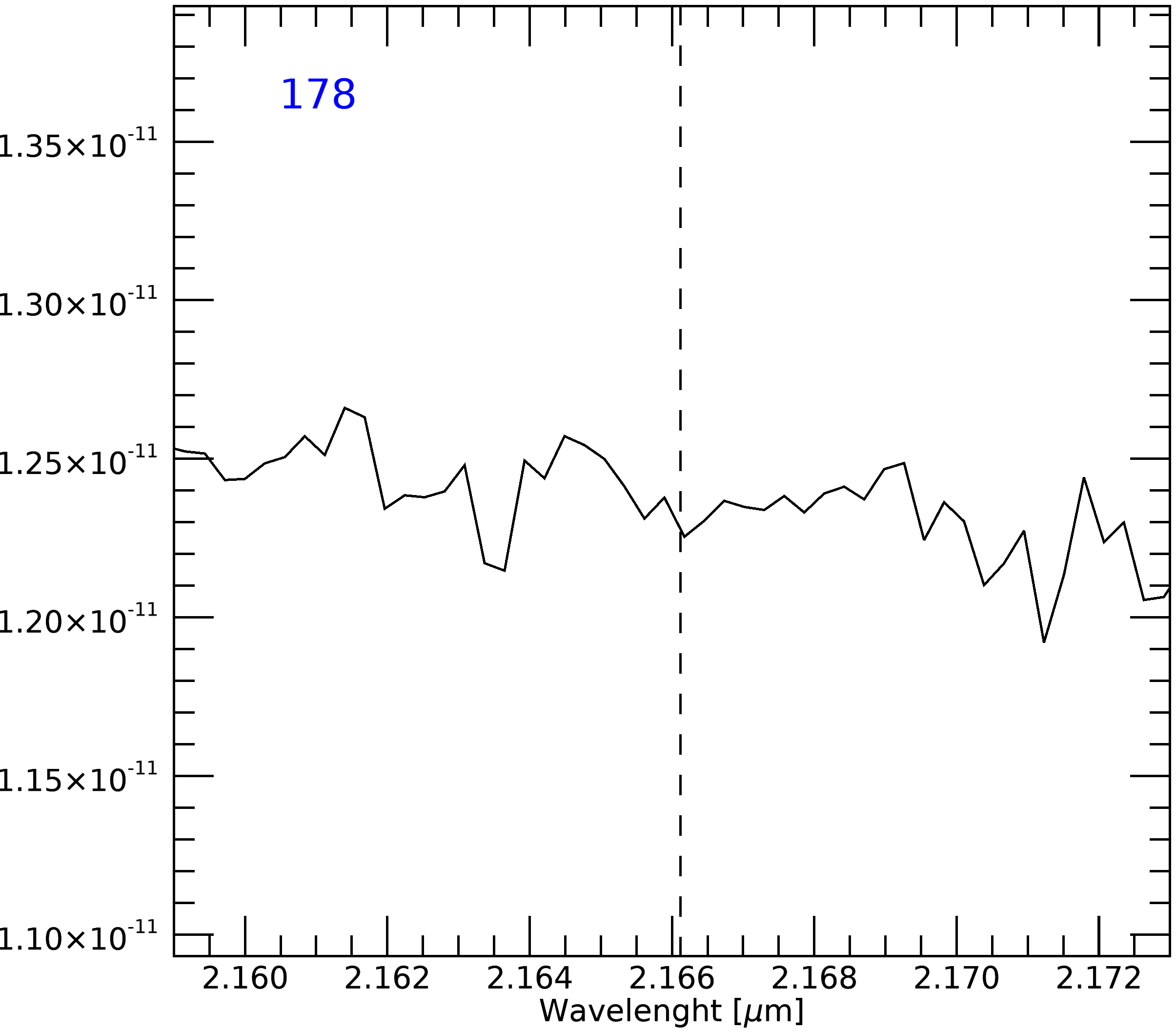}%
 \includegraphics[width=0.2\textwidth]{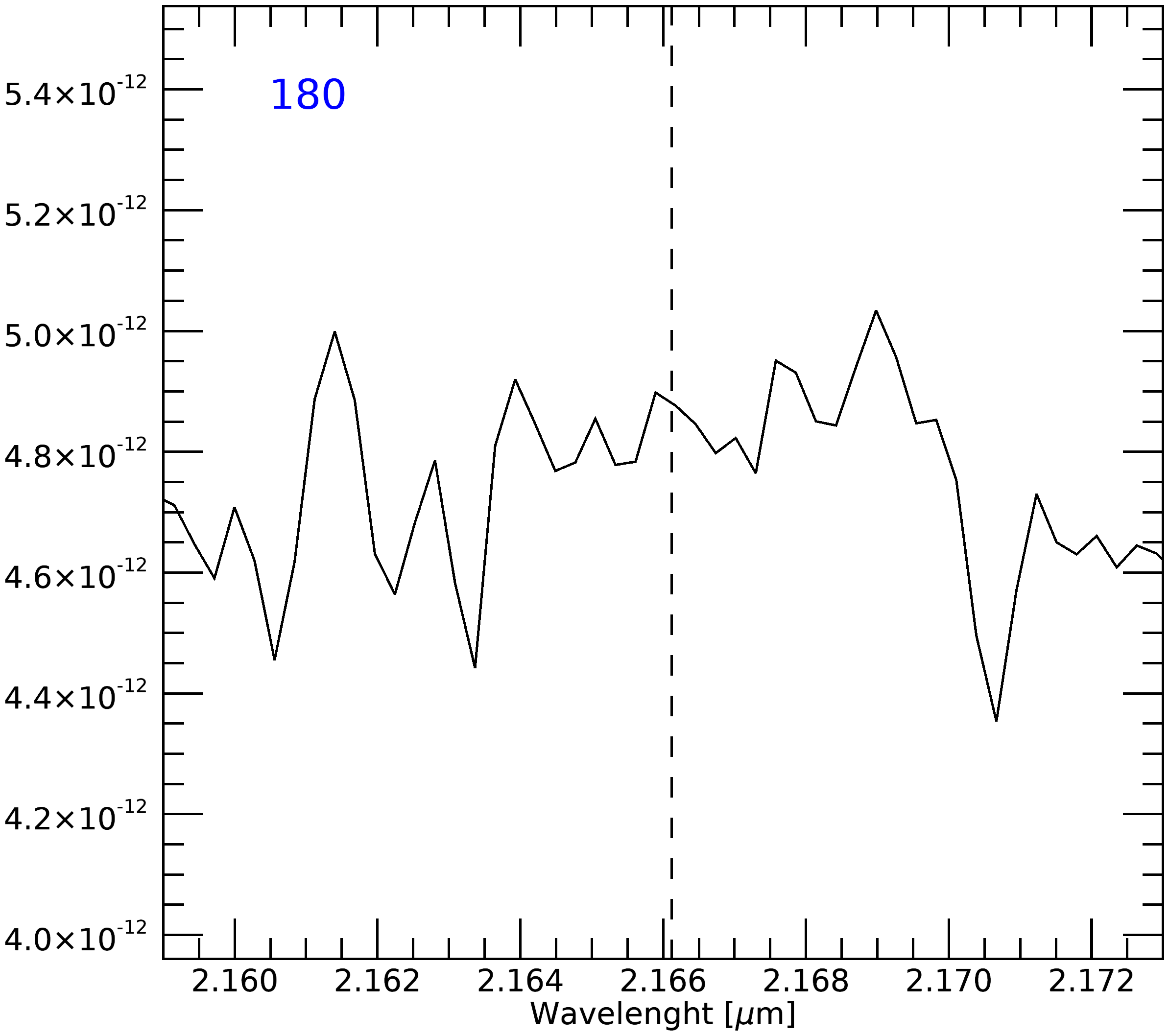}%
 
 \includegraphics[width=0.2\textwidth]{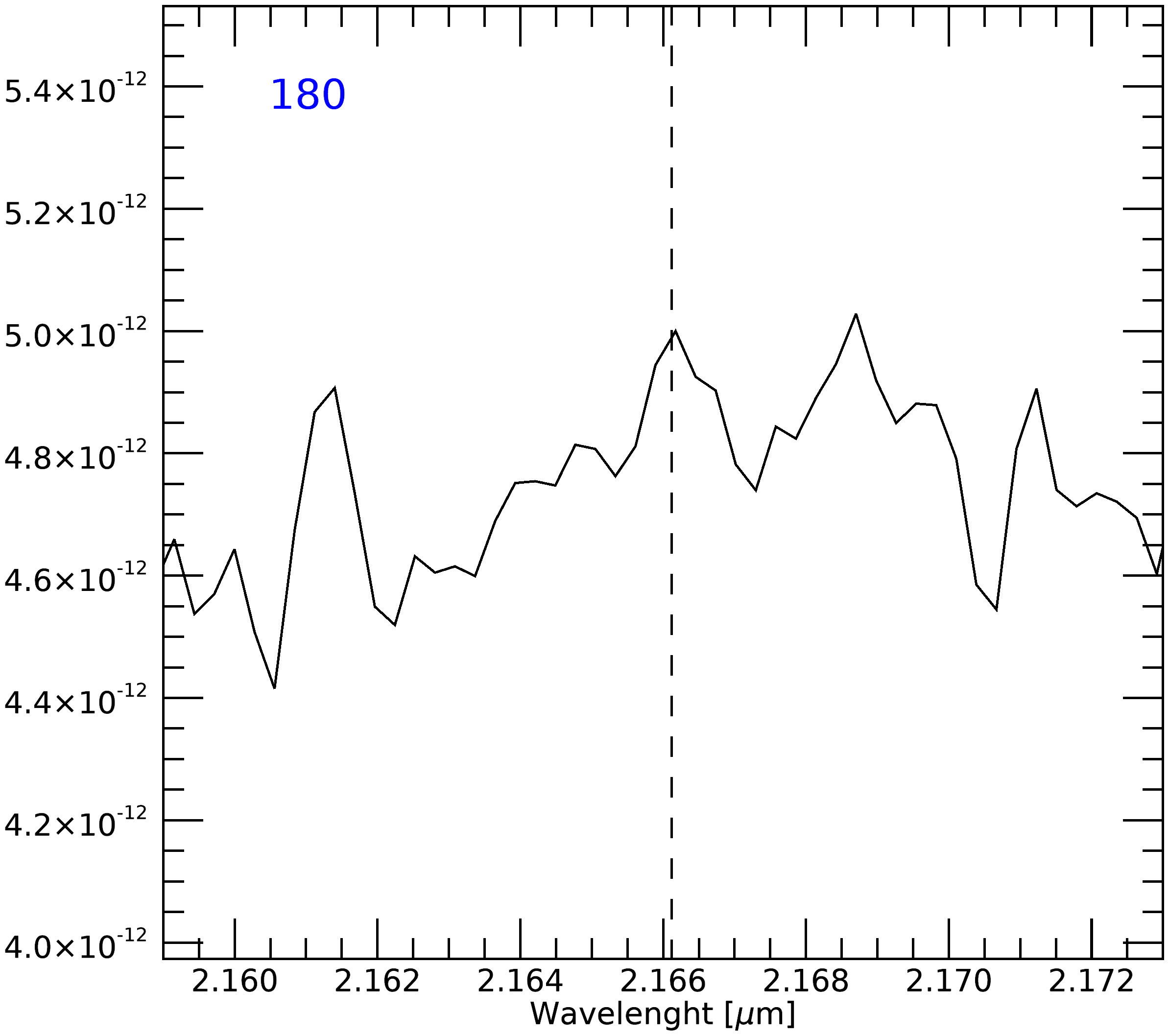}%
 \includegraphics[width=0.2\textwidth]{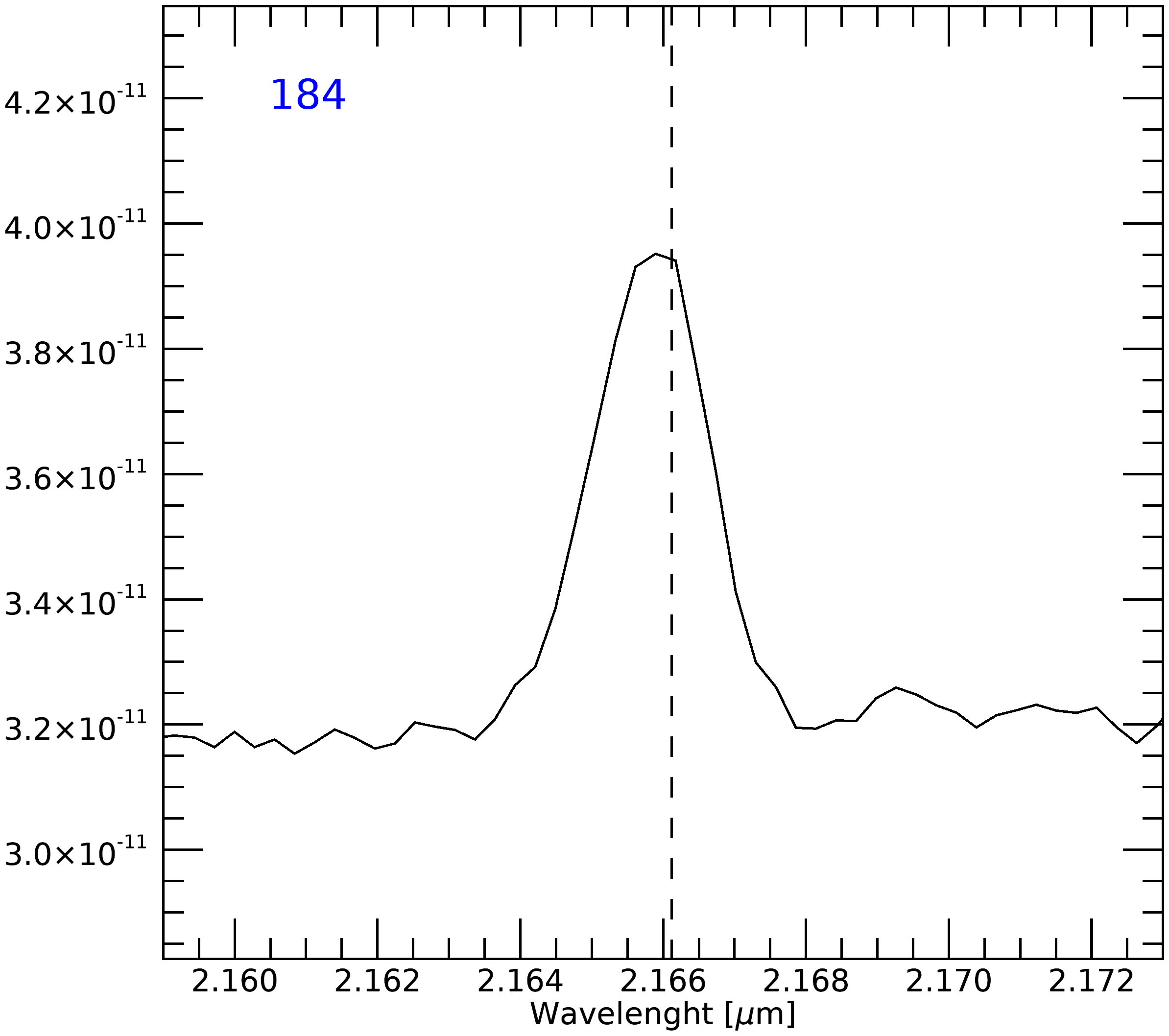}%
 \includegraphics[width=0.2\textwidth]{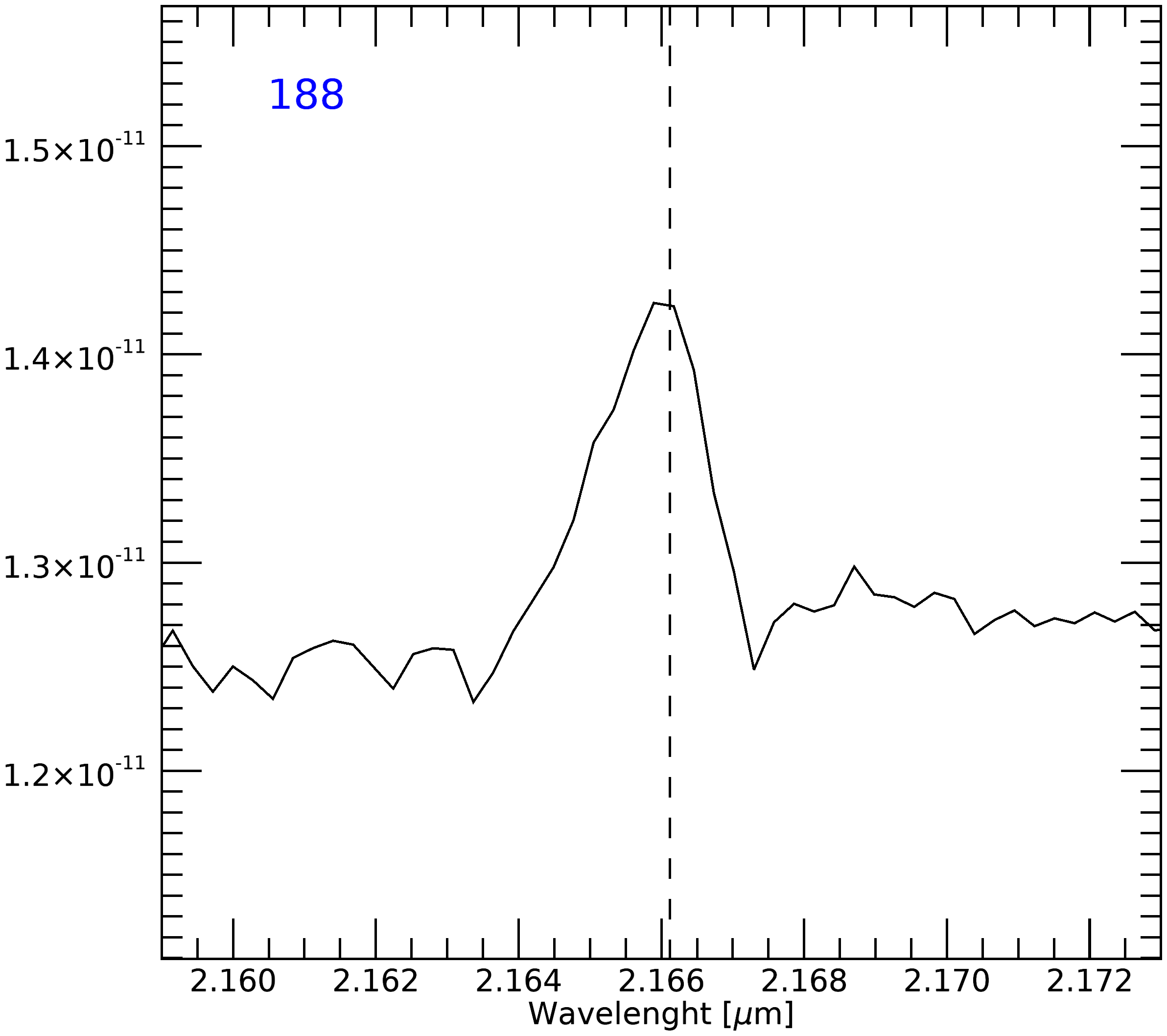}%
 \includegraphics[width=0.2\textwidth]{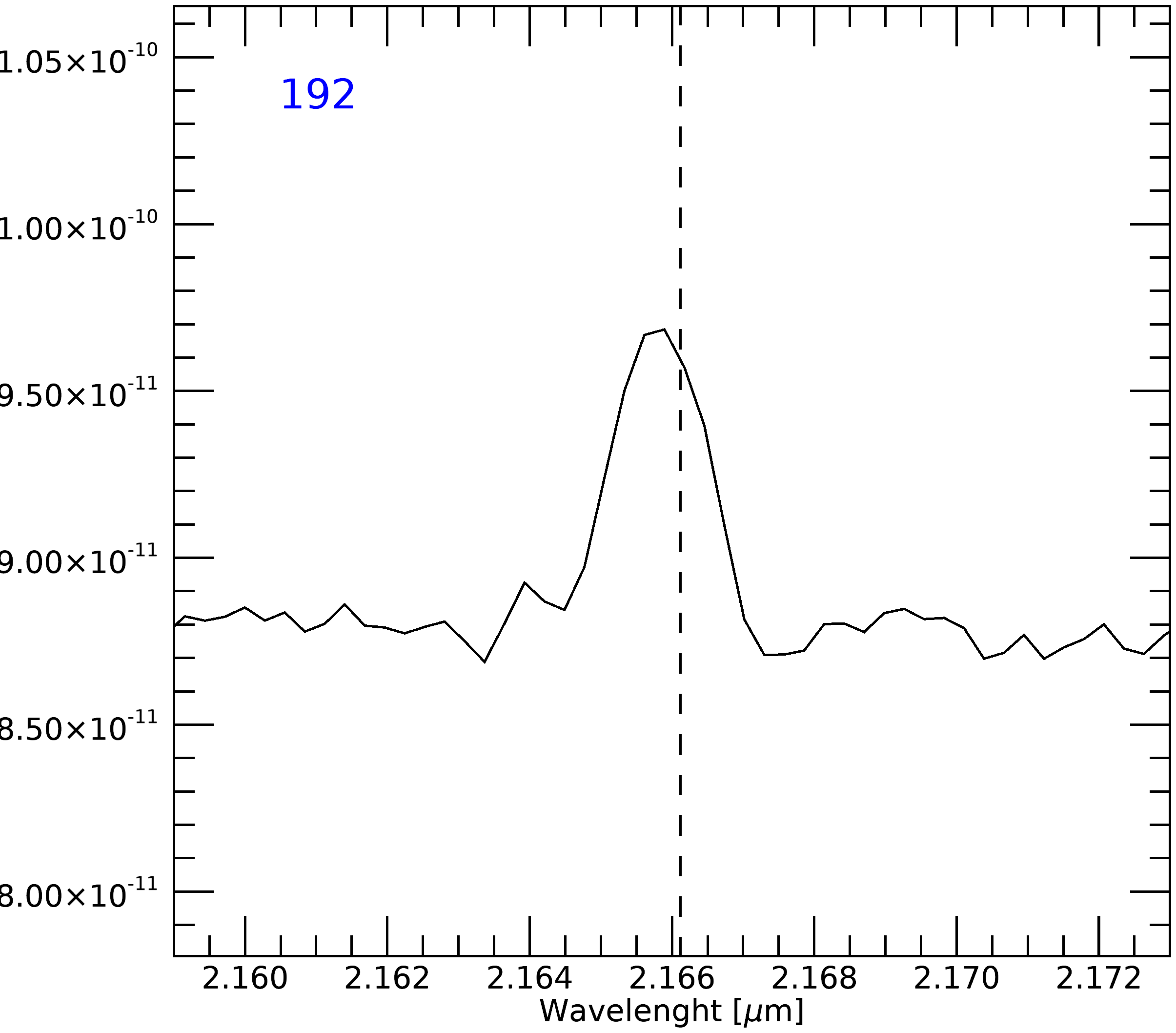}%
 \includegraphics[width=0.2\textwidth]{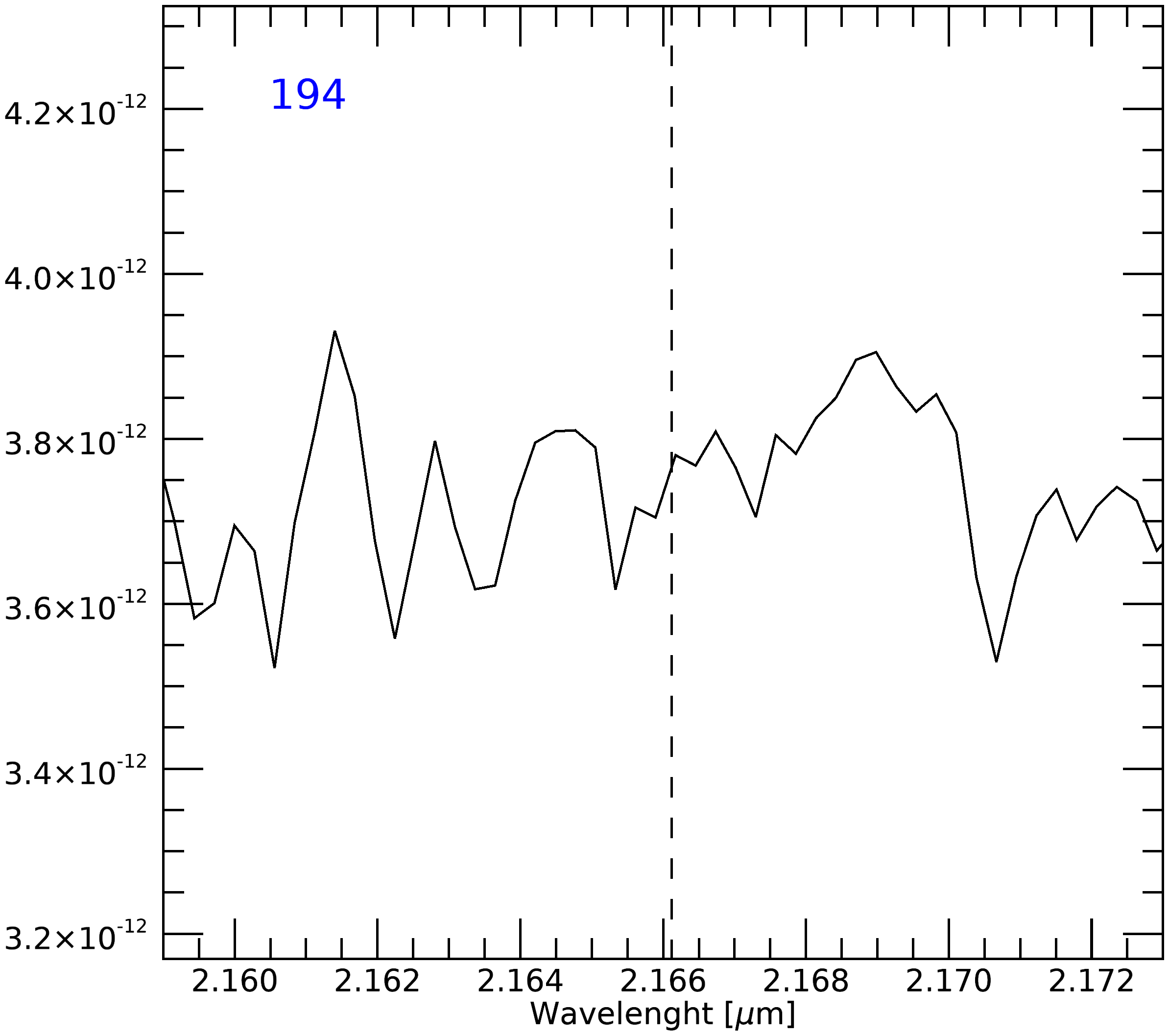}%
 
 \includegraphics[width=0.2\textwidth]{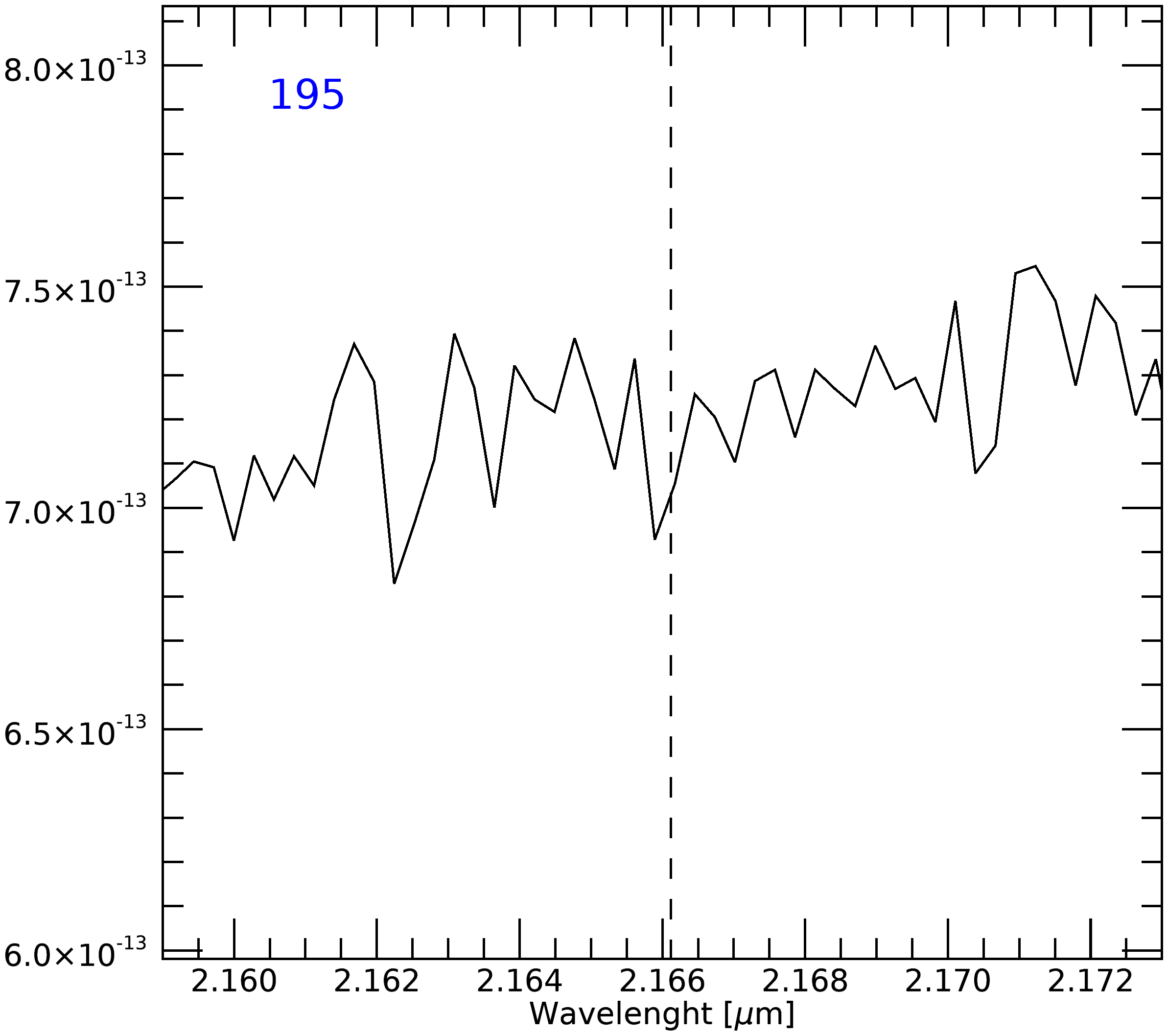}%
 \includegraphics[width=0.2\textwidth]{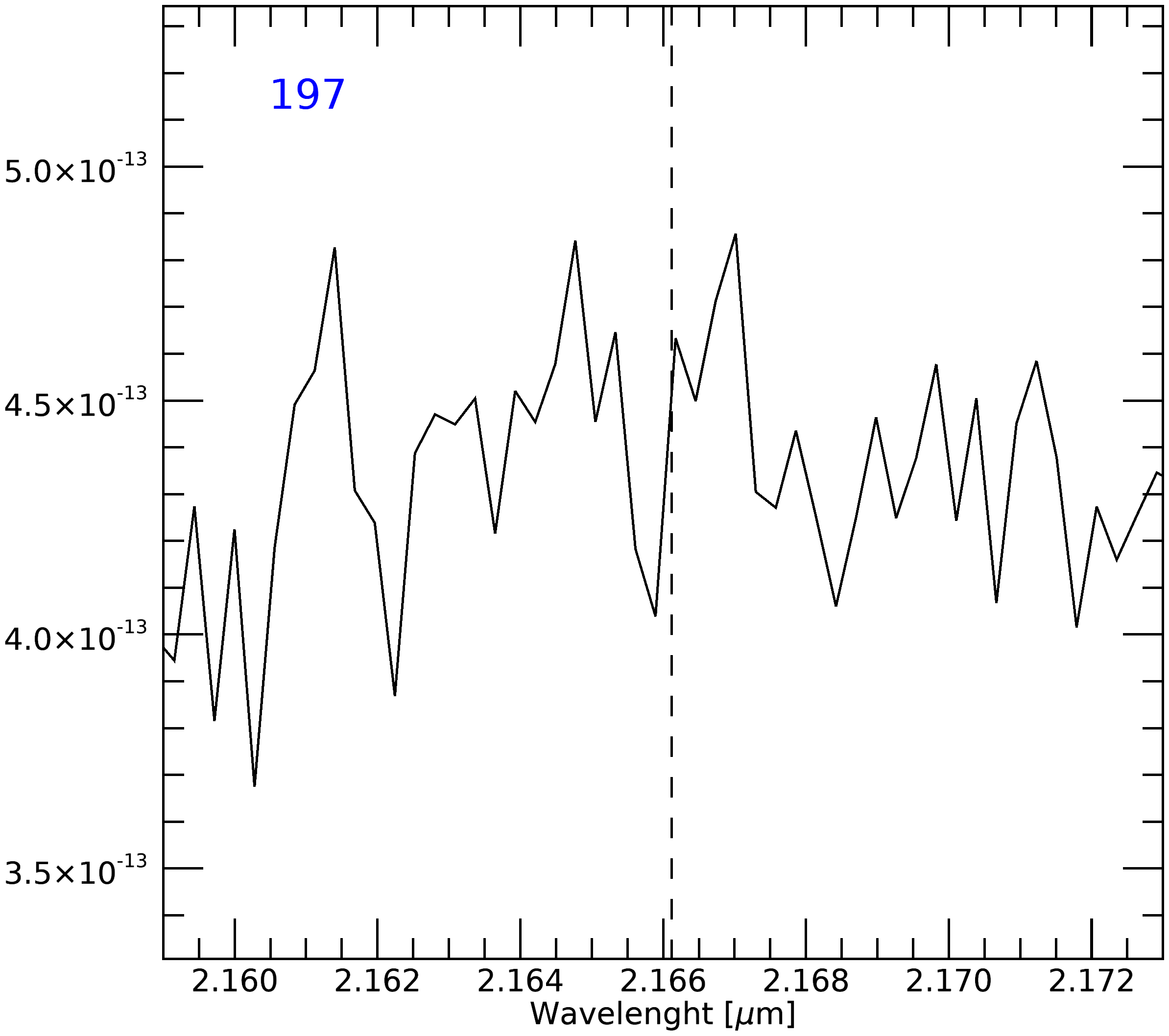}%
 \includegraphics[width=0.2\textwidth]{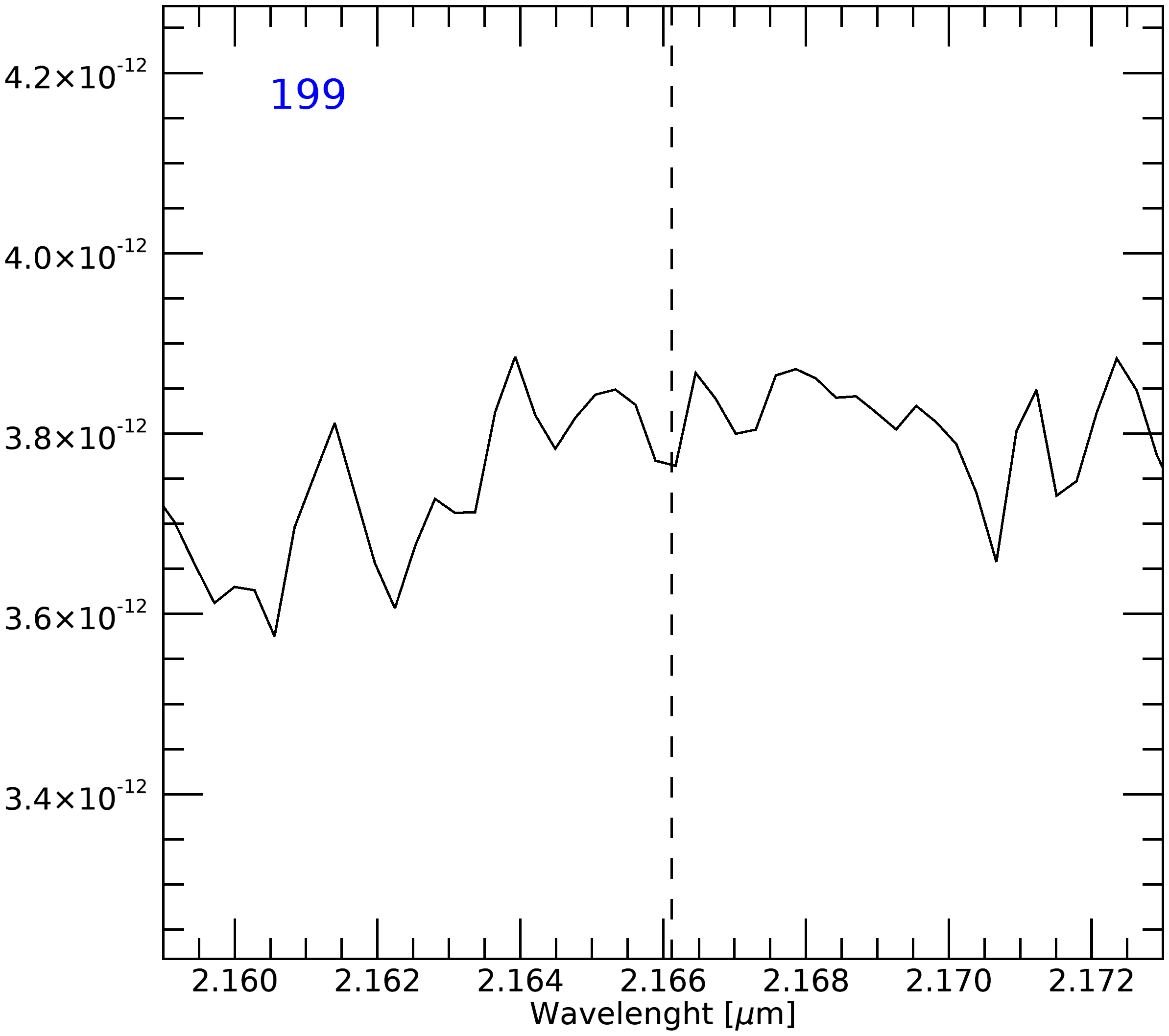}%
 \includegraphics[width=0.2\textwidth]{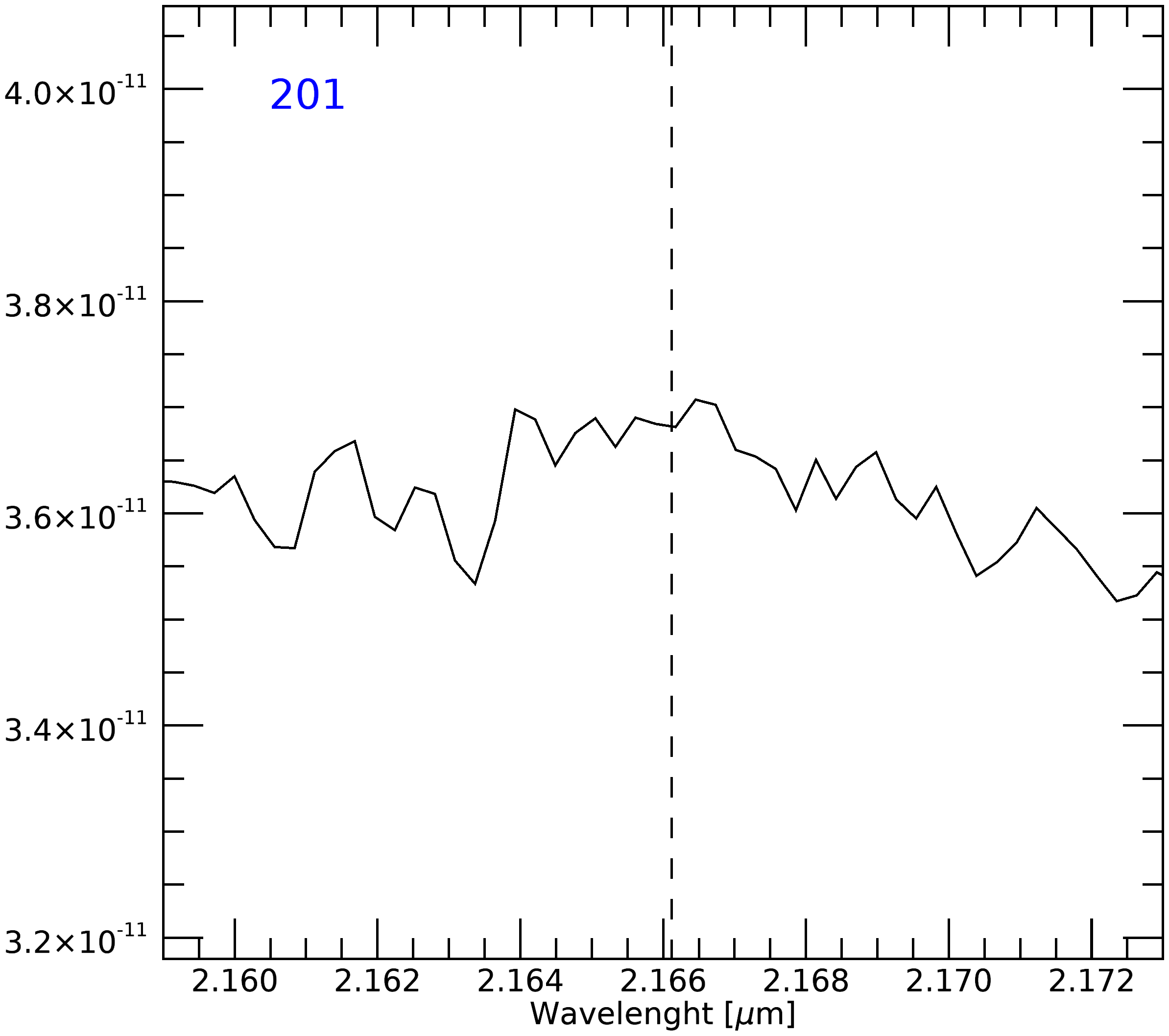}%
 \includegraphics[width=0.2\textwidth]{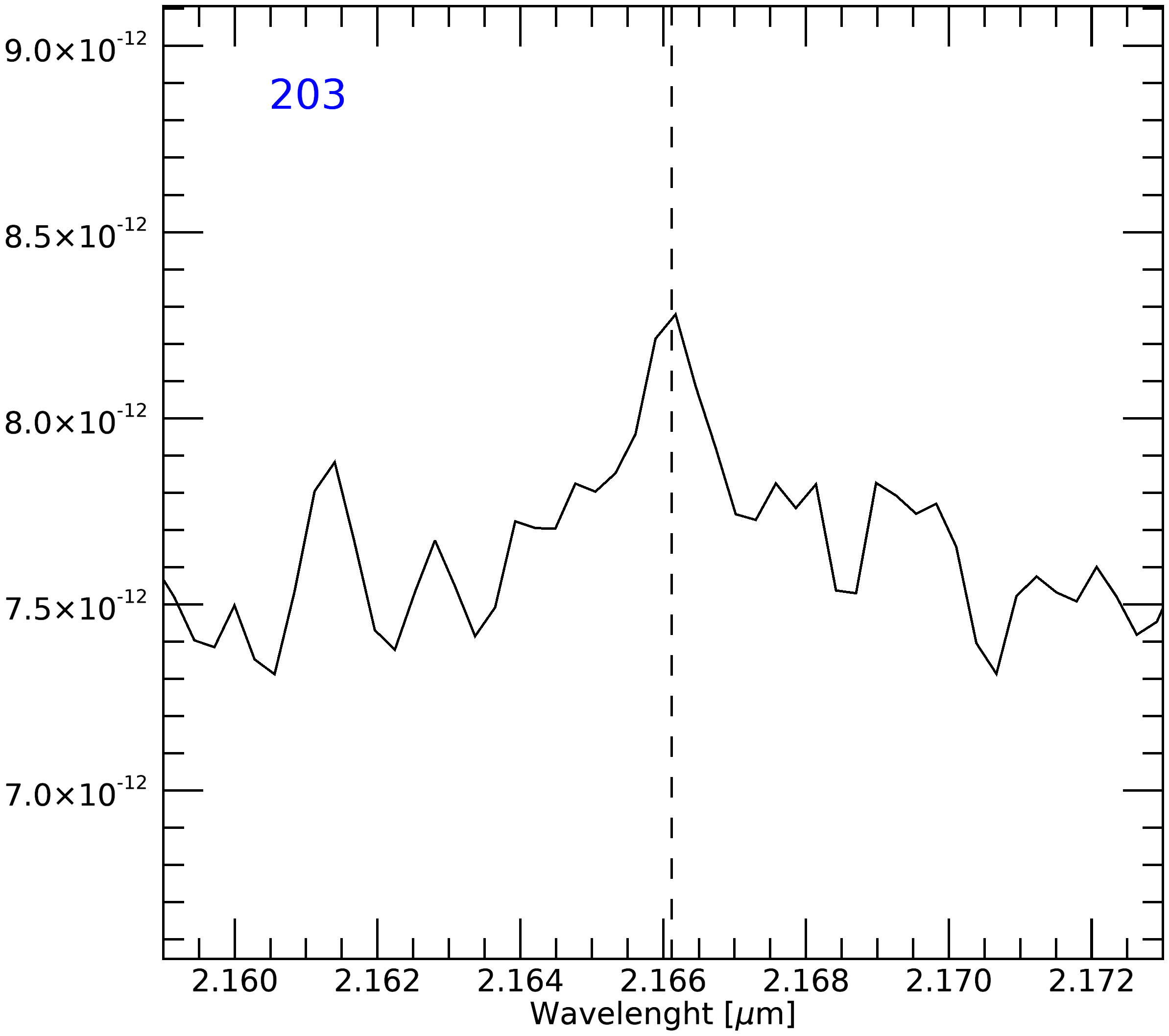}%
 
 \end{subfigure}
 \caption{\label{fig:lines2c}$\brg$ Class~II lines. The flux is in erg~s$^{-1}$cm$^{-2}\mu$m$^{-1}$.} 
\end{figure*}

\begin{figure*} 
\centering
 \begin{subfigure}{\textwidth}
 \centering
 \includegraphics[width=0.2\textwidth]{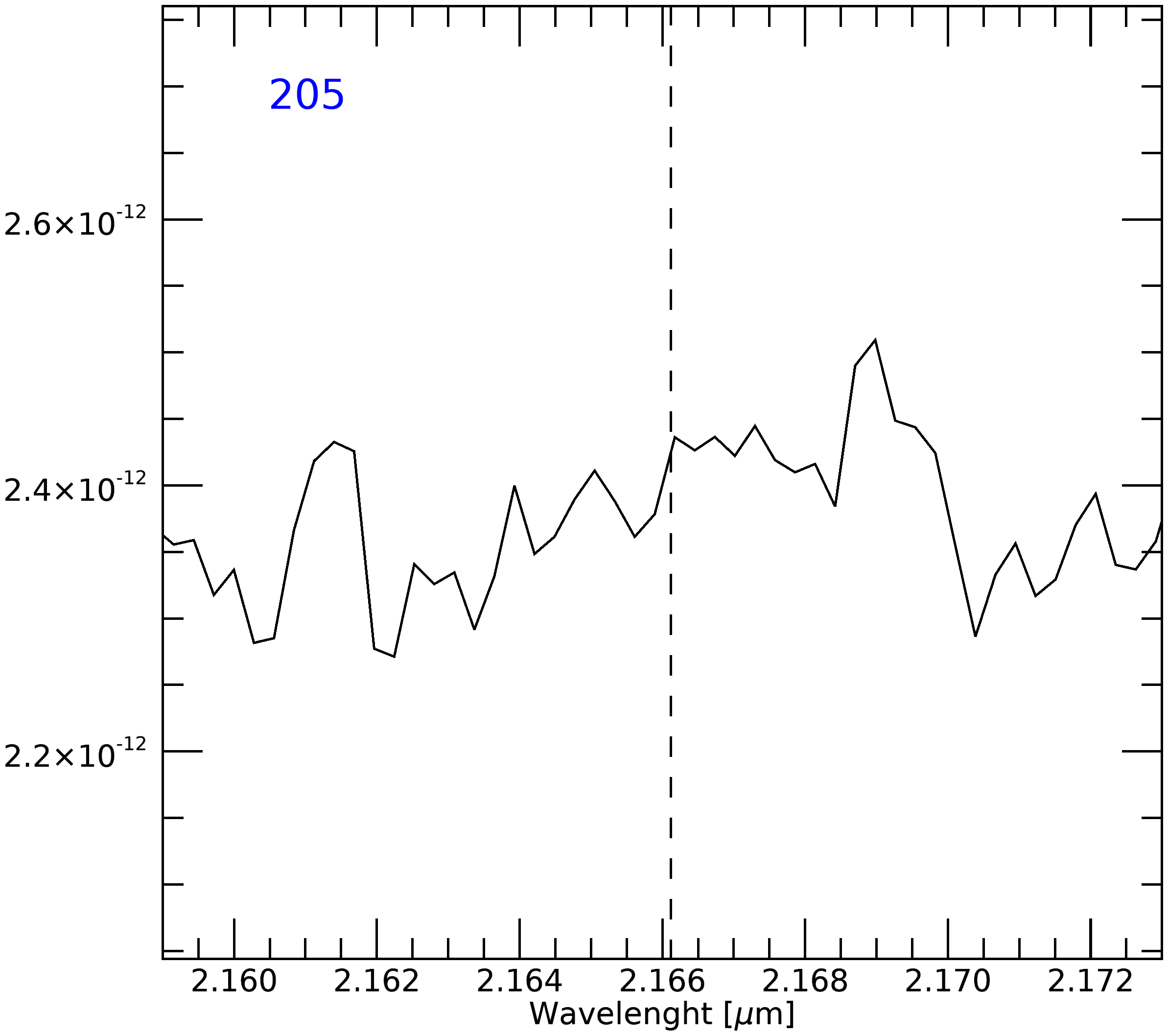}%
 \includegraphics[width=0.2\textwidth]{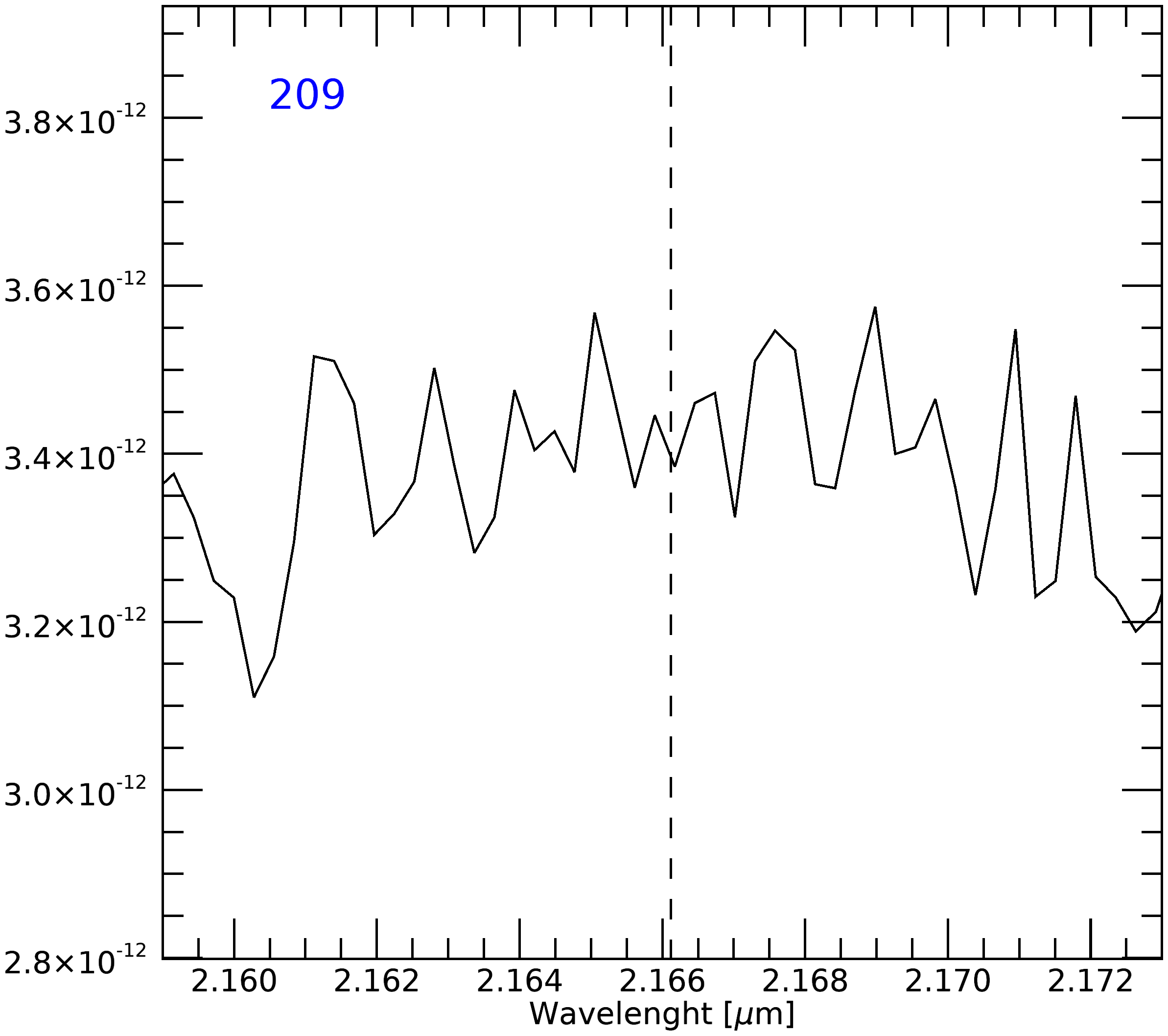}%
 \includegraphics[width=0.2\textwidth]{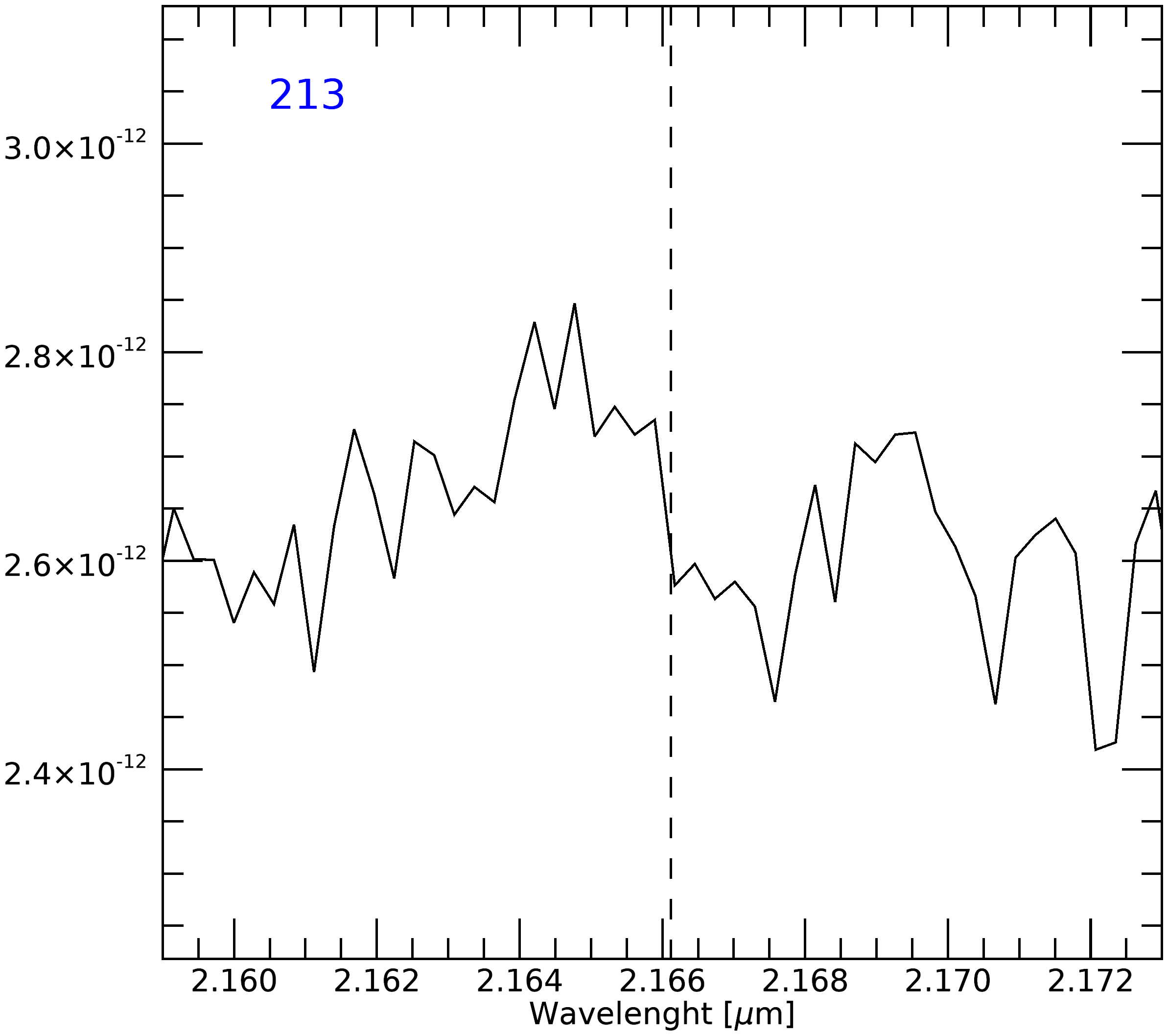}%
 \includegraphics[width=0.2\textwidth]{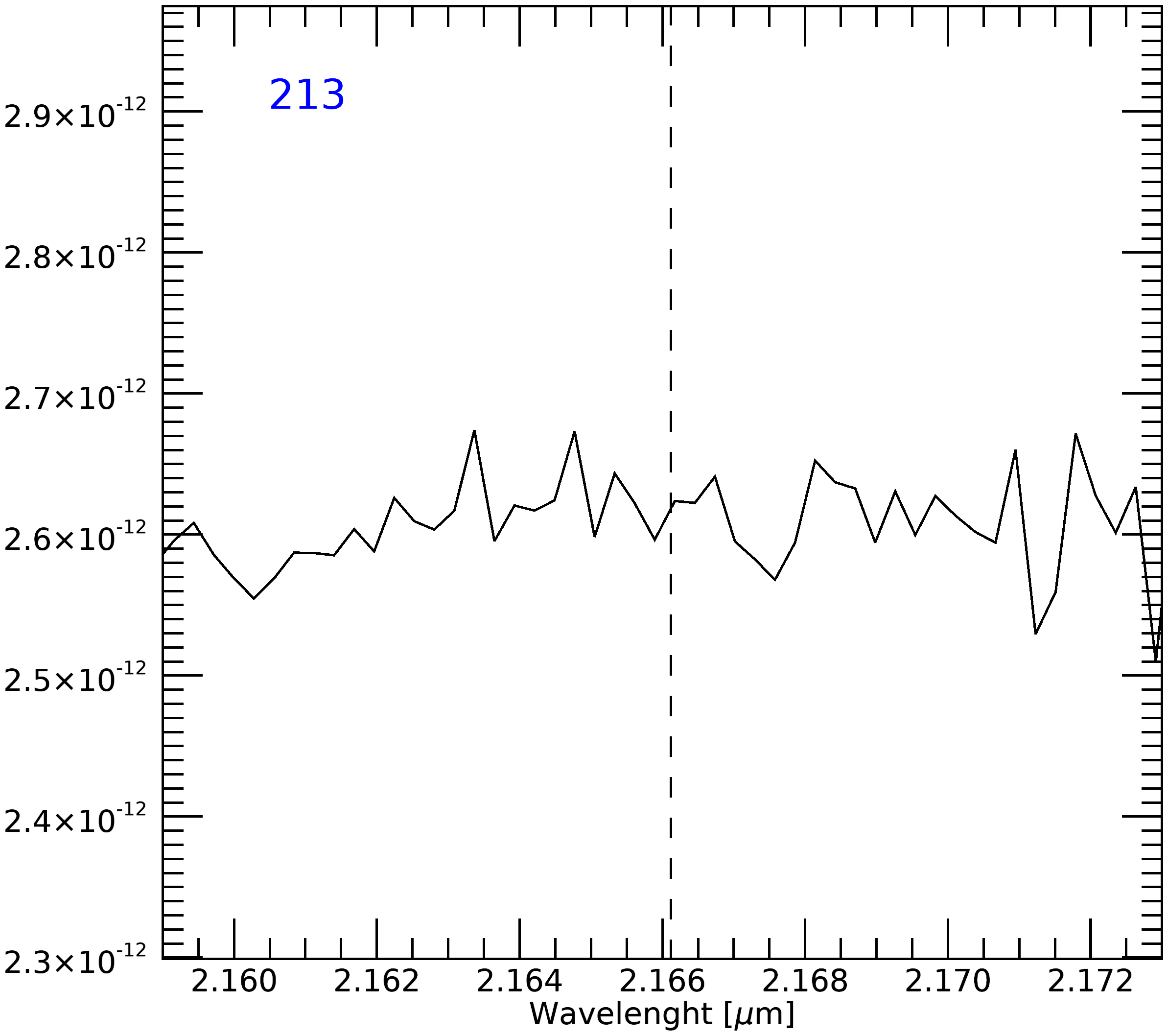}%
 \includegraphics[width=0.2\textwidth]{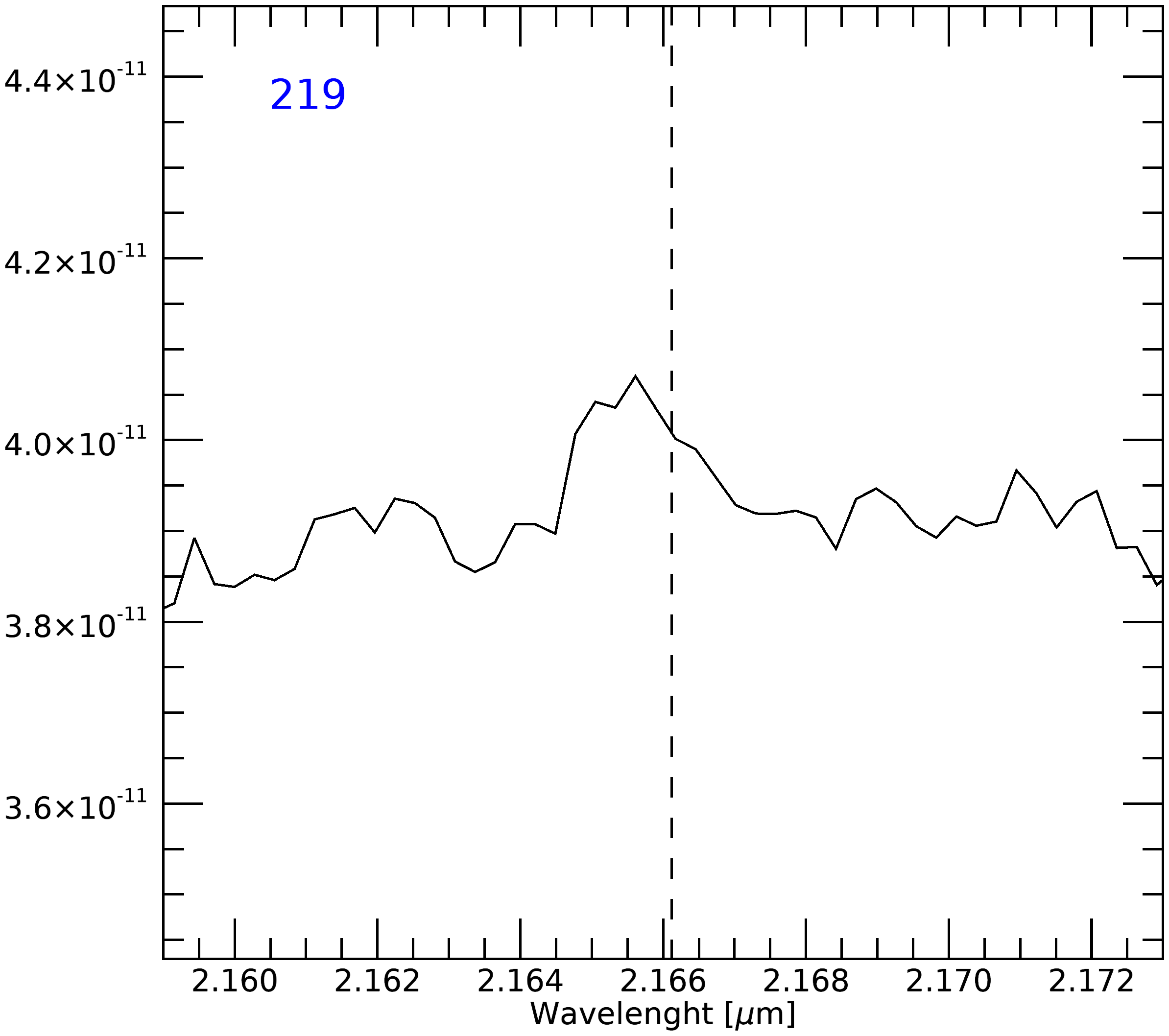}%
 
 \includegraphics[width=0.2\textwidth]{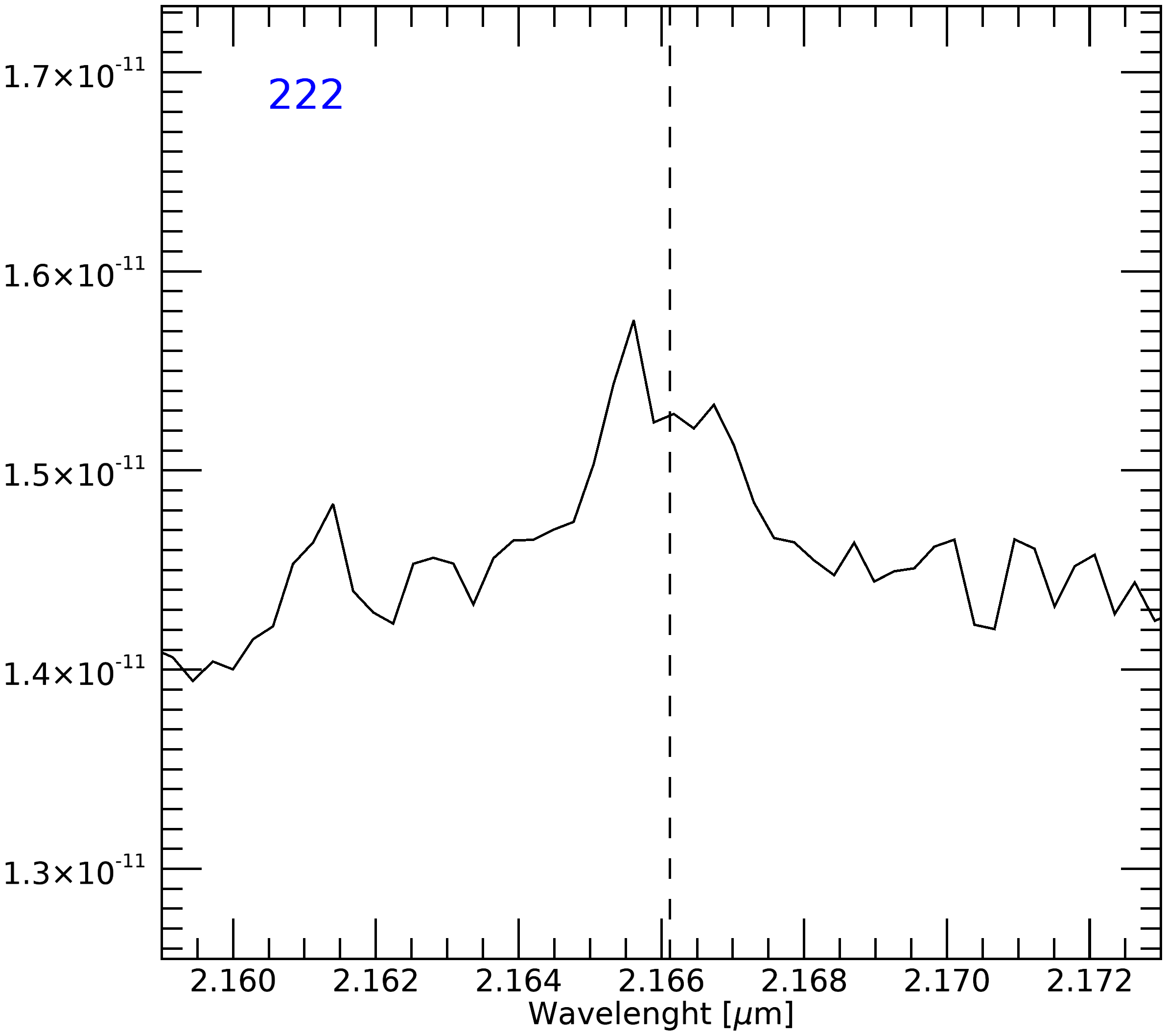}%
 \includegraphics[width=0.2\textwidth]{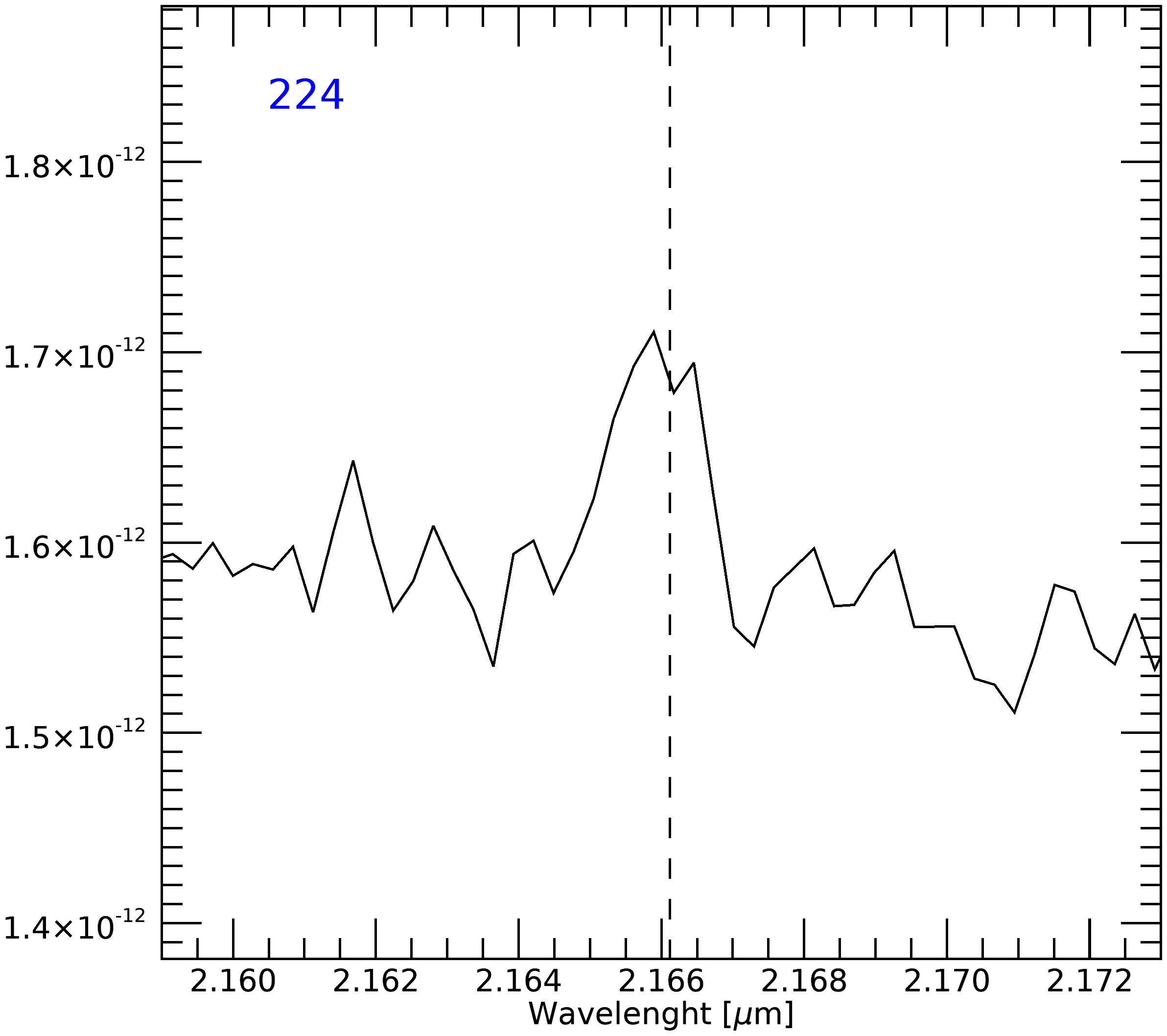}%
 \includegraphics[width=0.2\textwidth]{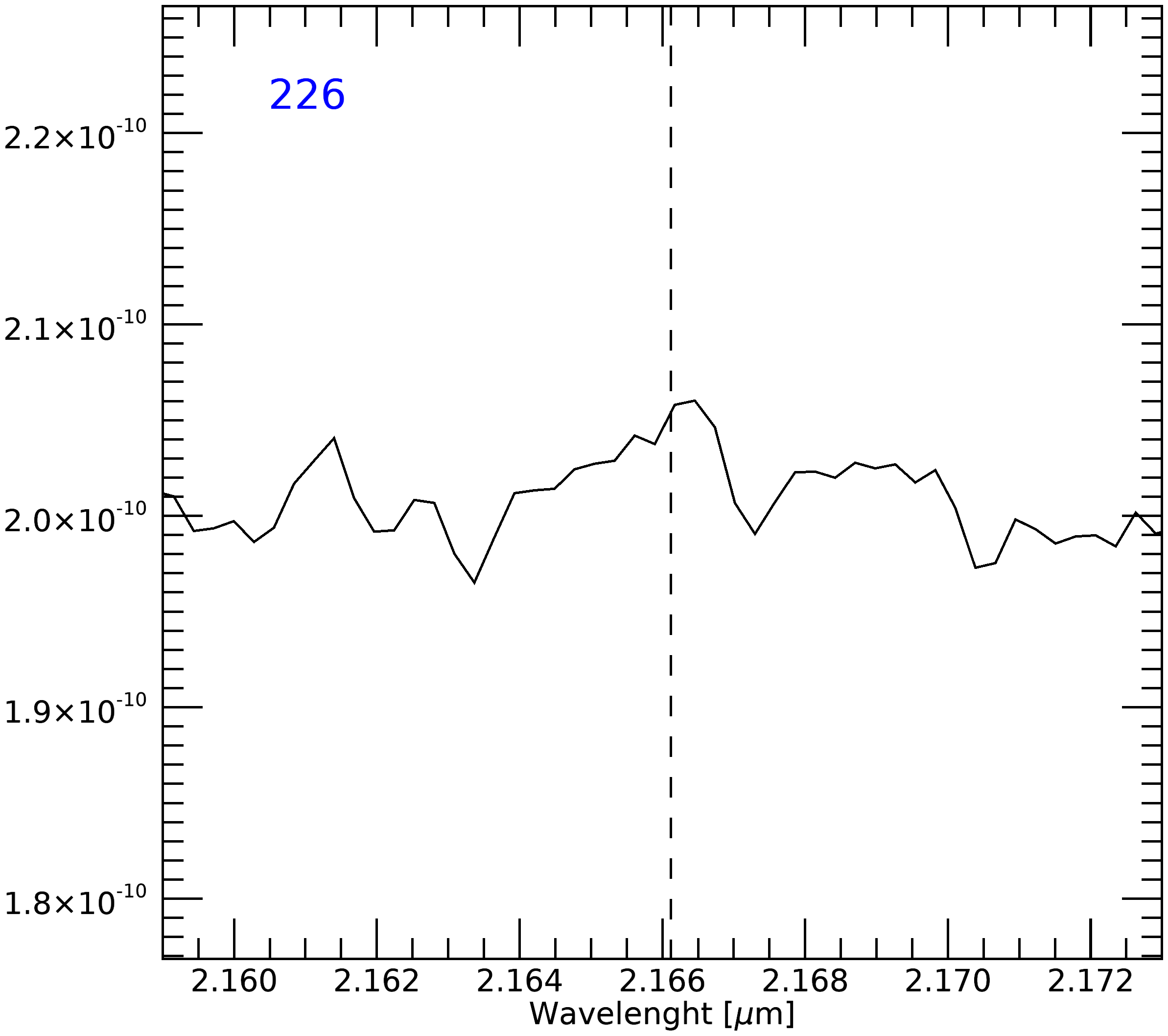}%
 \includegraphics[width=0.2\textwidth]{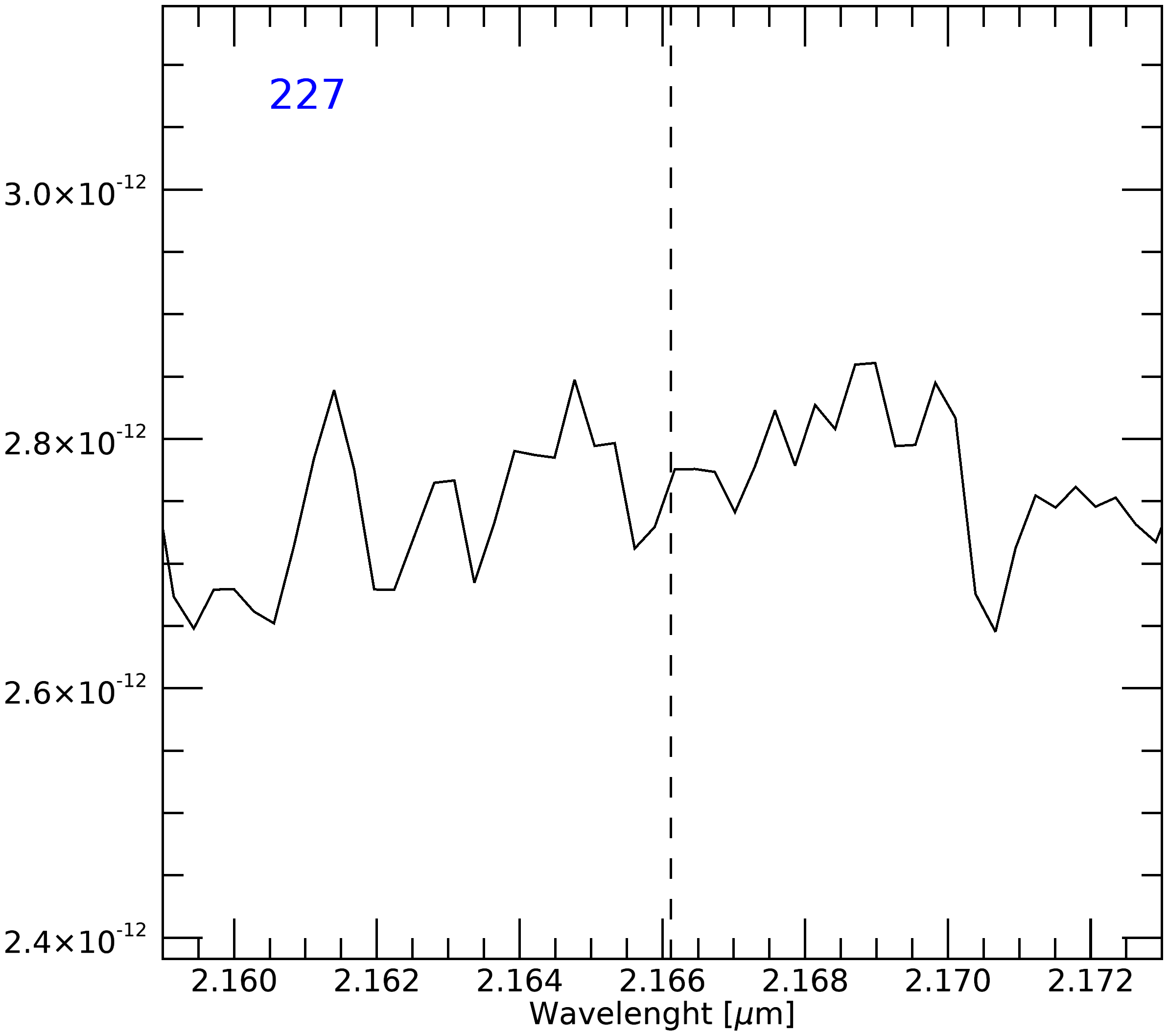}%
 \includegraphics[width=0.2\textwidth]{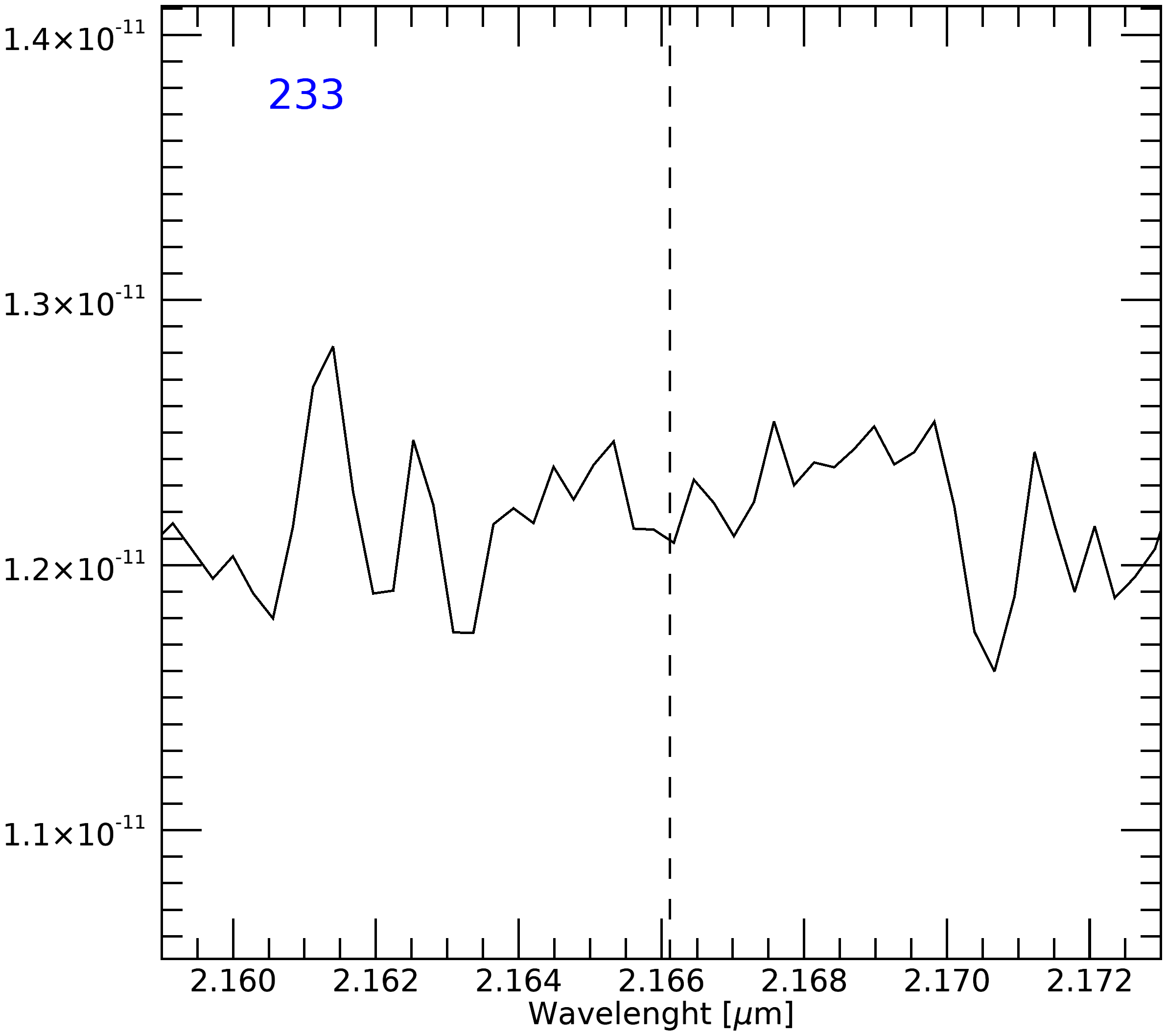}%
 
 \includegraphics[width=0.2\textwidth]{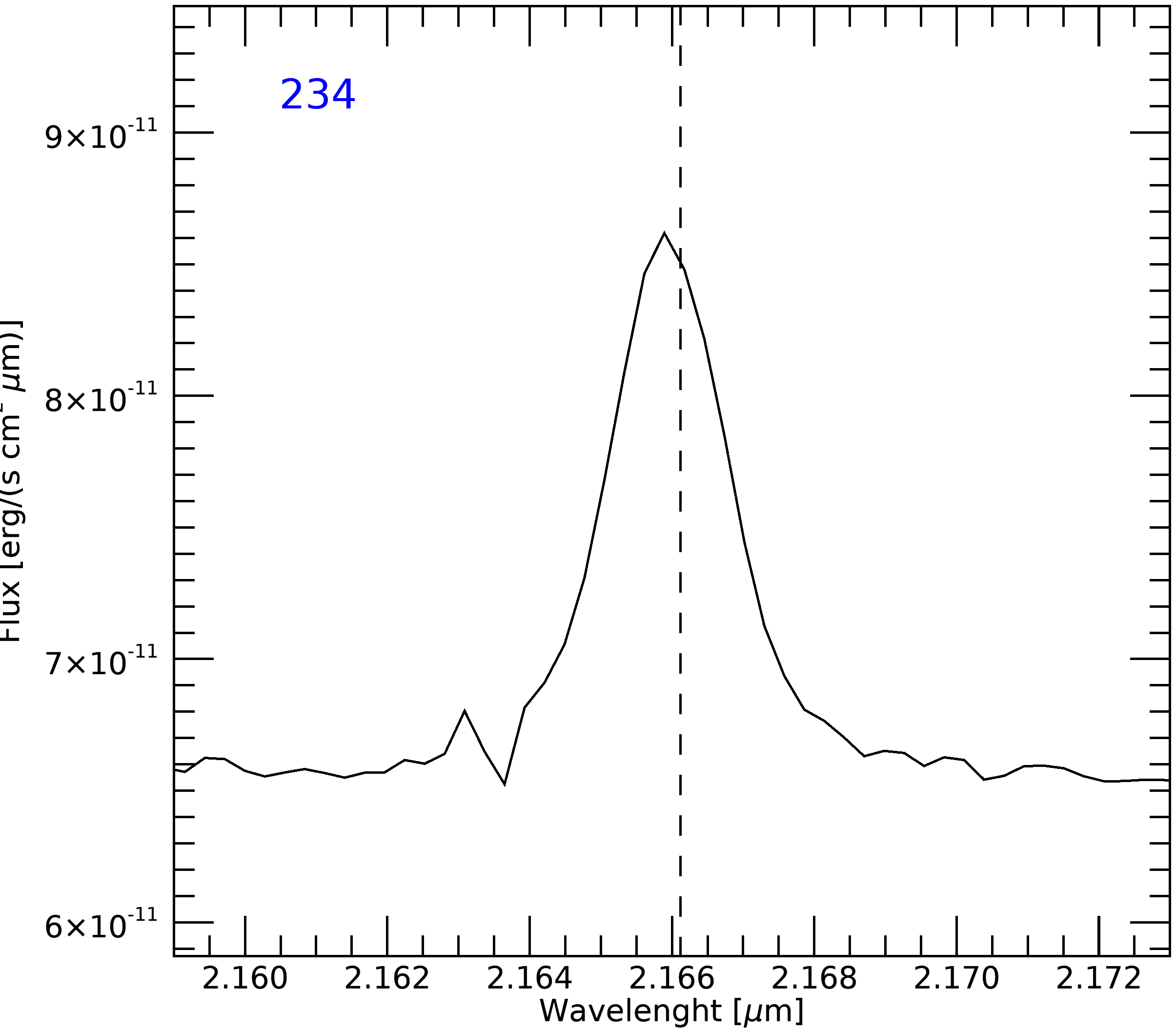}%
 \includegraphics[width=0.2\textwidth]{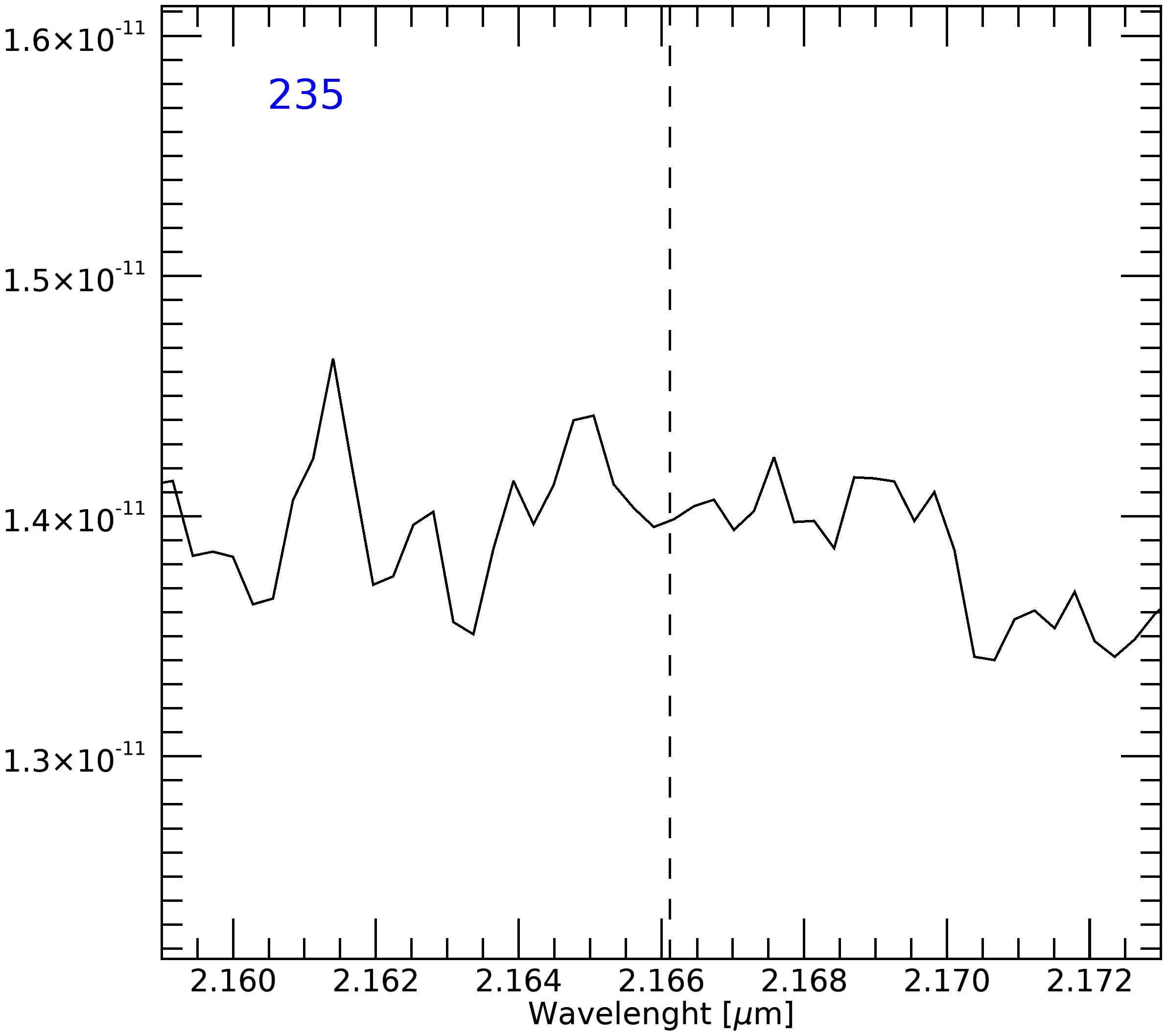}%
 \includegraphics[width=0.2\textwidth]{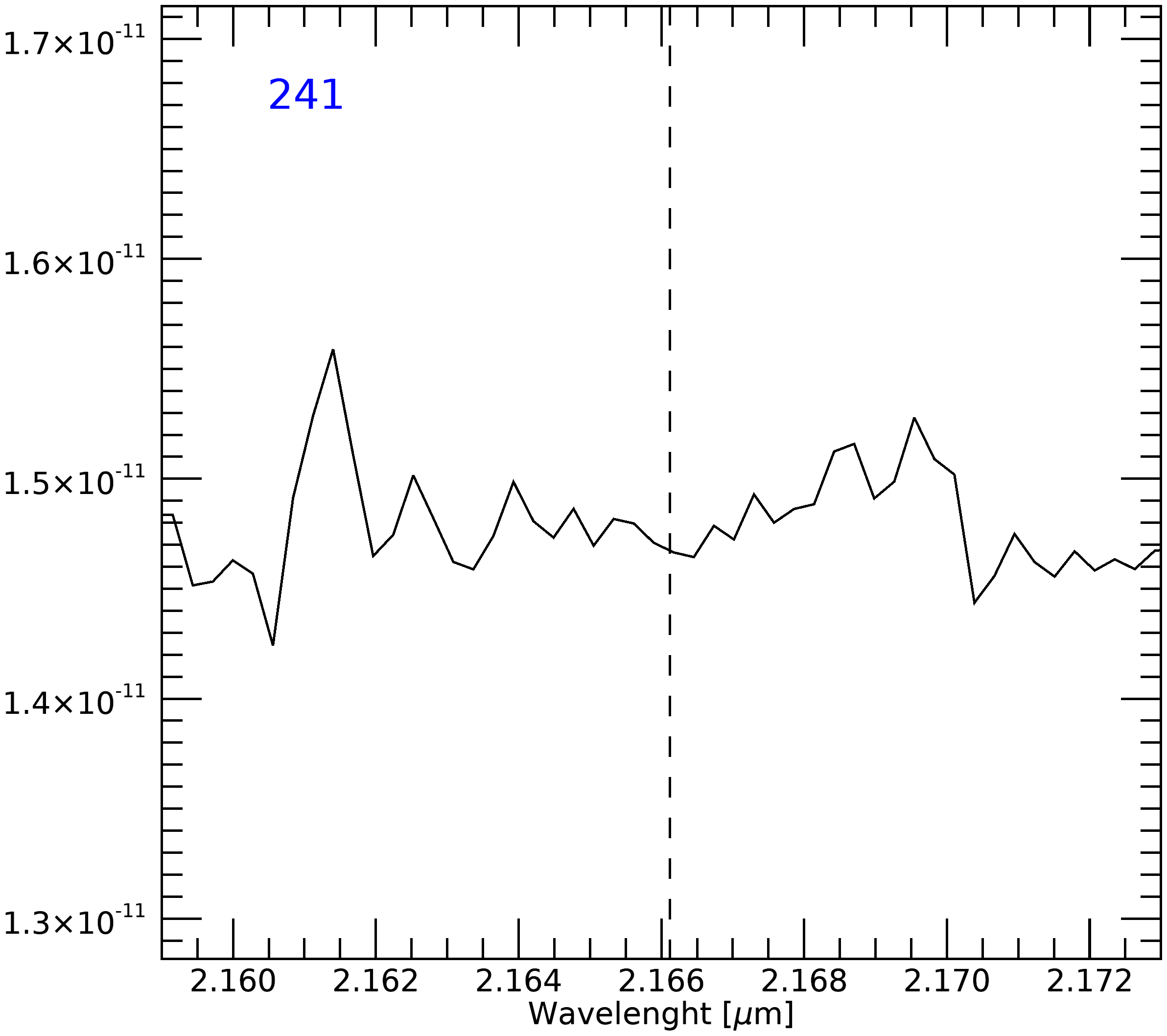}%
\end{subfigure}
\caption{\label{fig:lines2d}$\brg$ Class~II lines. The flux is in erg~s$^{-1}$cm$^{-2}\mu$m$^{-1}$.} 
\end{figure*}

\begin{figure*} 
\centering
 \begin{subfigure}{\textwidth}
 \centering
 \includegraphics[width=0.2\textwidth]{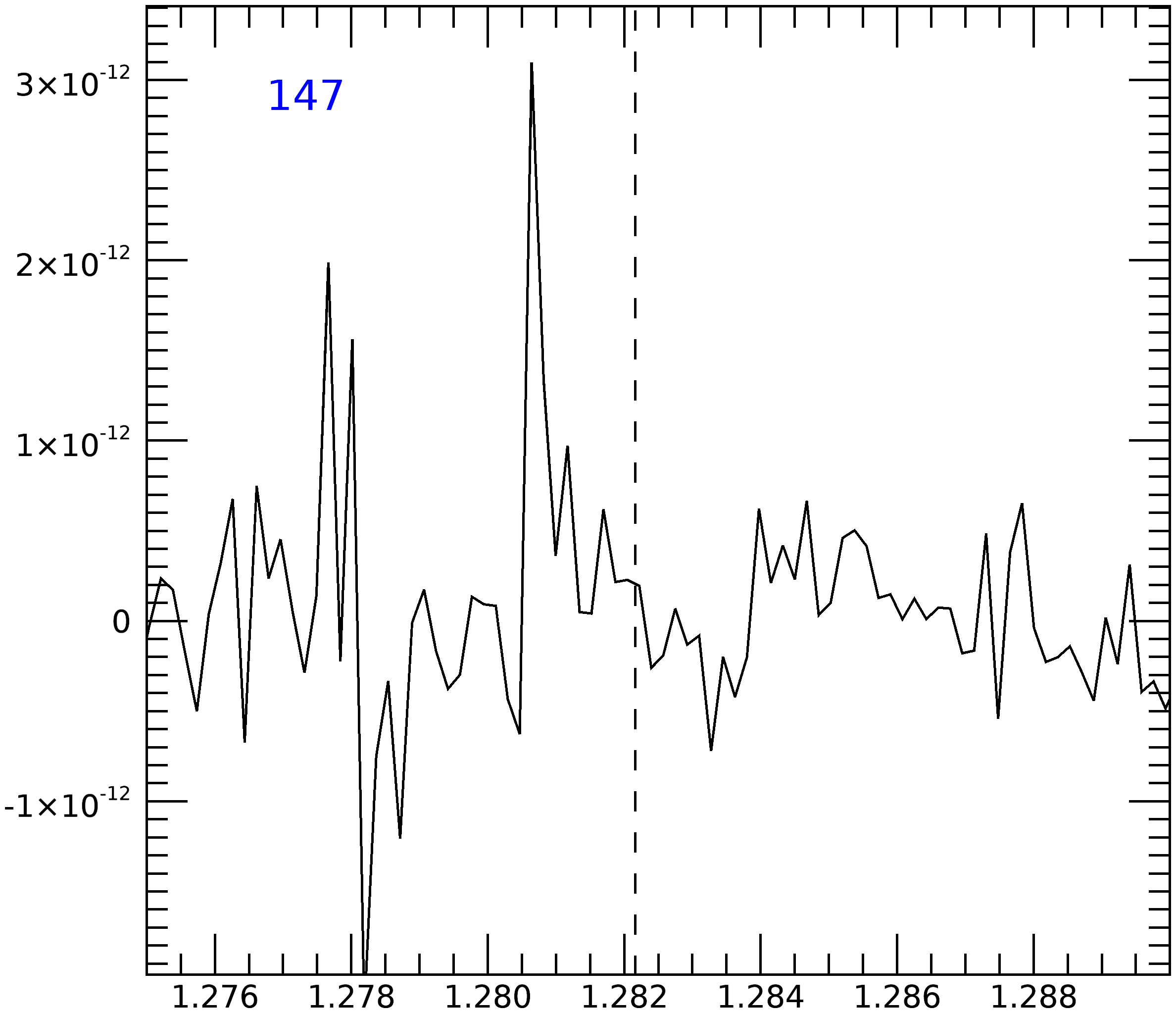}%
 \includegraphics[width=0.2\textwidth]{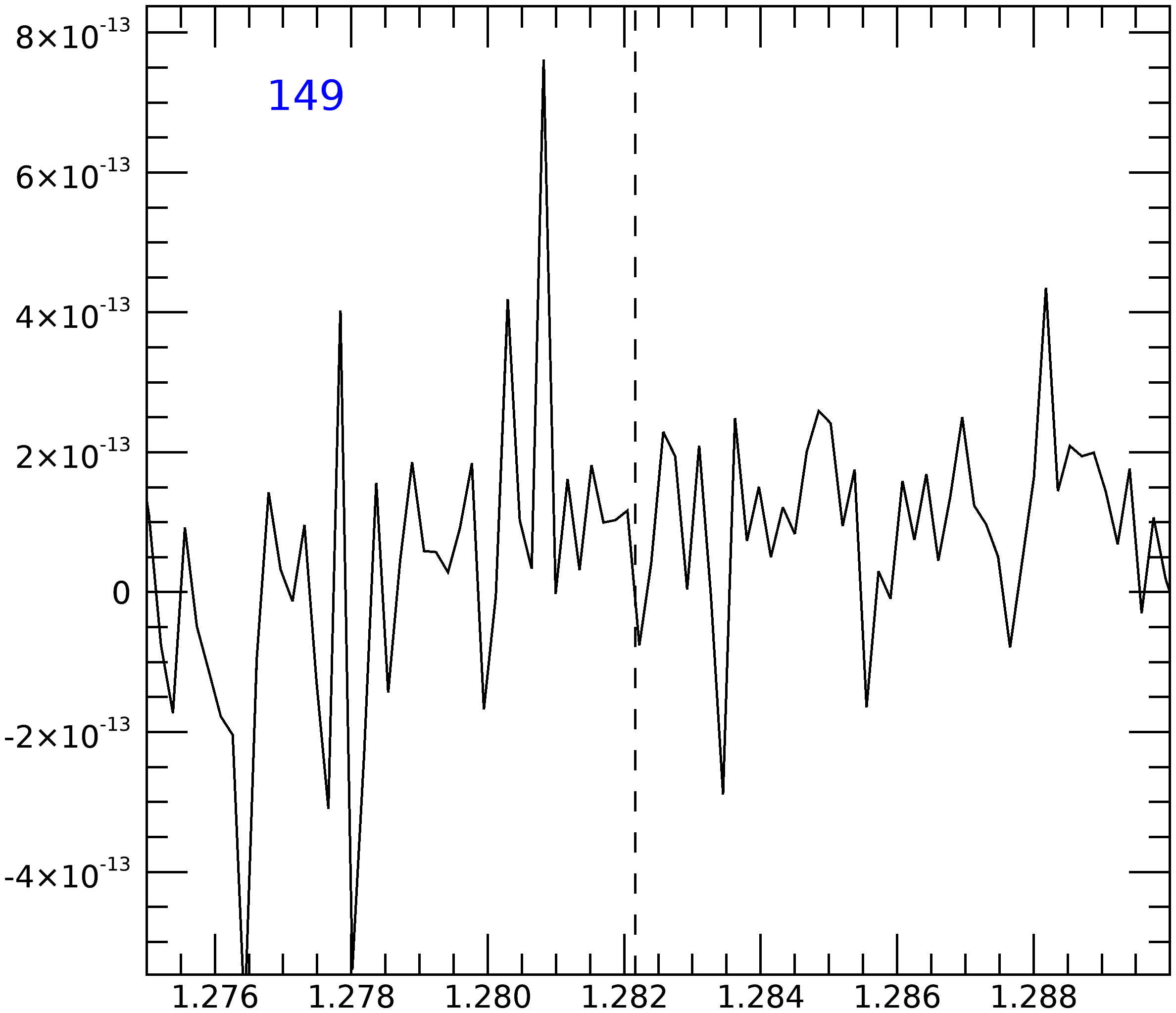}%
 \includegraphics[width=0.2\textwidth]{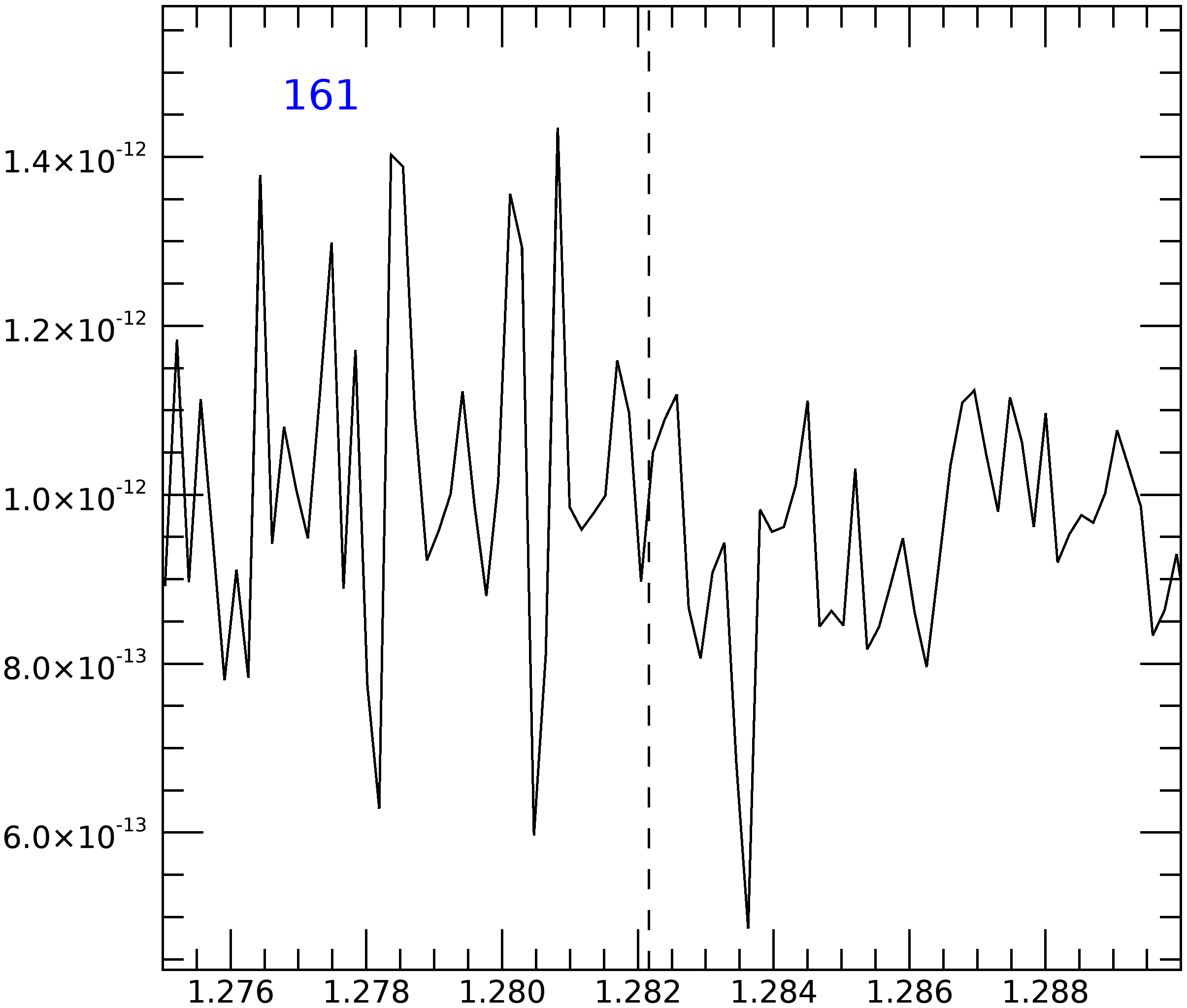}%
 \includegraphics[width=0.2\textwidth]{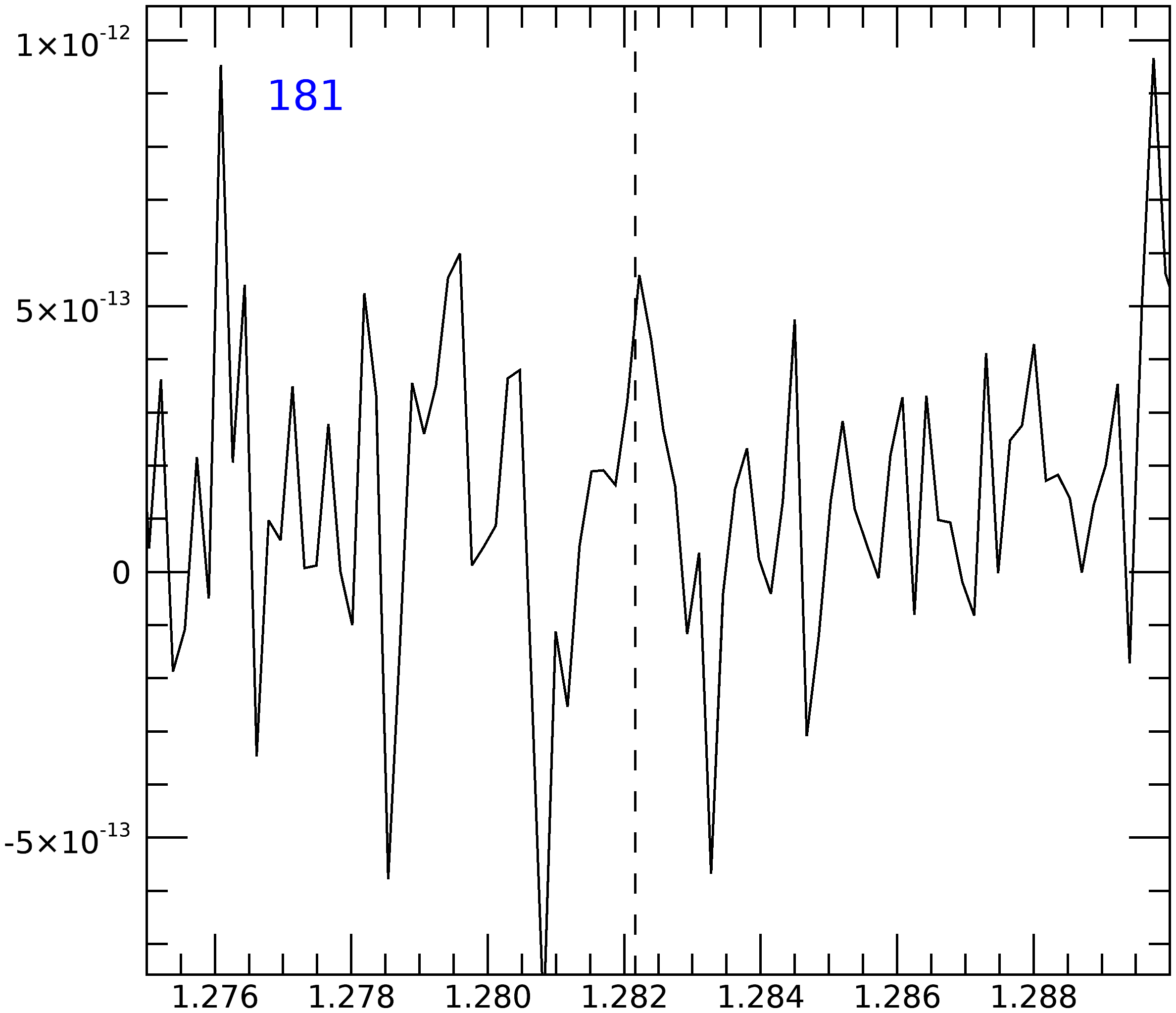}%
 \includegraphics[width=0.2\textwidth]{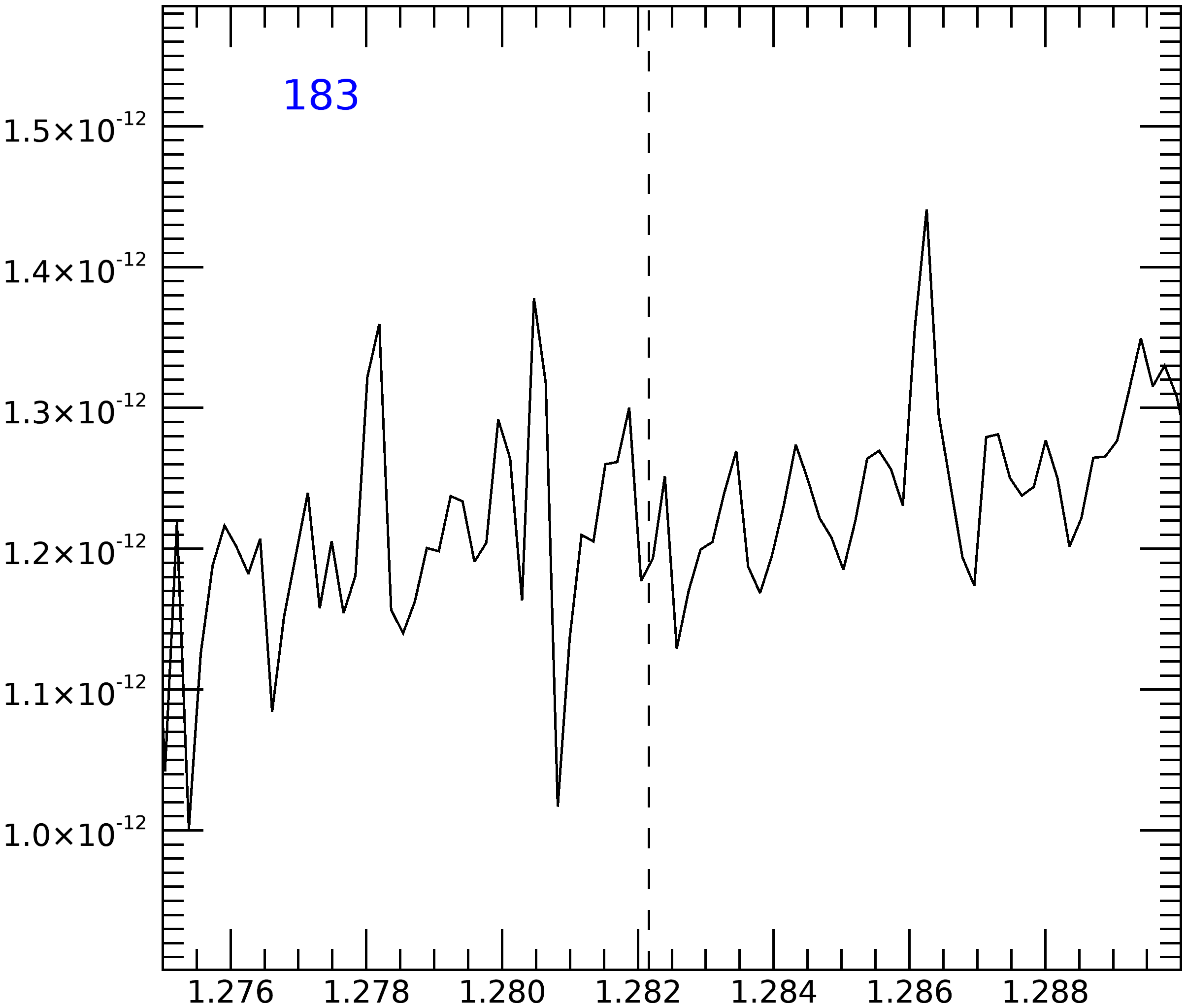}%
 
 \includegraphics[width=0.2\textwidth]{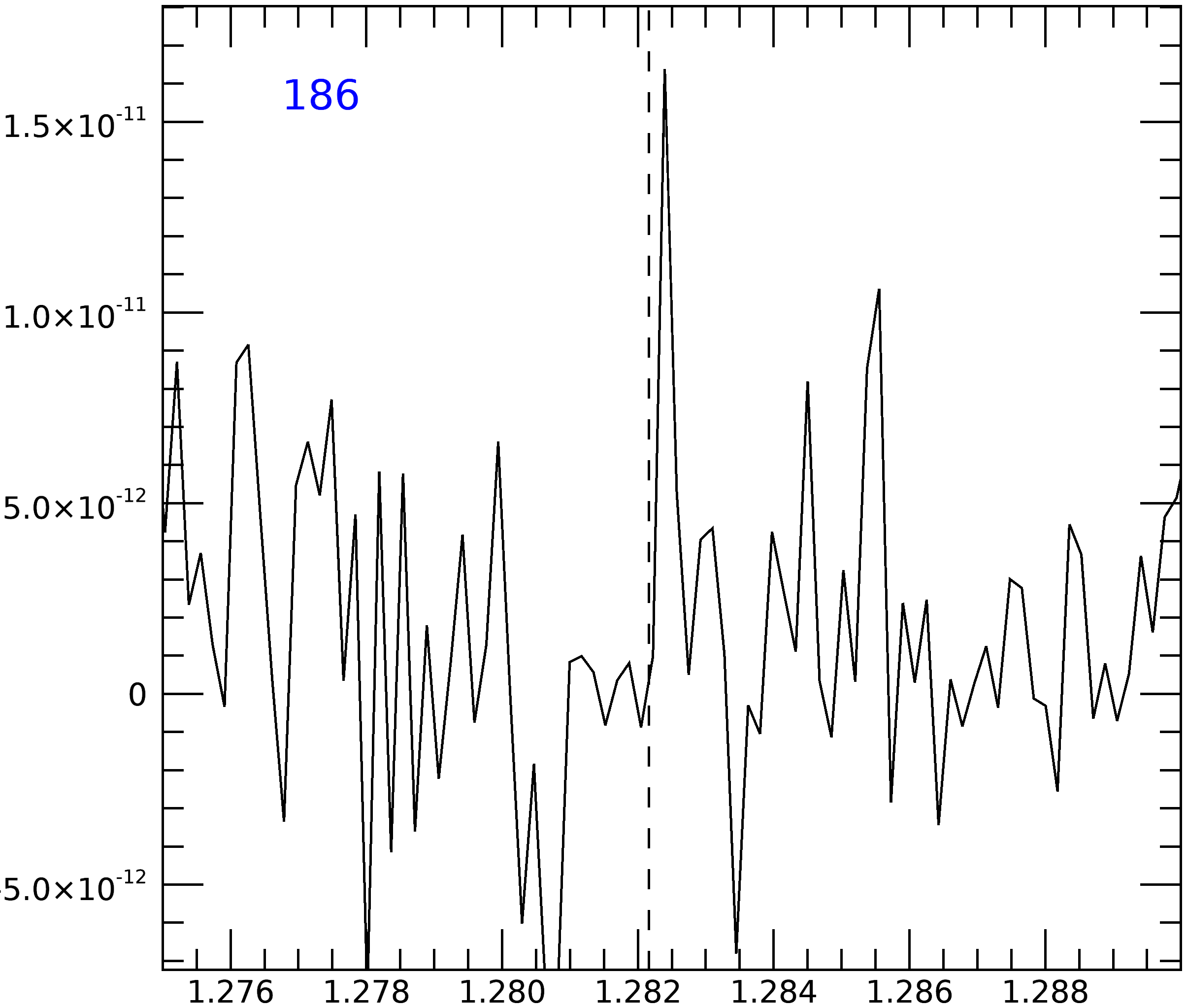}%
 \includegraphics[width=0.2\textwidth]{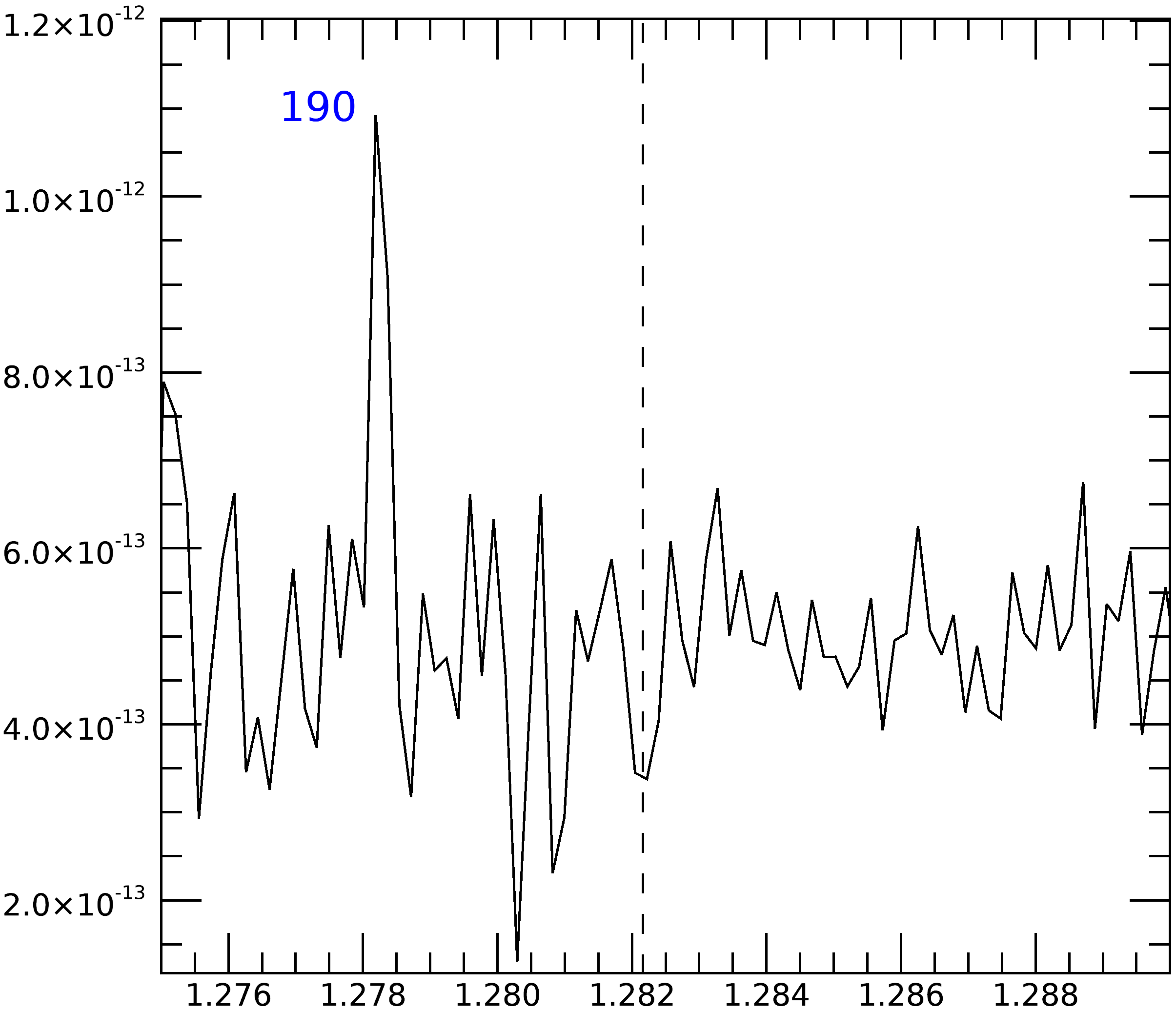}%
 \includegraphics[width=0.2\textwidth]{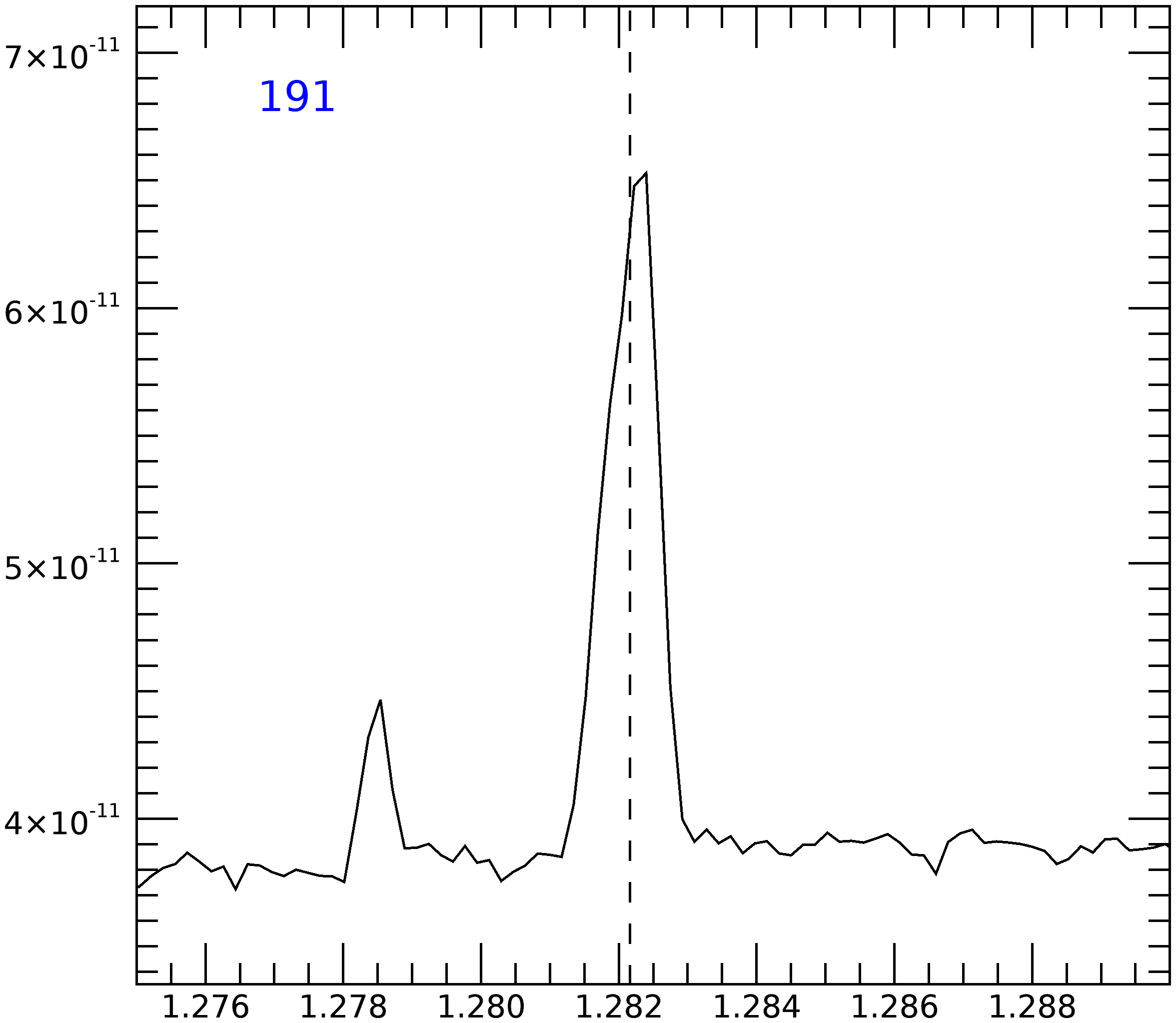}%
 \includegraphics[width=0.2\textwidth]{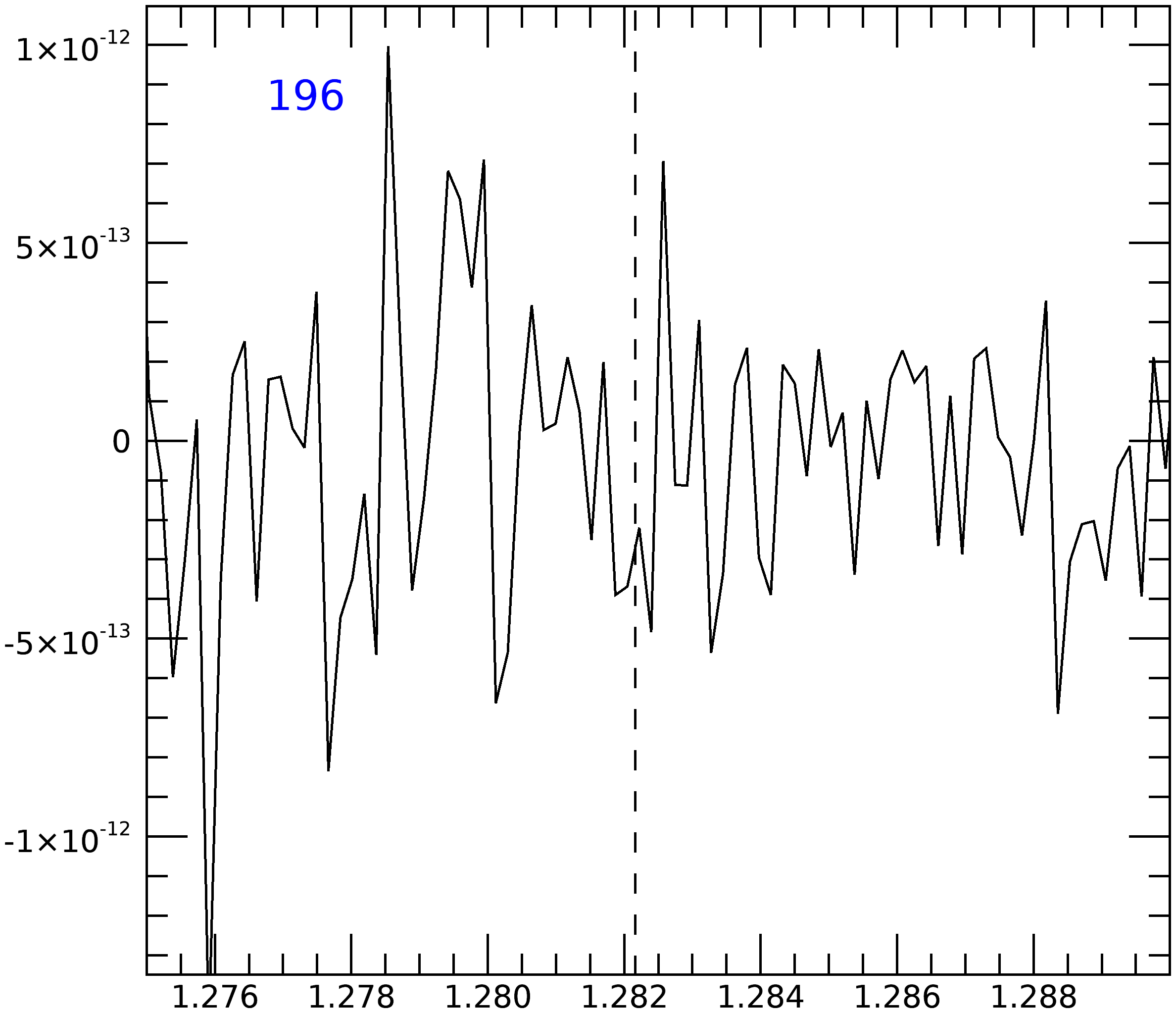}%
 \includegraphics[width=0.2\textwidth]{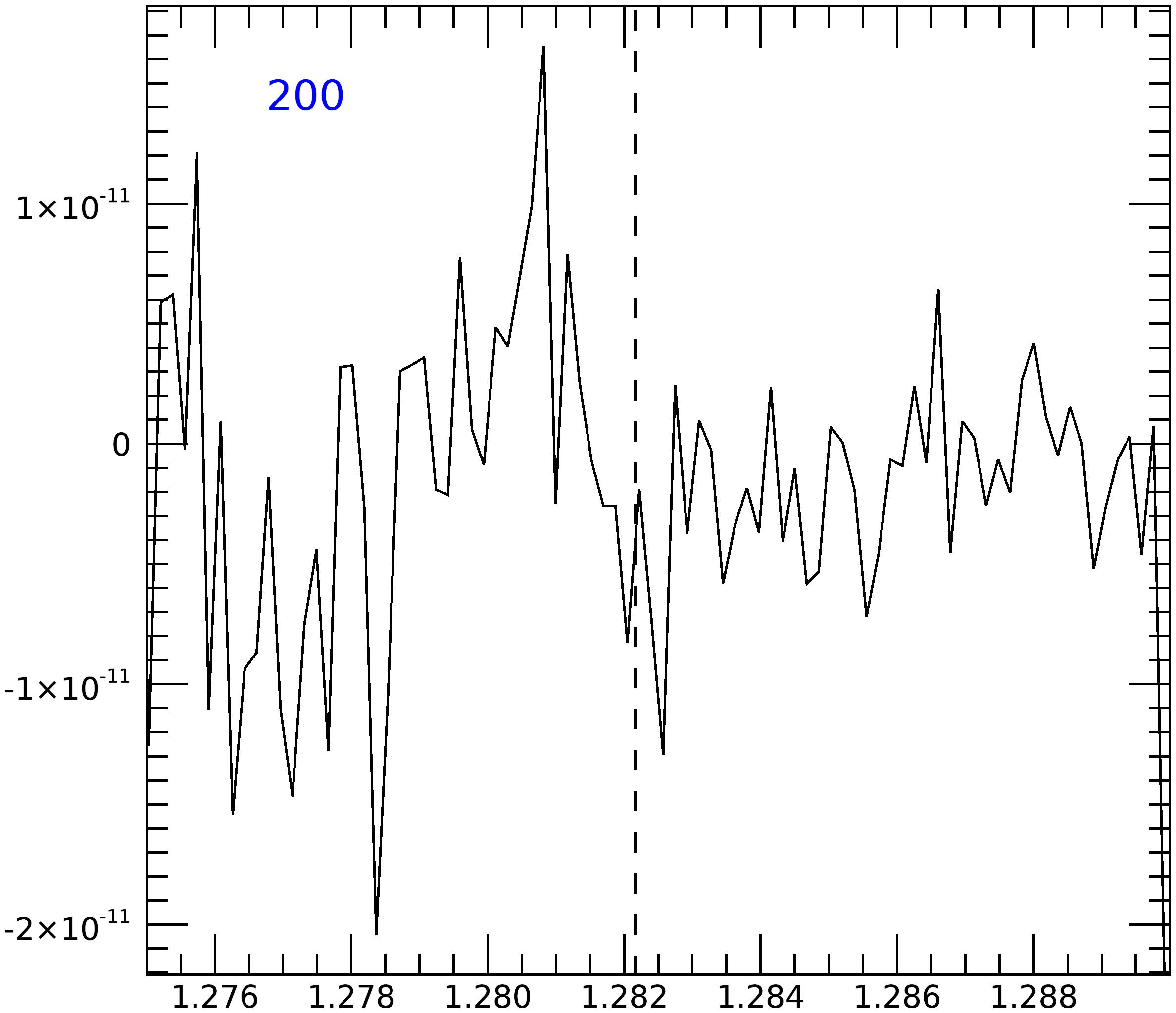}%
 
 \includegraphics[width=0.2\textwidth]{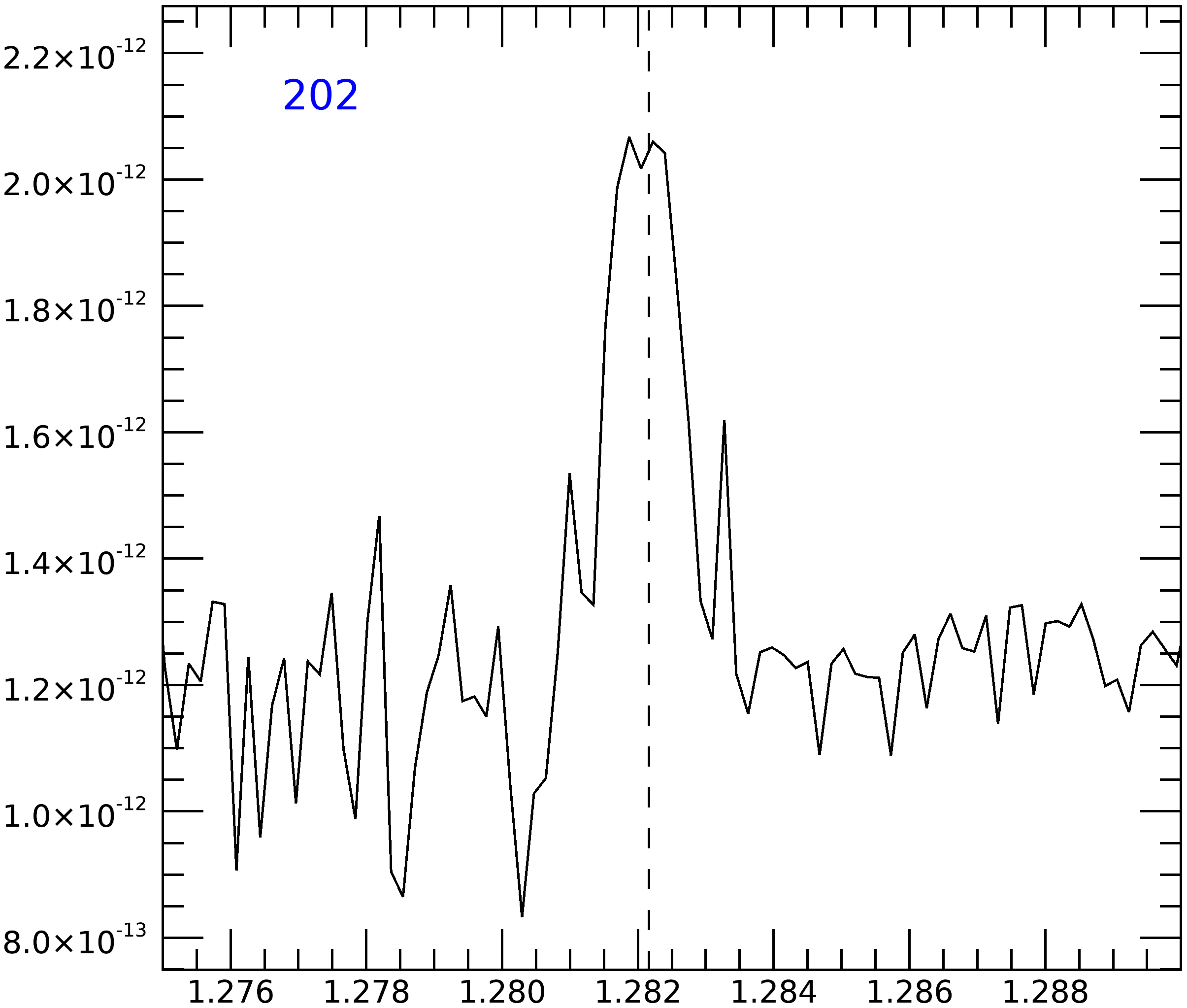}%
 \includegraphics[width=0.2\textwidth]{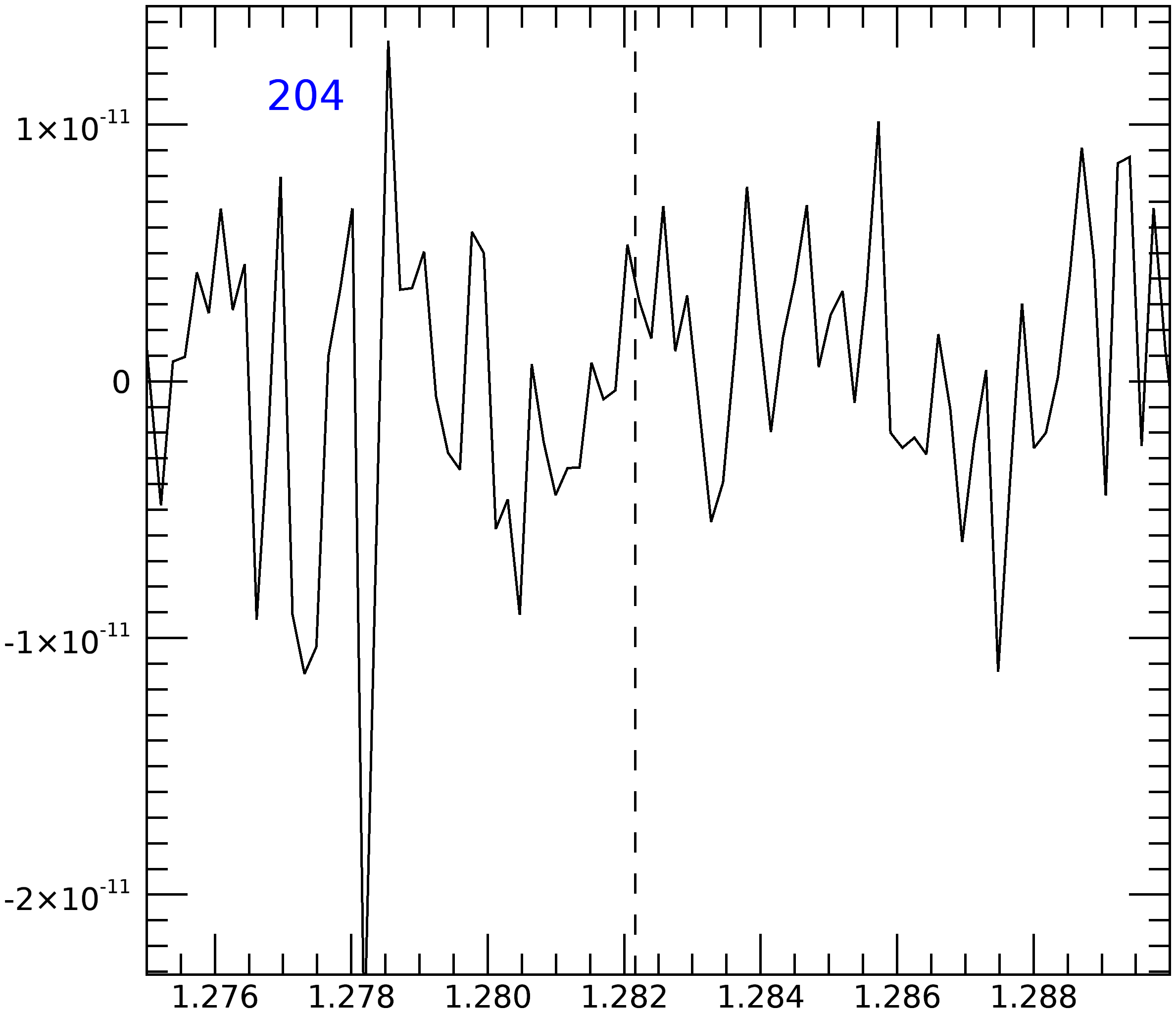}%
 \includegraphics[width=0.2\textwidth]{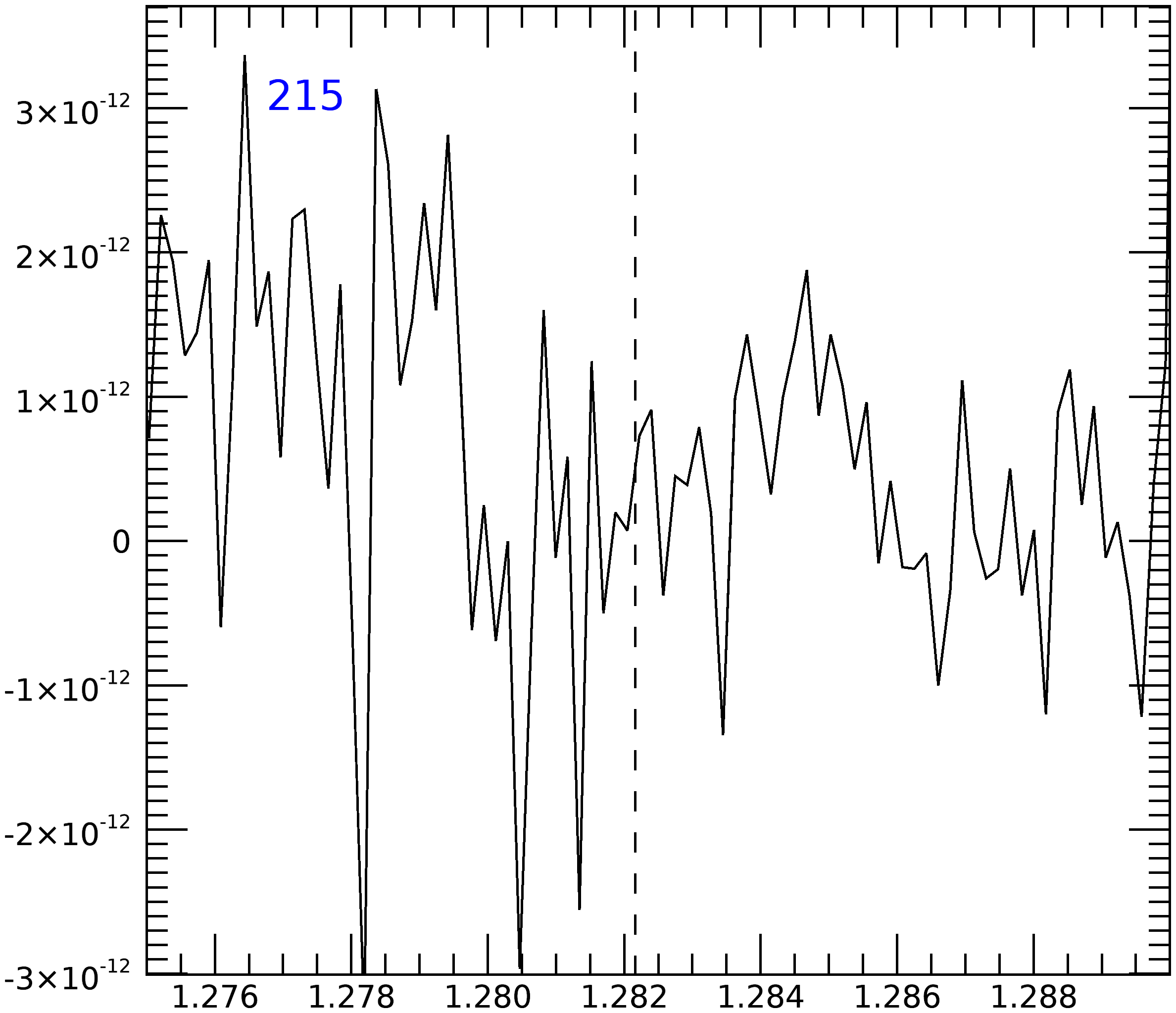}%
 \includegraphics[width=0.2\textwidth]{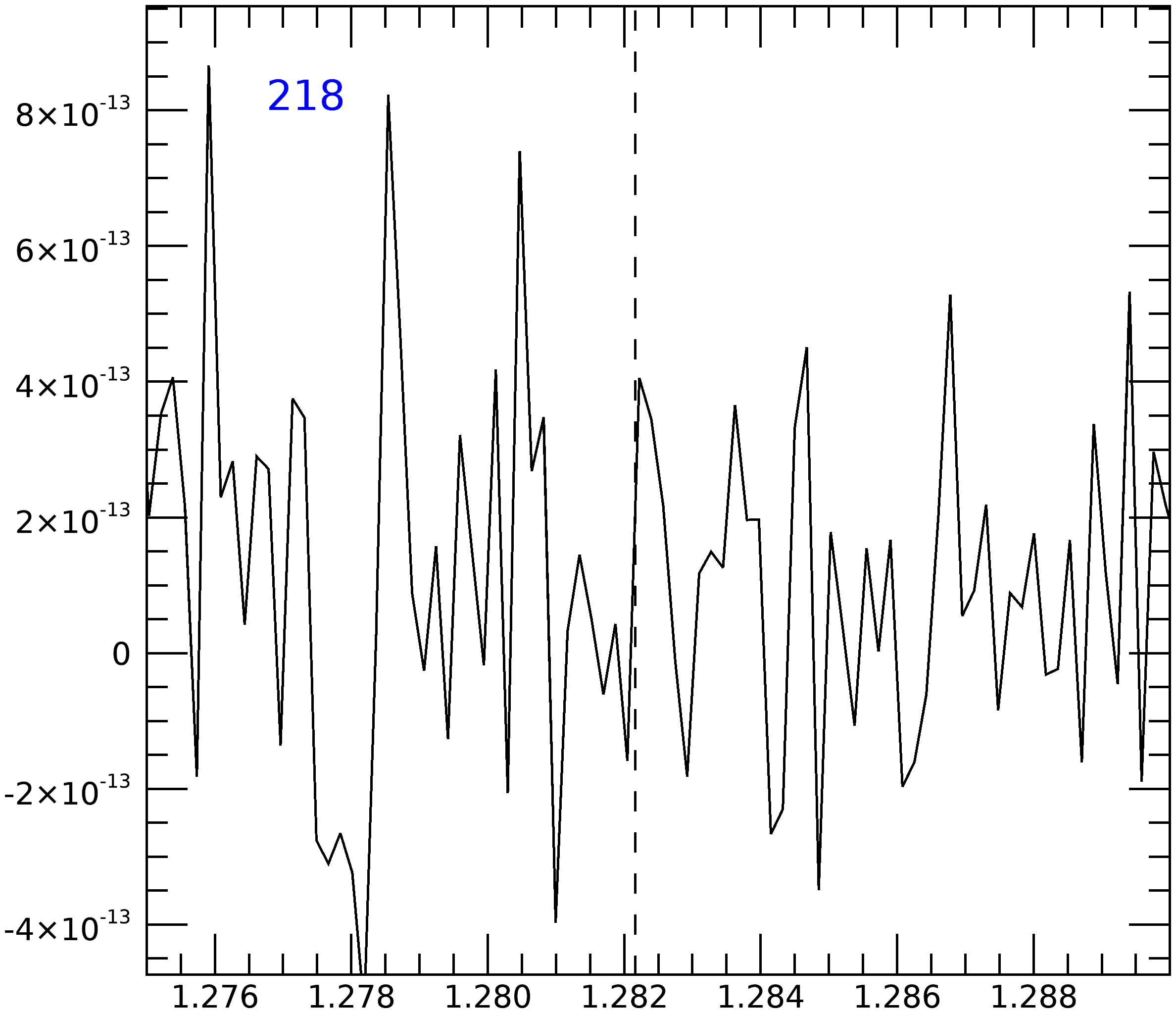}%
 \includegraphics[width=0.2\textwidth]{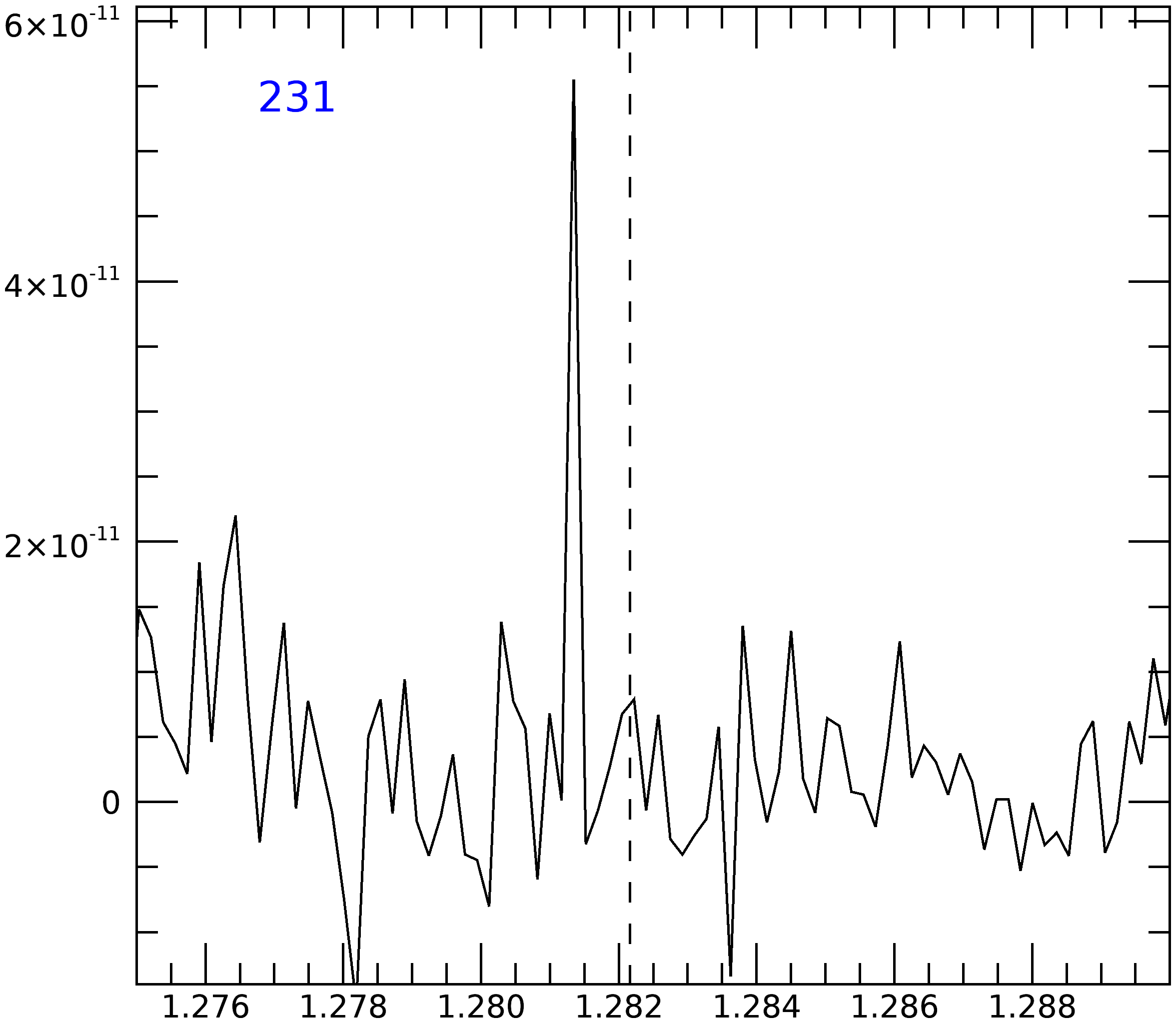}%
 
 \includegraphics[width=0.2\textwidth]{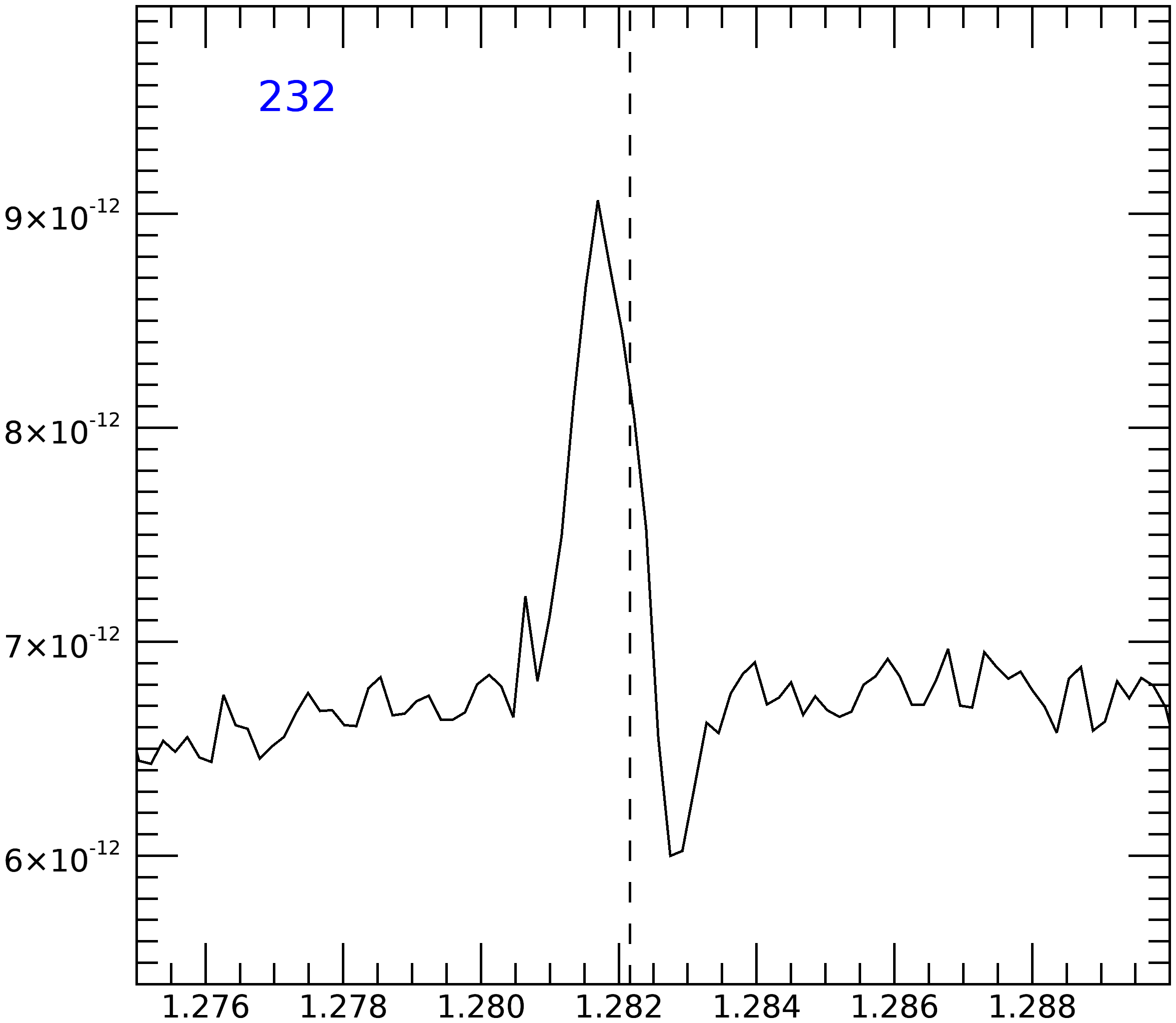}%
 \includegraphics[width=0.2\textwidth]{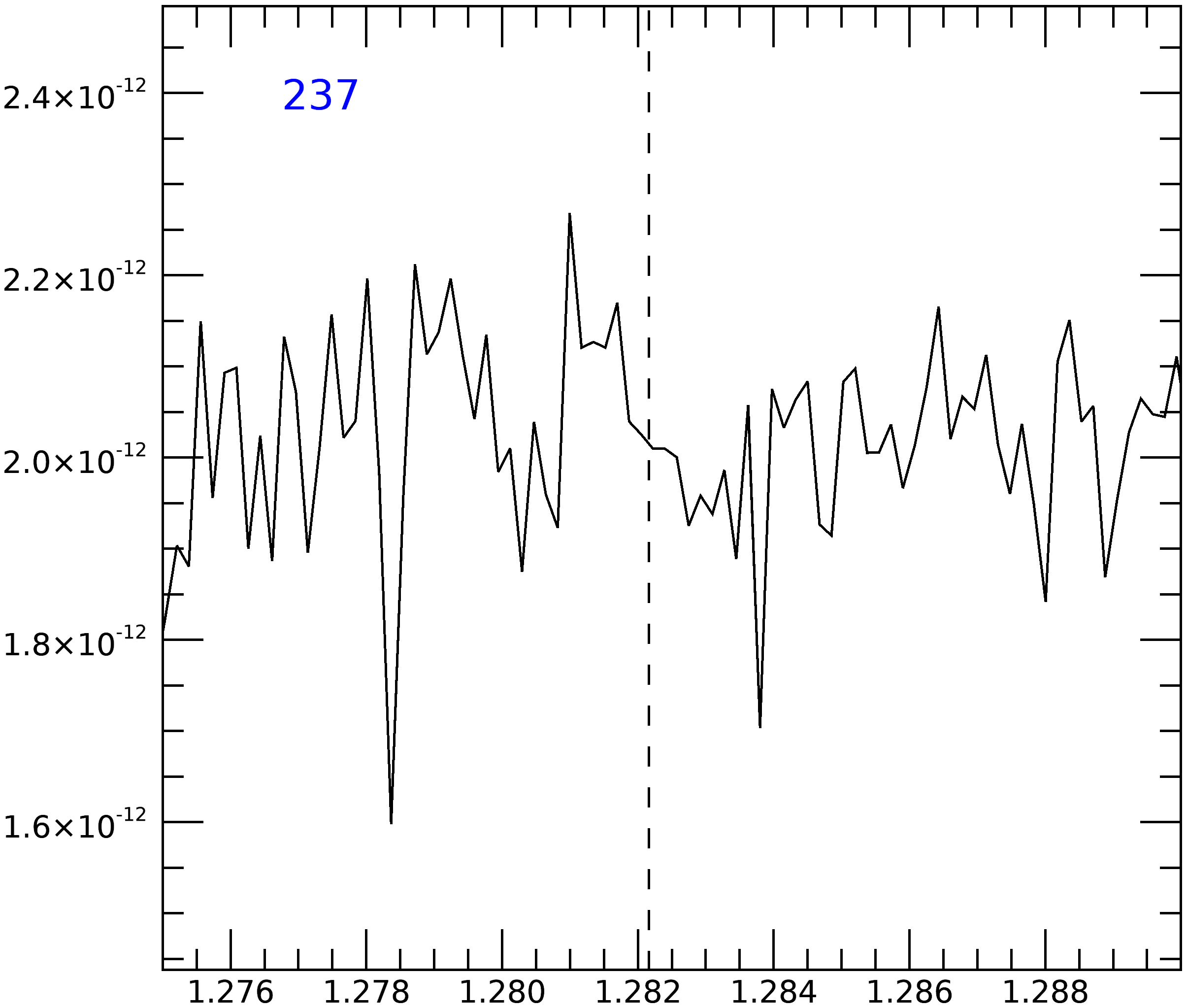}%

 \end{subfigure}
 \caption{\label{fig:linesIa}$\pab$ lines of Class~I objects. The flux is in erg~s$^{-1}$cm$^{-2}\mu$m$^{-1}$.} 
\end{figure*}

\begin{figure*} 
\centering
 \begin{subfigure}{\textwidth}
 \centering
 
 \includegraphics[width=0.2\textwidth]{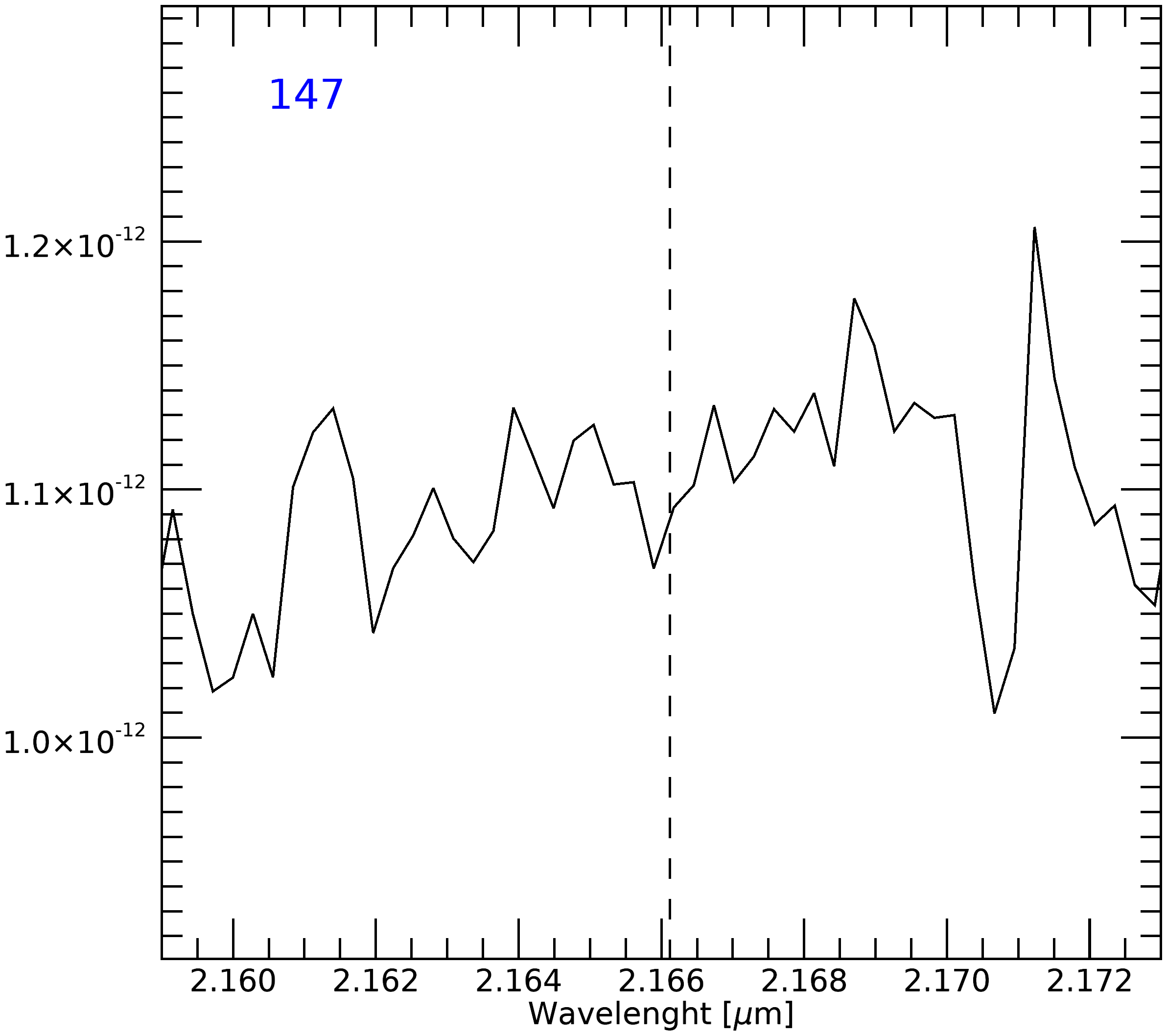}%
 \includegraphics[width=0.2\textwidth]{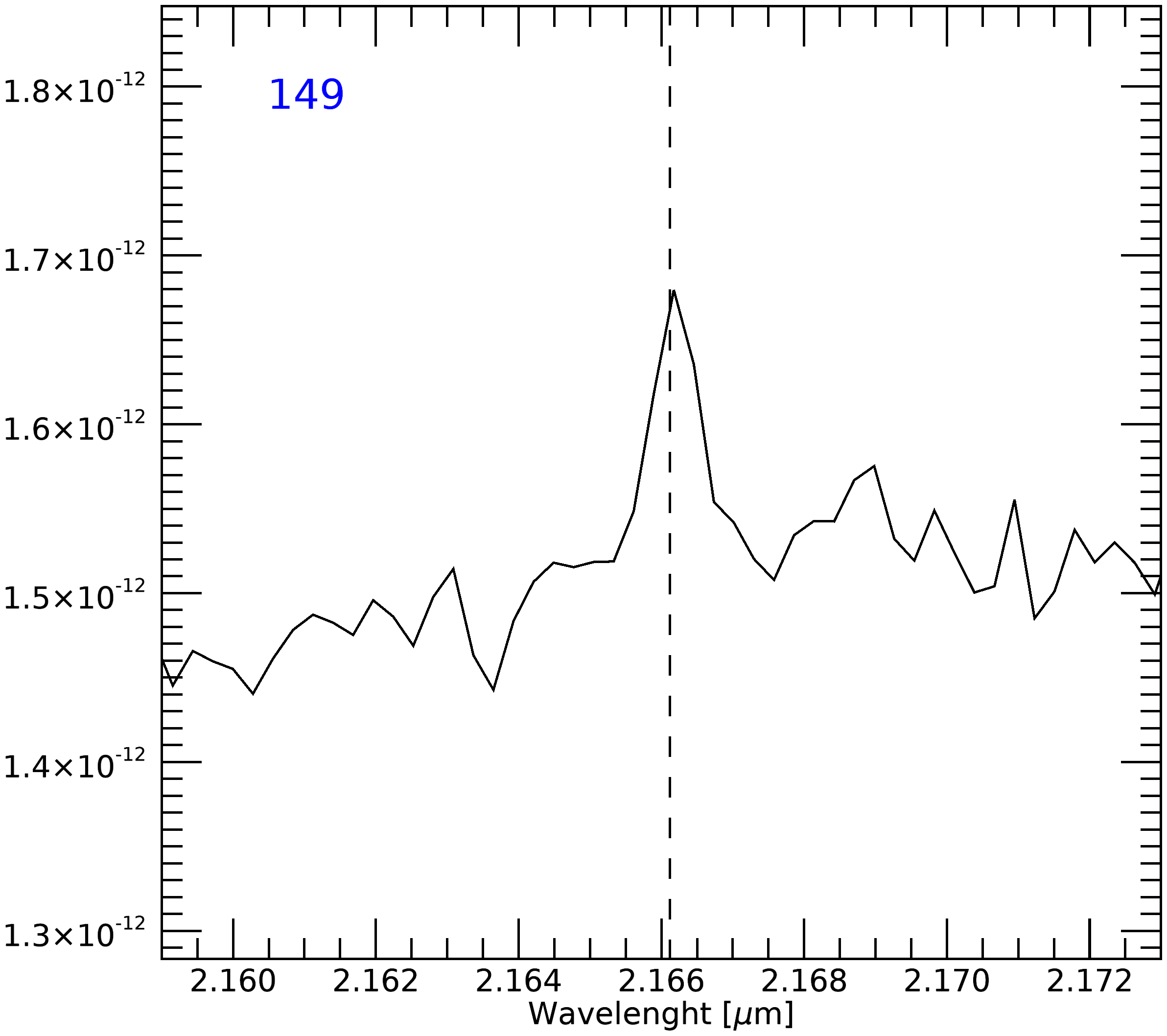}%
 \includegraphics[width=0.2\textwidth]{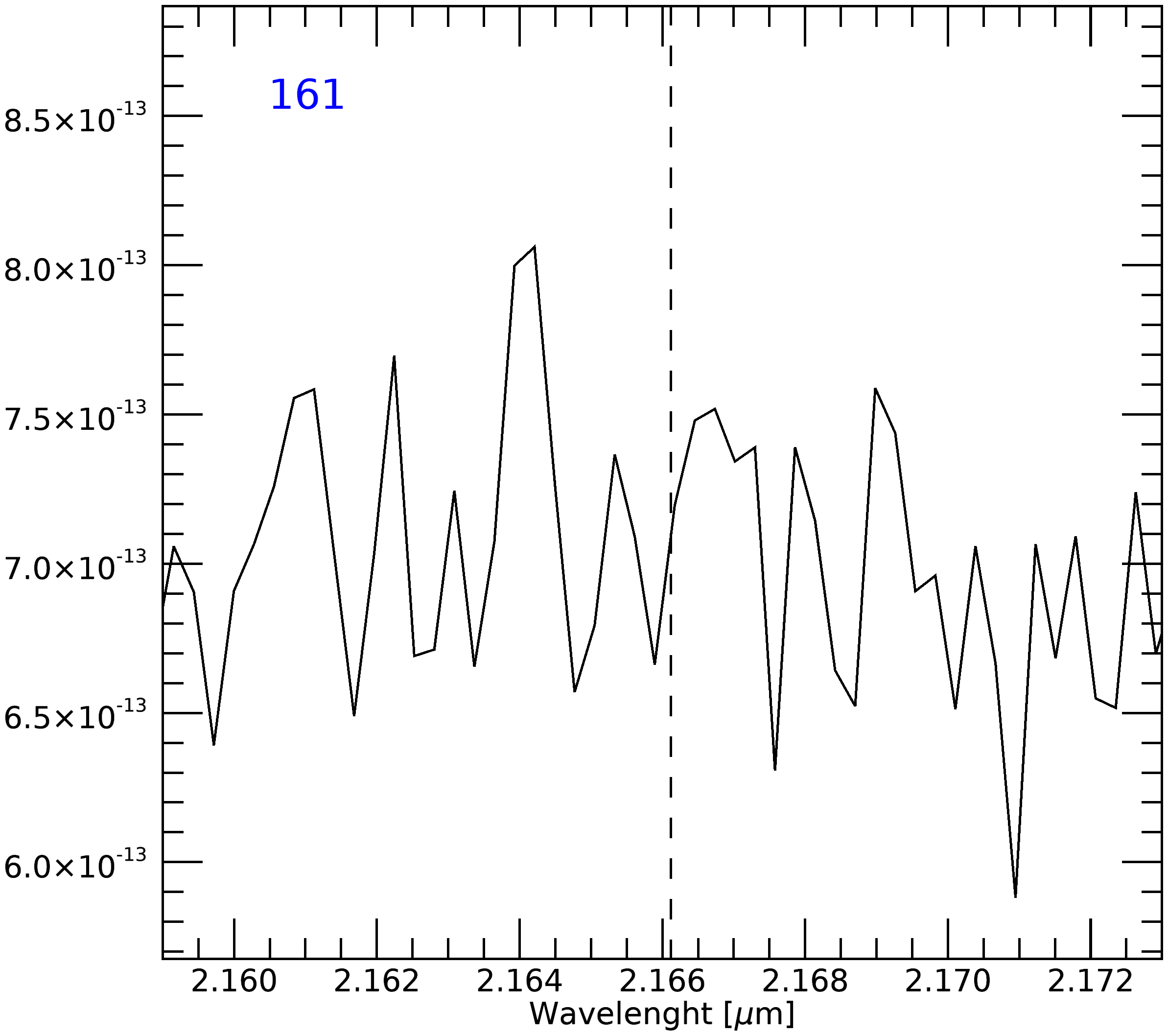}%
 \includegraphics[width=0.2\textwidth]{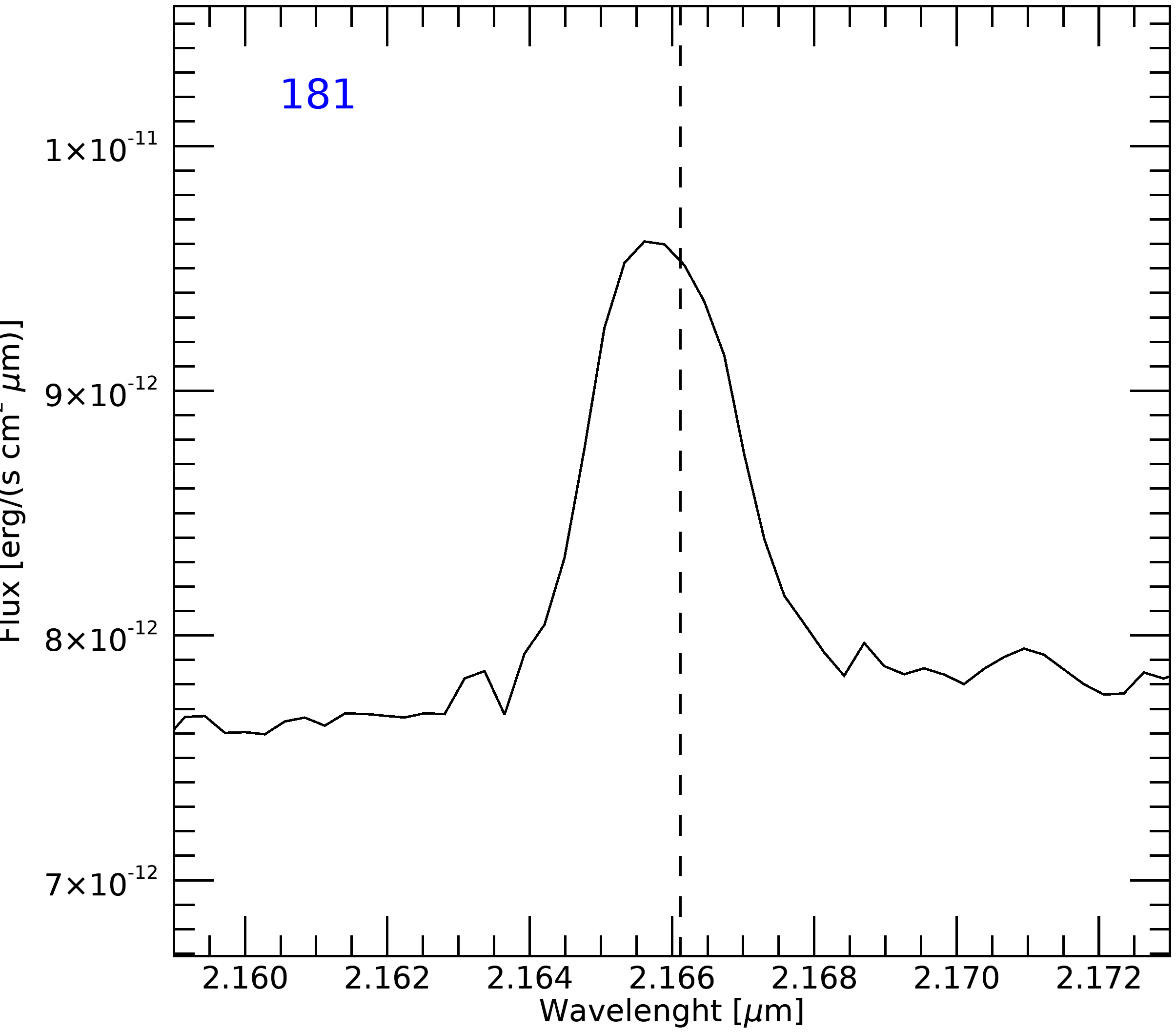}%
 \includegraphics[width=0.2\textwidth]{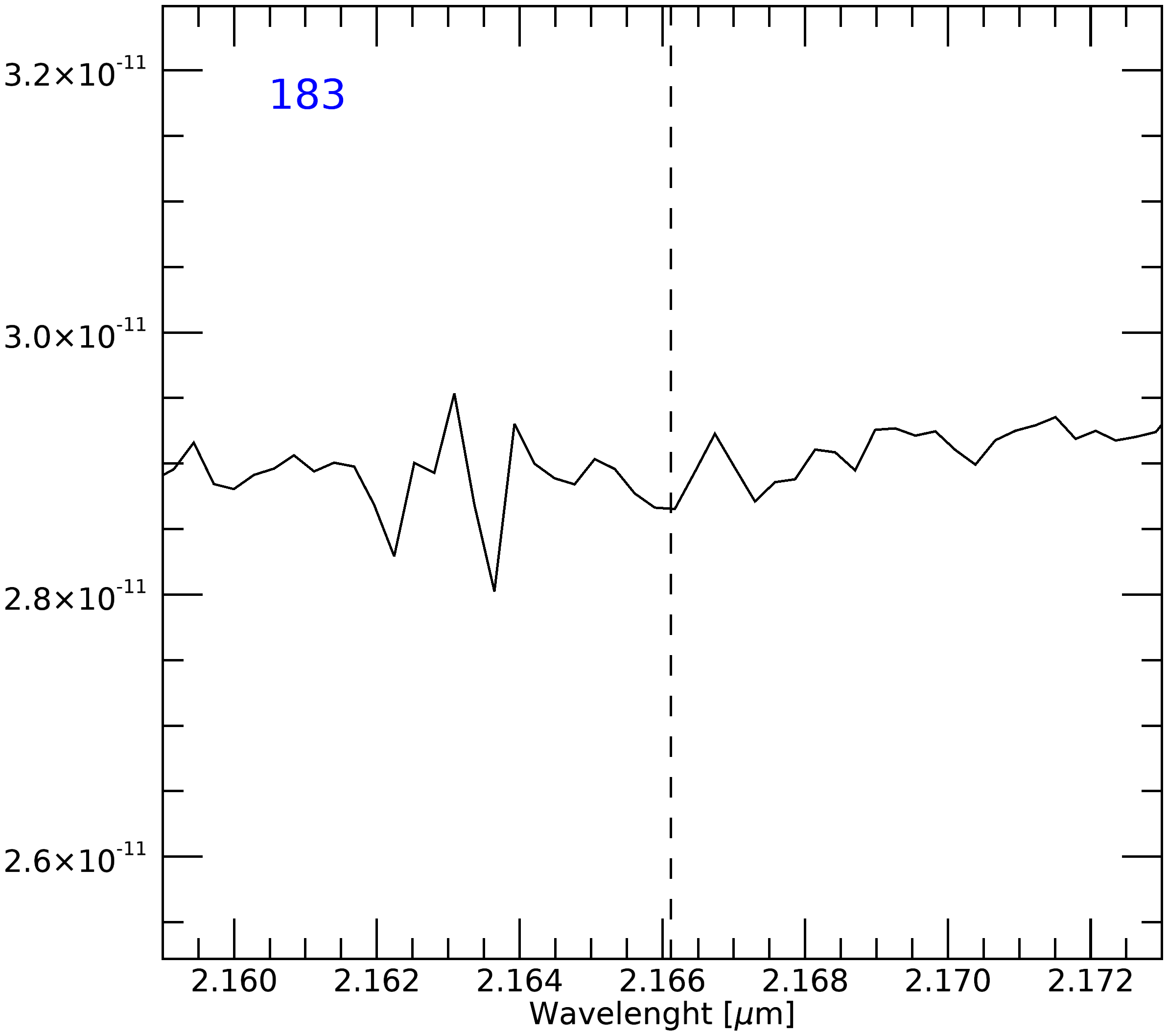}%
 
 \includegraphics[width=0.2\textwidth]{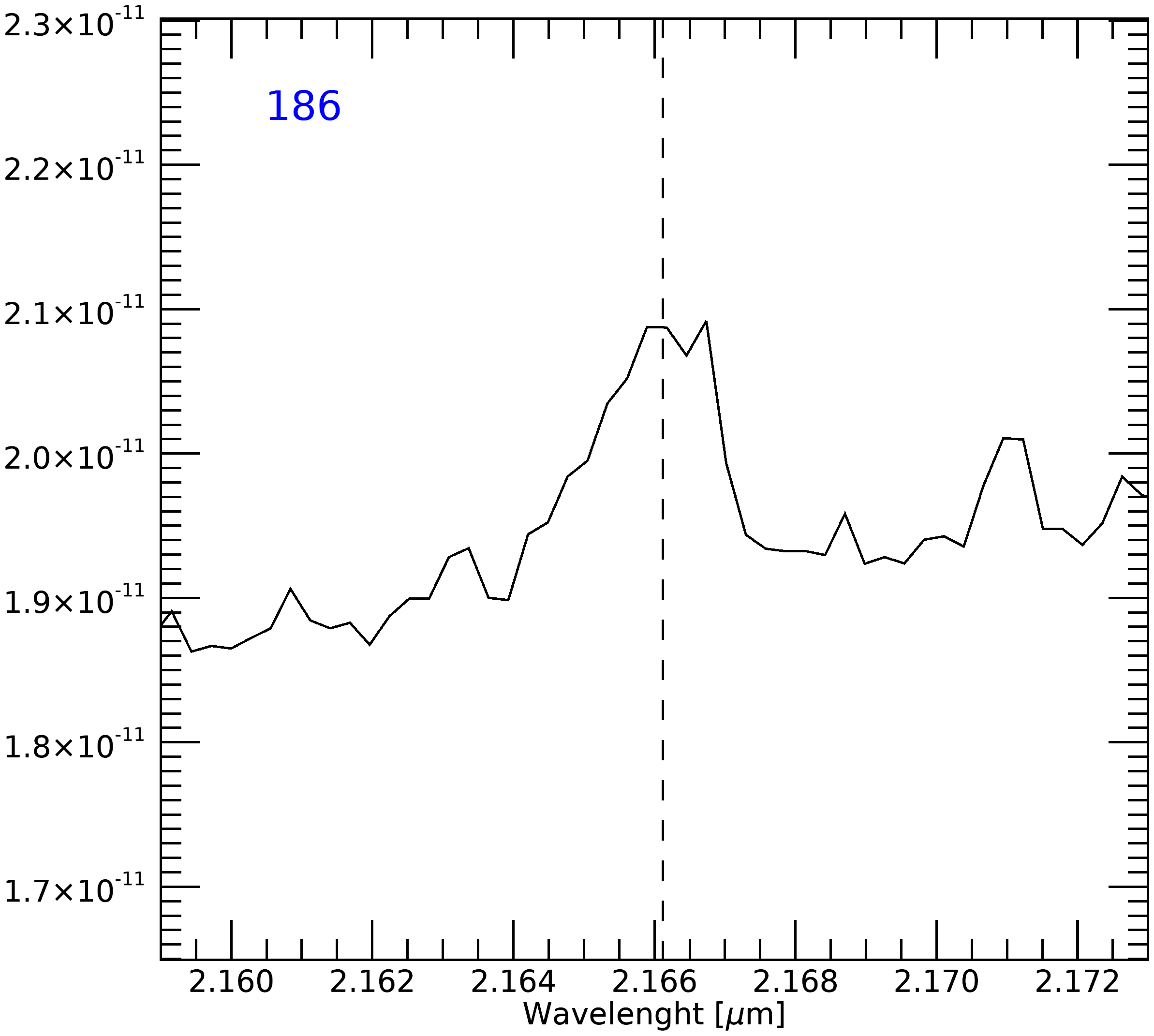}%
 \includegraphics[width=0.2\textwidth]{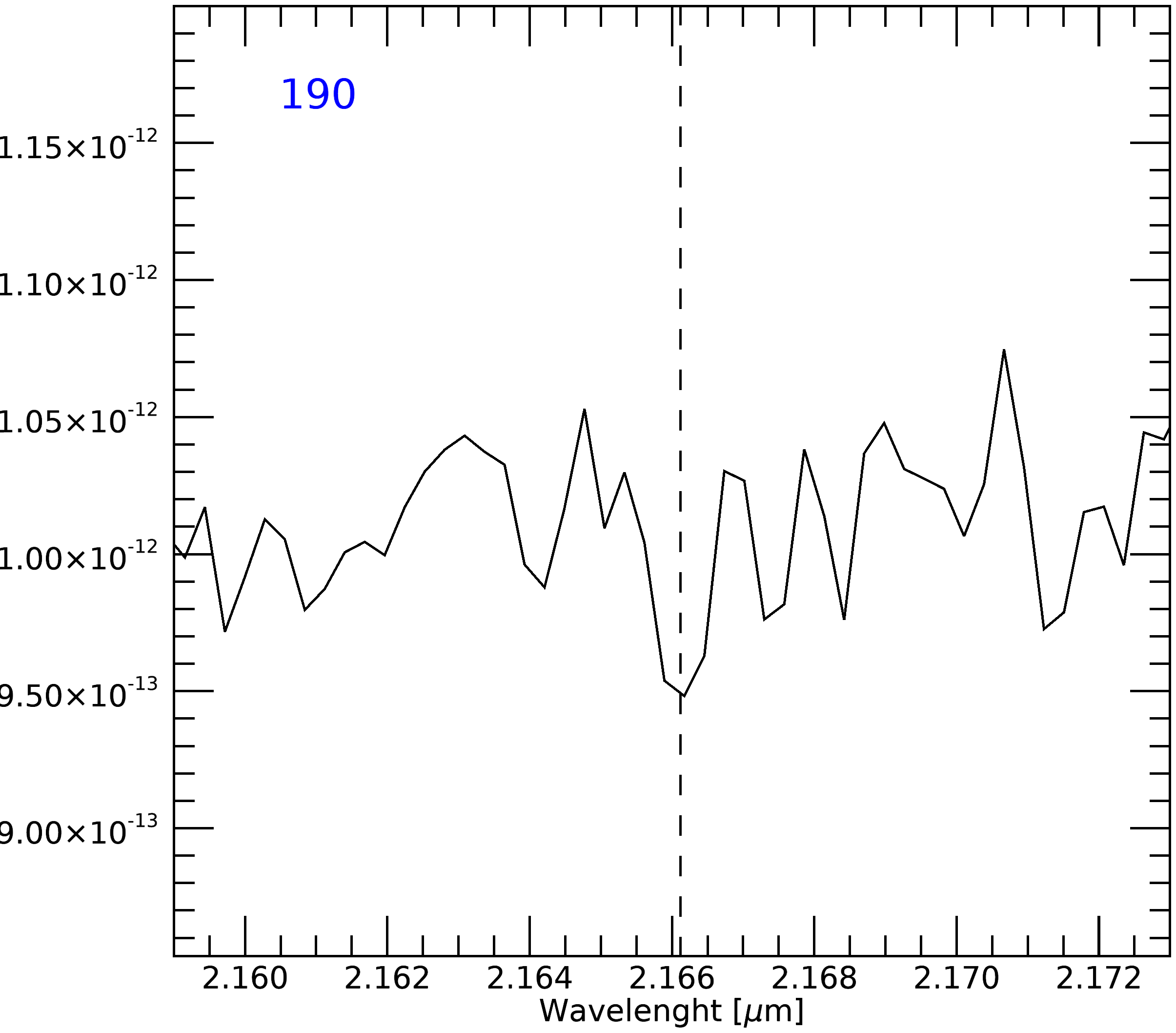}%
 \includegraphics[width=0.2\textwidth]{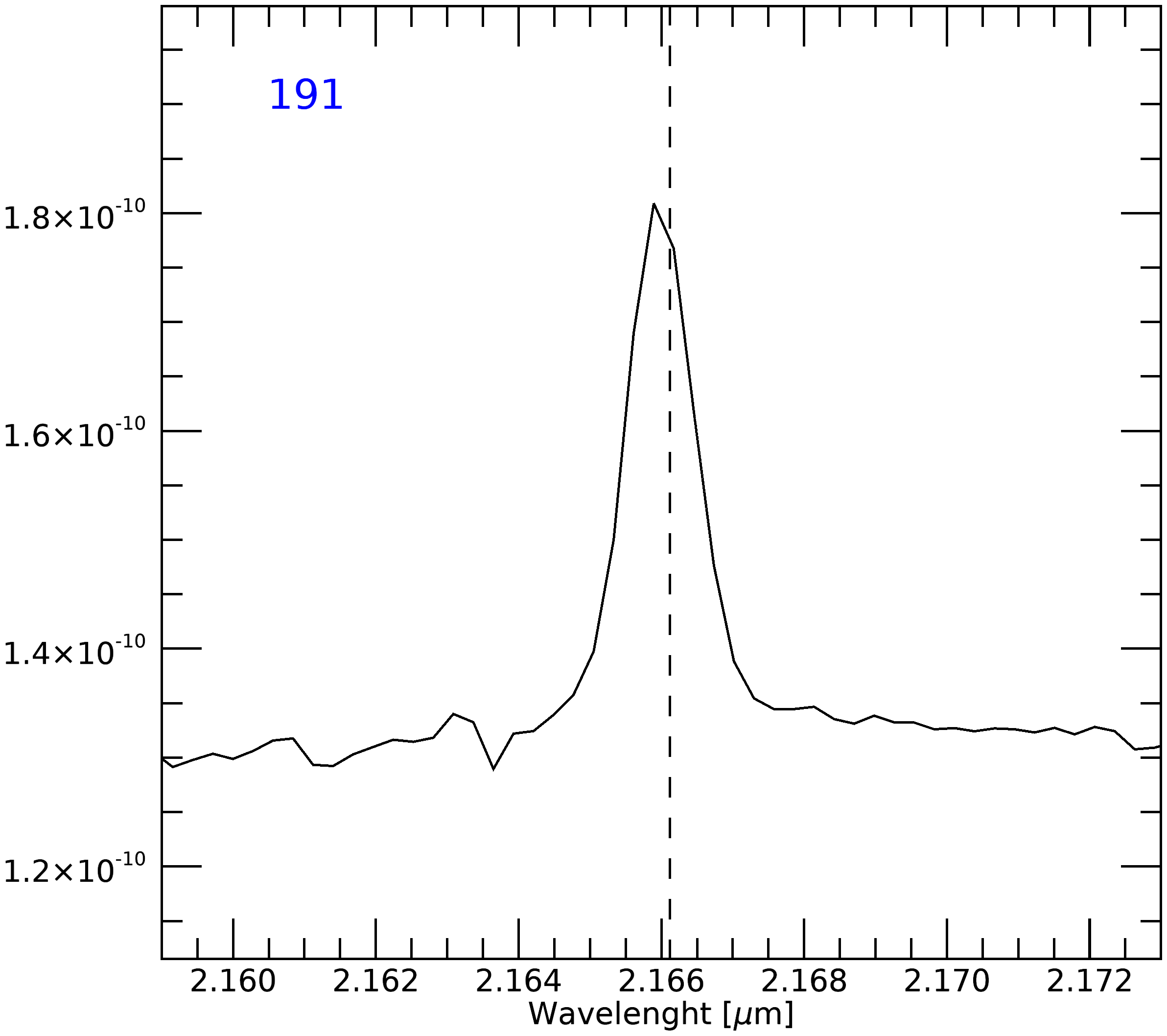}%
 \includegraphics[width=0.2\textwidth]{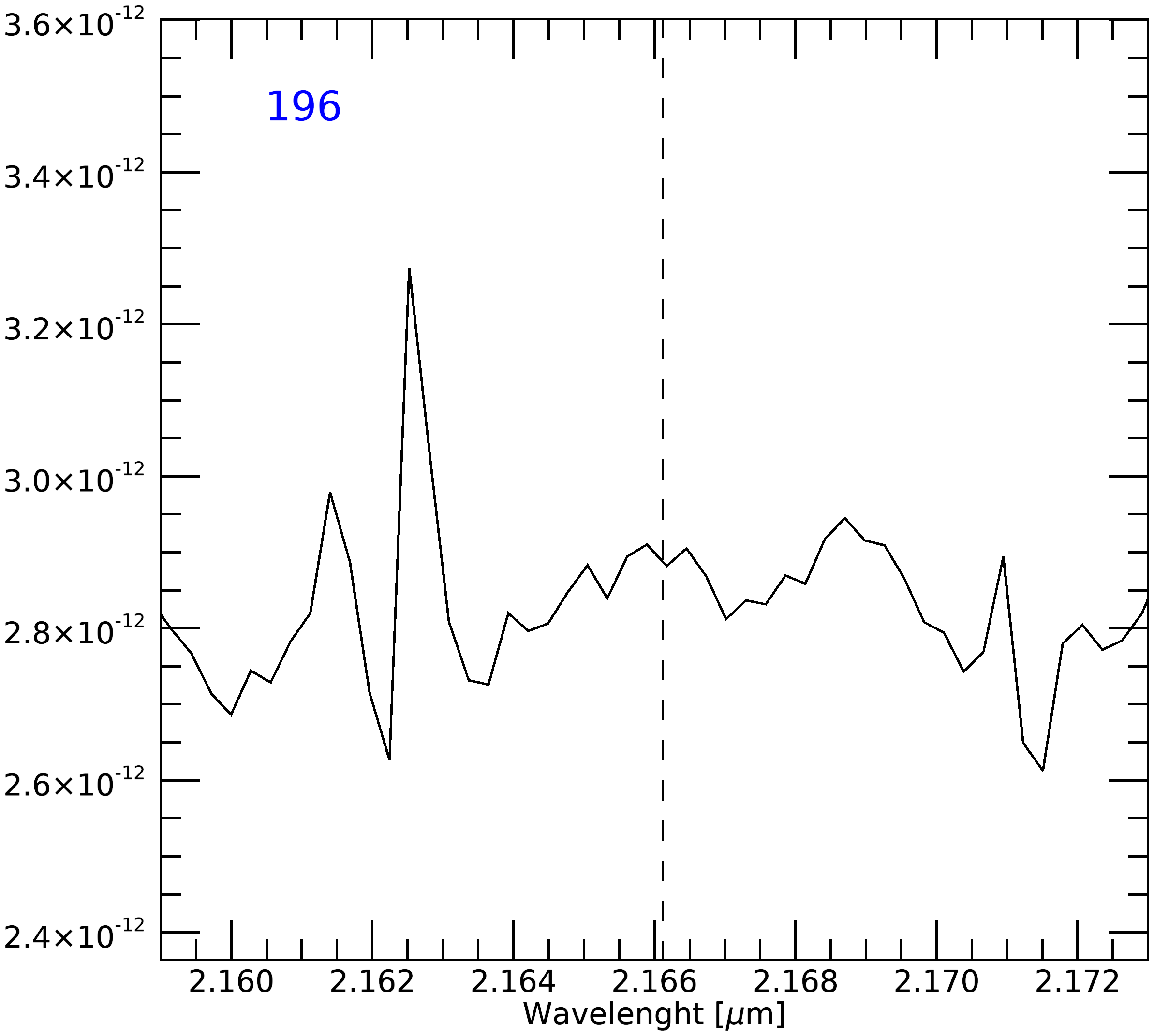}%
 \includegraphics[width=0.2\textwidth]{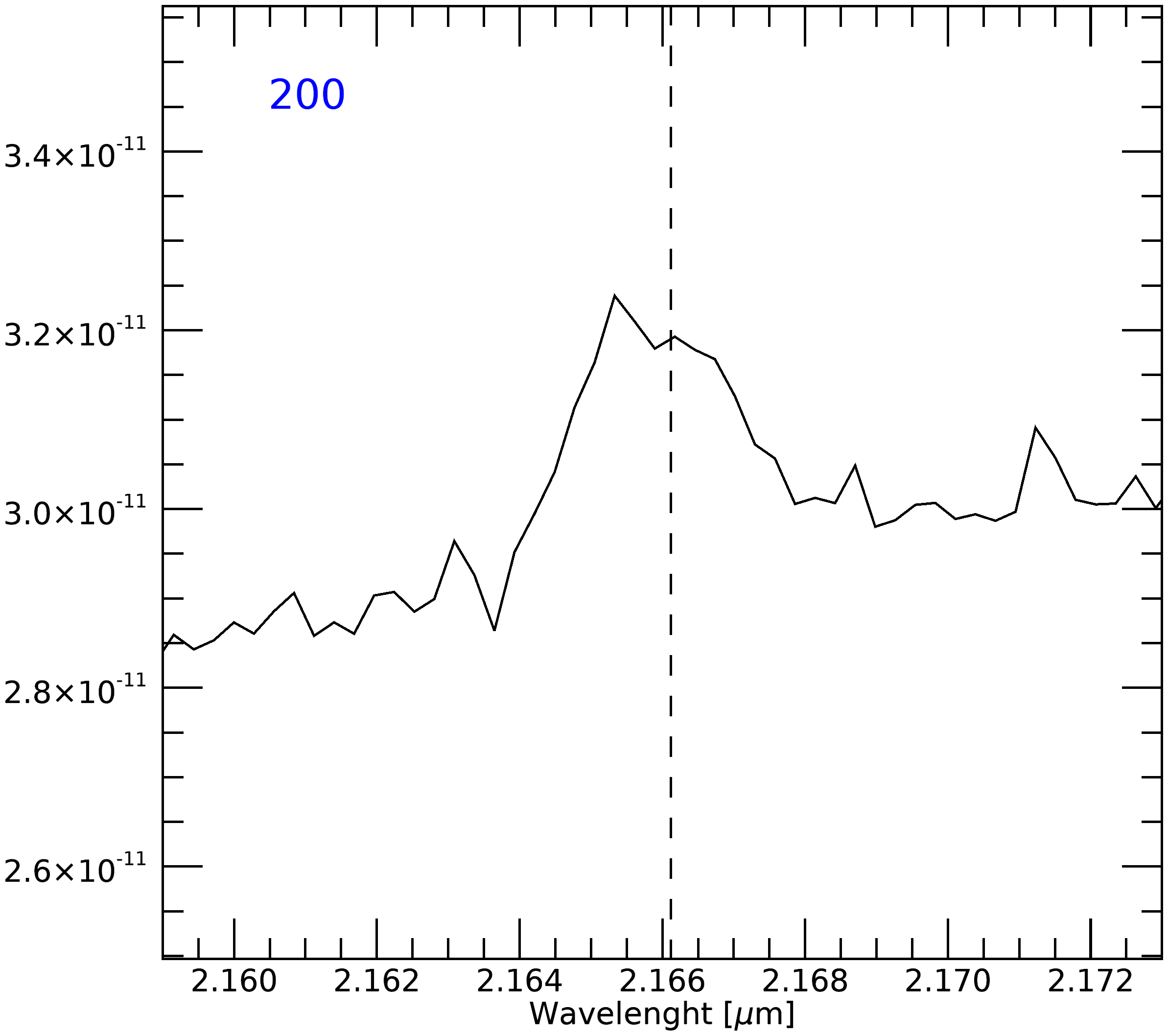}%
 
 \includegraphics[width=0.2\textwidth]{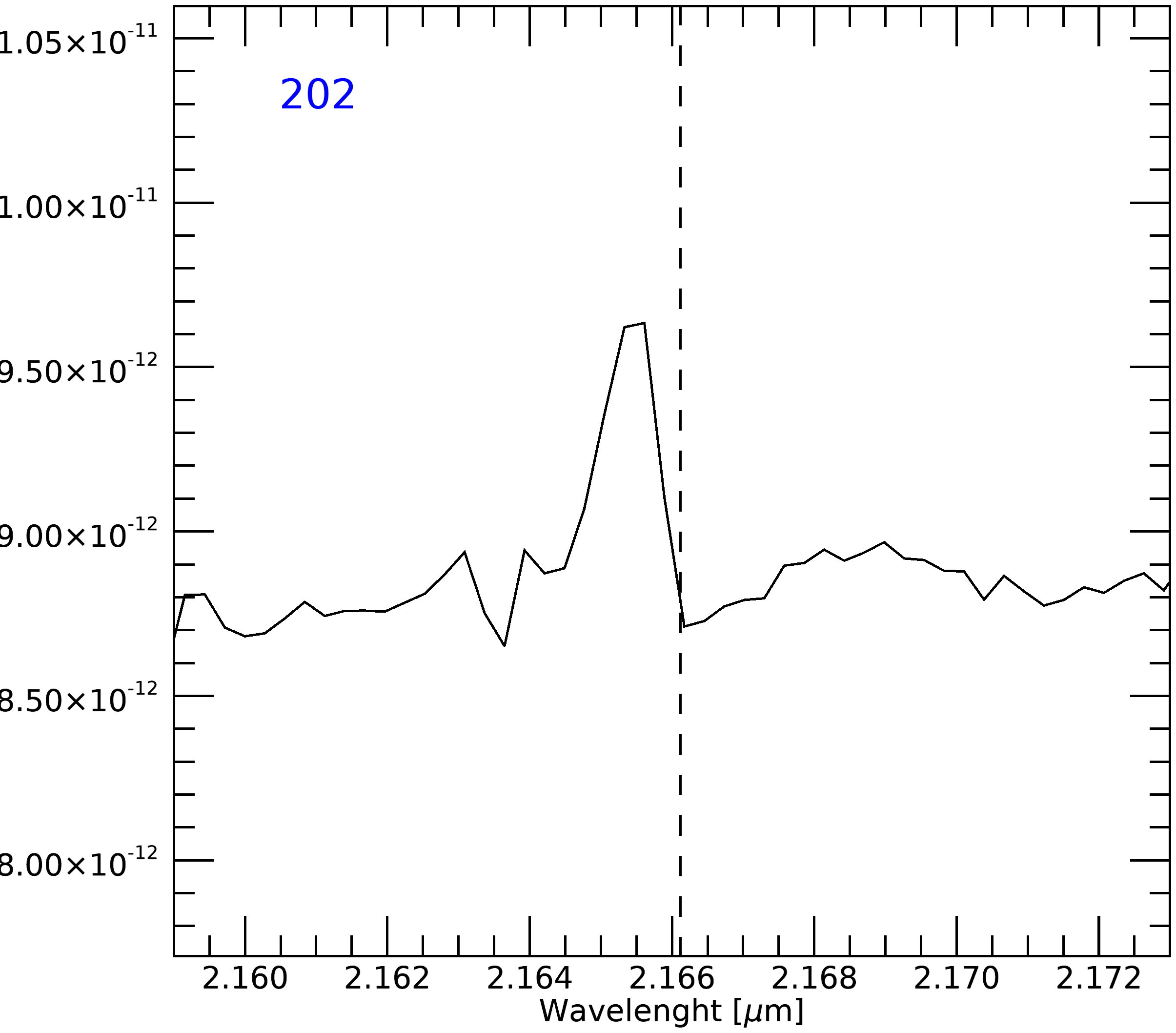}%
 \includegraphics[width=0.2\textwidth]{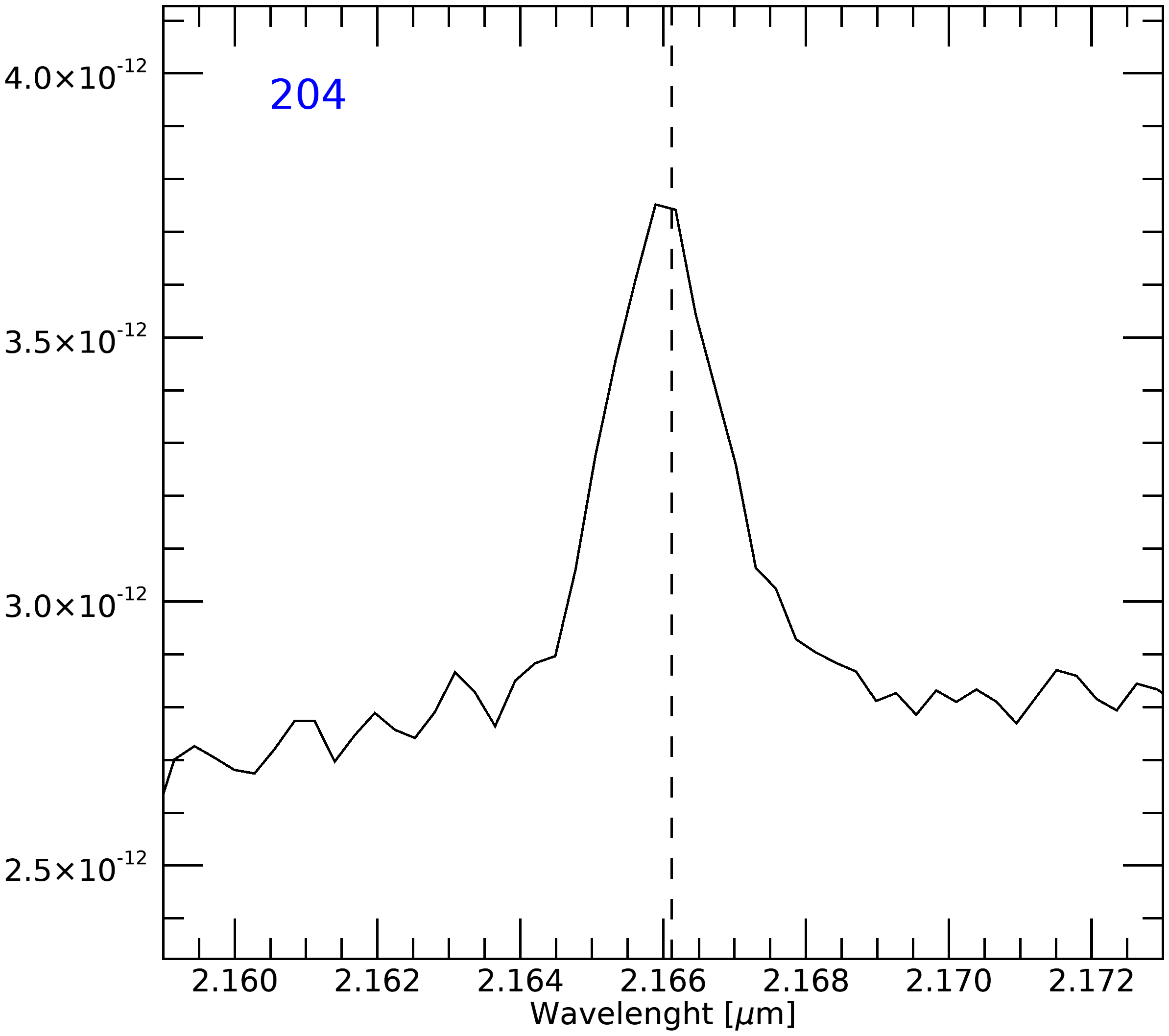}%
 \includegraphics[width=0.2\textwidth]{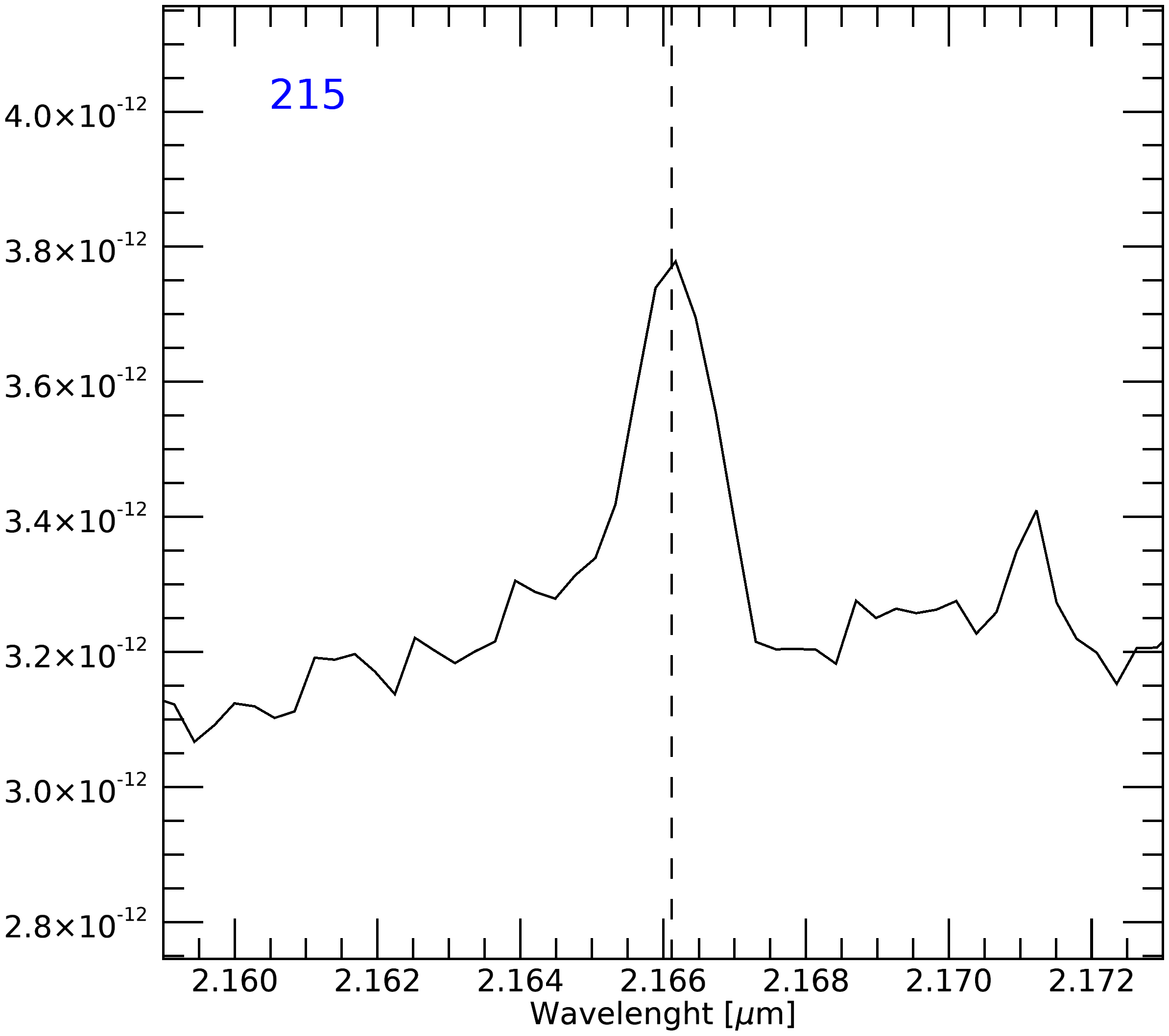}%
 \includegraphics[width=0.2\textwidth]{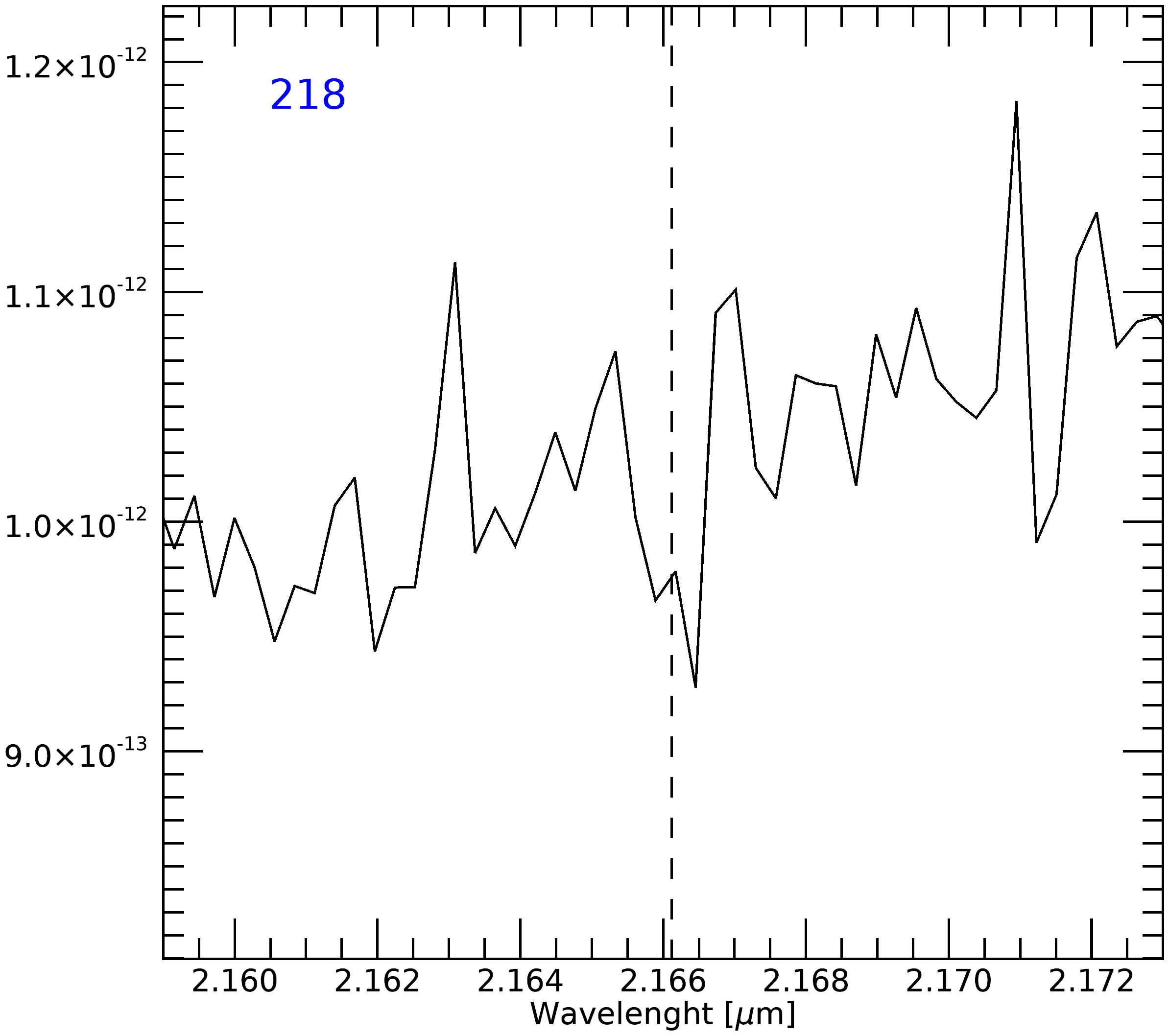}%
 \includegraphics[width=0.2\textwidth]{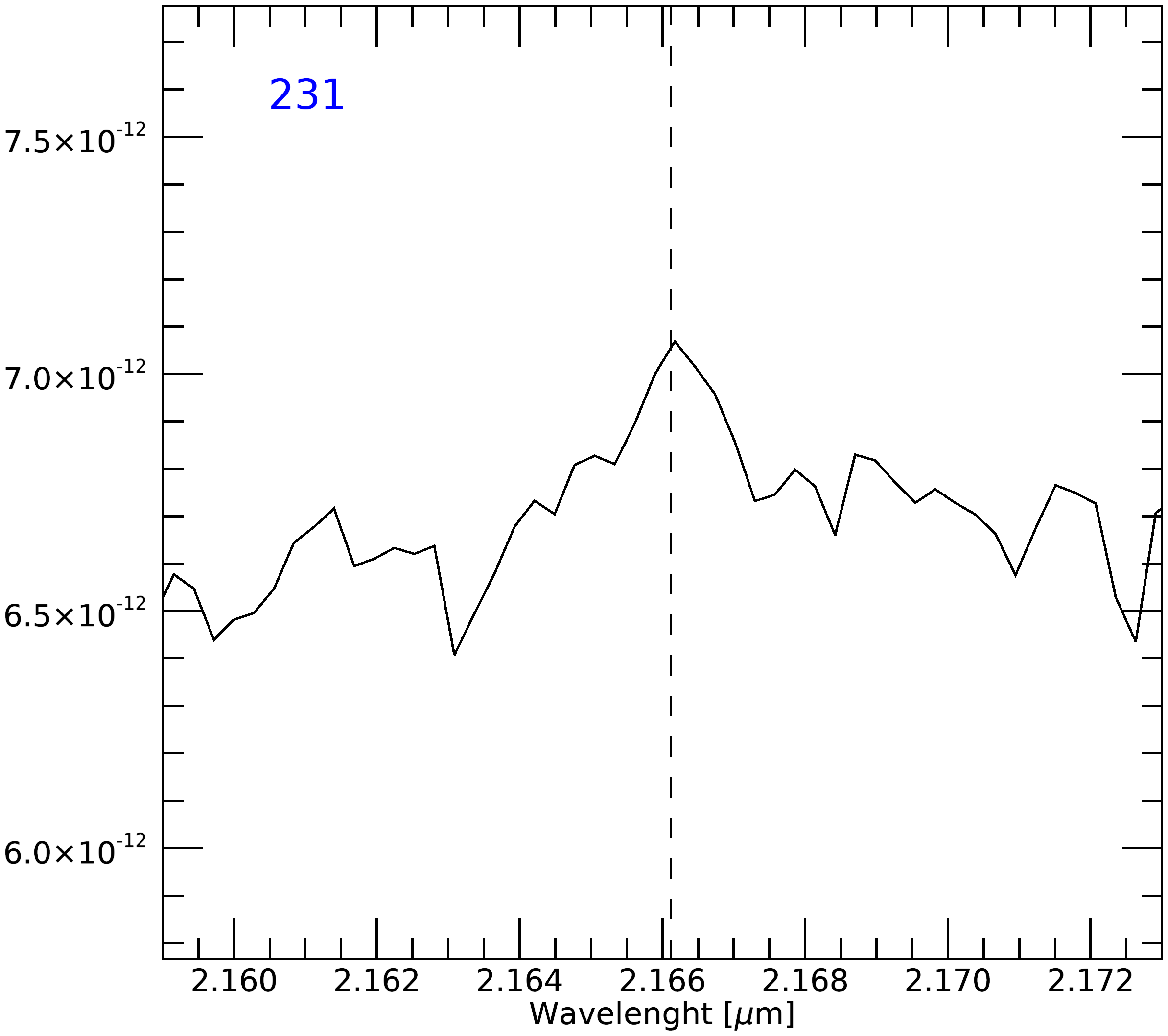}%
 
 \includegraphics[width=0.2\textwidth]{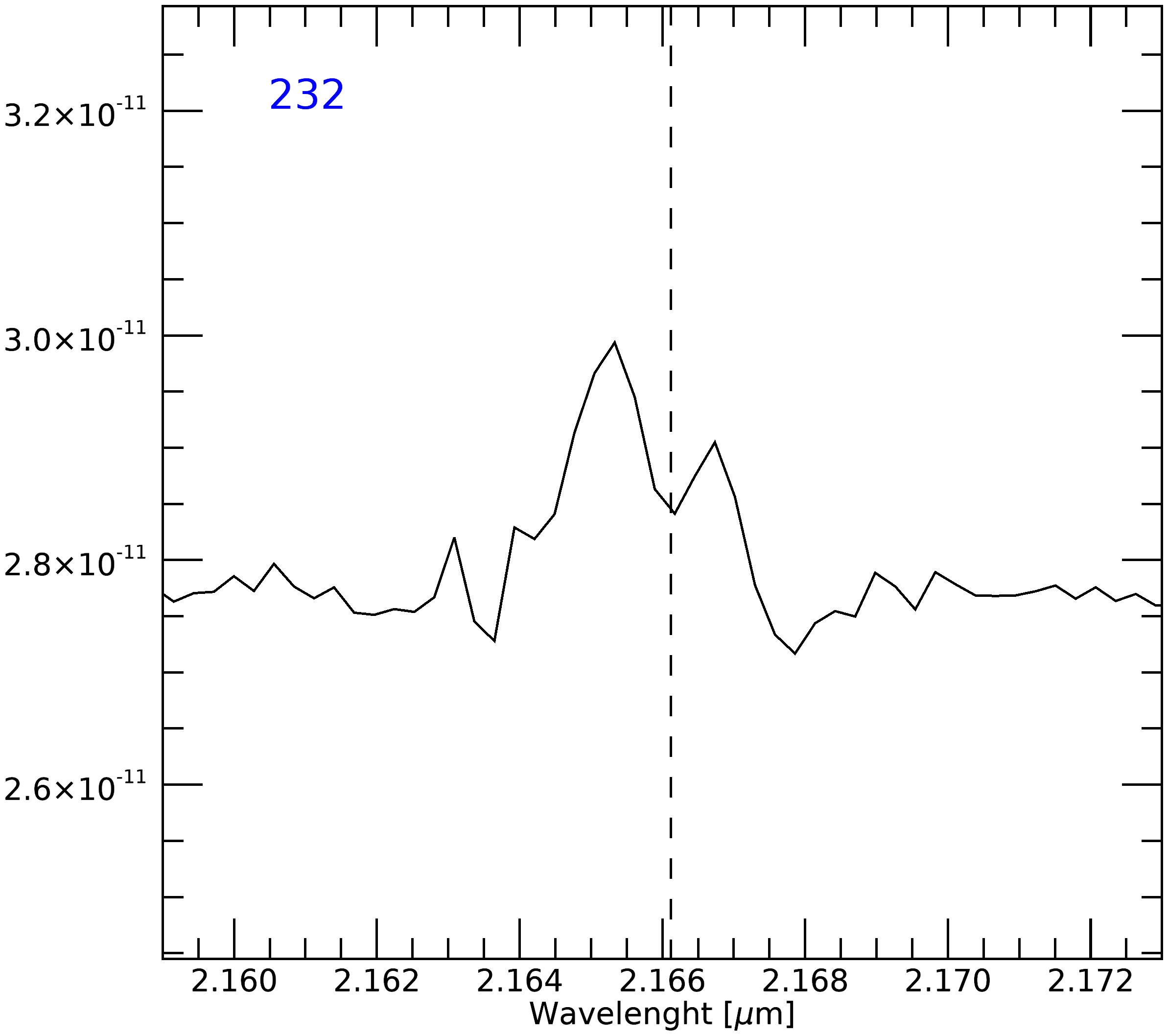}%
 \includegraphics[width=0.2\textwidth]{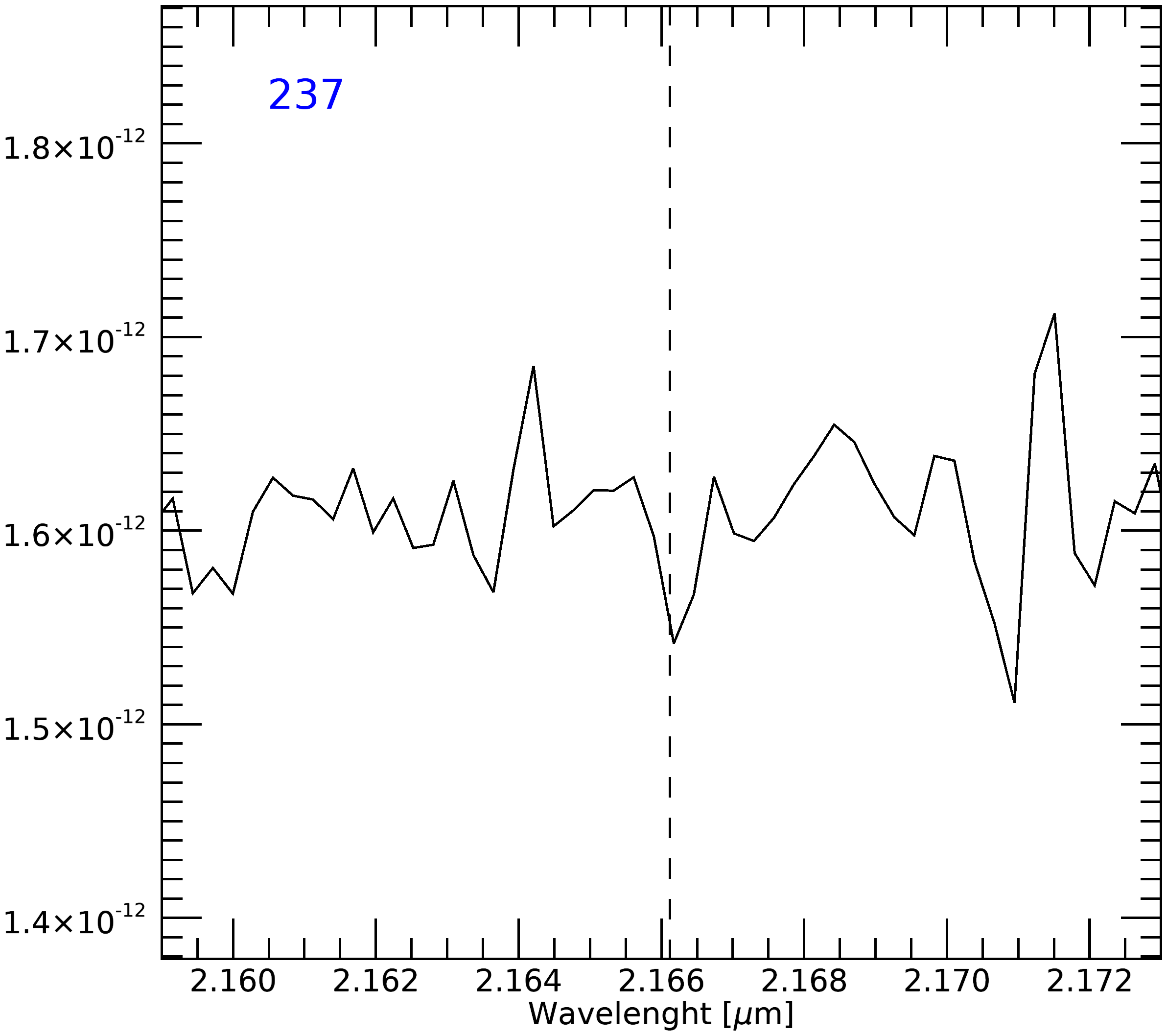}%
 
 \end{subfigure}
 \caption{\label{fig:linesIc}$\brg$ Class~I lines. The flux is in erg~s$^{-1}$cm$^{-2}\mu$m$^{-1}$.} 
\end{figure*}

Figs.~\ref{fig:lines2a} and \ref{fig:lines2b} show $\pab$ lines, while Figs.~\ref{fig:lines2c} and \ref{fig:lines2d} show $\brg$ lines for Class~II YSOs. 
Figs.~\ref{fig:linesIa} and \ref{fig:linesIc} show $\pab$ and $\brg$ lines, respectively, for Class~I YSOs. 
Some sources were observed two times (dim and bright OBs, see Table~\ref{obs}), but since there is no variability between them, we report only the spectra observed with a longer integration time.


\clearpage
\section{Other sources}
\label{appnoyso}
\begin{table*}[b]
 \centering
 \caption{\label{starparnoYSO} Observed objects  whose spectra are not compatible with the accreting YSOs ones.}
  \resizebox{\textwidth}{!}{%
  \begin{tabular}{lcccccrccc}
  \hline
  \hline
   ID&  E09&  2MASS name     &      RA   &    Dec   &Class  & $\alpha_{\rm IR}$ &   J              &     H          &     K        \\
     &      &                    &    &   &     &  & mag              & mag            &    mag  \\
    \hline
    \hline
 152 & 148  & J03284325+3117330  &  03:28:43.28 & +31:17:32.89 & Flat & $0.24$ & $12.59 \pm 0.02 $&$10.86 \pm 0.05$ &$  9.75 \pm 0.04$\\
 160 & 158  &J032851.20+311954.8 &  03:28:51.20 & +31:19:54.80 &   II  & $-1.00$ & $11.72 \pm  0.02$&$10.49 \pm 0.02$ &  $9.90 \pm 0.02$ \\
 176 & 175  &J032857.21+311419.1 &  03:28:57.21 & +31:14:19.10 &  I  & $1.40$ & $8.19  \pm  0.03$&$ 7.77 \pm 0.02$ &  $7.66 \pm 0.02$ \\
 198 & 202  &J032905.78+311639.6 &  03:29:05.78 & +31:16:39.61 &   II  & $-1.08$ & $14.49 \pm  0.06$&$11.61 \pm 0.04$ &  $9.93 \pm 0.03$   \\
 208 & 213  &J032909.65+312256.3 &  03:29:09.65 & +31:22:56.32 &   II  & $-1.60$ & $11.27 \pm 0.02 $&$10.16 \pm 0.03$ &  $9.53 \pm 0.02$ \\
 221 & 227  &J032914.40+311444.1 &  03:29:14.40 & +31:14:44.09 &   II  & $-0.70$ & $< 17.76 $       &$<16.02$         & $14.34 \pm 0.11$ \\
 223 & 229  &J032916.69+311618.2 &  03:29:16.69 & +31:16:18.19 &  III & $-2.43$  & $11.44 \pm 0.02$ &$10.75 \pm 0.03$ & $10.43 \pm 0.02$ \\
 228 & 235  &J032918.26+312319.9 &  03:29:18.26 & +31:23:19.90 &  I  & $1.09$ & $11.45 \pm 0.02$ &$10.68 \pm 0.03$ & $10.33 \pm 0.02$ \\
    \hline  
    \hline
  \end{tabular}  
  }
  \begin{quotation}
  \end{quotation}  
  \end{table*}

\begin{figure*}
    \centering
    \includegraphics[width=\textwidth]{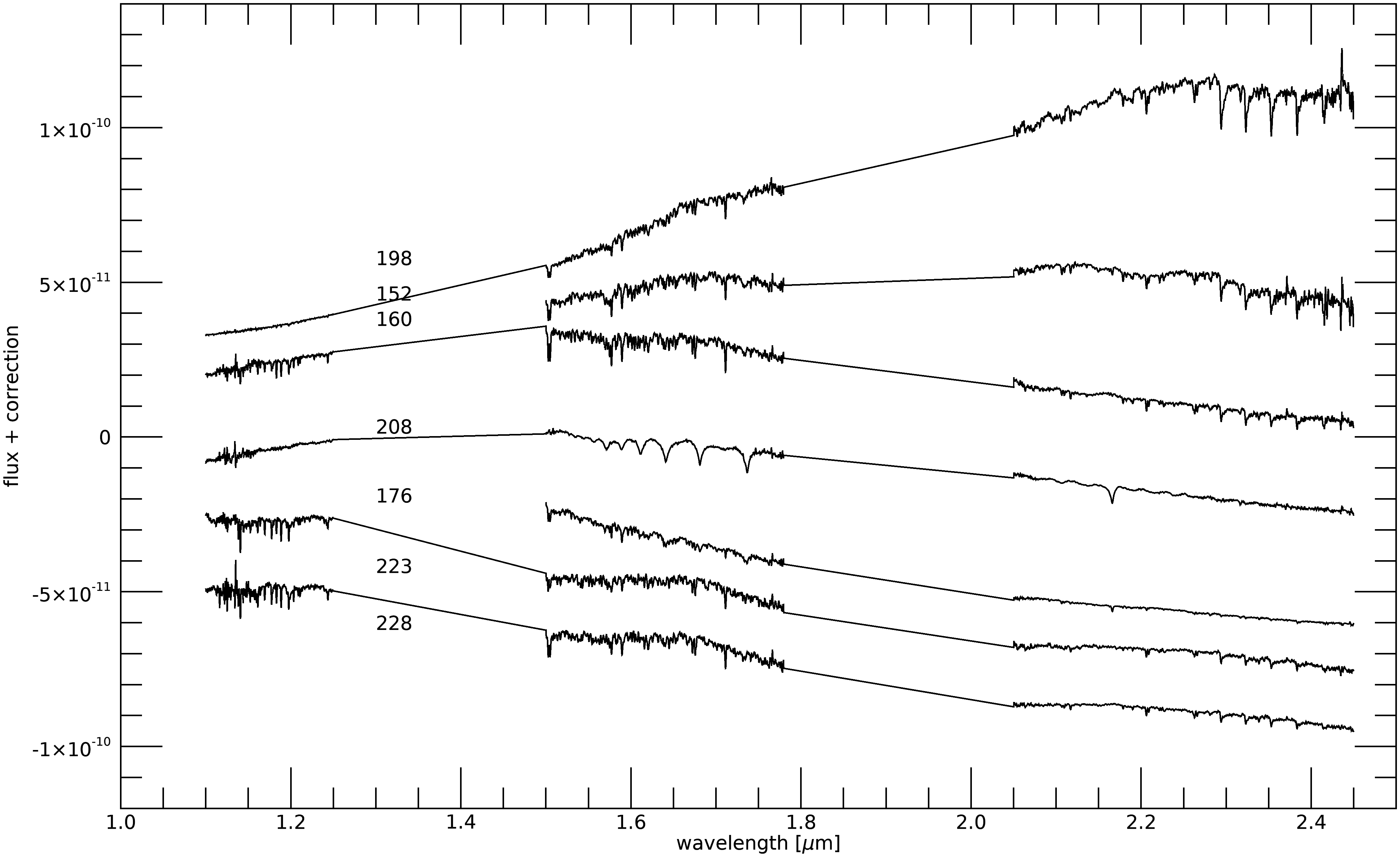}
    \caption{J, H, and K band spectra of removed sources. All the spectra are suitably shifted in flux for clarity.}
    \label{fig:spec_noyso}
\end{figure*}

Here, we report some information about the eight sources for which an estimate of the accretion parameters was not possible on the bases of their spectra (Table~\ref{starparnoYSO}). 
We removed these objects from our sample for two reasons: the spectrum shows at least one HI line ($\pab$ and/or $\brg$) in absorption, and/or it was not compatible with the one of YSOs.

In more in detail, the spectra of sources \#176 and \#208 are not compatible with the YSOs ones. Source \#176 shows HI lines in absorption and other absorption lines that are not compatible with the spectra of YSOs. 
Computing the extinction with the colour-colour diagram method (see Sect.~\ref{starpar}), we find this source is not extinct ($A_V = 0$). 
Looking on the Simbad database for further information on \#176, we found that its name is BD$+$30547, a G4V MS source \citep{fra18}. \newline Source \#208 shows absorption lines typical of evolved stars, as is clearly visible in Fig.~\ref{fig:spec_noyso}.
    
For what concerns sources \#152, \#160, \#198, \#223, and \#228, they all show HI lines in absorption, so they cannot be analysed by using the empirical relationships of \citet{alc17}. 
Source \#223 is a non-accreting Class~III YSO. 
For Class~II objects (\#160 and \#198), we conclude they are non-accreting YSOs as well. For objects classified as Class~I/Flat (see Tab.~\ref{starparnoYSO}), we checked the bolometric luminosity in the literature, finding $\lbol = 1.9$~$\lsun$ and $0.07$~$\lsun$ for \#152 and \#228, respectively \citep{eno09}.
Given the presence of HI in absorption and the low bolometric luminosity, we think these objects are non-accreting YSOs, suggesting they are not in the evolutionary stage expected for Class~I YSOs. 
As a matter of fact, source \#228 is spatially associated with ASR~7, a time-variable object exhibiting strong circular polarisation which suggests gyrosynchrotron emission 
\citep{rod99}, despite the $\alpha_{\rm IR}$ classifcation as a Class~I YSO. 

\clearpage

\section{HR diagram of Class~I}
\label{app:hrdiagrI}
\begin{figure}[t]
 \includegraphics[width=\columnwidth]{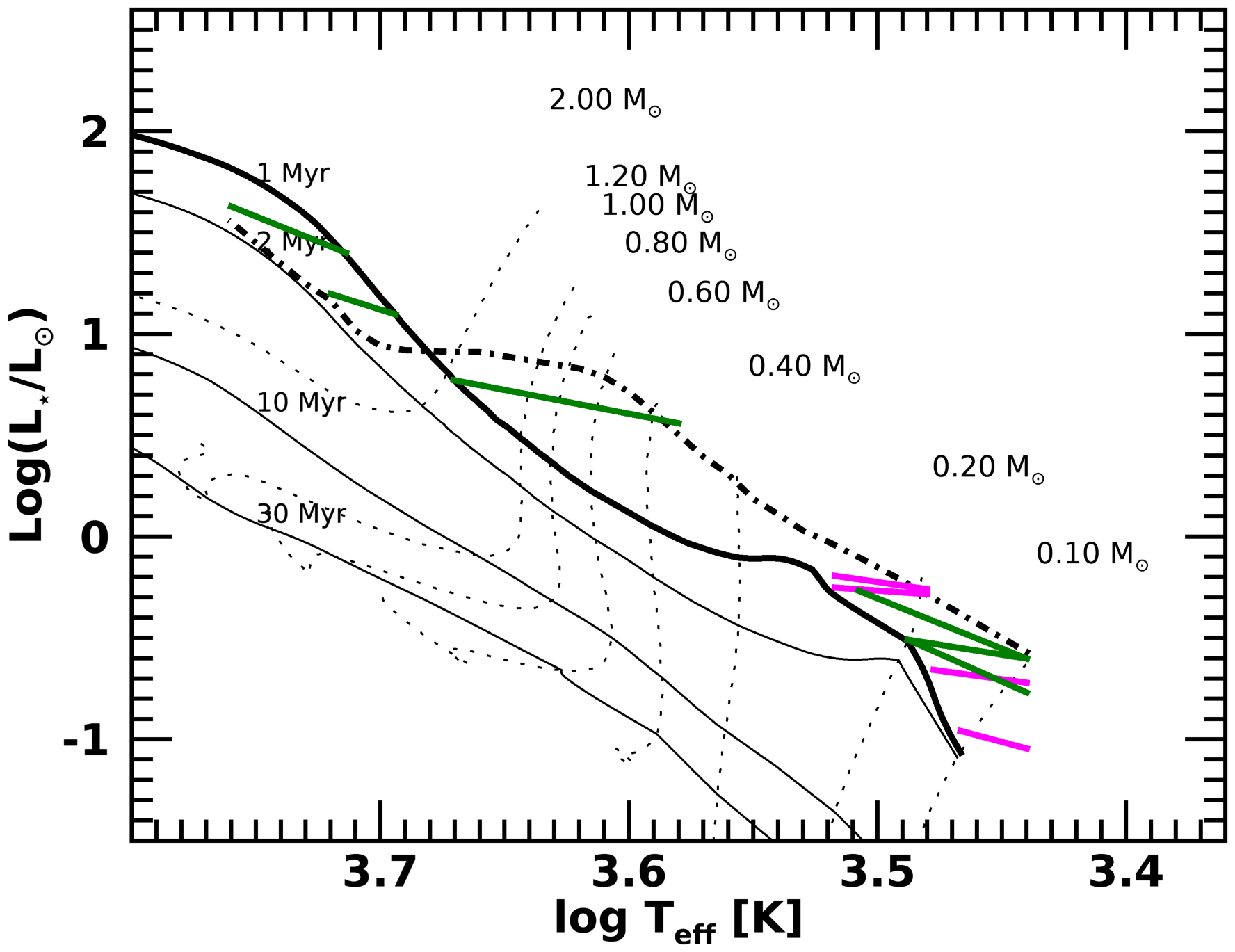}
  \caption{\label{hrdiag1}HR diagram of the Class~I YSOs. 
  Sources with veiling between 1 and 3 are plotted in magenta, while those with veiling between 3 and 8 are in green. 
  The birthline \citep{pal93} is plotted as a dot-dashed line. 
  Black solid and dashed lines show the isochrones and evolutionary tracks by \citet{sie00}.} 
\end{figure}

Fig.~\ref{hrdiag1} shows the position in the HR-diagram of the ten Class~I YSOs which show HI lines. 
It is our assumption that the age of these sources is between the birthline and 1~Myr (see Sect.~\ref{sect:ana1}). 

From this diagram, we can appreciate how the stellar mass, given the luminosity derived from our method, changes by varying the age from the birthline to the 1~Myr.
In particular, we note that sources \#191 and \#186 both remain in the intermediate-mass range if we put them on the birthline or in the 1~Myr isochrone, while source \#200 has a larger uncertainty on the mass due to the kink of the birthline curve in the luminosity range of this source. 

\section{Comparison with Luhman et al. 2016} \label{app:luh}

\citet{luh16} provide a census of YSOs and brown dwarfts (BDs) in the IC~348 and NGC~1333 clusters. 
In their work, the spectral characterisation of the sample is made through a comparison between the averages of dwarf and giant standards \citep{luh99} spectra with the low-resolution spectra ($R = 100 - 800$) of the sources taken with different spectrographs (SpeX@IRTF, IMACS@Magellan, GNIRS@Gemini, NIRC@Keck~I).
The extinction was estimated by dereddening the position of each source in the $[J- H]$~versus~$[I - K_s]$ diagram until it intersected the sequence of colours for dwarfs \citep{luh03}.

\begin{figure}
 \includegraphics[width=\columnwidth]{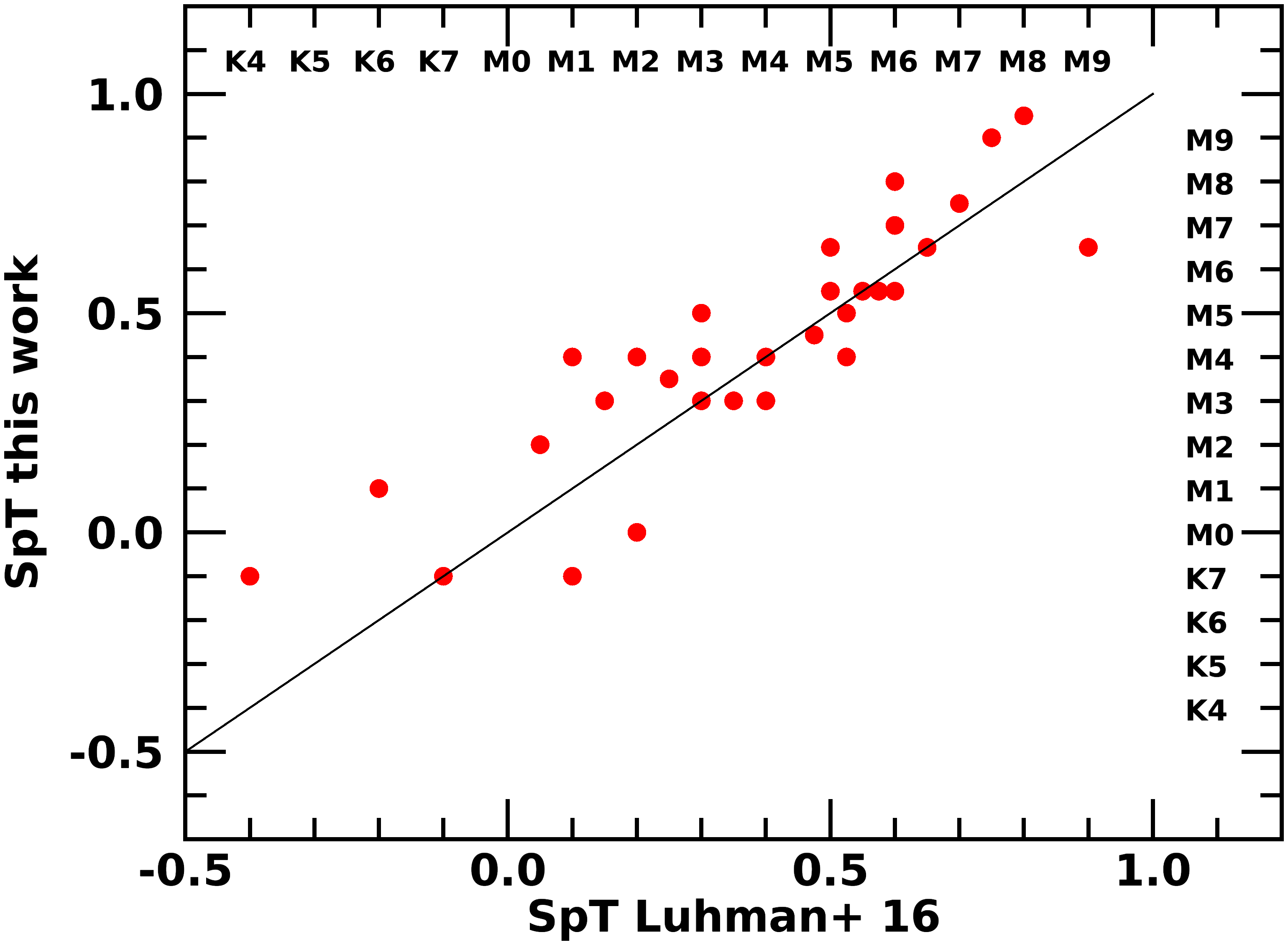}
 \caption{\label{spt} SpT of this work versus the spectral type from \citet{luh16} for the common sources.} 
\end{figure}

\begin{figure}
 \includegraphics[width=\columnwidth]{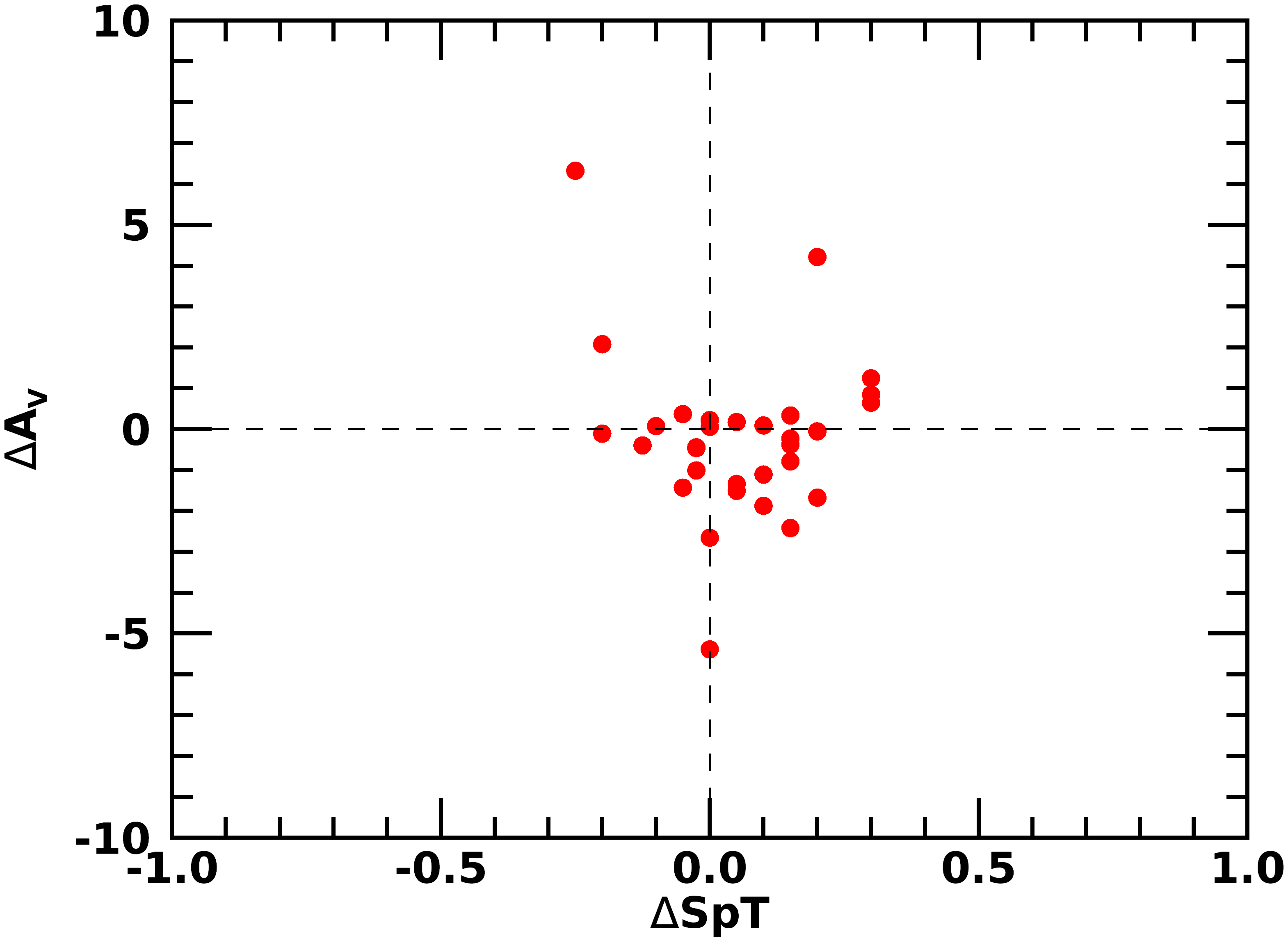}
 \caption{\label{deltaAvSpt}Variation between this work and the \citet{luh16} values of the extinction and the spectral type of our sample. We attributed an increasing number from early SpT to late SpT.
 }
\end{figure}

We matched the sample analysed by \citet{luh16} with ours, finding 40 sources in common. 
Of these, only 35 have both the extinction and the spectral type estimated by \citet{luh16} (see Table~\ref{starpar} and \ref{noHIclI}).
Fig.~\ref{spt} shows a comparison between the SpT estimated by \citet{luh16} and that derived in this paper. 
To be able to quantify the difference in the spectral type, we assigned a value in the following way to each spectral type: we set the M0 class as 0, then earlier types were associated to negative numbers with $0.1$ steps (i.e. K7 corresponds to $-0.1$, K6 to $-0.2$, and so on), and later types to positive numbers (i.e. M1 corresponds to $0.1$, M5 to $0.5$, etc.).
A general agreement is found between the two samples. 
As discussed in Sect.~\ref{sect:ana2}, the SpT and the extinction depend on each other. 
Therefore, in Fig.~\ref{deltaAvSpt}, we plotted the variation in SpT and $A_V$ between our results and those of \citet{luh16}.
For only two sources the discrepancy on $A_V$ is $> 5$ and all the sources have $\Delta$SpT contained within a spectral class. 
We also note that the most populated quadrant is the fourth (eight sources). 
This is a tentative trend, according to which, on average, \citet{luh16} classification prefers higher extinction and early-type spectral types. 
This can be due to the low resolution of their observations which were less sensitive to the molecular features. 

\section{Class~I SED's and $\lbol$}
 \label{app:sed}
Here, we present the spectral energy distributions of the Class~I YSOs and the derivation of their bolometric luminosity.  

To construct the SEDs, we used the IR photometry 2MASS and {\it Spitzer} of the \citet{you15} catalogue and, where possible, we added the measurements in FIR-submillimetre, from {\it Herschel} \citep{pez20}, and millimetre from {\it Bolocam} \citep{eno09}. The derived SEDs are displayed in Fig.~\ref{fig:sedI}.

In order to minimise the contamination from other stars in the relatively large Hershel FoV with respect to the Spitzer one, we inspected of the {\it Spitzer} and Hershel 70~$\mu$m maps and disregarded the Herschel fluxes when a clear contamination is found. 
This is the case of \#200.

We computed the bolometric luminosity (Table~\ref{starparI}) by integrating the SEDs with an ad hoc IDL\footnote{IDL (Interactive Data Language) is a registered trademark of Harris Corporation.} procedure, already used in \citet{ant08}.
The integration was performed starting from the J-band, and considering straight lines in the Log$(\lambda) - \mbox{Log}(\lambda F_\lambda)$ plan relatives to available SED points; a final correction at the longest wavelengths was applied assuming that the emission decreases as $1/\lambda^{2}$ after the last available observation.
The derived values are shown in Tables~\ref{starparI} and \ref{noHIclI}, and they range between 0.06~$\lsun$ to 58.8~$\lsun$. 

\begin{figure}
    \centering
    \includegraphics[width=\columnwidth]{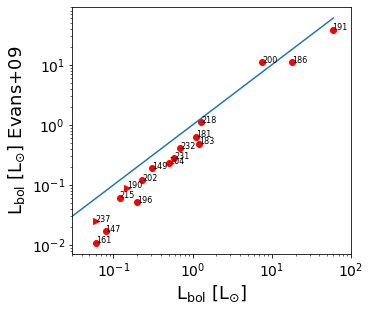}
    \caption{Comparison between the bolometric luminosity computed by \citet{eva09} and the one computed in this work.}
    \label{fig:Lbol_comparison}
\end{figure}

\begin{figure*}[b] 
\centering
 \begin{subfigure}{\textwidth}
 \centering
 \includegraphics[width=0.25\textwidth]{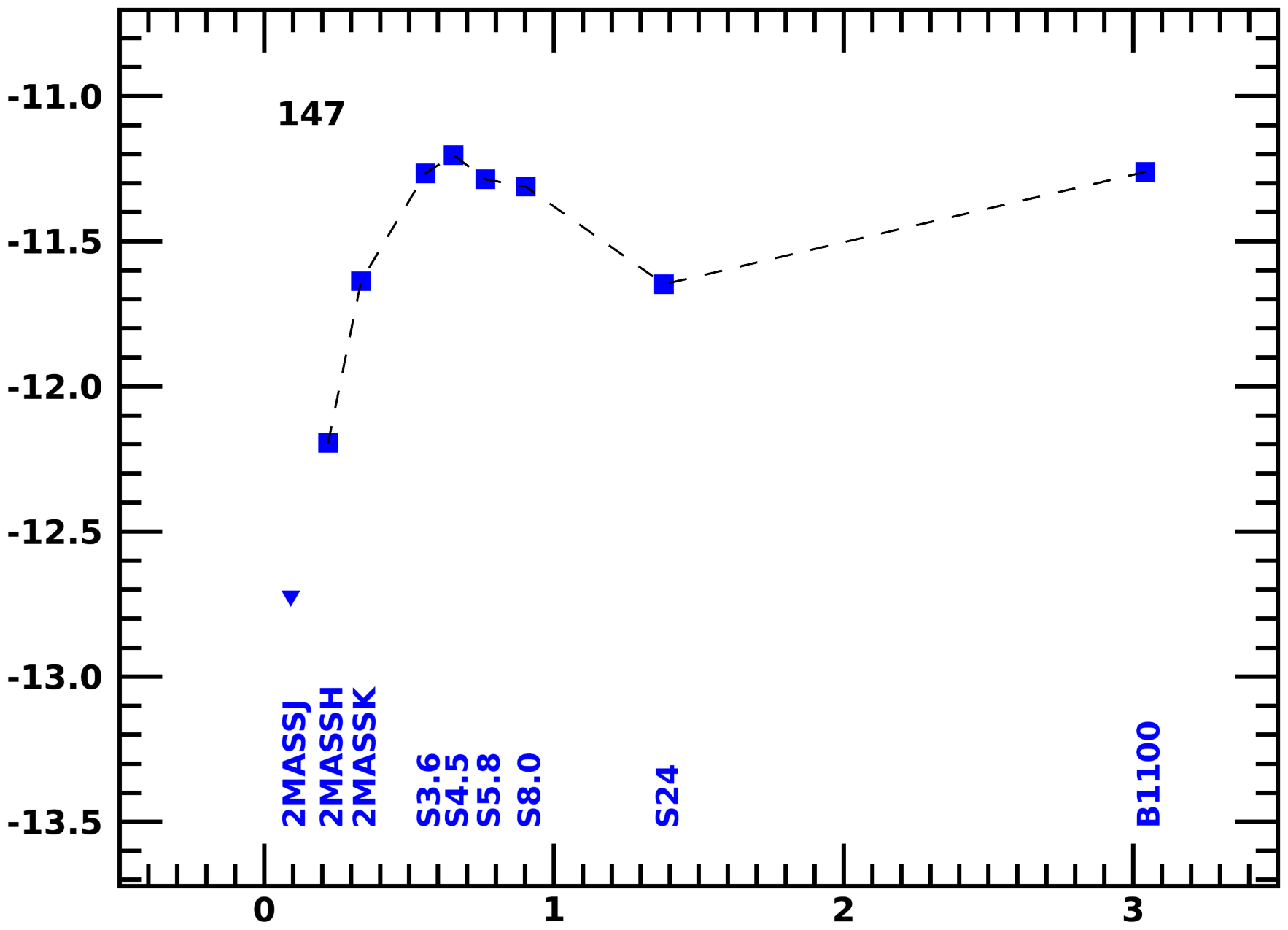}%
 \includegraphics[width=0.24\textwidth]{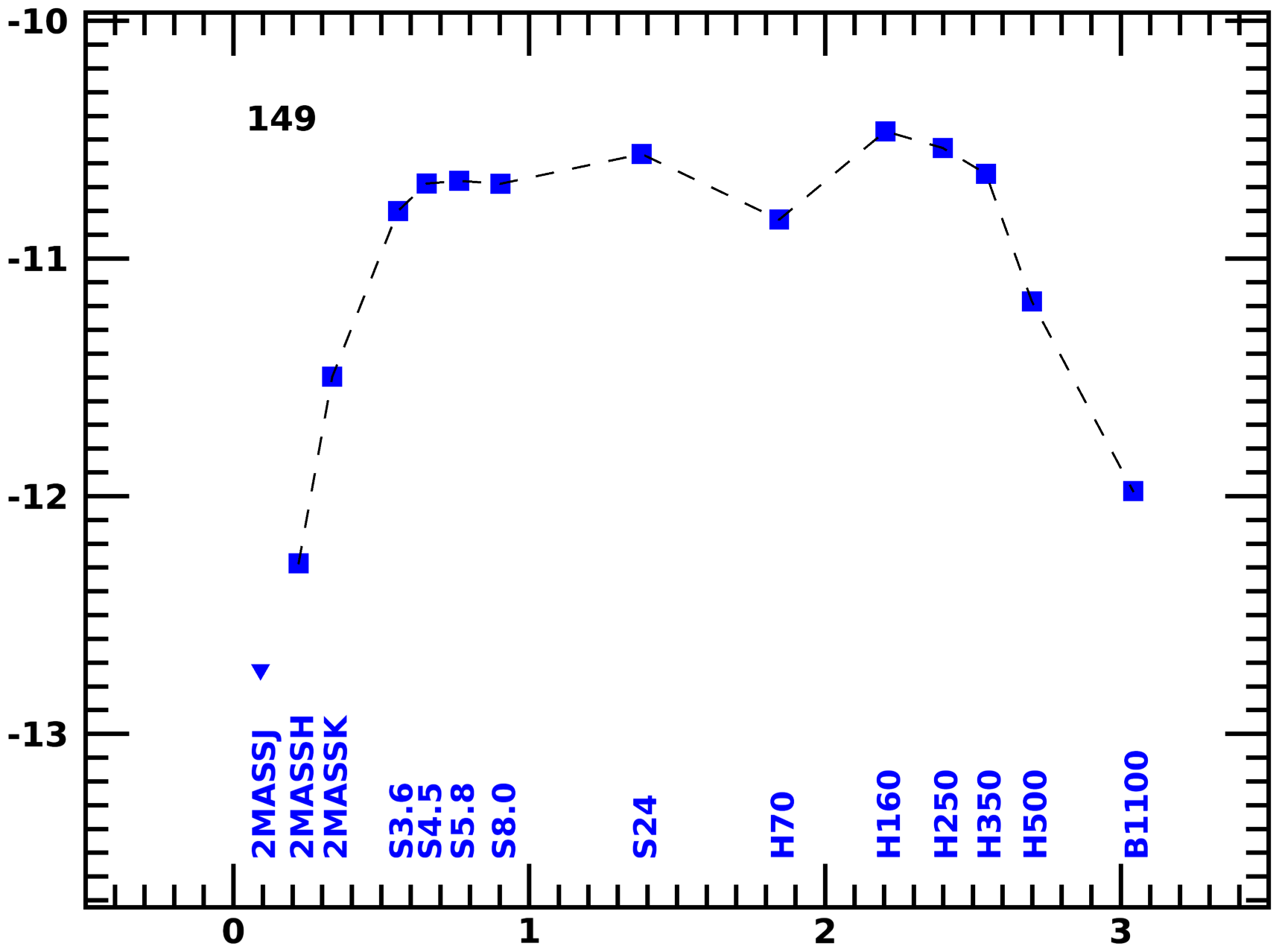}%
 \includegraphics[width=0.25\textwidth]{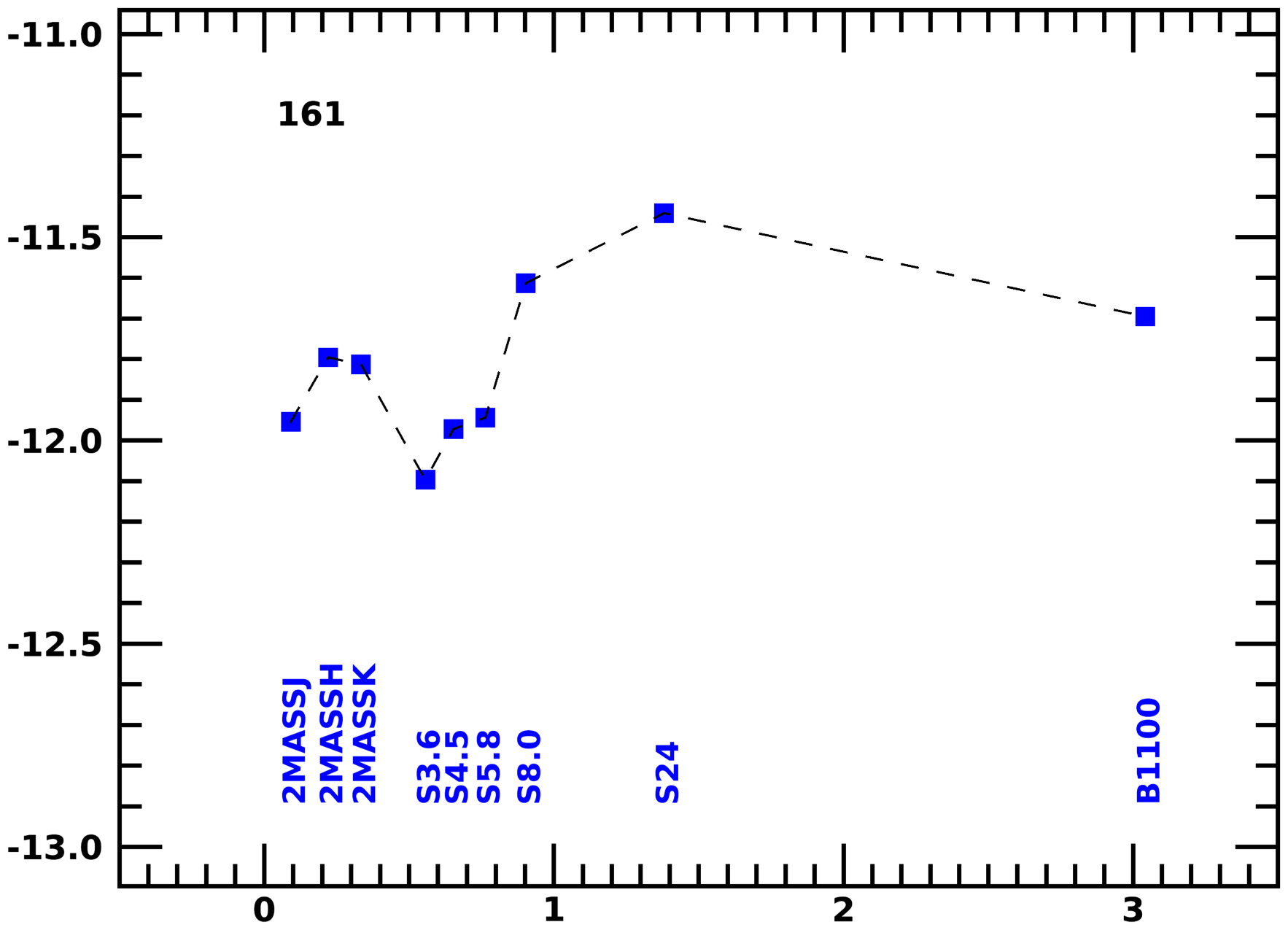}%
 \includegraphics[width=0.24\textwidth]{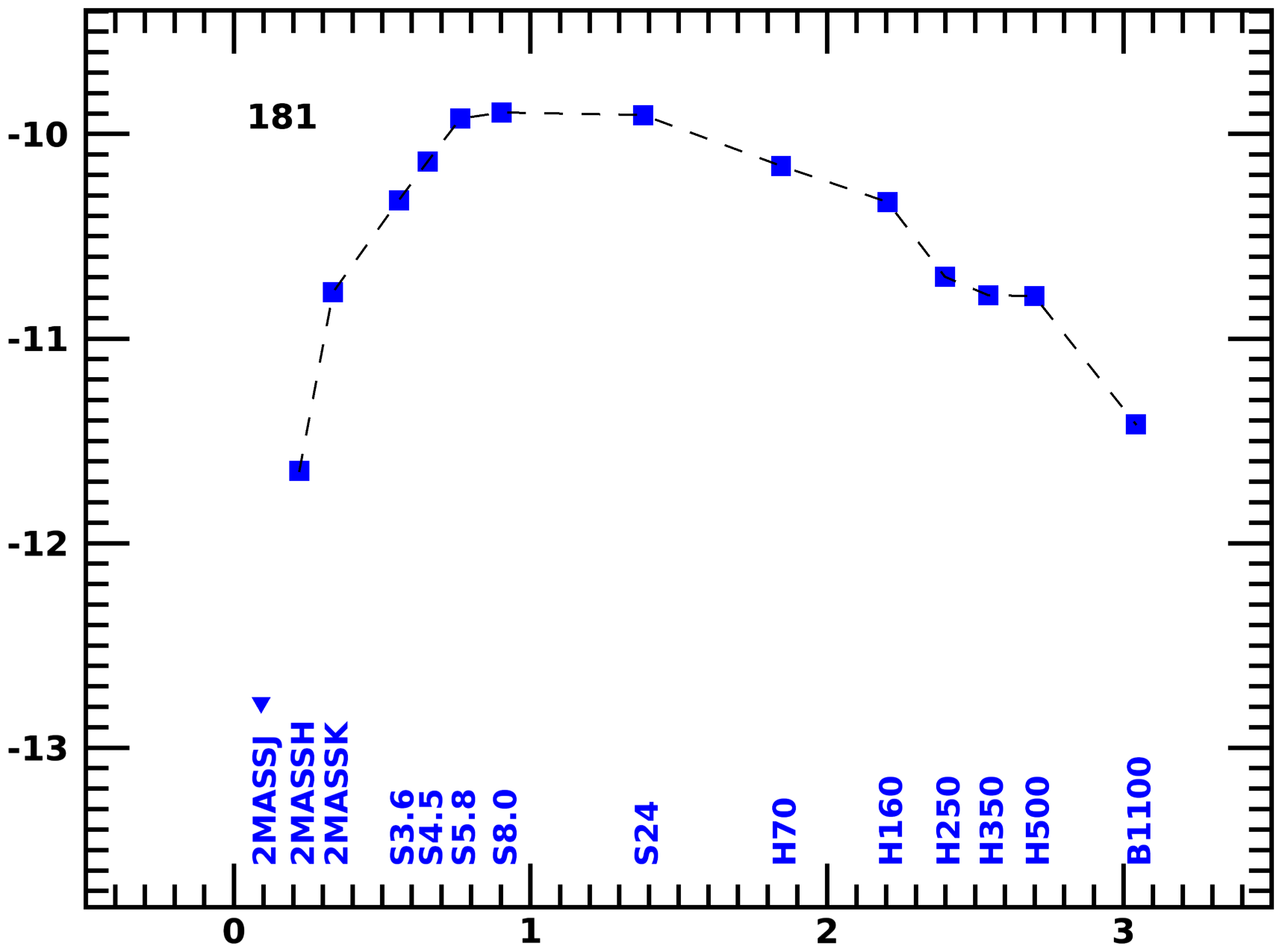}%
 
 \includegraphics[width=0.25\textwidth]{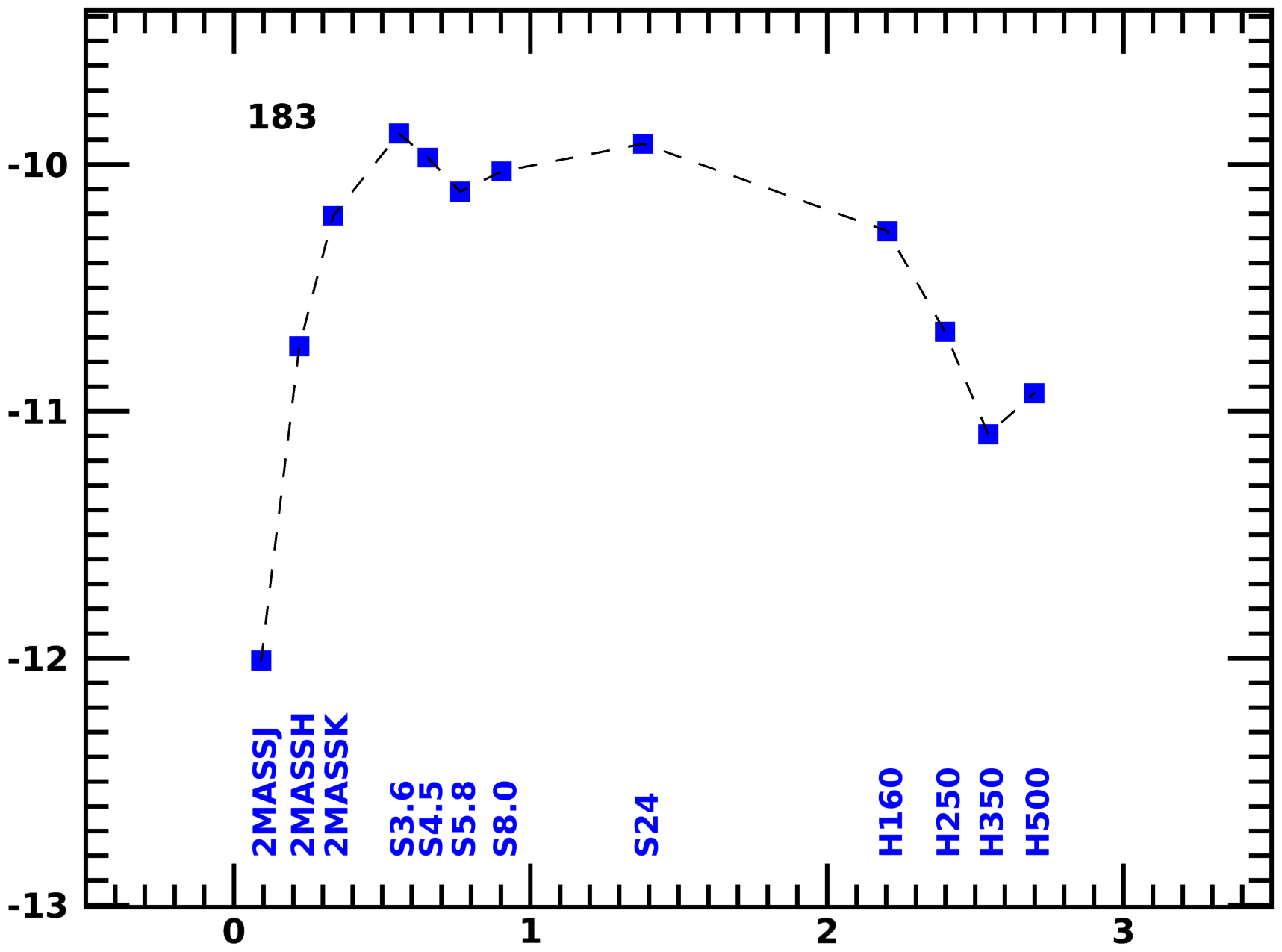}%
 \includegraphics[width=0.24\textwidth]{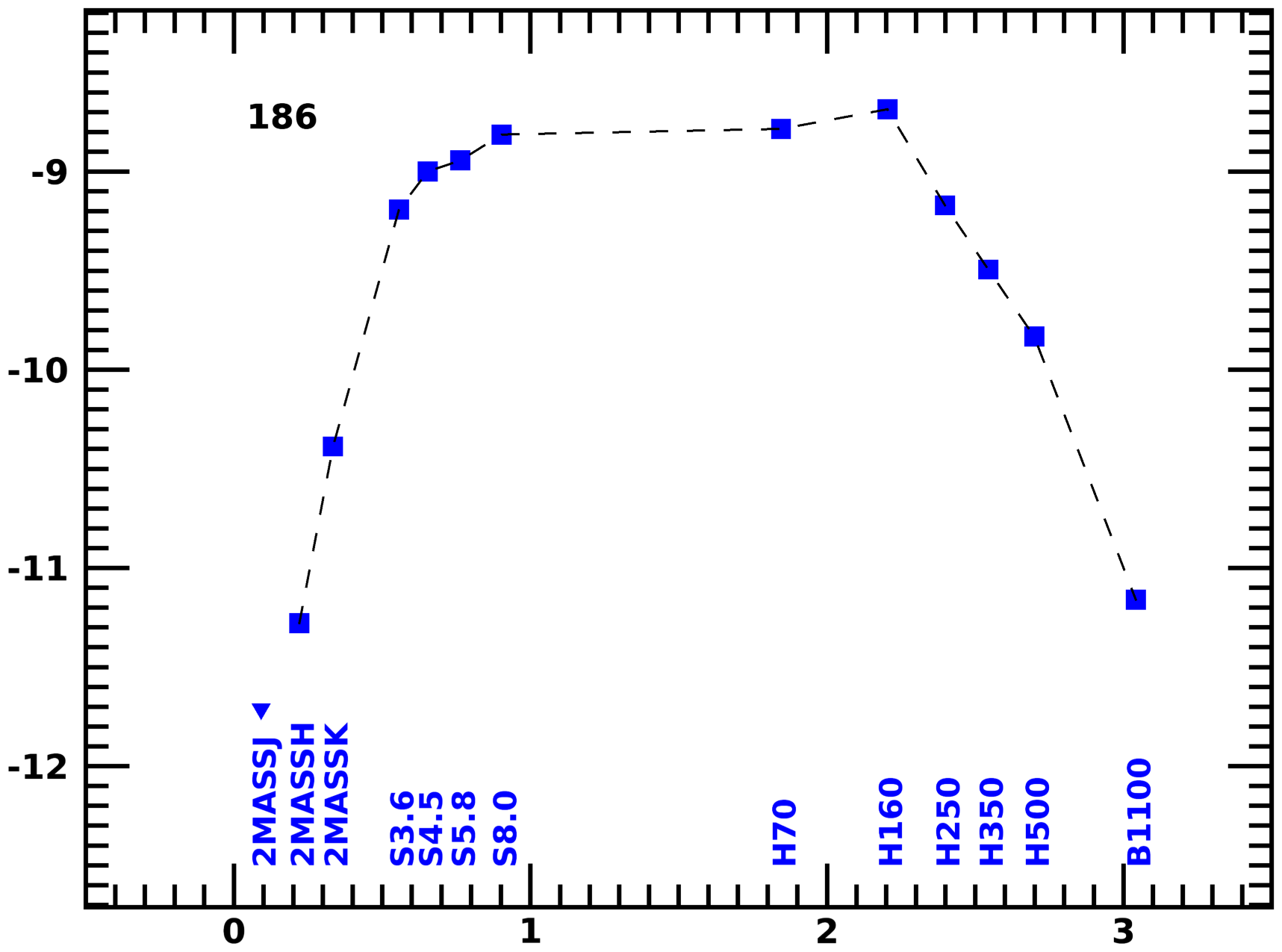}%
 \includegraphics[width=0.25\textwidth]{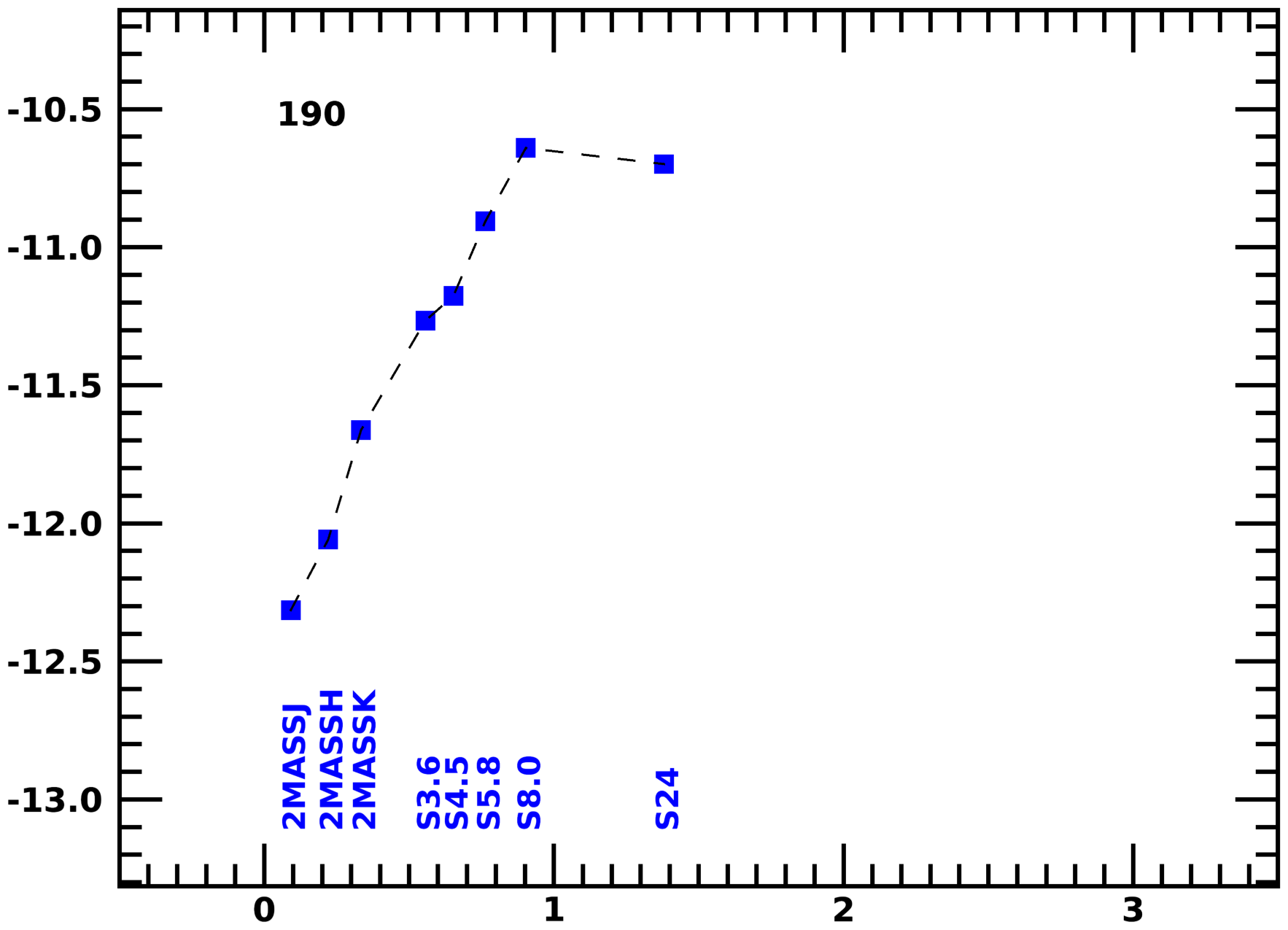}%
 \includegraphics[width=0.24\textwidth]{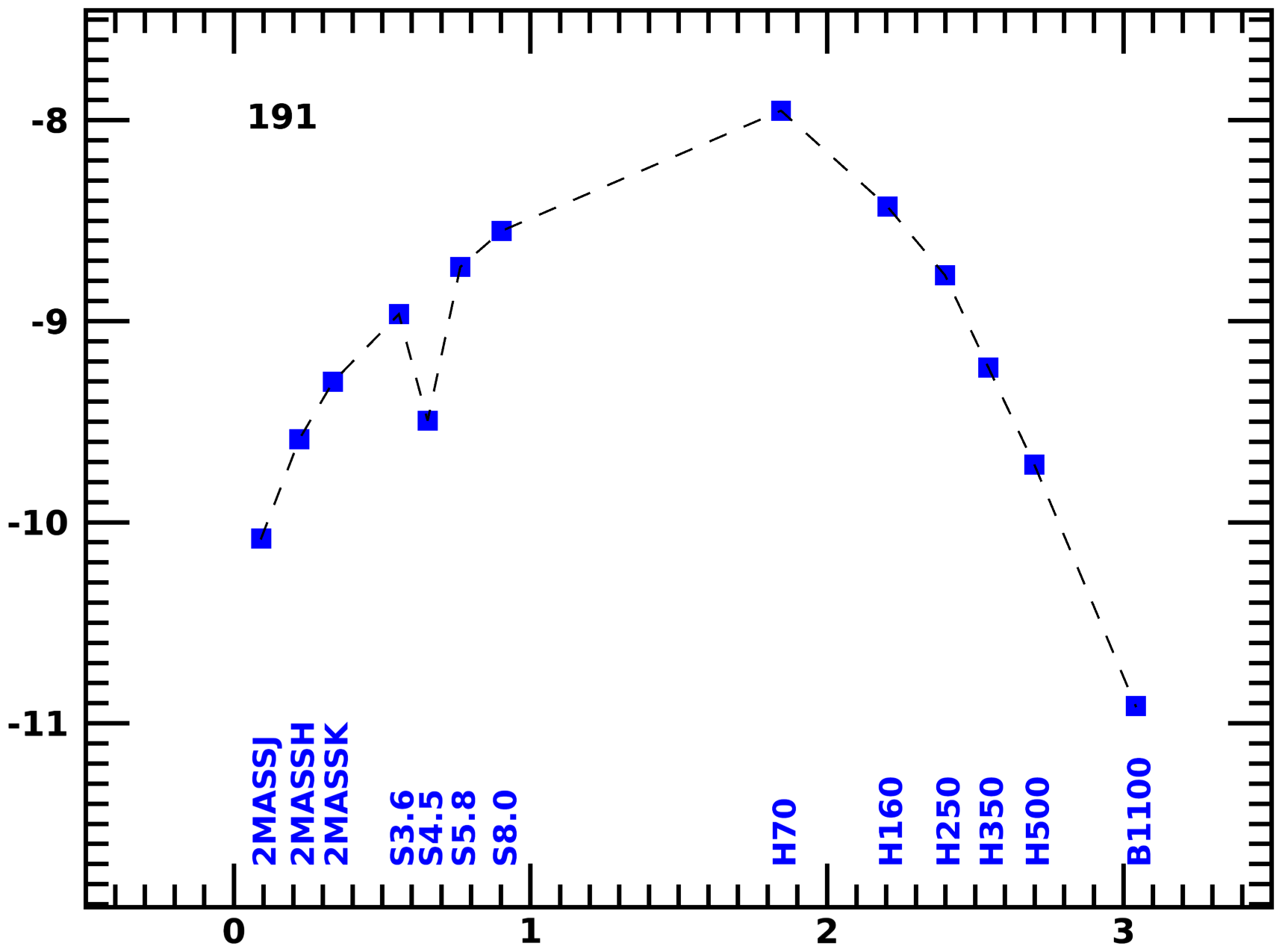}%
 
 \includegraphics[width=0.25\textwidth]{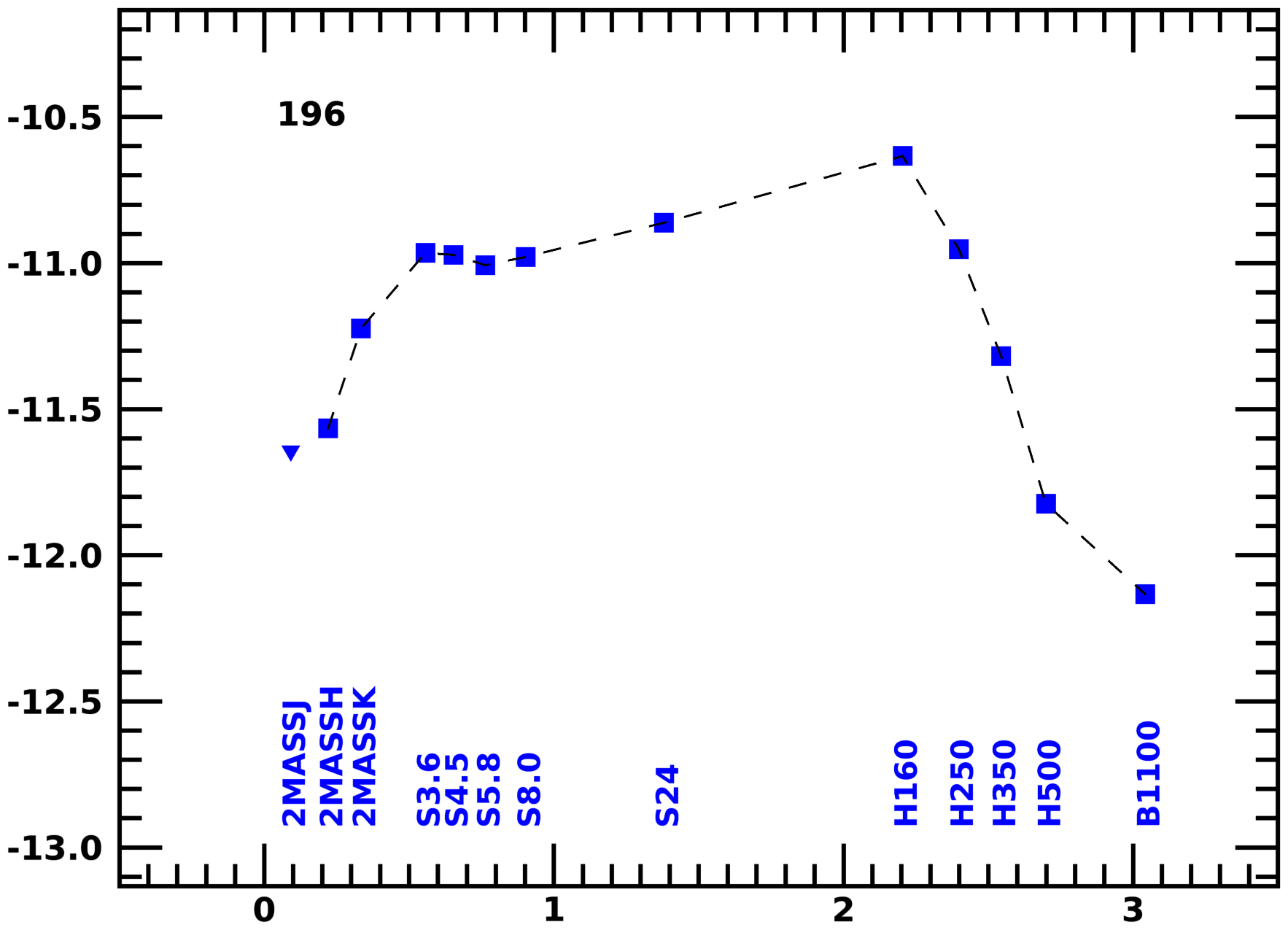}%
 \includegraphics[width=0.24\textwidth]{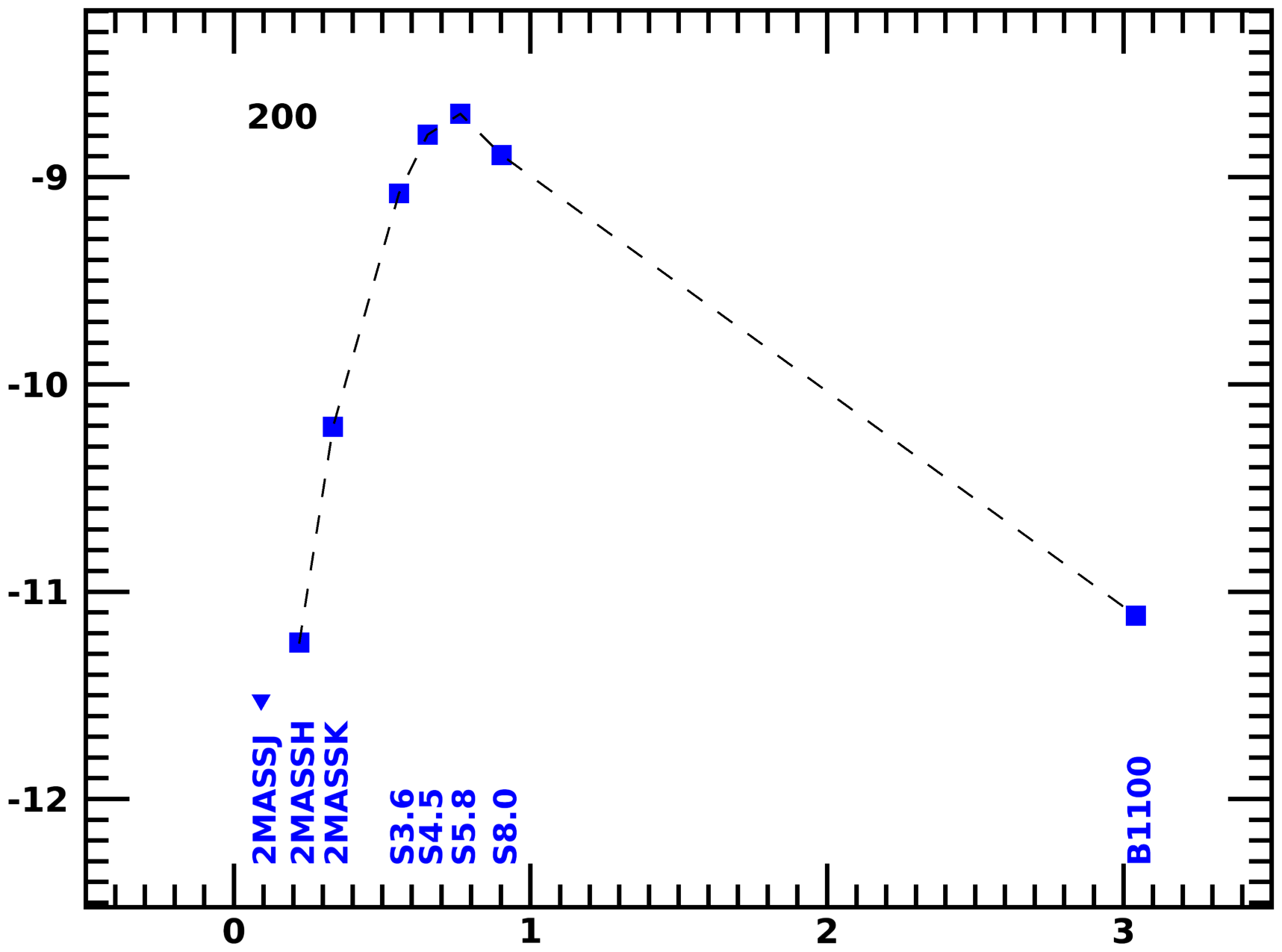}%
 \includegraphics[width=0.25\textwidth]{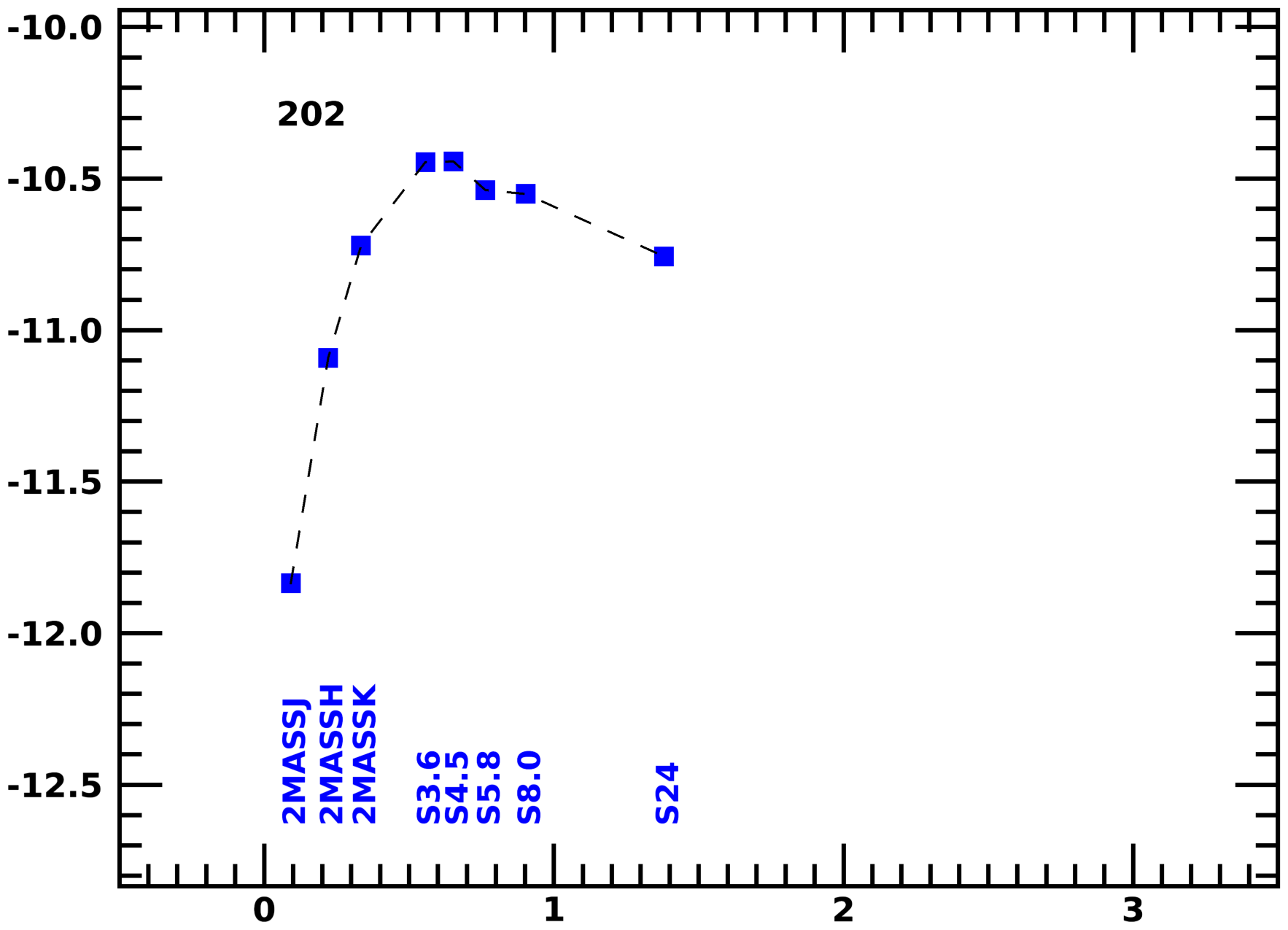}%
 \includegraphics[width=0.25\textwidth]{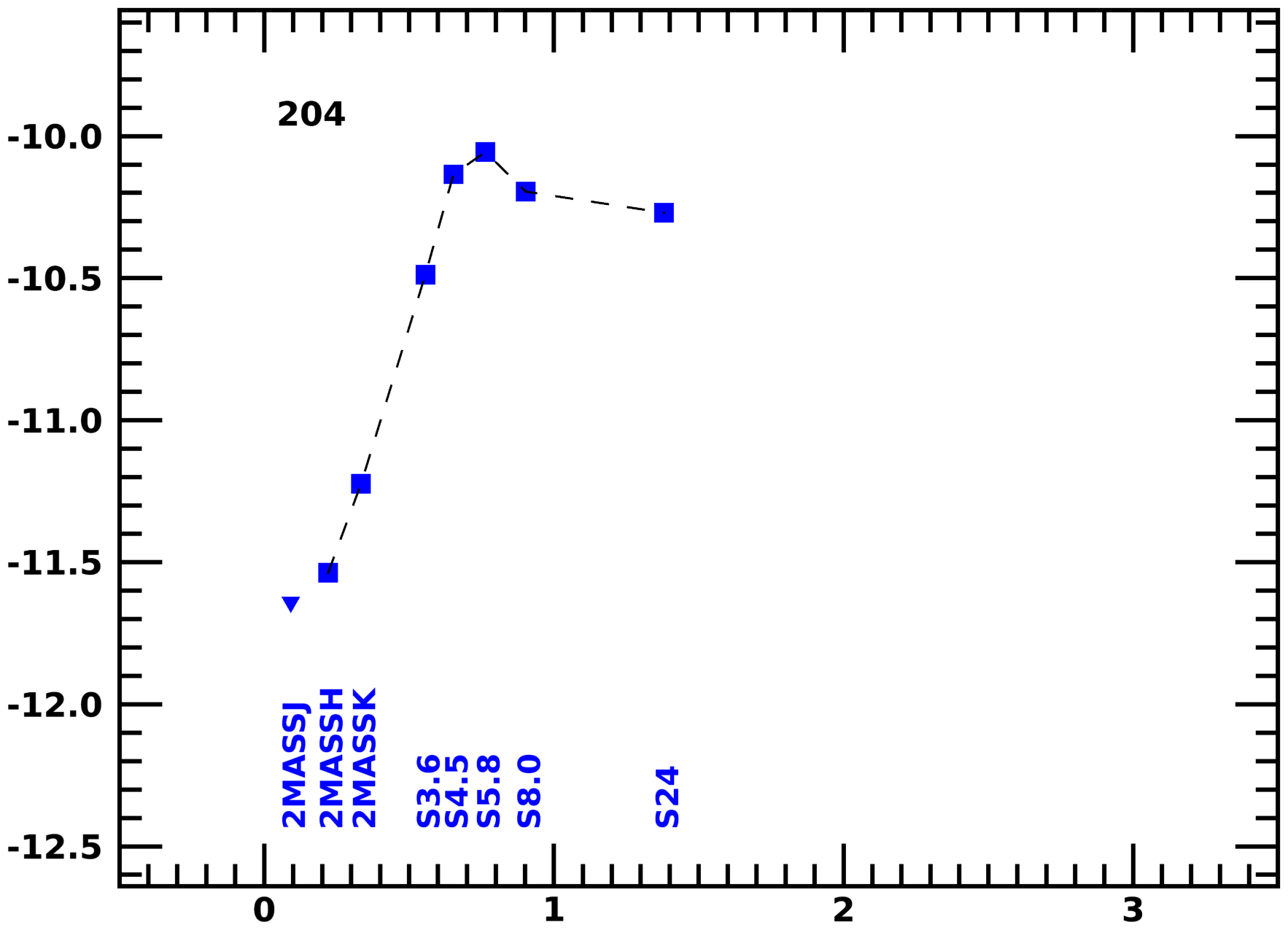}%
 
 \includegraphics[width=0.25\textwidth]{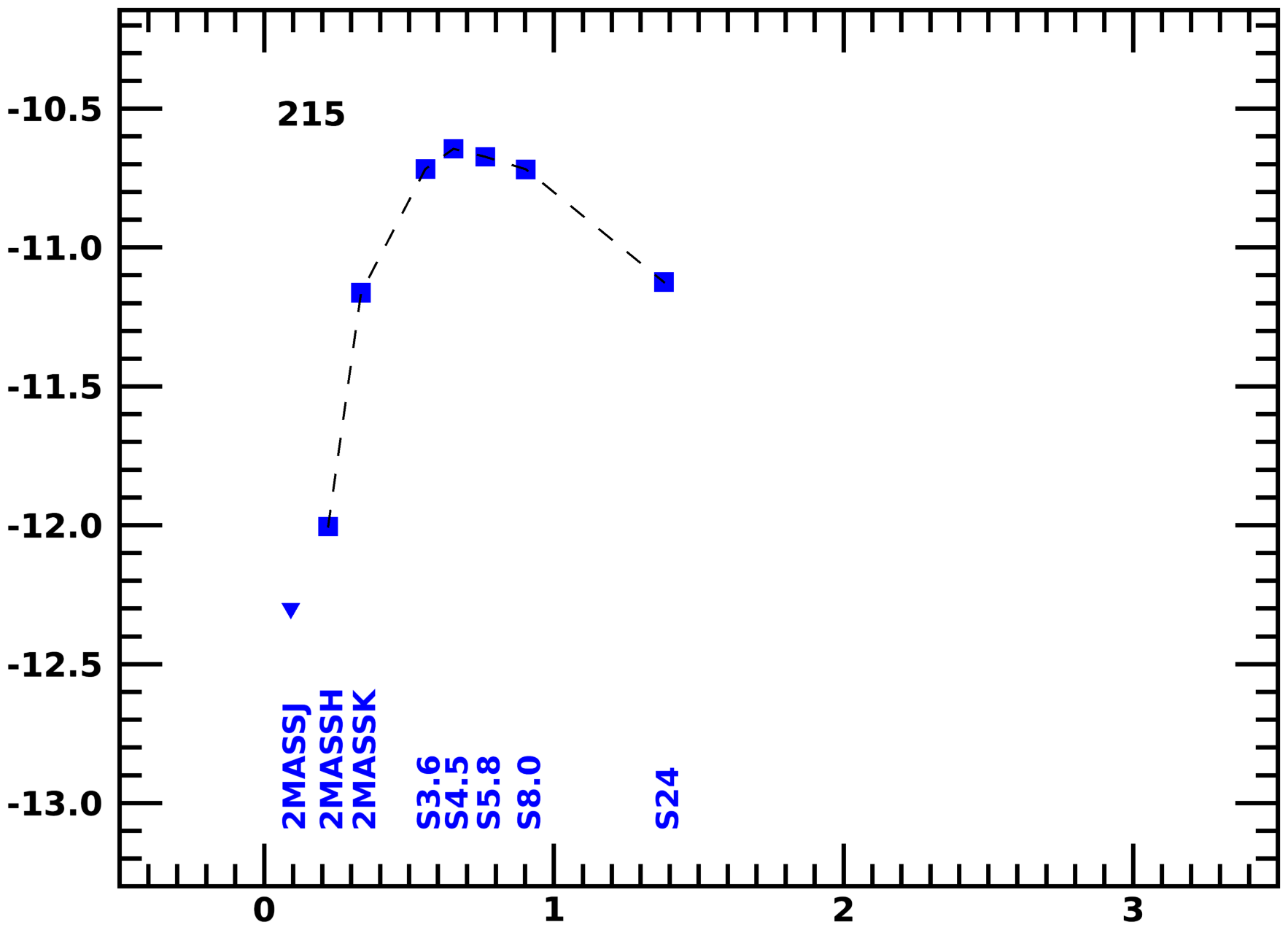}%
 \includegraphics[width=0.24\textwidth]{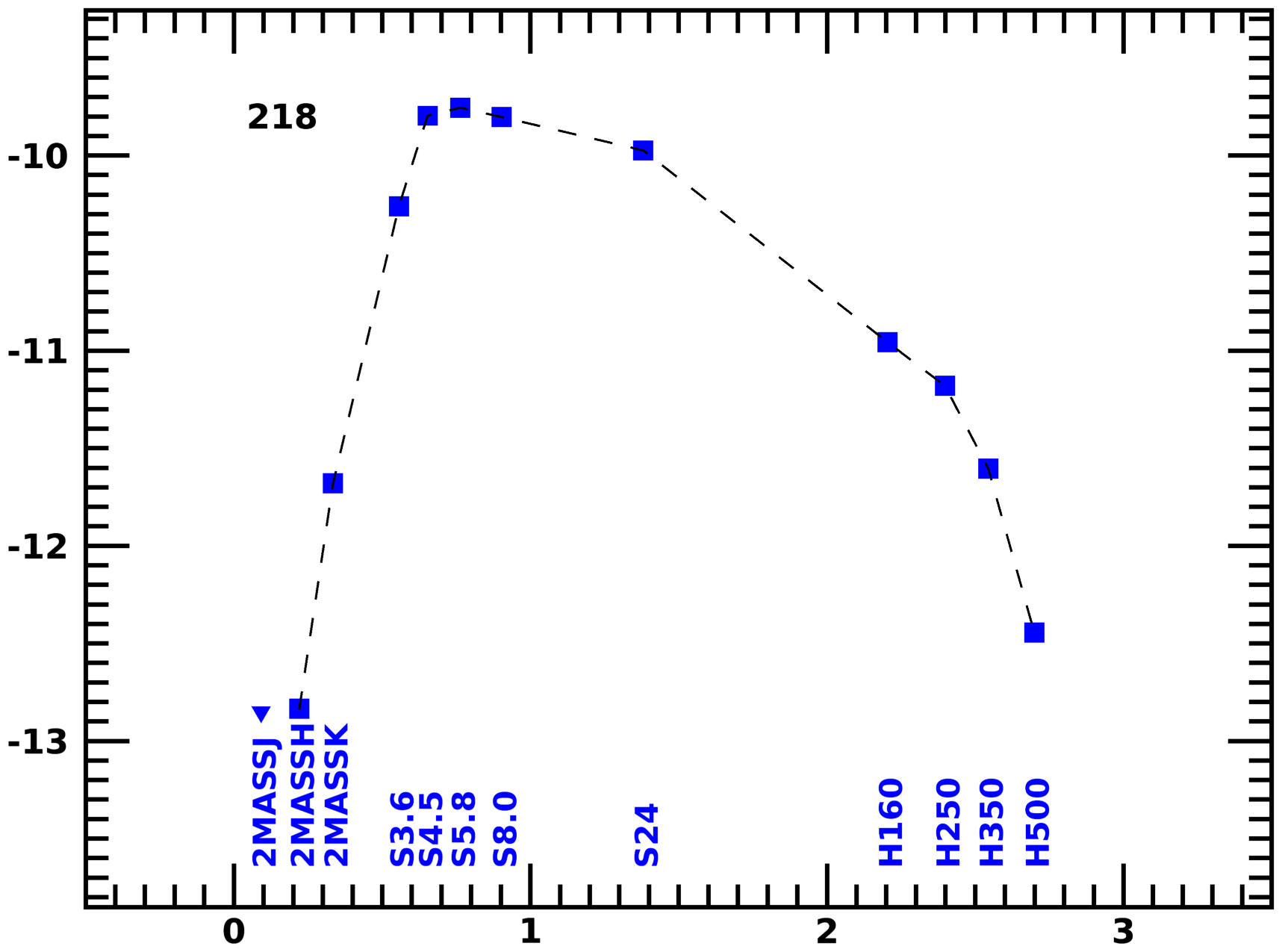}%
 \includegraphics[width=0.25\textwidth]{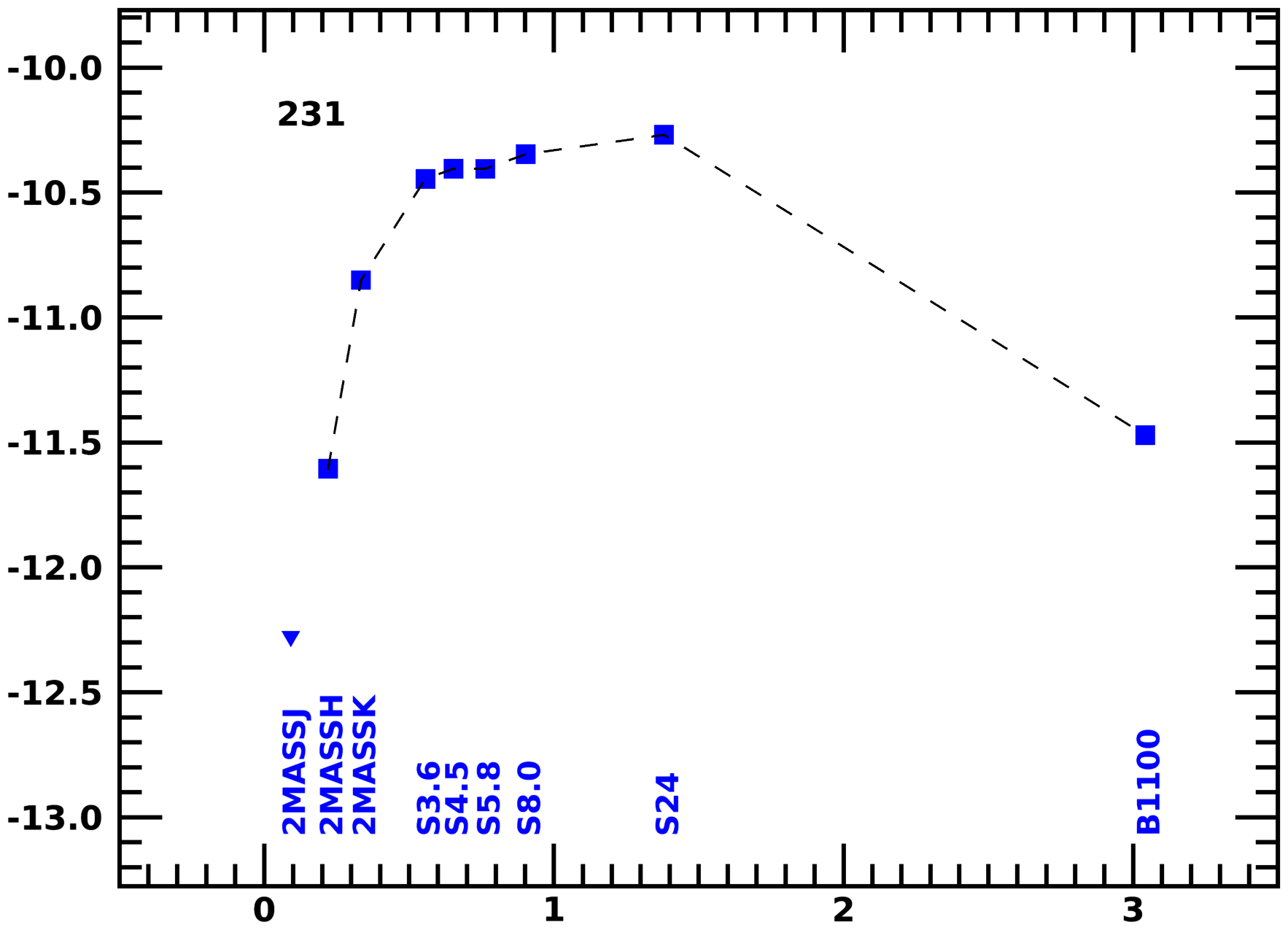}%
 \includegraphics[width=0.24\textwidth]{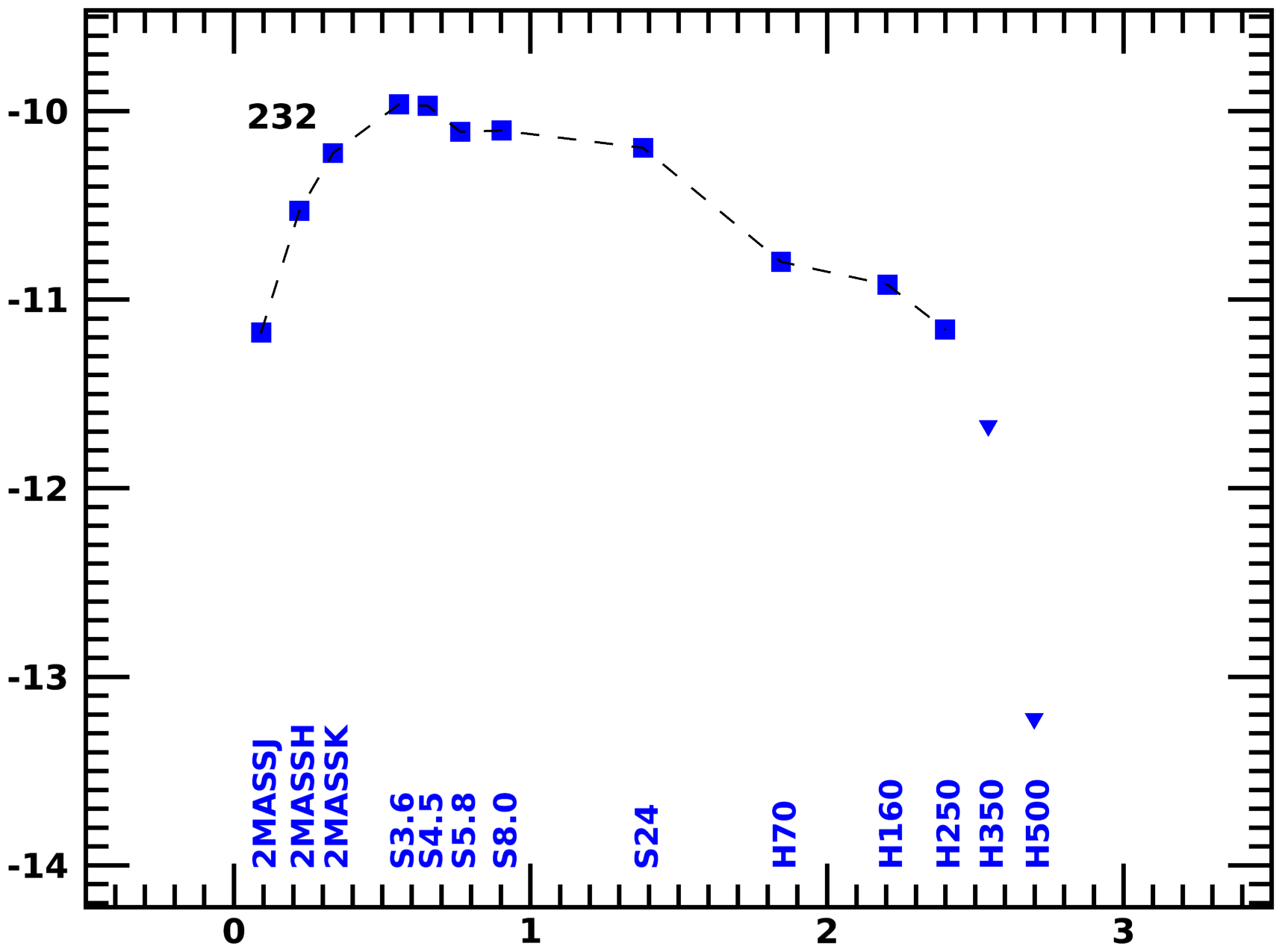}%
 
 \includegraphics[width=0.25\textwidth]{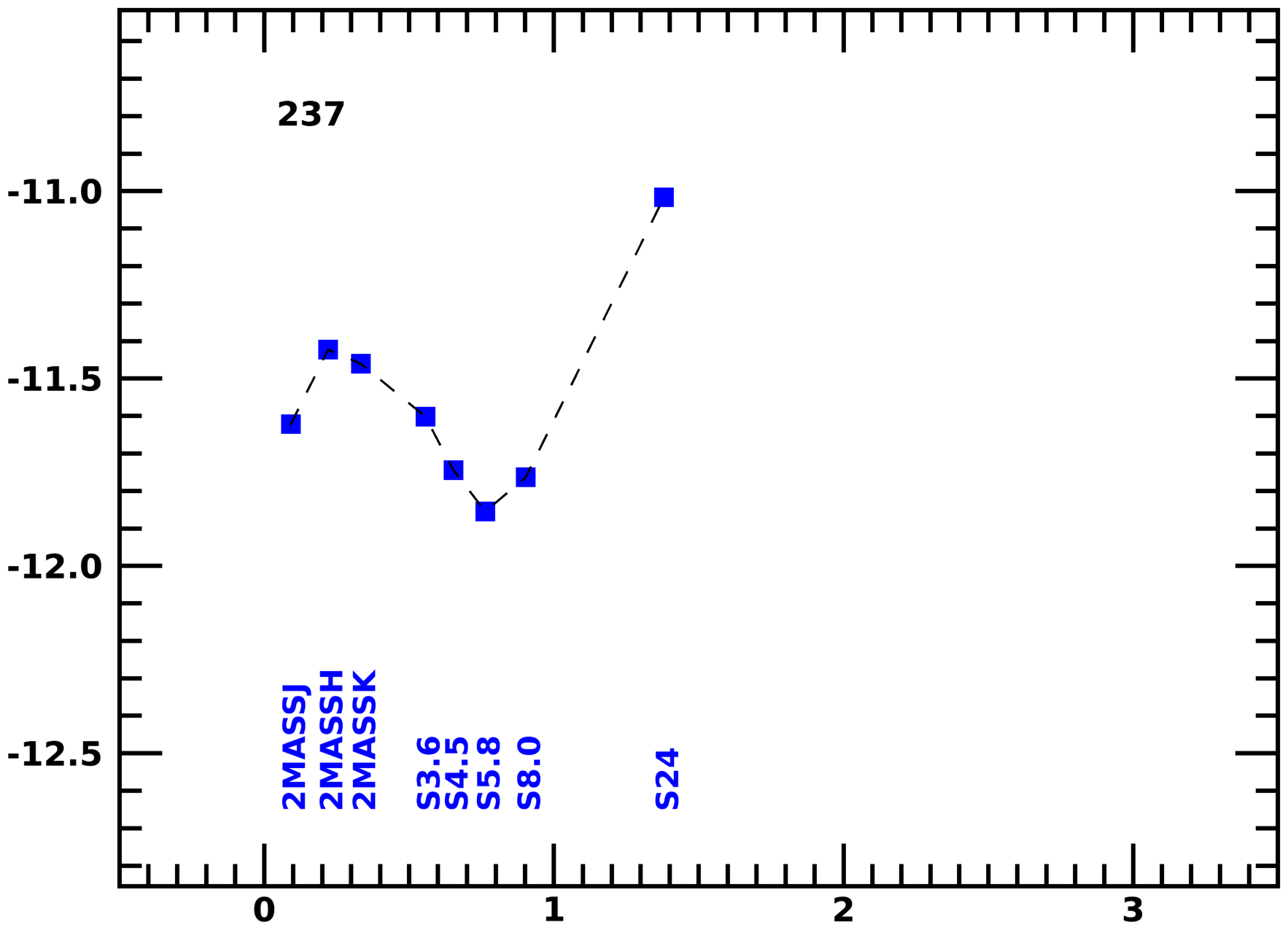}%
 
 \end{subfigure}
 \caption{\label{fig:sedI}Spectral energy distributions ($\lambda F_\lambda$ [erg s$^{-1}$cm$^{-2}$] vs. $\lambda$ [$\mu$m]) of the Class~I YSOs. We use the 2MASS, Spitzer and Herschel, and BOLOCAM photometry \citep{you15, pez20}.}
\end{figure*}

Fig.~\ref{fig:Lbol_comparison} shows the comparison between our estimates of $\lbol$ and the ones computed by \citet{eva09} for the Class~I sample. 
We find that for all the sources but \#200 the bolometric luminosity we estimated is compatible or larger with the one estimated by \citet{eva09}, as expected by adding {\it Herschel} fluxes in the FIR.
The results we show in this work are the first estimates of $\lbol$ which include the information of {\it Herschel} photometry. 

\end{document}